\documentclass[%
reprint,
superscriptaddress,
nofootinbib,
amsmath,amssymb,
aps,
prd,
]{revtex4-2}

\usepackage{xr-hyper}
\usepackage[]{graphicx}
\usepackage{xcolor}
\usepackage{dcolumn}
\usepackage{bm}
\usepackage{bold-extra} 

\definecolor{orange}{rgb}{0.8,0.35,0}

\usepackage[colorlinks=true
,urlcolor=blue
,anchorcolor=blue
,citecolor=blue
,filecolor=blue
,linkcolor=red
,menucolor=blue
,linktocpage=true
,pdfproducer=medialab
,pdfa=true
]{hyperref}

\usepackage{natbib}

\usepackage{array,mathtools,amssymb,booktabs}
\newcolumntype{C}{>{$}c<{$}}
\AtBeginDocument{
\heavyrulewidth=.08em
\lightrulewidth=.05em
\cmidrulewidth=.03em
\belowrulesep=.65ex
\belowbottomsep=0pt
\aboverulesep=.4ex
\abovetopsep=0pt
\cmidrulesep=\doublerulesep
\cmidrulekern=.5em
\defaultaddspace=.5em
}
\usepackage{multirow}

\defcitealias{List2020b}{Paper~I}

\begin{document}

\preprint{---}

\title{Dim but not entirely dark: \\
Extracting the Galactic Center Excess' source-count distribution with neural nets}

\author{Florian List}
\email{florian.list@univie.ac.at}
\affiliation{%
 Sydney Institute for Astronomy, School of Physics, A28, The University of Sydney, NSW 2006, Australia
}%
\affiliation{Department of Astrophysics, University of Vienna, T\"{u}rkenschanzstra{\ss}e 17, 1180 Vienna, Austria}
\author{Nicholas L. Rodd}
\affiliation{CERN, Theoretical Physics Department, Geneva 1211, Switzerland}
\author{Geraint F. Lewis}
\affiliation{%
 Sydney Institute for Astronomy, School of Physics, A28, The University of Sydney, NSW 2006, Australia
}%

\begin{abstract}
The two leading hypotheses for the Galactic Center Excess (GCE) in the \emph{Fermi} data are an unresolved population of faint millisecond pulsars (MSPs) and dark-matter (DM) annihilation. The dichotomy between these explanations is typically reflected by modeling them as two separate emission components. However, point-sources (PSs) such as MSPs become statistically degenerate with smooth Poisson emission in the ultra-faint limit (formally where each source is expected to contribute much less than one photon on average), leading to an ambiguity that can render questions such as whether the emission is PS-like or Poissonian in nature ill-defined. We present a conceptually new approach that describes the PS and Poisson emission in a unified manner and only afterwards derives constraints on the Poissonian component from the so obtained results. For the implementation of this approach, we leverage deep learning techniques, centered around a neural network-based method for histogram regression that expresses uncertainties in terms of quantiles. We demonstrate that our method is robust against a number of systematics that have plagued previous approaches, in particular DM / PS misattribution.
In the \emph{Fermi} data, we find a faint GCE described by a median source-count distribution (SCD) peaked at a flux of $\sim4 \times 10^{-11} \ \text{counts} \ \text{cm}^{-2} \ \text{s}^{-1}$ (corresponding to $\sim3 - 4$ expected counts per PS), which would require $N \sim \mathcal{O}(10^4)$ sources to explain the entire excess (median value $N = $ 29,300 across the sky). Although faint, this SCD allows us to derive the constraint $\eta_P \leq 66\%$ for the Poissonian fraction of the GCE flux $\eta_P$ at 95\% confidence, suggesting that a substantial amount of the GCE flux is due to PSs.
\end{abstract}

\maketitle

\section{Introduction}
There is strong evidence for the existence of dark matter (DM) in the Universe (see e.g. Ref.~\cite{Bertone2005} for a review), perhaps most notably thanks to the precise CMB measurements of the Planck satellite \cite{PlanckCollaboration2018}. Yet, the very nature of DM remains subject to speculation given the lack of a convincing detection. A promising avenue, which complements collider searches and direct detection efforts, is \emph{indirect detection}: the search for standard model particles resulting from the decay or annihilation of DM. An unexplained excess of $\gamma$-ray emission from the Galactic Center region in the data of the \emph{Fermi} space telescope, peaked at $\sim1 - 3$ GeV, has attracted much interest as it seems to be generally consistent with a signal originating from annihilating DM (for a recent review, see \textcite{Murgia:2020dzu}). This so-called Galactic Center Excess (GCE) extends $\sim10^\circ$ outwards from the Galactic Center and broadly follows the spatial profile expected for pair annihilation in a generalized NFW halo \cite{Calore2015, Daylan2016}. Possible DM explanations of the GCE have been extensively investigated \cite{Goodenough2009,Hooper:2010mq,Hooper:2011ti,Abazajian:2012pn,Hooper:2013rwa,Gordon:2013vta,Abazajian2014,Abazajian:2014hsa,Ajello2016,Linden2016,Macias:2016nev,Clark2018}, but other studies suggest an astrophysical origin such as a faint population of millisecond pulsars (MSPs) too dim to be individually resolved \cite{Hooper:2010mq, Abazajian:2012pn, Mirabal2013, Abazajian2014,Petrovic:2014xra,Yuan:2014yda, Brandt:2015ula}, young pulsars \cite{OLeary2015}, or cosmic-ray emission \cite{Carlson2014, Petrovic2014, Cholis2015}. Further, it has been argued that the spatial distribution of the excess follows the morphology of the stellar bulge more closely than the expected distribution of DM annihilation \cite{Macias:2016nev,Ploeg2017,Bartels:2017vsx, Macias:2019omb, Abazajian2020, Calore2021}, although a recent study in Refs.~\cite{DiMauro:2020rcr,DiMauro2021} found that with a different modeling of the background a shape more consistent with DM was preferred.
\par Most methods for the analysis of photon-count maps rely on template fitting, where the $\gamma$-ray sky is modeled as a linear combination of emission from different physical sources, each of which is associated with a spatial template. In addition, leading methods such as the Non-Poissonian Template Fit (NPTF; \cite{Lee2016,Mishra-Sharma2017}), 1pPDF \cite{Malyshev:2011zi}, or the Compound Poisson Generator (CPG; \cite{Collin:2021ufc}) harness the statistical differences between smooth (Poissonian) emission, which would arise from DM annihilation, and point-like (non-Poissonian) flux as in the case of emission from a population of astrophysical point-sources (PSs).
\par In 2016, \textcite{Lee2016} (see also Ref.~\cite{Lee:2014mza}) found strong evidence for a PS-like GCE using NPTF, and \textcite{Bartels2016} came to the same conclusion based on the application of a wavelet technique. However, re-analyses were presented more recently, which sound a note of caution on the interpretation of the 2016 results as definitive evidence against DM: Ref.~\cite{Zhong:2019ycb} showed that while the excess is still present when masking the bright sources of the updated \emph{Fermi} 4FGL source catalog \cite{Abdollahi2020}, the stacked power of the remaining bright PSs detected by the wavelet method in the \emph{Fermi} map is not enough to account for the entire excess, suggesting that the bulk of bright sources previously thought to explain the GCE forms part of the 4FGL catalog. As for the NPTF-based analysis, Ref.~\cite{Leane2019a} found that artificially injected DM flux was not correctly recovered from the \emph{Fermi} map, potentially hinting at a spurious preference for PSs due to mismodeling. This behavior was shown to be remedied by using an improved model of the diffuse foregrounds or harmonic marginalization \cite{Buschmann2020}. 
\par Yet, the worry that mismodeling might bias the analysis results remains: in Refs.~\cite{Leane2020, Leane2020a}, it was demonstrated that a mismatch between a spatial template and the true spatial distribution of the associated sources can produce an artificial preference for PSs with NPTF, even in the absence of any PS emission, as a PS model can more easily accommodate the observed larger variance caused by the mismodeling than a Poissonian model. Interestingly, when allowing for different normalizations for the GCE templates in the northern and southern hemisphere, Ref.~\cite{Leane2020} reported that within a region of interest (ROI) of $10^\circ$, the preference for PS emission vanishes, and NPTF favors a smooth asymmetric GCE. Thus, it is currently unclear to what extent the deficiencies in the modeling -- particularly of the diffuse Galactic foregrounds, which account for the majority of photon counts in the \emph{Fermi} map and constitute the largest source of uncertainty -- bias the analysis results. To counter this, different ways of endowing the spatial templates with additional degrees of freedom have been proposed, such as by using penalized likelihoods \cite{Storm2017}, expanding the diffuse template in a series of spherical harmonics \cite{Buschmann2020}, or Gaussian Processes \cite{Mishra-Sharma2020}.
\par An orthogonal approach to the problem is the development of new analysis methods, which might behave differently in the presence of shortcomings in the modeling. Recently, convolutional neural networks (CNNs) were used for the estimation of the DM vs. PS flux components of the GCE in the \emph{Fermi} map \cite{Caron2018}, and we showed in \textcite{List2020b} (henceforth \citetalias{List2020b}) that CNNs are able to learn the essential physics of template fitting, namely the accurate estimation of the flux fractions for \emph{all} the templates. Nevertheless, unlike existing template fitting methods, where the image likelihood is computed treating each pixel as statistically independent, CNNs base their judgment on properties of small patches in the photon-count maps. This leads to important differences in the case of mismodeling -- for example, CNNs seem to be fairly robust against a modest north-south asymmetry of the GCE flux (see \citetalias{List2020b}; Fig.~S8). We will later discuss this aspect in detail.
\par In \citetalias{List2020b}, we considered the task of estimating the flux fractions from $\gamma$-ray photon-count maps, treating (Poissonian) GCE DM and (non-Poissonian) GCE PS as two separate templates (albeit spatially identical, but associated with different photon-count statistics), as is also done in analyses using NPTF and CPG. However, an exact mathematical degeneracy between Poisson flux and PSs arises in the limit of infinitely faint PSs, resulting in an ambiguity in attempts to distinguish between the two templates. For illustration, consider the scenario of $N$ PSs with the same flux $\bar{f}$, giving a total flux of $F_\text{tot} = N \, \bar{f}$. In the hypothetical limit of infinitely many PSs $N \to \infty$ emitting an infinitely small flux $\bar{f} \to 0$, where the limit is formed in such a way that the total flux $F_\text{tot}$ remains constant, the PS emission becomes exactly degenerate with smooth Poisson emission. Thus, in this limit, a template fitting method such as a neural network (NN) should recognize that, assuming no preference for Poissonian / PS emission imposed by prior knowledge, any split of the flux into a Poissonian and a PS fraction is equally likely. Yet, this basic fact has not been accounted for in GCE analyses thus far. Indeed, the choice of priors adopted in existing NPTF analyses introduces a bias for either the Poissonian or the PS component, as recently demonstrated in Ref.~\cite{Collin:2021ufc}. The authors of that paper show that this issue can be overcome by reparameterizing the priors in a natural coordinate system. Although perfect degeneracy between the two flux regimes is only reached in the ultra-faint limit of infinitely many PSs, a partial degeneracy can be seen \emph{in practice} already for finite numbers of faint PSs, causing misattribution between Poissonian and PS-like flux, as has been shown to occur in NPTF analyses even when the templates perfectly describe the data (see Ref.~\cite{Chang2019}, Figs.~4 \& 5), while being further exacerbated in the presence of mismodeling (see Sec.~V in that paper). We also studied this phenomenon in Sec.~S4 of \citetalias{List2020b} for our NN-based method, where we analyzed the NN errors in the predicted flux fractions as a function of the PS brightness: as expected, the misattribution between bright PSs and Poisson emission is very small, but then gradually increases as the PSs become dimmer, and culminates in complete confusion as the source-count distribution (SCD) approaches a flux corresponding to roughly $1$ expected photon per PS. While the NN that we used in \citetalias{List2020b} yields estimates of the uncertainties inherent in the data (``aleatoric'') such as due to this very degeneracy, in addition to model-related (``epistemic'') uncertainties (and can even be trained to predict correlations in the uncertainties between multiple templates, see Sec.~S7F in \citetalias{List2020b}), the estimated distribution of the flux fraction for a PS template does not reveal any information about the SCD of the underlying PS population, for which reason it is not possible to judge how likely it is that PS and smooth emission might be confused.
\par Therefore, we present a more expressive deep learning-based approach in this paper: for training our NN, we assume the GCE to be entirely composed of PSs, where we make sure that our priors for the SCD allow for maps with PSs that are \emph{nearly} as faint as Poisson emission. In addition to the flux fraction of each template, we estimate the SCDs of the GCE and disk PS populations using a two-stage approach. To this aim, we first develop a histogram-based framework that makes use of a novel loss function, the \emph{Earth Mover's Pinball Loss}, which allows us to derive an estimate for the SCD and uncertainties on that estimate in a non-parametric way (in that we will derive the SCD without any assumption as to its functional form).\footnote{Although our SCD estimation is non-parametric, it should be expected that the prior functional forms used for the SCDs in the training data will be reproduced by the NN when evaluated on unseen data. For instance, a NN trained on unimodal SCDs will not be able to recover multimodal SCDs.} Second, we address the problem of constraining the Poissonian fraction $\eta_P$ of the GCE flux. While ultra-faint PSs are degenerate with Poisson emission, brighter PSs are not, and so to the extent the estimated SCD has support away from the ultra-faint regime, we can establish a limit on the fraction of the flux that is purely Poissonian. With this in mind, we determine a constraint on $\eta_P$ in a separate step. When evaluated on maps with a genuinely Poissonian GCE, our NN produces a faint SCD, reflecting the faint PS / Poisson degeneracy. By quantifying exactly how faint the SCDs estimated by our NN are for Poissonian emission with the help of \emph{another} NN, we obtain constraints on $\eta_P$ that become tighter as the brightness of the GCE PSs increases. 
\par For the GCE in the \emph{Fermi} map, our NN favors a faint SCD that would require $\mathcal{O}(10^4)$ PSs to explain $100 \%$ of the GCE emission. Whilst our less sophisticated framework presented in \citetalias{List2020b} attributed the entire GCE flux to the smooth GCE template, the SCD of the GCE PSs that we identify in the present work is faint enough for the above-mentioned confusion between PSs and Poissonian flux to explain this discrepancy.
\begin{figure}[t]
\centering
  \noindent
   \resizebox{1\columnwidth}{!}{
    \includegraphics{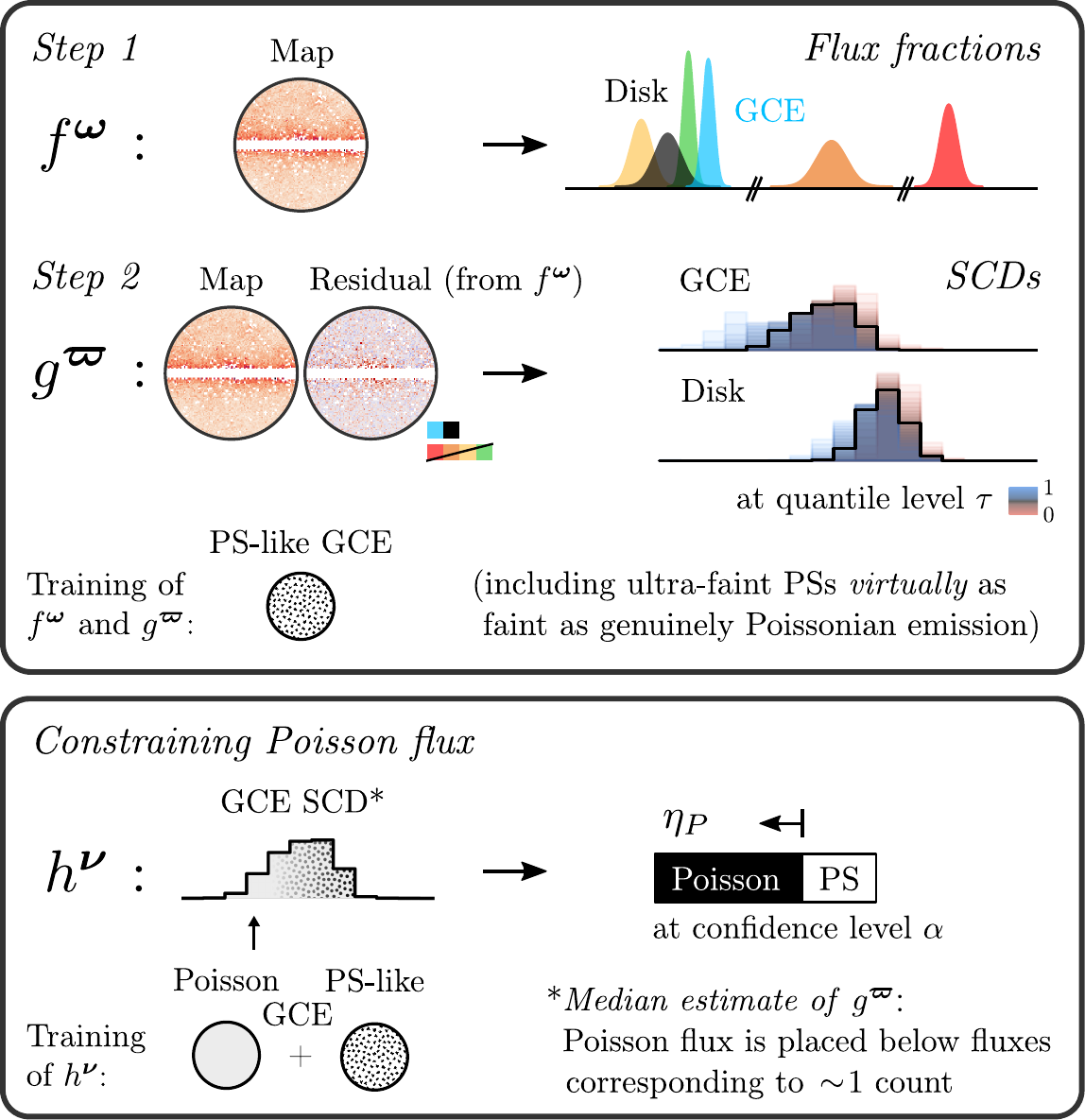}
    }
    \caption{A schematic depiction of the three NNs used in this work. In the upper panel we outline our two-step procedure for estimating the flux fractions of all emission components in the inner Galaxy (Step 1), followed by the SCDs for the GCE and disk (Step 2). These two steps are performed by sequential NNs $f^{\boldsymbol{\omega}}$ and $g^{\boldsymbol{\varpi}}$. When applying this procedure to the \emph{Fermi} data, we obtain the results shown in Fig.~\ref{fig:fermi_results}, finding a SCD for the GCE that is peaked just above a flux corresponding to a single photon. In the lower panel, we depict how we use a third NN $h^{\boldsymbol{\nu}}$ to estimate the fraction of the GCE flux consistent with Poisson emission, $\eta_P$, given the SCD determined by $g^{\boldsymbol{\varpi}}$. When $h^{\boldsymbol{\nu}}$ is applied to the \emph{Fermi} map, we obtain the results in Fig.~\ref{fig:fermi_constraints}, and in particular find that the NN estimates that at 95\% confidence, the GCE can be no more than $\sim 66\%$ Poisson emission. In all cases, on the left we show the inputs taken by each NN, and on the right the relevant outputs, with the types of GCE emission used for the training in each case shown below. Much more detail on each of these steps is provided in the text.}
    \label{fig:NN_sketch}
\end{figure}

\section*{Outline and Summary of Results}
\par Before we begin, let us outline in detail how the remainder of this work will be structured. As we do so, we will emphasize our key results in bold.
\par In Sec.~\ref{sec:cnns}, we briefly introduce CNNs, one of the fundamental tools our analysis makes use of, and then compare them to traditional likelihood-based analysis methods for $\gamma$-ray maps. We particularly discuss how mismodeling on large angular scales leads to differences in the results between our \emph{macroscale} CNN-based approach, which considers patches of the sky, and \emph{microscale} likelihood-based methods, which consider each pixel individually. A schematic example of this difference is shown in Fig.~\ref{fig:reshuffling_sketch}.
\par We introduce our two-stage approach for the NN-aided analysis of the $\gamma$-ray sky in Sec.~\ref{sec:NN_method}, the details of which are illustrated in the upper panel of Fig.~\ref{fig:NN_sketch}. We train a NN $f^{\boldsymbol{\omega}}$ to estimate the flux fraction of each template. For templates where we expect both a PS and Poisson contribution (such as the GCE), we only estimate the combined flux of both at this stage, with no attempt to distinguish whether the flux is more consistent with PSs or Poisson emission. Afterwards, the NN $g^{\boldsymbol{\varpi}}$ learns to recover the SCDs of the disk and the GCE populations, using the residuals of the maps after removing the best-fit emission of the other templates as judged by $f^{\boldsymbol{\omega}}$ as a second input channel. Importantly, \emph{for the training of both NNs, we only include a PS-like GCE}; however, our priors on the SCDs generated ensure that the training dataset contains maps with a PS-like GCE faint enough to be \emph{indistinguishable} from Poissonian flux.
\par As a first test, in Sec.~\ref{sec:toy} we consider the characterization of a single isotropic PS population in isolation. We demonstrate that we can recover the injected SCD (within uncertainties) even below fluxes where a PS would be expected to generate only a single photon, with examples shown in Fig.~\ref{fig:isotropic_results}. Further, in Fig.~\ref{fig:isotropic_poisson} we show that \textbf{genuine Poisson emission is reconstructed in the SCD well below the flux associated with 1 photon.}
\par We then turn toward the scenario of interest in Sec.~\ref{sec:fermi}, the real \emph{Fermi} map, where we include flux templates for all the sources that are expected to (potentially) contribute to the $\gamma$-ray sky; moreover, we account for the non-uniformity of the \emph{Fermi} exposure, and mask the known bright sources in the 3FGL catalog \cite{Acero2015}. Before considering the actual data, we validate our method on simulated \emph{Fermi} mock maps, showing in Figs.~\ref{fig:flux_fraction_errors} and \ref{fig:simulated_hist_plot} that we can accurately reconstruct the injected flux fractions and SCDs, respectively, for each template. In \textbf{Fig.~\ref{fig:fermi_results}}, we present the \textbf{main results} of our paper, namely our findings for the \emph{Fermi} data. \textbf{We infer a faint SCD for the GCE peaked at} $\boldsymbol{\sim4 \times 10^{-11} \ \textbf{counts} \ \textbf{cm}^{-2} \ \textbf{s}^{-1}}$ (yielding $\sim3 - 4$ expected counts per PS). Unlike in previous analyses, the SCD is used to account for both the Poissonian and PS flux, and a purely Poissonian GCE is expected to peak below fluxes corresponding to $1$ expected count per PS.
\par In Sec.~\ref{sec:constraining_Poisson}, we introduce a method for constraining the fraction of the flux that is consistent with purely Poissonian emission, $\eta_P$. To do so, we take the SCD predicted by $g^{\boldsymbol{\varpi}}$ as an input for \emph{another} NN $h^{\boldsymbol{\nu}}$, as illustrated in the bottom panel of Fig.~\ref{fig:NN_sketch}. We show that in a toy example where the exact likelihood can be calculated, our approach provides constraints on $\eta_P$ that are not much weaker than the frequentist constraints computed from the analytic likelihood, allowing us to exclude substantial Poissonian contributions in maps from PSs that on average emit less than one detected count each (see Fig.~\ref{fig:constraints_iso_no_PSF}). Afterwards, we apply this approach to the \emph{Fermi} map and derive constraints on the Poissonian GCE component as a function of confidence level and SCD. While the faint nature of the SCD identified in our analysis prevents us from excluding a Poisson-dominated GCE at high confidence, \textbf{we obtain a 95\%-confidence constraint on the Poissonian GCE flux fraction of $\boldsymbol{\eta_P \leq 66 \textbf{\%}}$} for our median SCD, suggesting the GCE cannot be entirely explained by Poissonian emission as predicted by DM annihilation, see \textbf{Fig.~\ref{fig:fermi_constraints}}.
\par Lastly, we test the robustness of our findings in Sec.~\ref{sec:robustness} against potential systematics. We show in Fig.~\ref{fig:comparison_w_best_fit} that for simulated \emph{Fermi}-like maps with a purely Poissonian GCE, we indeed obtain SCD estimates fainter than for the real \emph{Fermi} GCE. Then, we consider different sources of mismodeling in Fig.~\ref{fig:mismodelling}, showing for example the \textbf{robustness of our results against a north-south asymmetry of the GCE} that was found to cause a spurious PS preference with the NPTF in Ref.~\cite{Leane2020}, in addition to finding that \textbf{diffuse mismodeling could be absorbed in the GCE SCD, but is likely to do so at the lower fluxes characteristic of Poisson emission}. Notably, in our unified approach for the GCE, increasing mismodeling can be expected to gradually shift the SCD estimate instead of suddenly changing the PS vs. Poisson preference. Finally, we demonstrate in Fig.~\ref{fig:injection_FF} that both \textbf{Poissonian and PS-like GCE flux injected into the \emph{Fermi} map are accurately recovered}.

\section{Deep learning for $\gamma$-ray maps}
\label{sec:cnns}
We start this section with a brief introduction to CNNs \cite{Lecun1998}. In particular, we describe several particularities in the \texttt{DeepSphere} framework \cite{Perraudin2019a, Defferrard2020}, upon which we base our NN architecture, thereby avoiding the need for projecting the input maps to 2D images. Having introduced CNNs, we then contrast CNN-based inference with traditional template fitting methods, focusing on the effect of large-scale mismodeling.

\subsection{Convolutional neural networks}
Like most NNs, CNNs belong to the class of supervised learning methods. Thus, labeled training data $\mathcal{X}_L = \left(\mathbf{x}_l\right)_{l=1}^L$ is required, i.e. the true label $\mathcal{Y}_L = \left(\mathbf{y}_l\right)_{l=1}^L$ for each of the $L$ training samples must be available. Then, the task of the NN is to learn a mapping $f^{\boldsymbol{\omega}}: \Omega_X \to \Omega_Y$, $\mathbf{x} \mapsto \tilde{\mathbf{y}} = f^{\boldsymbol{\omega}}(\mathbf{x})$ from the input domain $\Omega_X$ to the target domain $\Omega_Y$, which approximates the true relation between inputs and outputs. Here and in what follows, we use a tilde to indicate estimated (and therefore approximate) quantities. Provided that the training set $\mathcal{X}_L \subset \Omega_X$ is a sufficiently large ``representative'' (discrete) subset of $\Omega_X$, one expects the NN output to be a good approximation of the (possibly unknown) true label $\mathbf{y} \in \Omega_Y$, that is $\tilde{\mathbf{y}} \approx \mathbf{y}$, even for samples $\mathbf{x} \in \Omega_X \setminus \mathcal{X}_L$ that the NN has not been trained on. The mapping $f^{\boldsymbol{\omega}}$ is defined by a series of operations (known as the NN \emph{layers}) that successively map each input $\mathbf{x}$ to an output $\tilde{\mathbf{y}}$. Some of these layers have trainable parameters, known as the \emph{weights} of the NN, which we gather in the vector $\boldsymbol{\omega}$. In order to assess the fidelity of the NN prediction with respect to the truth, one defines a loss function $\mathcal{L}: (\tilde{\mathbf{y}}, \mathbf{y}) \mapsto \mathcal{L}(\tilde{\mathbf{y}}, \mathbf{y}) \in \mathbb{R}$, which represents the optimization objective. Typical loss functions for regression problems are the mean absolute error ($l^1$) or the mean squared error ($l^2$). The NN ``training'' simply refers to the iterative minimization of the mean loss over the training set using a variant of a batch gradient descent method, which adjusts the weights $\boldsymbol{\omega}$ after each iteration step. Each \emph{batch} consists of a fixed number of samples that are simultaneously shown to the NN (as the entire training data $\mathcal{X}_L$ and labels $\mathcal{Y}_L$ do not usually fit in the memory, and a smaller batch size can improve the generalization from the training to testing dataset \cite{Smith2018}). 
\par Whilst the above concepts apply to many types of NNs, the distinctive operation of a CNN is the \emph{convolution}, which enables the extraction of salient spatial features from the data. Following \citetalias{List2020b}, we base our NN on the \texttt{DeepSphere} graph-CNN architecture \cite{Perraudin2019a, Defferrard2020}, which is particularly suitable for astrophysical and cosmological applications: in \texttt{DeepSphere}, the sphere is described by an edge-weighted, undirected graph, which leverages the \texttt{HEALPix} equal-area tessellation of the sphere \cite{Gorski2005}. Specifically, the center of each \texttt{HEALPix} pixel defines a vertex of the graph, and neighboring pixels are connected with an edge, leading to $7 - 8$ edges incident to each vertex. The edge weights determine how the influence between pixels decays with increasing distance. In this work, we use the new scheme for the edge weights proposed in Ref.~\cite{Defferrard2020}.
The trainable parameters of the convolutional layers are given by filters (or kernels) that detect specific patterns in the data, such as gradients or edges. These filters have a (user-defined) size, which determines the field of view or, in other words, the neighborhood of each pixel that affects the output of the convolution. For standard CNNs that operate on Euclidean domains, the convolution is performed by sliding the filters over the input image. In the context of graphs, the convolution can be defined in Fourier space using the graph Laplacian (see Ref.~\cite{Perraudin2019a} for additional details). To emphasize, for all filter sizes greater than 1, the convolution is inherently an \emph{inter-pixel} operation. In \texttt{DeepSphere}, the filters are restricted to be radially symmetric, which can be used to build NN architectures that are rotationally invariant (or more generally equivariant) on the sphere, which is useful for all-sky applications where the location on the sky should not matter, but which is not needed for our task at hand. However, we did not notice any detrimental effect of this specific form of the filters as compared to a standard 2D CNN applied to projected photon-count maps, for which reason we decided to use \texttt{DeepSphere} as it does not require projecting the maps to flat images.
Since \texttt{DeepSphere} supports partial maps, the input to our NN is only the relevant ROI, rather than the entire sphere. Besides the convolution operation, our CNN consists of maximum pooling layers, each of which reduces the spatial resolution by computing the maximum over blocks of 4 adjacent pixels (exploiting the hierarchy of the \texttt{HEALPix} tessellation, where each pixel contains 4 pixels at the next finer resolution level), activation functions, which introduce nonlinearity and enable the CNN to learn complex mappings, and batch normalization \cite{Ioffe2015} or instance normalization \cite{Ulyanov2016}, which have been shown to speed up the training process. The detailed NN architecture for each scenario is specified in App.~\ref{sec:NN_architectures}.

\subsection{Comparison with traditional methods}
In this section, we illustrate in a minimal scenario how the conceptual differences between CNN-based and likelihood-based inference may lead to different results in the presence of large-scale mismodeling, which can bias analyses of the \emph{Fermi} map and hence is a major hindrance to a conclusive resolution of the GCE. 
We also briefly comment on differences and similarities between our approach and the wavelet technique that was applied to the \emph{Fermi} map in Refs.~\cite{Bartels2016, McDermott:2015ydv,Balaji:2018rwz,Zhong:2019ycb}.
\par A challenge for any analysis of the \emph{Fermi} dataset is the treatment of cross-pixel correlations. One source of such correlations is the instrument point-spread function (PSF), which distributes incident photons among nearby pixels, in a statistically predictable manner. A second source arises from the mismodeling that results from using imperfect models for the spatial distribution of either Poissonian or PS flux, which is unavoidable given our present imperfect understanding of the $\gamma$-ray sky. If we ignore the effect of the PSF, then from the perspective of the true underlying distribution that the data is drawn from, each pixel represents an independent draw and is therefore uncorrelated. However, an analysis of that same data making use of imperfect models will induce apparent correlations over distances corresponding to the scale of mismodeling. For instance, these correlations are clearly noticeable in the structure observed in residual maps, where the best-fit model is subtracted from the data. In summary, both the PSF and template mismodeling imply that the observed values for the number of counts in nearby pixels are not independent at the level of the analysis.

Despite this, even with elaborate methods such as the recently introduced CPG framework \cite{Collin:2021ufc} that computes an individual instrumental and PSF correction for each pixel, a fully-consistent treatment of the cross-pixel correlation just from the PSF would involve solving heavy combinatorics in each likelihood evaluation to account for all possible combinations of counts being smeared from one pixel into another, which is computationally infeasible. Thus, the total image likelihood is ultimately calculated as the product over the individual pixel likelihoods, treating the pixels as being statistically independent. In practice, this often leads to the inferred posterior being narrower than it should be, as treating correlated pixels as independent artificially provides more information than is present in the data~\cite{Collin:2021ufc}. An important consequence of this product likelihood assumption is that all outputs of existing likelihood methods are invariant under a permutation of the pixel ordering (assuming the template values are also permuted accordingly).
\begin{figure}
\centering
  \noindent
   \resizebox{1\columnwidth}{!}{
    \includegraphics{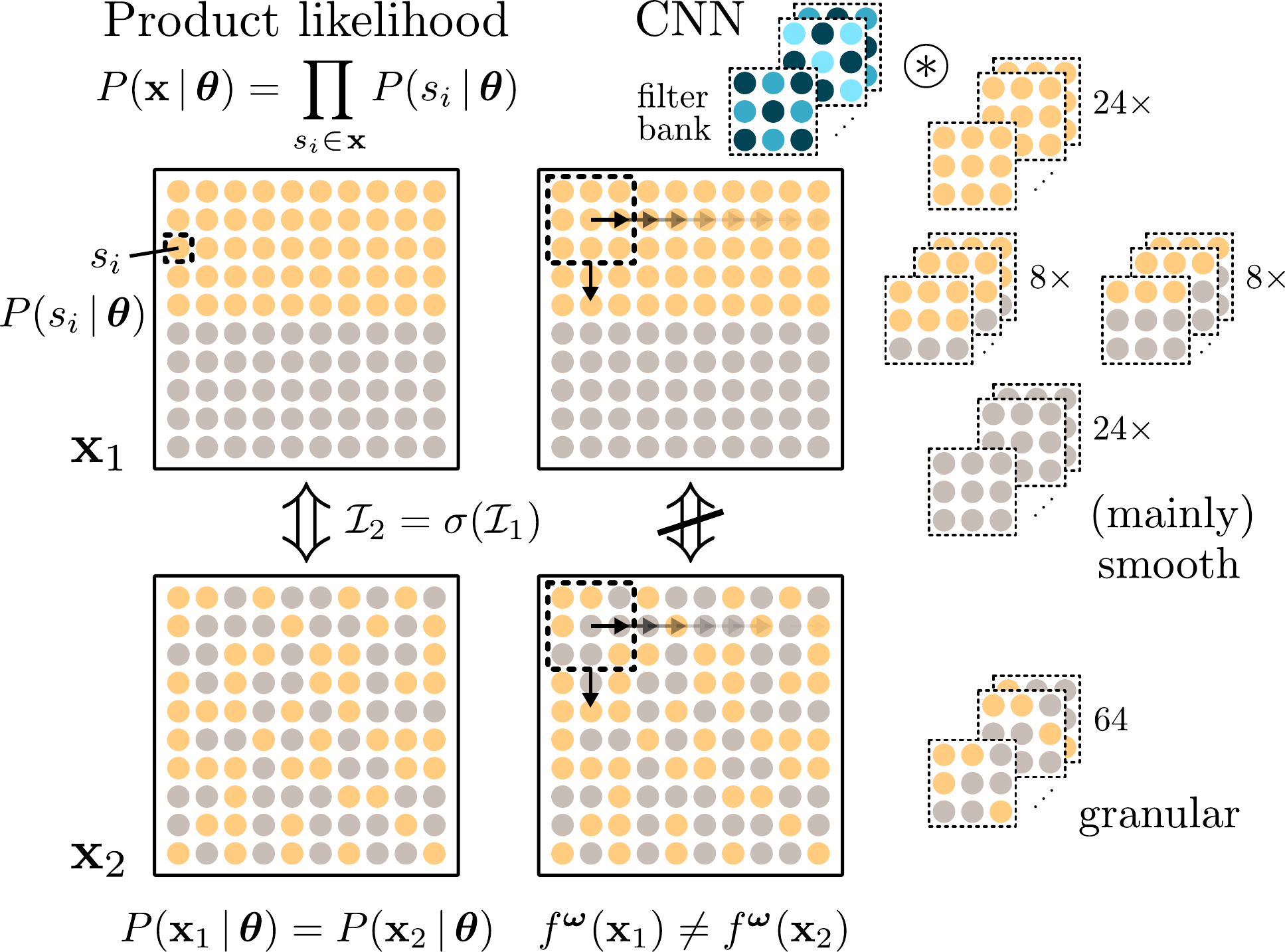}
    }
    \caption{Schematic of the inference process using likelihood-based methods (left) and CNNs (right). Map $\mathbf{x}_1$ (top) has a strong north-south asymmetry, expressed by the different colors in the northern and southern half of the map, where each dot represents a pixel (with orange pixels brighter than gray). In $\mathbf{x}_2 = \sigma(\mathbf{x}_1)$, the pixels are randomly shuffled by the permutation $\sigma$. If the \emph{expected} spatial distribution of the counts over the map is assumed to be homogeneous (and the asymmetry is hence ``unmodeled''), the shuffling leaves the product likelihood unaffected when there is no accounting of the inter-pixel correlations (as is the case for the NPTF and CPG). Thus, the smooth map with a single jump $\mathbf{x}_1$ is indistinguishable from the grainy map $\mathbf{x}_2$. For a more formal derivation of this effect, we refer to Ref.~\cite{Leane2020a}.
    In contrast, CNNs assess patches of a fixed size ($3 \times 3$ in this sketch) using filters that are convolved with the map. We neglect edge effects and padding here for simplicity. The large-scale mismodeling present in map $\mathbf{x}_1$ does not affect most of the patches, whereas the texture of map $\mathbf{x}_2$ strongly differs from an isotropic Poissonian map. Therefore, the NN outputs for the two maps will generally \emph{not} be the same. As each convolutional layer is followed by a pooling operation, the size of the patches considered by the CNN gradually increases with each layer, allowing the CNN to harness information on different scales.}
    \label{fig:reshuffling_sketch}
\end{figure}
\par Unlike CPG or NPTF, deep learning methods often do not rely on an explicit form of the image likelihood and therefore do not require such assumptions. In fact, CNNs draw much of their power from their ability to assess cross-pixel information such as image granularity. Accordingly, such methods are not invariant under a permutation of the pixelated data, and this has important consequences for the inference in the presence of mismodeling. We emphasize that although the inherently inter-pixel nature of CNNs could account for the correlations induced by the PSF, it could never fully account for those induced by mismodeling. Nevertheless, as the inference performed by the CNN is based on regions, rather than by extracting information from each pixel treated independently, its behavior in the presence of incorrect flux models can be dramatically different to likelihood approaches, as we now demonstrate.
\par For illustration, let us consider a simple toy example, inspired by the preference of NPTF for a GCE north-south asymmetry in the \emph{Fermi} data within a radius of $10^\circ$ around the Galactic Center that was identified by Refs.~\cite{Leane2020, Leane2020a}. We neglect the PSF such that inter-pixel correlations in the map are entirely caused by the flawed modeling. We consider purely Poissonian emission whose intensity in the northern and southern hemisphere differs, but is constant \emph{within} each hemisphere. For simplicity, we assume that the exposure is uniform. Such a map is sketched in Fig.~\ref{fig:reshuffling_sketch} (top), where the Poissonian scatter is not drawn for simplicity. Now, we consider the effect of incorrectly modeling the entire sky with an isotropic Poissonian and PS template. Whilst we qualitatively discuss and compare the different methods in this section, we explicitly perform this experiment for an example map in App.~\ref{sec:NS_asymmetry_example}.
\par For methods that compute the image likelihood as the product over the pixel likelihoods, this map $\mathbf{x}_1$ is indistinguishable from a map $\mathbf{x}_2$ in which the pixels are randomly reshuffled by a permutation $\mathbf{x}_2 = \sigma(\mathbf{x}_1)$, and their likelihoods are identical.\footnote{If the asymmetry is correctly modeled, and the permutation that transforms $\mathbf{x}_1 \mapsto \mathbf{x}_2$ is only applied to the data but not the asymmetric background template, then the situations are of course distinguishable. Nevertheless, note that even in this situation if we also permute the background model, then again the two maps will produce identical outputs.}
An example of such a permutation is provided in the bottom of Fig.~\ref{fig:reshuffling_sketch}. The permuted map exhibits large pixel-to-pixel variation that is suggestive of a population of sources, and indeed likelihood-based methods attribute the majority of the flux in these scenarios to PSs. However, given the invariance to permutations, these methods also predict that for $\mathbf{x}_1$ the asymmetry arises from PSs that are effectively all in the northern half of the map. This is reminiscent of the discussion in Ref.~\cite{Collin:2021ufc}, which evokes the analogy of gas molecules in a box: it would be completely unexpected to find all the molecules in just one half of the box; however, such a microstate is just as likely as any other configuration of the molecules. Similarly, if one expects isotropically distributed sources, the probability of them uniformly covering one hemisphere is identical to any other possible spatial distribution. This equivalency between the original and the shuffled case in terms of the resulting product likelihoods is depicted on the left hand side of Fig.~\ref{fig:reshuffling_sketch}. In view of the large pixel-to-pixel variance in the maps caused by the mismodeling, it is not surprising that non-Poissonian PS emission leads to a higher likelihood than smooth Poisson emission and is therefore preferred by NPTF (see also Ref.~\cite{Leane2020a} for a mathematical derivation of such a behavior). Note that while we consider an abrupt \emph{jump} in the flux intensity here, an unmodeled large-scale \emph{gradient} can be expected to induce a qualitatively similar behavior. Importantly, we point out that this equivalence of the two maps in terms of the resulting likelihoods is \emph{not} a flaw of the NPTF, but the consequence of the mismatch between the template and the true data, in conjunction with the \emph{microstate} (i.e. pixelwise) assessment of the maps by the NPTF.
\par CNNs, on the other hand, operate differently: rather than computing pixelwise likelihoods, trainable filters (illustrated in blue in Fig.~\ref{fig:reshuffling_sketch}) of a specified size -- $3 \times 3$ in the sketch -- are convolved with image \emph{patches}. These filters extract characteristic patterns, based on which the model parameters $\boldsymbol{\theta}$ (or their distribution) can be inferred for each input map. In practice, multiple convolutional layers are applied successively, enabling the CNN to distill more complex features from the data. The results of the convolution operations are further processed by nonlinearities and pooling operations, which is not essential for this discussion. Coming back to the original and randomly shuffled maps $\mathbf{x}_1$ and $\mathbf{x}_2$, respectively, the NN output (whose exact meaning is left unspecified for the moment) can be expected to be very different for these two maps: in map $\mathbf{x}_1$, all the patches save those containing the equator are constant up to Poisson scatter. In contrast, all the patches in $\mathbf{x}_2$ contain some pixels with many and others with few counts. Thus, the feature maps, i.e. the results of the convolution between the filters and the images, will generally not be identical for the two maps. For a realistic analysis of the GCE, the Galactic Plane is typically masked, such that the north-south transition region would not even be part of the considered ROI in this specific example.
Resorting to the analogy of molecules again, the CNN-based inference could be equated with an assessment of the molecule configuration within each of many small (overlapping) sub-boxes (or local \emph{macrostates}), which together make up the entire box. Since the majority of these sub-boxes look exactly as expected in the Poissonian case for map $\mathbf{x}_1$ (although they are not compatible with a single isotropic template) whereas their counterparts in $\mathbf{x}_2$ are granular, it is comprehensible that the CNN generally finds map $\mathbf{x}_1$ to be more ``Poissonian'' than map $\mathbf{x}_2$ (and in fact this occurs in practice, see Fig.~\ref{fig:asymmetry_comparison} in App.~\ref{sec:NS_asymmetry_example} for an example). Clearly, neither method can be expected to work perfectly in this situation, as the true model lies outside the space of models considered in the analysis. Finally, it is important to note that this example explicitly considers the effects of \emph{large-scale} mismodeling: the presence of \emph{small-scale} mismodeling, e.g. due to an overly smooth or grainy diffuse model on pixel-to-pixel scales, can be expected to introduce considerable biases with our CNN-based method (see Sec.~\ref{sec:mismodeling_experiment} for an assessment of the robustness of our results with respect to different sources of mismodeling).
\par At this point, let us also mention \emph{probabilistic cataloguing}, which rather than estimating the SCD, instead aims to resolve the location and intensity of each PS individually, even in crowded fields~\cite{Brewer2013, Daylan2017, Portillo2017}. The permutation invariance discussed for the NPTF and CPG using the example in Fig.~\ref{fig:reshuffling_sketch} does not apply to probabilistic cataloguing. More specifically, each possible number of PSs $N$ of a population defines a separate \emph{metamodel}, which itself comprises parameters for each of the $N$ PSs, leading to a large number of degrees of freedom of a few times $N$ (at fixed $N$). As $N$ is itself a parameter, a fundamental challenge is to ensure that transdimensional transitions occur efficiently in the Markov Chain Monte-Carlo runs (as changing $N$ varies the number of total model parameters). Moreover, for sufficiently crowded fields containing many sources in each pixel, the exact location and properties of all PSs may be of less interest than the global properties of the distribution encoded in the SCD. Hence, we will focus in this work on methods that describe PS populations \emph{globally} in terms of a SCD. For further discussion of this point, we refer to \textcite{Collin:2021ufc}.
\par As for CNNs, the convolution operation is also the crux of the wavelet technique \cite{Bartels2016, McDermott:2015ydv, Balaji:2018rwz}, but there are important differences. (1) For the wavelet technique, the convolution kernel needs to be manually specified, with the Mexican hat family being a popular choice. On the other hand, CNNs possess a large number of different filters, arranged in multiple layers, which are \emph{learned} by means of a stochastic gradient descent method. (2) The wavelet technique produces a signal-to-noise ratio map that reveals the location of detected bright sources in the map. The statistics of the identified peaks can then be compared to those expected in the purely Poissonian case in order to constrain the flux coming from smooth and PS emission (see Refs.~\cite{Bartels2016, Zhong:2019ycb}). In contrast, our CNN does not produce an output map, but rather infers \emph{global} properties such as template flux fractions and the SCDs of the PS populations. Another approach, which we defer to future work, would be the use of an encoder-decoder NN architecture such as a U-Net \cite{Ronneberger2015}, which allows for the inference of local (i.e., pixel-wise) quantities (see e.g. Ref.~\cite{Caron2021} for a recent application to the identification of PSs). (3) The wavelet technique does not attempt to disentangle the photon counts into multiple components that model different emission processes. Therefore, fully characterizing the emission typically requires a template fit (to determine the flux fractions of the templates) in addition to the wavelet analysis (to search for small-scale power), as done in Ref.~\cite{Zhong:2019ycb}. CNNs, just like NPTF and CPG, are able to simultaneously estimate flux fractions (or template normalizations) and other model parameters that describe the PS populations. In sum, CNNs combine certain aspects of both traditional template fitting methods and the wavelet technique, while providing an entirely independent way of analyzing photon-count maps, and the rapid progress in the development of new powerful deep learning techniques leaves significant room for further improvement going forward.

\section{A two-step approach for neural network-based inference}
\label{sec:NN_method}
In \citetalias{List2020b}, we included both a PS-like non-Poissonian component and a smooth Poissonian component of the GCE by modeling them as two separate templates, each associated with an individual flux fraction, similar to NPTF-based analyses. However, this simple approach neglects the inherent degeneracy between PS and Poisson emission that arises gradually as the PS brightness tends to zero. 
\par Therefore, we present an improved version of our NN in this work, which characterizes the flux associated with each (potentially) non-Poissonian template by means of a \emph{histogram} that expresses the discretized SCD of the PS population.
We introduce a two-step approach for the fully-supervised deep learning-based analysis of $\gamma$-ray maps, where the flux fractions are determined in Step 1, followed by the estimation of brightness histograms in Step 2. Importantly, we estimate a single flux fraction for the Poissonian and the PS-like component associated with a spatial template, and we will then use the SCD estimate to distinguish between the two. In what follows, we will describe the two steps in detail.

\subsection{Step 1: Estimating flux fractions}
\label{sec:NN_method_step_1}
Since the flux fraction estimation follows the ideas presented in \citetalias{List2020b}, we only summarize the key points here. Let $f^{\boldsymbol{\omega}}$ be a NN with trainable parameters $\boldsymbol{\omega}$. The task of this NN is to predict the vector of flux fractions $\mathbf{y} = (y_t)_{t=1}^T \in \Delta^{T - 1}$ for $T$ templates given an input map $\mathbf{x}$. Here, $\Delta^{T - 1}$ is the $(T - 1)$-dimensional standard simplex, namely the set of all $\mathbf{a} = (a_t)_{t=1}^T \in \mathbb{R}^{T}$ such that $a_t \geq 0$ for all $t \in \{1, \ldots, T\}$ and $\sum_{t=1}^T a_t = 1$.
Making the simplifying assumption that the flux fraction of each template $t$ can be modeled independently by a Gaussian distribution with standard deviation $\sigma_t$, the negative maximum log-likelihood for the NN prediction is given by
\begin{equation}
    \mathcal{L}\left(f^{\boldsymbol{\omega}}(\mathbf{x}), \mathbf{y}\right) = \sum_{t=1}^{T}\left(\frac{1}{2 \sigma_{t}^{2}(\mathbf{x})}\left(f^{\boldsymbol{\omega}}(\mathbf{x})_{t}-y_{t}\right)^{2}+\frac{\ln \sigma_{t}^{2}(\mathbf{x})}{2}\right),
\label{eq:loss_step_1}
\end{equation}
where we omit the constant term $T/2 \,\ln(2 \pi)$. We do not assume the standard deviations $\sigma_t$ to be known a \emph{priori}, but rather train the NN to predict them in addition to the mean flux fractions, using the negative maximum log-likelihood in Eq.~\eqref{eq:loss_step_1} as the loss function. Note that the first and the second term of the loss function penalize too small and too large values of $\sigma_t$, respectively. Thus, for $T$ templates, the NN output has dimension $2 \times T$ and contains $\{(f^{\boldsymbol{\omega}}(\mathbf{x})_t, \sigma_t(\mathbf{x}))\}_{t=1}^T$, where $f^{\boldsymbol{\omega}}(\mathbf{x})_t = \tilde{y}_t \approx y_t$, and $\sigma_t$ expresses the data-inherent (aleatoric) uncertainties. Since we found the model-related (epistemic) uncertainties of the trained NN to be comparatively small in \citetalias{List2020b}, we omit them in this work.
We enforce that the estimated flux fractions sum up to unity by applying a softmax activation function to the means $f^{\boldsymbol{\omega}}(\mathbf{x})$ after the last NN layer, which normalizes a vector $\mathbf{a} = (a_t)_{t=1}^T \in \mathbb{R}^T$ as follows:
\begin{equation}
\operatorname{softmax}(\mathbf{a})_t = \frac{\exp(a_t)}{\sum_{s=1}^T \exp(a_s)}.
\label{eq:softmax}
\end{equation}
We guarantee the positivity of the variance by estimating the log-variance, $\ln(\sigma_t^2)$.
The important difference as compared to \citetalias{List2020b} is that we now describe the GCE with a \emph{single} template instead of treating Poissonian and PS-like GCE emission as separate templates. This simplifies the task of the NN as the total number of templates is reduced by one and, more importantly, the above discussed degeneracy between smooth and PS emission for one and the same spatial template is eliminated, and only spatially distinct (albeit not disjunct) templates remain. A side effect of this unified approach is that the assumption of Gaussian uncertainties for the GCE flux fraction becomes more justifiable: whereas an error distribution of the flux fractions skewed away from zero is natural for templates with a very small flux fraction (see e.g. Figs.~S4 and S6 in \citetalias{List2020b} for this effect occurring for GCE DM and PS, respectively), the error distribution of the \emph{total} GCE flux can be well approximated by Gaussians (see the ``Total GCE'' column in the same figures). Of course, the most interesting question as to the nature of the GCE has been ignored until now, but we will address this in the second step.

\subsection{Step 2: Estimating source-count distributions}
\label{sec:NN_method_step_2}
We now present the second part of our approach, which enables us to characterize the underlying PS populations in terms of the SCD. As is customary, we model the SCD via a function $dN/dF$, which expresses the differential number of PSs $dN$ that fall within an infinitesimal flux interval $[F, F + dF]$. Note that this function specifies a probability density function (PDF) $P(F)$ via
\begin{equation}
    \frac{dN}{dF} = N \, P(F),
\end{equation}
where $N$ is the expected number of sources. For each individual PS, the probability of observing $s_i$ counts in a pixel $i$ depends on (1) the probability for the PS to emit a certain flux $F$ as described by $dN/dF$, (2) the probability distribution for the \emph{expected} observed counts given a flux $F$, which depends on detector effects such as exposure time, effective area, and the PSF, and (3) the Poisson probability for the actually \emph{observed} number of counts given the expected number of counts. Additionally, the observed number of PSs itself is a random variable that can be modeled with a Poisson distribution. 
\par Different avenues could be pursued for estimating the SCDs of PS populations using NNs. For instance, a versatile framework for the estimation of arbitrary probability distributions, which has recently found its way into cosmology (e.g. Refs.~\cite{Jeffrey2020, Hortua2020a, Mishra-Sharma2020}), is given by Normalizing Flows \cite{Trippe2018, Kobyzev2019, Papamakarios2021}. Another interesting approach, rooted in contrastive learning, considers the task of likelihood-to-evidence ratio estimation and frames it as a classification problem \cite{Miller2020}. In that framework, the trained NN outputs an approximation of the (marginalized) likelihood of each model parameter. For these approaches, the SCD function $dN/dF$ could be parameterized, e.g. as a multiply broken power law in log-space as usually done for NPTF analyses, with model parameters $\boldsymbol{\theta}$.
\par In this work, we opt for a different approach and use a \emph{binned} source-count function instead. Thus, arbitrary shapes of $dN/dF$ can be accounted for, and no explicit parametrization of $dN/dF$ is needed. A binned $dN/dF$ has also been considered for the analysis of the GCE in the context of NPTF by Ref.~\cite[Fig.~S14]{Lee2016}. Whilst obtaining posterior distributions with the above-mentioned methods typically requires sampling points and propagating them through the NN, we represent the distribution of possible SCD histograms in terms of their \emph{quantiles}, as will be explained further below. Specifically, we estimate the quantity 
\[
F \, \frac{dN}{d \log_{10} F} \propto F^2 \, \frac{dN}{dF},
\]
implying that the histogram values are proportional to flux $F$ when using log-spaced flux bins (or \emph{relative} flux after normalizing the histograms as described below).\footnote{In comparison, when binning $dN / d (\log_{10} F)$ into log-spaced bins, the histogram values are proportional to the number of PSs, which comparatively suppresses the importance of bright PSs. For example, consider a map containing 2,000 counts, 1,000 of which come from a single bright PS while the other 1,000 originate from 1,000 faint PSs each responsible for 1 count. Assuming uniform exposure, the bars for the fluxes corresponding to 1 count and 1,000 counts are equal when binning $F \, dN / d (\log_{10} F)$ because the PSs in both bins contribute the same flux to the map. In contrast, binning $dN / d (\log_{10} F)$ causes the bar for the faint PSs to be 1,000 times larger than that for the bright PS.}
Therefore, integrating this quantity over log-spaced flux bins yields the total flux of the PS population,\footnote{We remark that whenever we write $\log_{10}(F)$ or $\log_{10}\left(F \ [\text{counts} \ \text{cm}^{-2} \ \text{s}^{-1}]\right)$, this should be interpreted as $\log_{10}\left((F \ [\text{counts} \ \text{cm}^{-2} \ \text{s}^{-1}]) \, / \, (\text{counts} \ \text{cm}^{-2} \ \text{s}^{-1})\right)$ such that the logarithm is applied to a nondimensional quantity.} 
\begin{equation}
    F_\text{tot} = \int F \, \frac{dN}{dF}(F) \, dF = \int F \, \frac{dN}{d \log_{10} F}(F) \, d \log_{10} F.
\end{equation}
Instead of regressing a flux-based quantity, one could also consider the prediction of count-based histograms, e.g. by binning the counts according to the number of total counts detected from each PS (see \textcite{List2021}). Then, the labels would include the Poisson scatter that arises from drawing the number of observed counts given the expected number of counts, which would slightly simplify the task of the NN. However, since flux is the physical quantity that characterizes a PS, we choose a flux-based approach in this work, which leads to labels that are immune to the non-uniformity of the \emph{Fermi} exposure map and facilitates the comparison with conventional methods such as the NPTF.
\par In what follows, we introduce the notation that we will need for the definition of the loss function. Let $\mathbf{u} = (u_j)_{j=1}^M \in \Delta^{M - 1}$ be the true histogram that discretizes the normalized $F \, dN / d (\log_{10} F)$ into $M$ bins, such that each bin $j$ collects the relative flux $F / F_\text{tot}$ from all those PSs whose individual flux lies within the associated logarithmic flux range $(\Delta \log_{10} F)_j$. As above, $\Delta^{M - 1}$ denotes the $(M - 1)$-dimensional standard simplex. For example, for a population of identical PSs that each emit a fixed flux $\bar{f}$, we have $u_j = 1$ in the single bin $j$ for which $\log_{10} \bar{f} \in (\Delta \log_{10} F)_j$ and $u_m = 0$ for $m \neq j$. The motivation for dividing by the total flux of the PS population $F_\text{tot}$ is that $F_\text{tot}$ can simply be recovered from the flux fraction estimated for the template in Step 1, together with the known total flux in the map. Therefore, it is sufficient for the histograms to express the \emph{relative} amount of flux $F / F_\text{tot}$ coming from PSs within each logarithmic flux interval.
\par We define $g^{\boldsymbol{\varpi}}$ to be the NN for the task of the SCD estimation, with trainable parameters $\boldsymbol{\varpi}$. Again, a suitable loss function needs to be specified, now for comparing the true and estimated SCD histograms. A naive approach would be to compute the loss in each histogram bin (e.g. $l^1$, $l^2$, or cross-entropy loss) and to sum over the losses in the individual bins. However, this would ignore the natural ordering of the histogram bins: for example, the loss between a true histogram $\mathbf{u} = [1, 0, \ldots, 0]$ and an approximation $\tilde{\mathbf{u}}_{1} = [0, \ldots, 0, 1]$ would be the same as between $\mathbf{u}$ and $\tilde{\mathbf{u}}_{2} = [0, 1, 0, \ldots, 0]$, although a NN that predicts $\tilde{\mathbf{u}}_{2}$ is clearly preferable. In order to instill this logic into our NN, we utilize the loss function for histogram regression recently introduced in \textcite{List2021}, which incorporates cross-bin information and enables the estimation of the entire \emph{distribution} of possible histograms in terms of their quantiles. 

\subsubsection{The Earth Mover's Distance (in 1D)}
A natural way of including cross-bin information is to consider a loss function that acts with respect to the \emph{cumulative} rather than the density histograms. In fact, it can be shown \cite{Ramdas2017} that in the 1D case with equally-sized bins and normalized histograms, the $l^1$ distance applied to the cumulative histogram is a special case of the Earth Mover's Distance (EMD) \cite{Rubner2000} in Transportation Theory: the EMD measures the amount of work required in order to transform one probability distribution (or histogram in the discrete case) into another when using the optimal transport plan. In statistics, this metric is known as the Wasserstein metric, Kantorovich--Rubinstein metric, or Mallows distance. While determining the optimal transport plan is generally a challenging task, the problem is substantially simplified in 1D, where the EMD between histograms $\tilde{\mathbf{u}}$ and $\mathbf{u}$ is simply given by
\begin{equation}
    \mathcal{L}_\text{EMD}(\tilde{\mathbf{u}}, \mathbf{u}) = \frac{1}{M} \sum_{j = 1}^M | \tilde{U}_j - U_j |,
    \label{eq:EMD}
\end{equation}
with $\tilde{U}_j = \sum_{m = 1}^j \tilde{u}_m$ and similarly for $U_j$. This implies that the NN loss grows as it places probability mass in bins further away from the true bin, and $\mathcal{L}_\text{EMD}(\tilde{\mathbf{u}}_2, \mathbf{u}) < \mathcal{L}_\text{EMD}(\tilde{\mathbf{u}}_1, \mathbf{u})$ in the example above. In particular, this means that when the NN estimate is far away from the truth, the gradient of the EMD does not vanish, unlike for distances such as the Kullback--Leibler divergence -- a fact that in the context of deep learning has been exploited in other applications, most prominently in Wasserstein GANs \cite{Arjovsky2017}. The (squared) EMD has also been proposed as a loss function for NN-based ordered classification such as age estimation with ordinal labels ``baby'', ``child'', and ``adult'' \cite{Hou2016}. For these problems, a ground distance needs to be specified (or learned), which sets the ``distance'' in the notion of ``work'' required to transport probability mass between classes (e.g., the distance between ``baby'' and ``child'' might be different from that between ``child'' and ``adult'').
However, for histogram data like in our case, the definition of the bins induces a natural distance when defining the EMD as in Eq.~\eqref{eq:EMD}: this formulation implicitly assumes an underlying ground distance $d_{ij} \propto |i - j|$ proportional to the absolute difference between the bin indices $i$ and $j$. Throughout this paper, we use flux bins that are uniformly spaced with respect to $\log_{10}(F)$; therefore, the work required for transporting probability mass is proportional to this quantity.

\subsubsection{Quantile regression with the pinball loss}
\par Rather than regressing a single ``average'' histogram, we are interested in the entire \emph{distribution} of possible histograms so that we can quantify the uncertainties. Therefore, we extend the EMD loss function by harnessing ideas from \emph{quantile regression} \cite{Koenker1978, Koenker2001}. Recall that just as the mean \emph{squared} ($l^2$) error is minimized by the \emph{mean}, the mean \emph{absolute} ($l^1$) error is minimized by the \emph{median} (or more precisely \emph{any} median, given that it does not need to be unique), i.e. for a real-valued random variable $Y$, the median solves $c^* = \operatorname{argmin}_{c} \, \mathbb{E}_Y \left[| c - Y | \right]$.
While the median is the $0.5$-quantile by definition, an analogous result can be obtained for arbitrary quantiles, where the $\tau$-th quantile of $Y$ is defined as
\begin{equation}
    Q_Y(\tau) = F^{-1}_Y(\tau) = \inf \{y : F_{Y}(y) \geq \tau \},
\end{equation}
with $F_{Y}(y)$ denoting the cumulative distribution function (CDF) of $Y$. Let $\tilde{y}$ be an approximation of the true quantile function $Q_Y(\tau)$. The pinball loss function \cite{Fox1964, Koenker1978, Koenker2001, ferguson2014mathematical} compares $\tilde{y}$ with observed values $y$ as
\begin{equation}
    \begin{aligned}
    \mathcal{L}^\tau_\text{pin}(\tilde{y}, y) &= (y - \tilde{y}) \left(\tau - \mathbb{I}\left[y < \tilde{y}\right]\right) \\ &=
    \begin{cases}
    \tau (y - \tilde{y}), & \text{if } y \geq \tilde{y}, \\
    (\tau - 1) (y - \tilde{y}), & \text{if } y < \tilde{y}.
    \end{cases}
    \end{aligned}
    \label{eq:pinball_loss}
\end{equation}
Here, $\mathbb{I}\left[C\right]$ is the indicator function, which is $1$ if the condition $C$ is true and $0$ otherwise. One can then show that the expected pinball loss function is minimized by the $\tau$-th quantile, i.e. $Q_Y(\tau)$ solves $c^* = \operatorname{argmin}_{c} \, \mathbb{E}_Y \left[\mathcal{L}^\tau_\text{pin}(c, Y) \right]$. In particular, for the median ($\tau = 0.5$), the pinball loss function is equivalent to the $l^1$ distance (up to the factor of $1/2$).

\subsubsection{Earth Mover's Pinball Loss}
We now combine the idea of the pinball loss in Eq.~\eqref{eq:pinball_loss} with the EMD in Eq.~\eqref{eq:EMD}. This yields the loss function presented in Ref.~\cite{List2021} that allows us to estimate \emph{arbitrary} quantiles of the \emph{cumulative} histogram in each bin $j \in \{1, \ldots, M\}$, given by
\begin{equation}
    \mathcal{L}_\text{EMPL}^\tau(\tilde{\mathbf{u}}, \mathbf{u}) = \frac{1}{M} \sum_{j=1}^{M} \left[ \left(\tilde{U}_j - U_j\right) \left(\tau - \mathbb{I}\left[\tilde{U}_j < U_j\right]\right) \right],
    \label{eq:EM_pinball_loss_discrete}
\end{equation}
where EMPL stands for \emph{Earth Mover's Pinball Loss}. Thus, for each map $\mathbf{x}$ and quantile level $\tau \in (0, 1)$, a NN $g^{\boldsymbol{\varpi}}$ trained using the EMPL provides an estimate of the $\tau$-th quantile of the cumulative histogram in each bin, conditional on the input $\mathbf{x}$:
\begin{equation}
   g^{\boldsymbol{\varpi}}(\mathbf{x}; \tau) = \tilde{Q}^{\boldsymbol{\varpi}}(\mathbf{x}; \tau) \approx Q_\mathbf{U}(\tau \, | \, \mathbf{x}),
\end{equation}
where $Q_\mathbf{U}(\tau \, | \, \mathbf{x}) = \left(Q_{U_1}(\tau \, | \, \mathbf{x}), \ldots, Q_{U_M}(\tau \, | \, \mathbf{x})\right) \in [0, 1]^{M}$ is the vector that gathers the quantiles of the true cumulative histogram $\mathbf{U} = (U_j)_{j=1}^M$ in all bins.
\par We simultaneously train our NN for arbitrary quantile levels $\tau \in (0, 1)$ by randomly drawing an individual value $\tau \sim U([0, 1])$ for each training map, which greatly reduces quantile crossing for scalar quantile regression as compared to training separate NNs for different quantile levels, as shown in Ref.~\cite{Tagasovska2018}. Since all the operations involved are (almost everywhere) differentiable with respect to the NN weights $\boldsymbol{\varpi}$, the weights can be optimized iteratively by following the negative gradient $-\partial \mathcal{L}_\text{EMPL}^\tau / \partial \boldsymbol{\varpi}$. In practice, we use a slightly smoothed version of the EMPL (see App.~\ref{sec:NN_architectures}). To ensure the monotonicity and the normalization of the histograms, i.e. $\tilde{U}_{j+1} \geq \tilde{U}_j$ and $\tilde{U}_M = 1$ for each fixed quantile level $\tau$, we proceed as follows: first, we estimate the \emph{density} histogram $\tilde{\mathbf{u}}$. In terms of $\tilde{\mathbf{u}}$, the normalization condition becomes $\sum_{j = 1}^M \tilde{u}_j = 1$, which we enforce using a ``normalized softplus'' activation function after the last layer (used in another context in Ref.~\cite{Shridhar2018}), given by
\begin{equation}
    \overline{\operatorname{softplus}}(\mathbf{a})_j = \frac{\ln\left(1 + \exp(a_j)\right)}{\sum_{m=1}^M \ln\left(1 + \exp(a_m)\right)}.
\label{eq:softplus_norm}
\end{equation} 
Note the similarity to the softmax activation function in Eq.~\eqref{eq:softmax} that we use for the normalization of the flux fractions. Indeed, both functions map $\mathbb{R}^M$ to the standard simplex $\Delta^{M - 1}$, and their limit behavior as $a_j \to -\infty$ is identical; however, the activation function in Eq.~\eqref{eq:softplus_norm} grows linearly for $a_j \to \infty$ rather than exponentially, which resulted in a more stable training and slightly improved accuracy in our experiments. The \emph{cumulative} histogram is obtained as the cumulative sum over the normalized density histogram (i.e., the softplus output), which is then used for the computation of the EMPL in Eq.~\eqref{eq:EM_pinball_loss_discrete}. The monotonicity of the quantiles \emph{within each bin} with respect to the quantile level $\tau$ is not strictly guaranteed, but it is strongly encouraged by the definition of the EMPL in Eq.~\eqref{eq:EM_pinball_loss_discrete}. We verified that quantile crossing by more than physically negligible relative fluxes $\ll 1\%$ rarely ever occurs in practice once the NN is trained. For a detailed description of the EMPL loss function and applications to other problems, we refer the interested reader to Ref.~\cite{List2021}.

\subsection{The combined framework}
To obtain the flux fractions as well as the SCDs of the PS populations, we combine the above two steps. In the first step, we train the NN $f^{\boldsymbol{\omega}}$ to estimate the flux fractions using the maximum likelihood loss function in Eq.~\eqref{eq:loss_step_1}. Once trained, we freeze the weights $\boldsymbol{\omega}$ and turn toward the estimation of the SCD in the second step. For the training of the second NN, $g^{\boldsymbol{\varpi}}$, we exploit the predictions of the first part and use a two-channel input, with the raw photon-count map $\mathbf{x}$ in the first channel and the residual $\mathbf{x}_\text{res}^{\boldsymbol{\omega}}$ after removing the estimated flux of the templates that we assume to be purely Poissonian (all but GCE and disk) as determined by $f^{\boldsymbol{\omega}}$ in the second channel. Thus, for perfectly correct flux fractions $f^{\boldsymbol{\omega}}$, the residual map $\mathbf{x}_\text{res}^{\boldsymbol{\omega}}$ would only contain photon counts from the (potentially) non-Poissonian templates plus Poisson scatter from the other templates. In our experiments, this additional residual channel led to a modest improvement in the NN accuracy. We train $g^{\boldsymbol{\varpi}}$ for the same number of batch iterations as $f^{\boldsymbol{\omega}}$ using the EMPL (Eq.~\eqref{eq:EM_pinball_loss_discrete}) and then freeze the weights $\boldsymbol{\varpi}$, yielding a trained ``double NN'' that produces estimates of flux fractions as well as the SCDs of the PS templates.

\section{Proof-of-concept example: isotropic point-source population}
\label{sec:toy}
\begin{figure*}
\centering
  \noindent
   \resizebox{1\textwidth}{!}{
    \includegraphics{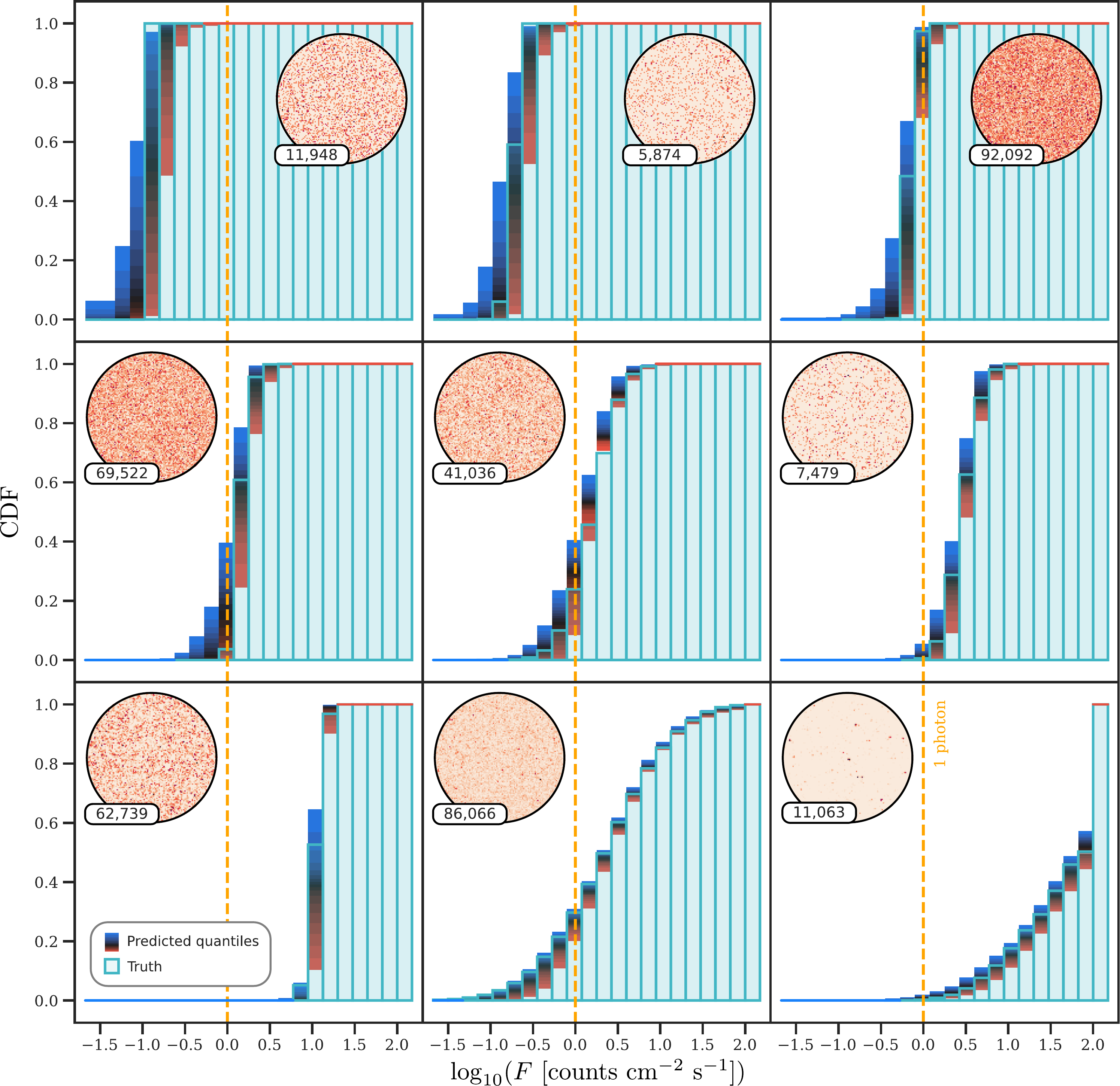}
    }
    \caption{NN predictions for 9 randomly selected maps from the test dataset for the isotropic proof-of-concept example: true cumulative $F \, dN / d (\log_{10} F)$ (light blue) and predicted quantiles (colored regions, 5 $-$ 95\% in steps of 5\%), sorted by the brightness of the PS population from very faint (top left) to very bright (bottom right). Specifically, the sorting criterion is the index where the true cumulative histogram $\mathbf{U}(\mathbf{x})$ surpasses $0.95$. The corresponding photon-count maps (i.e., the NN inputs $\mathbf{x}$) are shown in the inset plots, together with the total number of counts in the map. The colormap is normalized for each map, from $0$ to the maximum number of counts over all pixels. The flux range covers roughly three orders of magnitudes, with the faintest (brightest) PSs emitting on average $\sim0.1$ ($100$) counts. The flux associated with 1 expected count is indicated by the dashed orange line for orientation. From these results we see that the NN is able to accurately recover the true histogram in a wide variety of scenarios.}
    \label{fig:isotropic_results}
\end{figure*}
\begin{figure}
\centering
  \noindent
   \resizebox{1\columnwidth}{!}{
    \includegraphics{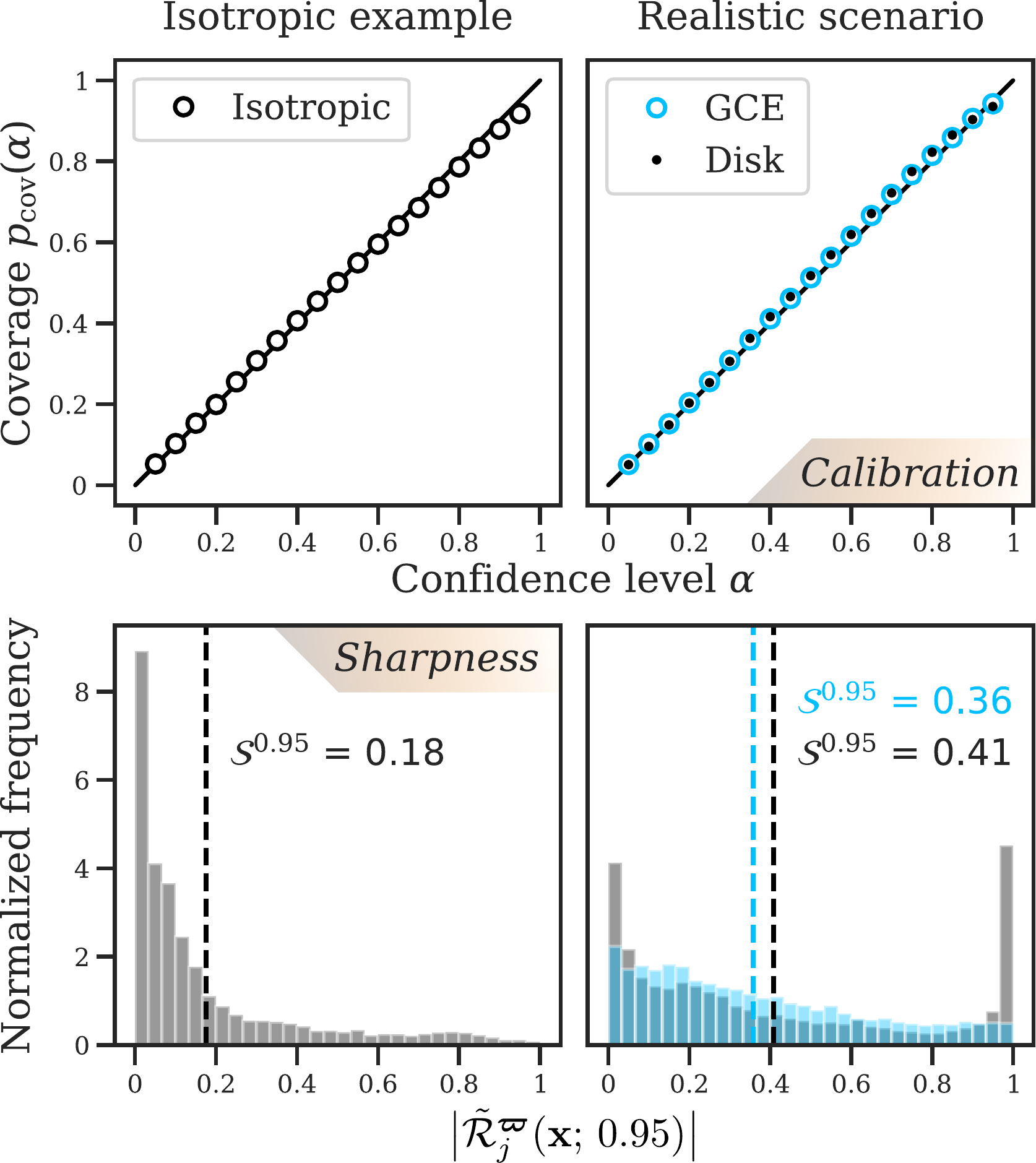}
    }
    \caption{Quantification of the uncertainty estimates of the NN $g^{\boldsymbol{\varpi}}$, for the isotropic proof-of-concept example considered in Sec.~\ref{sec:toy} (left) and the realistic scenario from Sec.~\ref{sec:fermi} (right). The upper panels show the average calibration of the uncertainties: for each confidence level $\alpha \in [0, 1]$, the coverage $p_\text{cov}(\alpha)$ is computed as the fraction of samples and bins for which the truth lies within the symmetric $\alpha$-interquantile range (IQR) around the median (see main text). Perfect calibration implies $p_\text{cov}(\alpha) = \alpha$, which is indicated by a solid line. In all the cases we consider, the uncertainties are well calibrated, which means that the uncertainties are approximately consistent with the errors in an average sense. The lower panels show how the size of the $95\%$-IQR is distributed. Here, it becomes apparent that the realistic scenario is much more difficult than the isotropic example, reflected by large uncertainties occurring more frequently. The dashed vertical lines are located at the mean size of $95\%$-IQR (average over maps and bins), which we define as the $95\%$-sharpness $\mathcal{S}^{0.95}$ (see Eq.~\eqref{eq:sharpness}). Very small and very large uncertainties are more common for the disk template than for the GCE in the realistic scenario.}
    \label{fig:UQ_plot}
\end{figure}
As a first test case for our SCD estimation method, we consider a simple scenario, where only a single isotropically distributed PS population is present (and Step 1 is therefore unnecessary). In this proof-of-concept example, we take the exposure to be 1 $\text{cm}^{2} ~\text{s}$ throughout our circular ROI, which is delimited by an outer radius of $25^\circ$ around the Galactic Center. Thus, the notions of flux $F$ and counts $S$, which are related via $F = S / E$ with the exposure $E$ in each pixel, are interchangeable in this example. We use a \texttt{HEALPix} resolution parameter of $n_\text{side} = 256$, corresponding to a pixel size of $13.7'$, and apply the \emph{Fermi} instrument PSF at $2$ GeV, modeled as the linear combination of two King functions.\footnote{For details of the \emph{Fermi} PSF, see \url{https://fermi.gsfc.nasa.gov/ssc/data/analysis/documentation/Cicerone/Cicerone_LAT_IRFs/IRF_PSF.html}.} Despite the fact that the standard deviation of the \emph{Fermi} PSF at this energy level is roughly twice the pixel size, training our NN with $n_\text{nside} = 256$ maps led to an improvement in accuracy over $n_\text{nside} = 128$ in our experiments, indicating that the NN is able to leverage information below the PSF scale.
\par We generate $1.5 \times 10^6$ maps and use $1.25 \times 10^6$ of them for training our CNN, while keeping the rest for testing. Throughout this work, when generating Monte Carlo (MC) data, we model $dN/dF$ as a skew normal distribution with respect to $\log_{10} F$, with randomly drawn parameters for location, scale, and skewness (see App.~\ref{sec:NN_priors}). In this example, our priors for the SCD result in the expected number of counts per PS to fall in the range $[0.1, 55]$ for the majority of PSs ($\sim95\%$). We take the total expected flux in the map to be uniformly distributed over $[1,$ 100,000$]$. For the discretization of the SCD, we take $M = 22$ bins, uniformly spaced in terms of $\log_{10} F$ from $\log_{10}\left(F \, / \, (\text{counts} \ \text{cm}^{-2} \ \text{s}^{-1})\right) = -1.5$ to $2$. The detailed NN architecture is provided in App.~\ref{sec:NN_architectures}. We train our CNN for 25,000 batch iterations at a batch size of $256$ on a single GPU on the supercomputer Gadi located in Canberra, which is part of the National Computational Infrastructure (NCI). We use an Adam optimizer \cite{Kingma2014} with learning rate $5 \times 10^{-4}$, which exponentially decays at a rate of $-1.5 \times 10^{-4}$ after each batch iteration.
\par Figure~\ref{fig:isotropic_results} shows the predictions of our CNN for 9 randomly selected maps from the test dataset that span a wide range of PS brightness, from a very faint PS population (top left) to a population with some very bright PSs (bottom right). We evaluate our CNN for quantile levels $\tau$ from $5\%$ to $95\%$ in steps of $5\%$, represented by the colored regions (from red to blue). The true cumulative $F \, dN / d (\log_{10} F)$ histograms are given by the light blue bars. The CNN has learned to recover the SCD of the underlying PS population, and the predicted histograms agree well with their true counterparts. Regressing the entire \emph{distribution} of possible histograms, expressed in terms of quantiles, allows us to draw conclusions about the uncertainties in the NN prediction. The quantile ranges at the low flux end of faint SCDs are generally large. For the first map, for instance, which contains PSs with $\ll 1$ count expected from each, the NN is uncertain about the exact brightness of the faintest PSs. Also, rather uniform PS populations with a steeply increasing CDF tend to produce higher uncertainties in the relevant bins than heterogeneous populations whose CDFs rise more gently over multiple flux magnitudes.
\par We now quantify the \emph{calibration} (or reliability) of our CNN on a more representative set of maps by means of a calibration plot. Specifically, we test how often the true value for the cumulative histogram in a given bin falls within the predicted quantiles -- ideally, we would expect that 90\% of true values would fall within our predicted 5 $-$ 95\% range. In detail, for every confidence level $\alpha \in [0, 1]$, we define the bin-averaged \emph{coverage probability} as 
\begin{equation}
    p_\text{cov}(\alpha) = \left\langle \frac{1}{|\mathcal{B}_\varepsilon(\mathbf{x})|} \sum_{j \in \mathcal{B}_\varepsilon(\mathbf{x})} \mathbb{I}\left[U_j(\mathbf{x}) \in \tilde{\mathcal{R}}^\varpi_j(\mathbf{x}; \, \alpha)\right] \right\rangle_{\mathbf{x}},
\end{equation}
where $\langle \cdot \rangle_\mathbf{x}$ denotes the average over samples and
\begin{equation}
    \tilde{\mathcal{R}}^\varpi_j(\mathbf{x}; \alpha) = \left[\tilde{Q}^{\boldsymbol{\varpi}}\left(\mathbf{x}; \frac{1 - \alpha}{2}\right), \tilde{Q}^{\boldsymbol{\varpi}}\left(\mathbf{x}; \frac{1 + \alpha}{2}\right)\right]
\end{equation}
is the predicted $\alpha$-interquantile range (IQR) symmetrically around the median. In the average over the bins, we exclude the bins in which the cumulative histogram is outside $[\varepsilon, 1 - \varepsilon]$ and only consider the subset 
\begin{equation}
    \mathcal{B}_\varepsilon(\mathbf{x}) = \left\{j \in \{1, \ldots, M\} \mid U_j(\mathbf{x}) \in [\varepsilon, 1 - \varepsilon] \right\}.
\end{equation}
This is to prevent bias arising from the bins where all the quantiles are very close to $0$ or $1$, and numerical inaccuracies far below the physically relevant magnitudes determine whether or not the true value lies within the estimated quantile range. We choose $\varepsilon = 10^{-5}$, but have confirmed that the results are not sensitive to the exact cutoff $\varepsilon$.
In other words, we compute the coverage probability as the fraction of bins for which the true cumulative histogram value $U_j$ falls within the predicted $\alpha$-IQR, averaged over a large number of maps. For perfectly calibrated quantiles, the coverage probability would be given by the identity $p_\text{cov}(\alpha) = \alpha$. Note that this notion of calibration thus assesses the \emph{average} reliability of the NN when evaluated on maps from the test dataset whose model parameters are randomly drawn from our priors.
\par Figure~\ref{fig:UQ_plot} (top left) shows the coverage probability $p_\text{cov}(\alpha)$ as a function of the confidence level $\alpha$, evaluated on 1,024 maps from the test dataset. For all confidence levels $\alpha \leq 0.65$, the deviation from perfect calibration is less than a percent, i.e. $|p_\text{cov}(\alpha) - \alpha| < 0.01$. For larger $\alpha$, the coverage lies slightly below the identity line, which means that our CNN on average underestimates the uncertainties; however, the deviations are small. The largest deviation among the considered confidence levels occurs at $p_\text{cov}(0.95) = 0.918$, implying there are $8.2\%$ outliers outside the $95\%$ confidence interval, while $5\%$ are expected. 
\par Whilst calibration is critical in order to avoid systematic biases, it is not sufficient to guarantee the usefulness of the estimates: for example, a NN that entirely ignores its input and always predicts the same true quantiles of the \emph{marginalized} distribution yields calibrated but quite useless predictions (e.g. Ref.~\cite{Kuleshov2018}, Fig.~4). An additional desideratum is therefore \emph{sharpness} of the uncertainties: for each uncertainty level $\alpha \in [0, 1]$, we define the $\alpha$-sharpness as the average size of the predicted $\alpha$-IQR, averaged over many maps and (relevant) bins:
\begin{equation}
    \mathcal{S}^{\alpha} = \left\langle \frac{1}{|\mathcal{B}_\varepsilon(\mathbf{x})|} \sum_{j \in \mathcal{B}_\varepsilon(\mathbf{x})} \big{|}\tilde{\mathcal{R}}^\varpi_j(\mathbf{x}; \, \alpha)\big{|} \right\rangle_{\mathbf{x}}.
    \label{eq:sharpness}
\end{equation}
Smaller values of $\mathcal{S}^{\alpha}$ indicate lower average uncertainties, as this corresponds to quantiles tightly grouped around the median prediction. In Fig.~\ref{fig:UQ_plot} (bottom left), we plot the distribution of $\big{|}\tilde{\mathcal{R}}^\varpi_j(\mathbf{x}; \, 0.95)\big{|}$ (the size of the predicted 95\%-IQR) over 1,024 test maps and the relevant bins $j \in \mathcal{B}_\varepsilon(\mathbf{x})$. A value of $1$ in this distribution means that at 95\% confidence, the value of the cumulative histogram in the respective bin cannot be confined to any proper subinterval of $[0, 1]$ by the NN. The dashed line indicates the mean of this distribution that defines the sharpness according to Eq.~\eqref{eq:sharpness}, which for this isotropic proof-of-concept example is given by $\mathcal{S}^{0.95} = 0.18$. The distribution of $\big{|}\tilde{\mathcal{R}}^\varpi_j(\mathbf{x}; \, 0.95)\big{|}$ is heavily right-skewed, and small uncertainties expressed by 95\%-IQRs $\lesssim 0.1$ occur frequently. The right-hand side in both rows of this figure quantifies the performance in a realistic scenario -- i.e. more representative of the actual \emph{Fermi} data -- that will be discussed in the following section.
\par Finally, we report the mean EMD between the median prediction and the true histogram over the 1,024 test maps (see Eq.~\eqref{eq:EMD}), given by $\mathcal{L}_\text{EMD} = 0.32$. This can be interpreted as the average amount of work required for transporting the median histogram to the truth in units of ``bins'' $\times$ ``probability mass'' (note that the total probability mass equals one because the histograms are normalized). For example, the EMD between the histograms $\mathbf{u} = [1, 0, \ldots, 0]$ and $\tilde{\mathbf{u}}_2 = [0, 1, 0, \ldots, 0]$ mentioned at the end of Sec.~\ref{sec:NN_method_step_2} is $1$ as the entire probability mass needs to be moved by one bin, namely from the second to the first. Converting from bins to flux, one finds that the mean EMD corresponds to a multiplicative factor of $1.14$ in flux space.
 
\begin{figure}
\centering
  \noindent
   \resizebox{0.9\columnwidth}{!}{
    \includegraphics{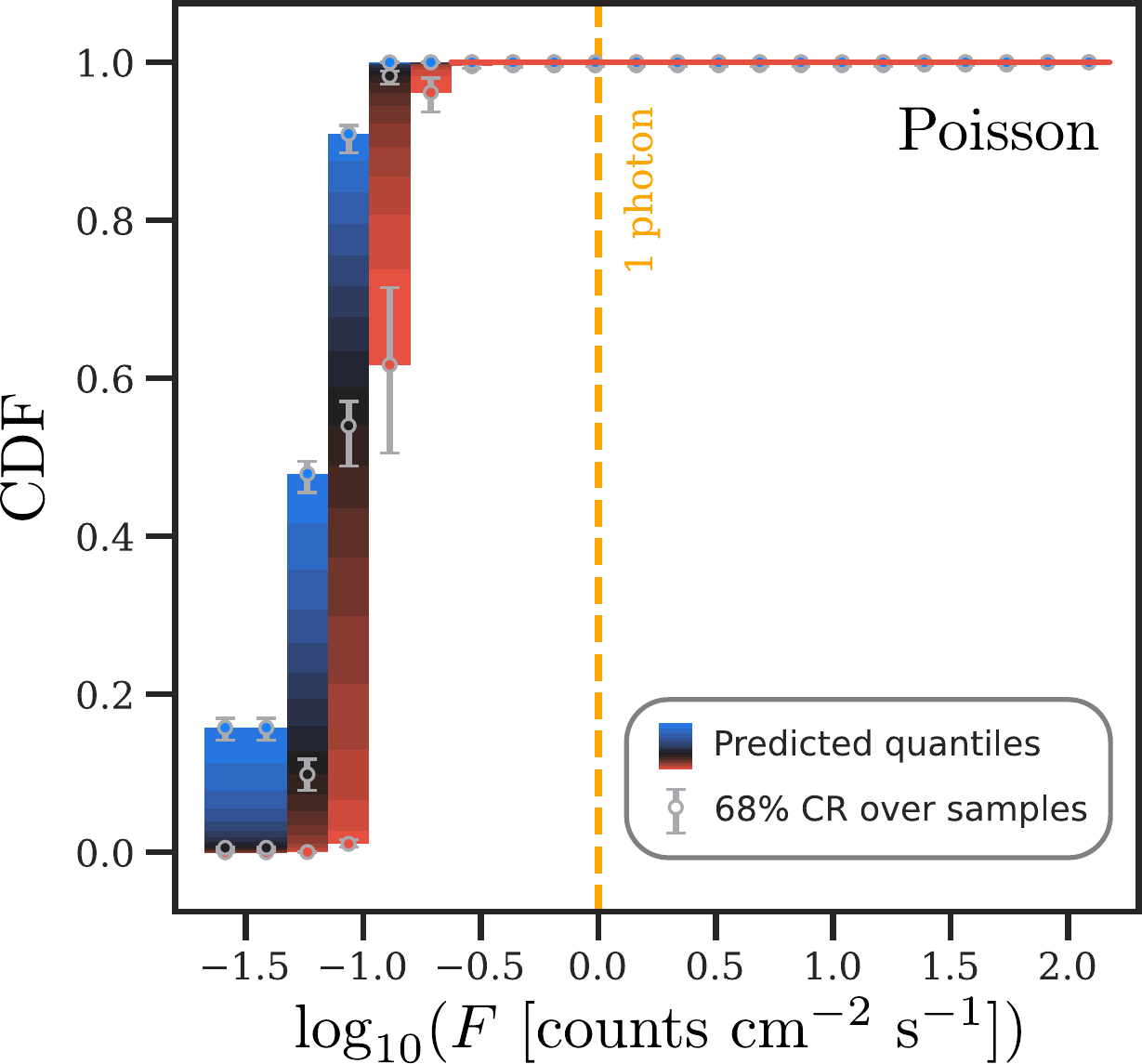}
    }
    \caption{
    NN prediction for the isotropic example when evaluated on purely Poissonian maps. The colored regions show the median of the different quantiles (5 $-$ 95\%) computed over 1,024 randomly generated maps. For the $5\%, 50\%$, and $95\%$ quantiles, the errorbars indicate the $68\%$ scatter over the maps. Whilst the NN has only seen (non-Poissonian) PS maps during its training, the variance of faint PSs only very slightly exceeds that of genuinely Poissonian emission and hence, the extrapolation effort required of the NN is small. As expected, the NN places the Poissonian flux far below the 1-photon line.}
    \label{fig:isotropic_poisson}
\end{figure}
\par Now, let us discuss how purely Poissonian emission is accommodated within our analysis framework. As already mentioned in the introduction, a central theme in this work is to describe Poissonian and PS-like emission associated with the same spatial template in a unified manner. (Note that we only apply this approach to emission components that are potentially PS-like; for purely Poissonian templates such as the diffuse foregrounds, we simply estimate the flux fraction as described in Sec.~\ref{sec:NN_method_step_1}.) Strictly speaking, annihilating DM can just as well be viewed as a huge collection of extremely faint PSs, where each PS corresponds to the location where a pair of DM particles annihilate. Clearly, modeling the resulting emission as Poissonian is justified, however, as the number of DM particles expected in each pixel is gargantuan for WIMP-like candidates. But even faint astrophysical PSs may strongly resemble Poisson emission: consider a population with an expected number of $N$ PSs, each of which produces $\bar{S}$ counts on average, such that the expected number of total counts is $\mu = N \bar{S}$. The variance of the counts for this population is given by $\sigma_\text{NP}^2 = N \bar{S} (1 + \bar{S}) = \mu (1 + \bar{S})$, compared with $\sigma_\text{P}^2 = \mu$ for Poisson emission with the same expected number of counts. Thus, $\sigma_\text{NP}^2 = (1 + \bar{S}) \, \sigma_\text{P}^2$, implying $\sigma_\text{NP}^2 > \sigma_\text{P}^2$ with $\sigma_\text{NP}^2 \to \sigma_\text{P}^2$ as $\bar{S} \to 0$. Hence, for the faintest populations considered in this example with $\sim0.1$ expected counts per PS, the variance exceeds that of Poisson emission only by $\sim10 \%$.\footnote{This argument ignores the PSF, which makes PS maps even smoother.} We can therefore expect our NN to locate the $F \, dN / d (\log_{10} F)$ at the very low flux end when applied to purely Poissonian maps -- even though \emph{truly} Poissonian maps were never shown to the NN during the training.
\par Figure~\ref{fig:isotropic_poisson} reveals that this is indeed the case: we plot the median prediction (same quantiles as in Fig.~\ref{fig:isotropic_results}) over 1,024 random Poissonian realizations with expected counts uniformly drawn from $[1$, 100,000$]$ as for the PS maps. For $\tau = 0.05, 0.5$, and $0.95$, the errorbars indicate the $68\%$ scatter over the samples. Compared to the prediction for the faintest PS map in Fig.~\ref{fig:isotropic_results} (top left), the estimated SCD for the Poissonian maps is even fainter, and the presence of PSs that emit more than $\approx 10^{-0.5} = 0.3$ expected counts is excluded at high confidence (see also Sec.~\ref{sec:constraining_Poisson}, where we consider how the Poissonian flux fraction can be constrained based on the estimated SCD histogram). Thus, it is justifiable to train our NN only with PS flux for the templates whose emission might be either smooth or PS-like -- provided that the dataset contains faint PS populations deep in the (partially) degenerate regime. Altogether, this experiment demonstrates that our CNN is able to accurately recover the underlying PS distribution as described by the SCD $F \, dN / d (\log_{10} F)$, and Poisson emission is placed at the low flux end far below the 1-photon line.

\section{Application to the \emph{Fermi} map}
\label{sec:fermi}
Now, we turn toward the realistic scenario, where we model all the components of the emission present in the inner Galaxy region of the \emph{Fermi} map. First, we describe the dataset that we use in this work and detail our modeling. Then, we briefly summarize the generation of training data and the NN training. Afterwards, we evaluate our CNN on simulated maps and finally present and discuss our results for the real \emph{Fermi} dataset.

\subsection{\emph{Fermi} data}
\label{sec:fermi_data}
\begin{figure*}
\centering
  \noindent
   \resizebox{1\textwidth}{!}{
    \includegraphics{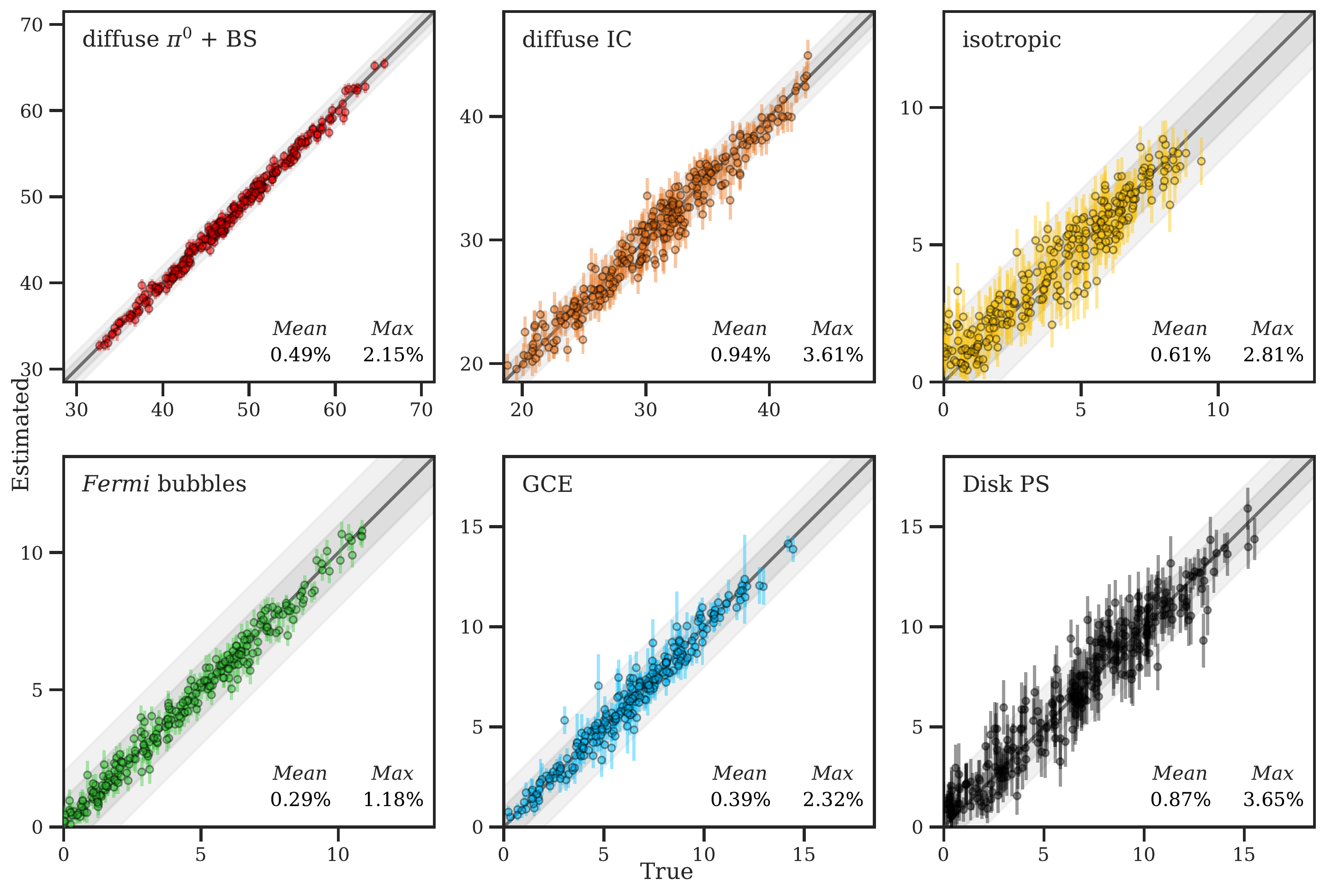}
    }
    \caption{
    True vs. estimated flux fractions produced by the NN $f^{\boldsymbol{\omega}}$ (in $\%$), for 256 randomly selected MC maps from the test dataset. Note that we zoom into the relevant flux region that arises from our flux priors for each template, for which reason the axes for the different templates have individual scales. The dark (light) gray stripes are included for orientation, and depict errors of $\pm$1\% (2\%). The inset values state the mean and maximum error over the maps for each template. For all the templates, the mean error lies below one percent. More accurate predictions generally come with smaller uncertainty estimates (compare e.g. the \emph{Fermi} bubbles to the disk). 
    }
    \label{fig:flux_fraction_errors}
\end{figure*}

\begin{figure*}
\centering
  \noindent
   \resizebox{1\textwidth}{!}{
    \includegraphics{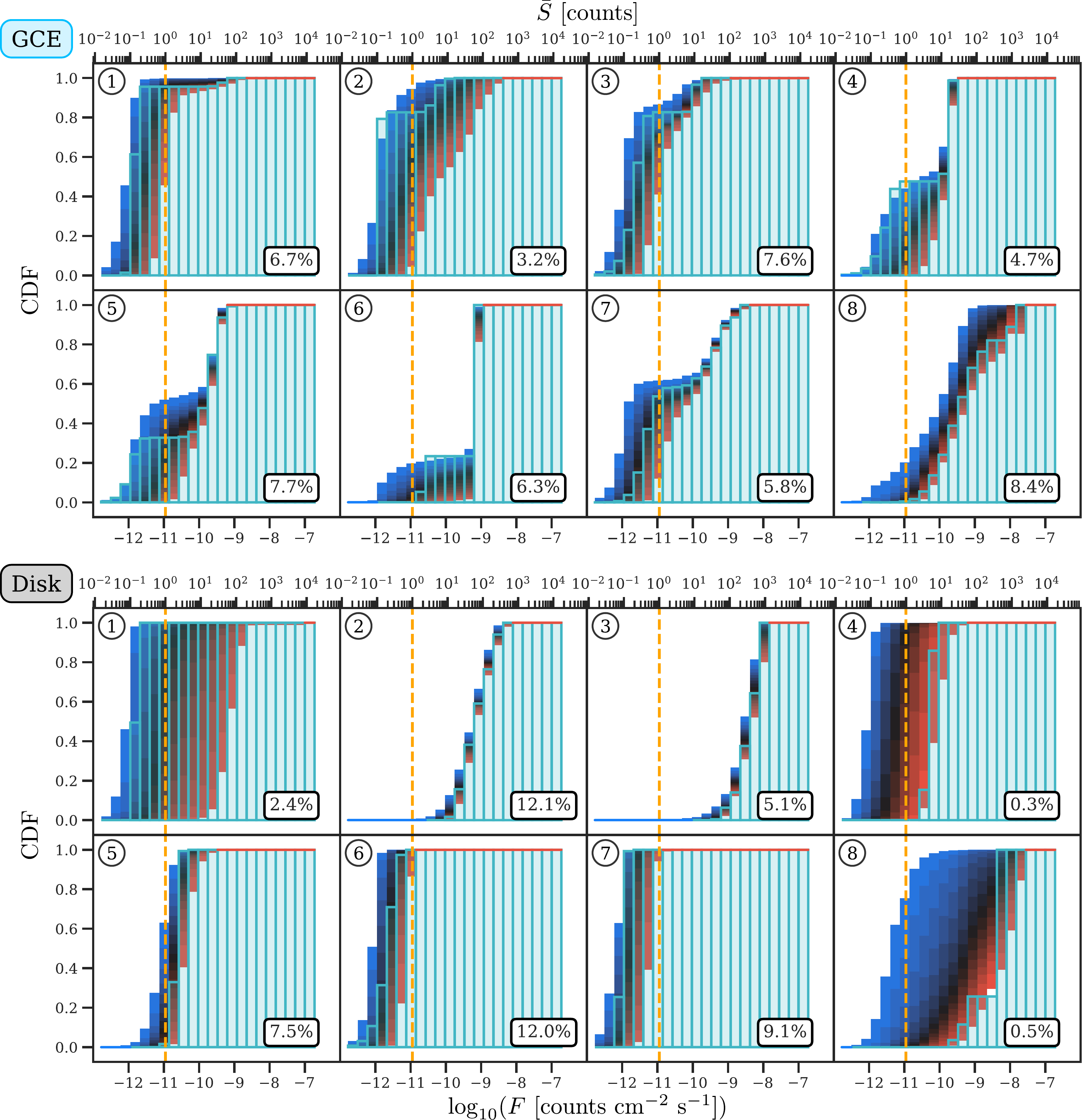}
    }
    \caption{Predictions of the NN $g^{\boldsymbol{\varpi}}$ for 8 randomly selected MC maps from the test dataset for the realistic scenario: true cumulative $F \, dN / d (\log_{10} F)$ (light blue) and predicted quantiles (colored regions, $5 - 95\%$ in steps of $5\%$), sorted by the brightness of the GCE PS population (top), from very faint to very bright.
    The disk PS histogram (true and predicted) for each of the maps is plotted below. The percentages in the lower right corners state the flux fraction of the respective template (GCE / disk). For some of the maps, the two distinct GCE populations can be clearly identified in the histograms. The NN has learned to distill the two PS populations from the smooth background emission and to recover the underlying SCDs, although sharp kinks are sometimes slightly smoothed out. The histograms for the disk PSs reveal the correlation between larger flux fractions and sharper uncertainty estimates (e.g., compare the uncertainties for maps 1 and 2 with 2.4\% and 12.1\% disk PS flux, respectively). 
    } 
    \label{fig:simulated_hist_plot}
\end{figure*}

To construct our data, we begin with all photons collected by \emph{Fermi} in the PASS 8 dataset between 4 August 2008 and 19 June 2019, which corresponds to almost 11 years of data.
To minimize background contamination from charged cosmic-rays, we use events in the \texttt{UltracleanVeto} class. 
Further, to reduce the diffuse background to PS searches, we keep only the top quartile of $\gamma$-rays as graded by the quality of reconstruction of their incident direction.\footnote{We remark that the recent works \cite{Leane2020, Leane2020a} considered the three best-graded quartiles, rather than only the top one. Whilst this leads to three times more photon counts, it also increases the radius of the PSF. We leave a comparative study of different data selection criteria to future work.}
Finally, to ensure we only consider data that was collected during good time intervals, when the instrument was operating in science configuration, and that is uncontaminated by emission from the Earth's limb, we apply the conventional quality cuts \texttt{DATA\_QUAL==1}, \texttt{LAT\_CONFIG==1}, and zenith angle $< 90^{\circ}$, respectively.

After applying these criteria, we are left with a list of photons labeled by two angles corresponding to their reconstructed origin on the celestial sphere, and their reconstructed energy.
We remove the energy information by combining the data into a single bin of events between 2 and 20 GeV, in order to capture the region where the GCE is expected to peak over backgrounds.
As for the isotropic example in Sec.~\ref{sec:toy}, we bin the resulting list of photons into \texttt{HEALPix}-discretized input maps at a resolution of $n_\text{side} = 256$. In our experiments, we did not achieve substantial improvements by increasing the resolution to $n_\text{side} = 512$. However, it might be possible to exploit the additional information contained in higher resolution maps by using more complex NN architectures (see e.g. Ref.~\cite{He2016}). We leave an in-depth study in this direction to future work. We consider a circular ROI of radius of $25^\circ$ around the Galactic Center, and then mask the inner $|b| \leq 2^\circ$ around the Galactic Plane as well as the pixels that are within the 95\% containment radius at 2 GeV ($\approx 0.47^{\circ}$ for these quality cuts) of any source in the 3FGL catalog.\footnote{In more detail, to construct the PS mask we start with an $n_\text{side} = $ 2,048 map, and mask any pixel with center within 95\% containment radius of a source.
This map is then downgraded to $n_\text{side} = 256$, and if more than half of the parent pixels were masked, we mask the pixel in the lower resolution map.}

\subsection{Flux templates}
In line with previous analyses (e.g. Refs.~\cite{Lee2016, Leane2019a, Buschmann2020, Leane2020, Leane2020a}), we include templates modeling the following physical processes: (1) Galactic diffuse foregrounds from decay of neutral pions ($\pi^0$) together with bremsstrahlung (BS), both of which originate from the interaction of cosmic rays with the interstellar gas, for cosmic-ray protons and electrons, respectively, (2) Galactic diffuse foregrounds from photons of the interstellar radiation field, which are up-scattered by cosmic-ray electrons to $\gamma$-ray energies via the inverse Compton (IC) effect, (3) extragalactic emission, described by a spatially uniform template, (4) the \emph{Fermi} bubbles \cite{Su2010}, a large-scale structure in the $\gamma$-rays stretching to the north and south of the Galactic Plane, (5) emission from PSs associated with the Galactic Disk, which we model with a doubly-exponential disk with scale height $z_s = 0.3 \ \text{kpc}$ and scale radius $R_s = 5 \ \text{kpc}$, and (6) a template for the GCE, given by the line-of-sight integral of a squared generalized NFW profile \cite{Navarro1997} with slope parameter $\gamma = 1.2$. Further, we assume that templates 1 $-$ 4 are purely Poissonian; i.e., isotropic PSs are not included as their impact has been found to be very small~\cite{Buschmann2020}, nor do we consider hypothetical PSs associated with the \emph{Fermi} bubbles as evoked in a proof-of-concept example in \textcite{Leane2019a}. Templates 5 and 6 are hence the only PS-like templates used in our analysis. For the diffuse Galactic foreground emission, we choose Model~O, which was introduced in \textcite{Buschmann2020} (building on Refs.~\cite{Macias:2016nev, Macias:2019omb}), and provides a much better fit at low energies as compared to the official \emph{Fermi} model \texttt{p6v11} (see e.g. Fig.~17 in Ref.~\cite{Buschmann2020}). As we include more data than considered in Ref.~\cite{Buschmann2020} and \citetalias{List2020b}, we refit the components used to construct Model~O to our maps, using the same procedure described in the former work.

\subsection{Data generation and neural network training}
For training and testing our NN, we generate $1.5 \times 10^6$ maps in total, $10^5$ of which we set aside for testing while using the remaining $1.4 \times 10^6$ maps for training $f^{\boldsymbol{\omega}}$ and $g^{\boldsymbol{\varpi}}$. For the four Poissonian templates, the counts in each map are drawn from a Poissonian distribution with pixel means given by the product of the template normalization $A$ and the respective spatial template. Whilst we chose wide priors in the main body of \citetalias{List2020b} to present CNNs as a general template fitting method for $\gamma$-ray maps, our priors for the template normalizations cover a much tighter range around the expected values for the \emph{Fermi} map in this work, so as to maximize the performance in this region of the parameter space. The exact prior ranges are tabulated in App.~\ref{sec:NN_priors}. For the PS templates, we take the SCD functions $dN/dF$ to be skew normal distributions, whose parameters for location, scale, and skewness are randomly drawn. For each map, the PSs are distributed across the map in accordance with the spatial template, a Poisson draw is performed for each PS to determine the number of counts, and the \emph{Fermi} PSF correction is applied. In order to allow for more complex SCDs and, more importantly, to include maps with both a bright and a very faint GCE population that together model a mixed PS + (nearly) Poissonian GCE, we generate twice as many template maps for the GCE ($3 \times 10^6$) and add them pairwise such that each combined count map contains \emph{two} individual GCE populations. For the disk PSs, we assume a single population. Our more flexible modeling for the SCD of the GCE could lead to comparably more robust results for the GCE than those for the disk -- justifiably given the GCE is our primary concern -- however, further improvement of the disk modeling would be an interesting future direction. The labels for each map are given by the flux fractions of each template for $f^{\boldsymbol{\omega}}$ and by the discretized (relative) $F \, dN / d (\log_{10} F)$ for $g^{\boldsymbol{\varpi}}$, where the bin edges range from $\log_{10}\left(F \, / \, (\text{counts} \ \text{cm}^{-2} \ \text{s}^{-1})\right) = -12.5$ to $-7$ in steps of $0.275$, resulting in 22 equally-spaced flux bins with respect to $\log_{10}(F)$.
\par We train our NN using the two-step procedure outlined in Sec.~\ref{sec:NN_method} for the two NN parts $f^{\boldsymbol{\omega}}$ and $g^{\boldsymbol{\varpi}}$, both times minimizing the respective loss function for 30,000 batch iterations at batch size $256$. For both steps, we use an Adam optimizer with the same hyperparameters as in the isotropic example, resetting the learning rate to its original value before starting the training of $g^{\boldsymbol{\varpi}}$. 

\subsection{Results for simulated data}
\label{sec:fermi_example_results_simulated}
First, we discuss the flux fraction estimation using the NN $f^{\boldsymbol{\omega}}$ (Step 1). We evaluate our trained NN on 256 randomly selected maps from the test dataset. The true vs. estimated flux fractions for these maps are plotted in Fig.~\ref{fig:flux_fraction_errors} (in \%), zoomed into the relevant range for each template. For orientation, the dark (light) gray bands delimit errors of $\pm$1\% (2\%). Compared to the NN errors for the realistic scenario in \citetalias{List2020b}, the NN errors are generally smaller, which can be explained by a combination of (1) the fact that GCE DM and PS are modeled by a \emph{joint} template, (2) more training data, (3) the higher data resolution ($n_\text{side} = 256$ instead of $128$), (4) narrower prior ranges (except for Fig.~S26 in \citetalias{List2020b}, where we also used narrow priors around the \emph{Fermi} values), and (5) we consider a fixed ROI radius of $25^\circ$ in this work instead of varying between $15 - 25^\circ$. On the other hand, the SCD of the GCE PSs is more complex now as the GCE PS counts are the sum of two individual template maps. For all the templates, our NN recovers the flux fractions on average well within percent accuracy. In particular, for the GCE template, the mean error is $< 0.5\%$. Large errors are generally accompanied by large uncertainties, suggesting that the NN recognizes which maps and templates are difficult to predict. The flux fraction predictions are least accurate for the diffuse IC and disk PS templates: both templates have smooth emission that is correlated with the disk of the Milky Way, for which reason there might be confusion between faint disk PSs (which, recall, are indistinguishable from Poisson emission) and diffuse IC emission. In Fig.~S20 in \citetalias{List2020b}, where we considered a full uncertainty covariance matrix, this is reflected by a large negative correlation between the flux fractions of these two templates (Pearson correlation coefficient $r = -0.3$).
\par Now, we consider the SCD prediction with the NN component $g^{\boldsymbol{\varpi}}$ (Step 2). Figure~\ref{fig:simulated_hist_plot} shows the true cumulative SCDs (GCE and disk) for 8 randomly selected maps from our test dataset, together with the NN estimates. As compared to the isotropic proof-of-concept example, the SCD estimation becomes considerably more difficult now as GCE PSs and disk PSs each only make up $\sim0 - 15$\% of the counts in the map. Nonetheless, the NN has learned to provide accurate uncertainty regions for the SCDs of both templates that trace the true histograms. As the flux fraction of a PS template (given in the lower right corner) approaches zero, the uncertainties for the associated SCD diverge, indicating that the NN becomes aware that tight constraints on the SCD can no longer be derived in this situation. The GCE histograms typically have more complex shapes than those for the disk due to the two distinct GCE populations present in each map, which is generally well reproduced by the NN (see, e.g., the varying slopes of the histograms for maps 1 and 7).
\par As in the isotropic example, we analyze the calibration and the sharpness of the uncertainty estimates based on 1,024 randomly selected test maps, as shown in Fig.~\ref{fig:UQ_plot} on the right-hand side. Also for the realistic scenario, the uncertainties are very well calibrated for both PS templates. Rather than causing overconfident or underconfident predictions that would be reflected by large deviations from the identity line in the calibration plot, the increased difficulty of the problem affects the \emph{sharpness} of the uncertainties: the sharpness with respect to the $95\%$-IQR increases from $\mathcal{S}^{0.95} = 0.18$ in the isotropic case to $0.36$ and $0.41$ for GCE and disk PSs, respectively. Interestingly, the distribution of $|\tilde{\mathcal{R}}_j^{\boldsymbol{\varpi}}(\mathbf{x}; 0.95)|$ for the disk PS template is bimodal and peaks at zero and one, whereas it decreases roughly monotonically for the GCE PS template. This difference in behavior between the PS models can be traced to the fact that each map contains two GCE PS template maps, but only one for the disk. Accordingly, the disk SCD will be unimodal, whereas for the GCE the PSs will typically be associated with a wider distribution in flux (see Fig.~\ref{fig:simulated_hist_plot}). The bimodal distribution of $|\tilde{\mathcal{R}}_j^{\boldsymbol{\varpi}}(\mathbf{x}; 0.95)|$ for the disk is then associated with the lowest flux bins: if the disk PSs are bright, then the NN can be confident there are no low flux sources (as it was trained on a unimodal SCD), whereas if the disk sources are dim, then determining the exact peak of the distribution is challenging, resulting in large uncertainties. Note that another consequence of the different treatment of the two PS templates in the generation of the maps is that the distribution of the total flux of the PS templates over the maps follows a triangular distribution for the GCE, but a uniform distribution for the disk. However, we confirmed this difference is not a significant driver in the different shapes of the $|\tilde{\mathcal{R}}_j^{\boldsymbol{\varpi}}(\mathbf{x}; 0.95)|$ distribution between the two models: when restricting the testing dataset to maps in which the respective template has a flux fraction $\geq 5\%$, the distribution of the $95\%$-IQR size for the disk PS template remains bimodal, although the height of the peak at one is reduced, as the disk SCD can be determined more accurately in maps where disk PSs contribute more total flux.
\par We emphasize that even in the case of large uncertainties within one or multiple bins, it can be possible to obtain tight constraints on the SCD: for example, if all the quantiles of the predicted cumulative histogram are identically zero in bins $\leq j - 1$ and $1$ in bins $\geq j + 1$ (assuming the NN estimate is correct, all these bins are excluded from the set $B_\varepsilon$ and are hence not considered in our computation of the sharpness), but span the entire possible range $[0, 1]$ in bin $j$, we have $|\tilde{\mathcal{R}}_j^{\boldsymbol{\varpi}}(\mathbf{x}; 0.95)| = 1$; however, we know that the SCD can be non-zero only in bins $j$ and $j + 1$. 
\par The mean EMD between the predicted median and the true SCD histogram is now $\mathcal{L}_\text{EMD} = 0.90$ and $0.99$ for GCE and disk. We remark that these values are affected by maps where the flux fraction of the respective template is very small and the median SCD lies several bins away from the truth -- which the NN accounts for by producing uncertainties that span multiple orders of magnitudes in terms of flux (e.g. for the disk PSs in maps 1 and 8 in Fig.~\ref{fig:simulated_hist_plot}). Therefore, we also quote the median EMD, which is more representative of a typical map, given by $\mathcal{L}_\text{EMD} = 0.71$ and $0.56$ for GCE and disk, respectively, yielding multiplicative factors of $1.58$ and $1.42$ in terms of flux.

\subsection{Results for the \emph{Fermi} map}
\label{sec:results_for_fermi_map}
\begin{figure*}
\centering
  \noindent
   \resizebox{1\textwidth}{!}{
    \includegraphics{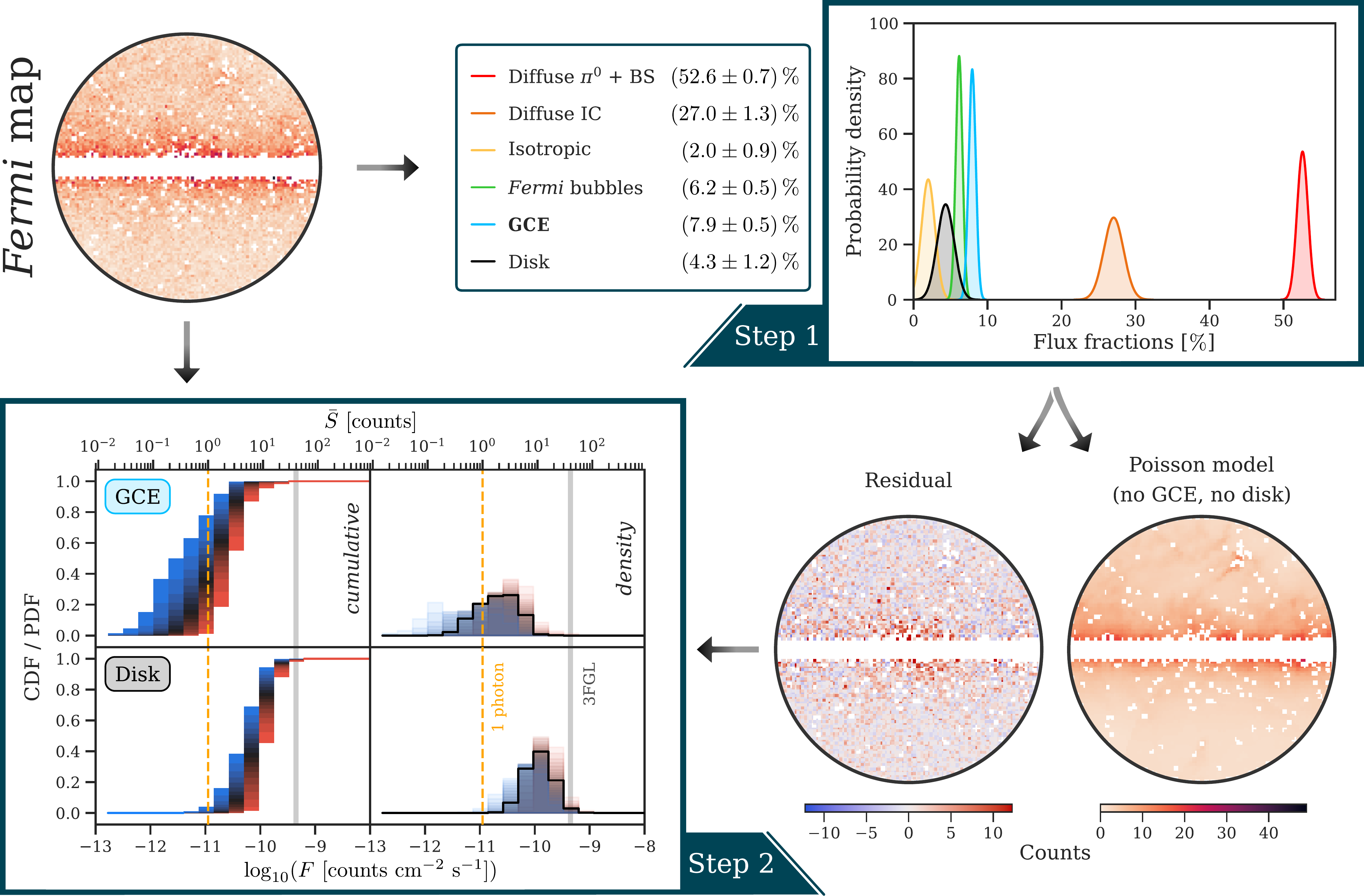}
    }
    \caption{Results for the \emph{Fermi} map. The flux fraction estimates of $f^{\boldsymbol{\omega}}$ in Step 1 are shown in the upper right panel. Our NN identifies $(7.9 \pm 0.5)$\% GCE emission within our ROI ($25^\circ$ around the Galactic Center, $|b| \leq 2^\circ$ and 3FGL sources masked). The resulting Poisson model, which accounts for all the templates except for GCE and disk PS, as well as the residual after subtracting the model from the \emph{Fermi} map are depicted on the lower right, where we use identical colormap limits for the counts in the \emph{Fermi} map and the Poisson model. This residual map, together with the original \emph{Fermi} map, form the input for $g^{\boldsymbol{\varpi}}$, which predicts the (normalized) SCD function $F \, dN / d (\log_{10} F)$ in Step 2 (lower left). We plot the cumulative histograms and the corresponding density histograms for the GCE and the disk PSs, where the colors illustrate the estimated $5 - 95\%$ quantiles in steps of $5\%$ (from red to blue). In the density histogram axes, the black lines show the median predictions. The gray vertical bars mark the location of the 3FGL threshold at $F \approx (4 - 5) \times 10^{-10} \ \text{counts} \ \text{cm}^{-2} \ \text{s}^{-1}$, above which PSs can be expected to be individually resolved. The upper $x$-axis indicates the expected number of counts $\bar{S}$ associated with the logarithm of the flux $F$ on the lower $x$-axis. The predicted GCE median histogram peaks at $\bar{S} \approx 3 - 4$ counts and ranges below the 1-photon line, with substantial uncertainty at the lower end. Nearly the entire GCE flux is attributed to PSs emitting less than $10$ counts. A much brighter SCD is preferred by the NN for the disk PSs, which is roughly delineated by the 1-photon line and the 3FGL threshold at the faint and bright end, respectively.}
    \label{fig:fermi_results}
\end{figure*}

Having confirmed that our method produces reliable estimates for both the flux fractions and the SCDs for simulated \emph{Fermi}-like photon-count maps, we now evaluate our NNs on the real \emph{Fermi} map (again, we refer to Sec.~\ref{sec:fermi_data} for the specific dataset considered in this work). 
\par In Fig.~\ref{fig:fermi_results}, we present our results for the \emph{Fermi} map (shown in the upper left corner within our ROI). The NN $f^{\boldsymbol{\omega}}$ assigns $(7.9 \pm 0.5)\%$ of the flux to the GCE template. Generally, the flux fraction estimates are similar to our findings in \citetalias{List2020b} (note that work used $\sim 8$ years of \emph{Fermi} data, whereas here we use $\sim 11$ years) and consistent with those of the NPTF implementation \texttt{NPTFit} in the same ROI (see App.~\ref{sec:comparison_nptfit}). Based on the estimated flux fractions for the purely Poissonian templates (all but GCE and disk), the best-fit Poisson model is determined, and the residual count map is provided as an input to the NN $g^{\boldsymbol{\varpi}}$ alongside the original \emph{Fermi} map for the SCD estimation in Step 2. The GCE is visible near the Galactic Center in the residual map. The resulting SCD estimates for the GCE and the disk are plotted in the lower left corner, where the different colors again correspond to quantile levels from $\tau = 0.05$ to $0.95$ in steps of $0.05$. We show the cumulative histograms on the left and the density histograms on the right, where the solid black lines mark the median predictions. The NN places $72$\% of the GCE flux in the three bins corresponding to a flux of $F = (0.8 - 5.0) \times 10^{-11} \ \text{counts} \ \text{cm}^{-2} \ \text{s}^{-1}$ (or equivalently $\bar{S} = 0.7 - 4.5$ expected counts) for the median prediction, and less than 1\% ($\approx 13\%$ for a quantile level of $\tau = 0.05$) is assigned to PSs brighter than $F = 9.4 \times 10^{-11}  \ \text{counts} \ \text{cm}^{-2} \ \text{s}^{-1}$ (or $\bar{S} \geq 8.4$ expected counts). Below the 1-photon line, there is substantial uncertainty and for $\tau \gtrsim 0.9$ (i.e., with an expected probability of $\sim10$\%), more than half of the GCE flux is attributed to PSs that on average even contribute less than $\sim1$ count to the \emph{Fermi} map. Qualitatively, the SCD predicted by our NN provides no indication of two distinct GCE components present in the \emph{Fermi} map such as e.g. a Poissonian and a PS component.
\par We now put our SCD estimate for the GCE into the context of previous NPTF-based studies: as pointed out in Ref.~\cite{Leane2020a}, most NPTF analyses identify a rather steep SCD for the GCE population, implying that most of the GCE flux would originate from sources close to the break $F_b$ of the broken power-law that commonly describes the SCD in NPTF analyses. For a more extensive discussion of SCDs found in NPTF analyses of the GCE than the one presented here, we refer to Ref.~\cite[Sec. VII A 1]{Leane2020a}. Using the \emph{Fermi} diffuse model \texttt{p6v11}, the first analysis of the GCE with NPTF conducted by \textcite{Lee2016} reported a value of $F_b = 1.76^{+0.44}_{-0.35} \left(1.62^{+0.45}_{-0.32}\right) \times 10^{-10} \ \text{counts} \ \text{cm}^{-2} \ \text{s}^{-1}$ for their analysis with unmasked (masked) 3FGL sources, and Ref.~\cite{Leane2019a} found $F_b = 1.94^{+0.34}_{-0.30} \times 10^{-10} \ \text{counts} \ \text{cm}^{-2} \ \text{s}^{-1}$ in their masked analysis within a $30^\circ$ radius. (Note that the first of these analyses used a different energy range than considered in the present work, although the difference due to this should be smaller than the other uncertainties on inferring properties of the SCD.) An unmasked analysis by Ref.~\cite{Leane2019a} with Model~A identified a lower value of $F_b = 1.07^{+0.20}_{-0.16} \times 10^{-10} \ \text{counts} \ \text{cm}^{-2} \ \text{s}^{-1}$ (which is still twice as large as the peak of the GCE SCD preferred by our NN with Model~O). Ref.~\cite{Leane2020a} obtained $F_b = 7.9^{+1.5}_{-1.3} \times 10^{-11} \ \text{counts} \ \text{cm}^{-2} \ \text{s}^{-1}$ in their baseline analysis with a narrow prior range of $[2.05, 5]$ for the negative slope of the SCD above the break ($n_1$), which prevents a sharp cutoff. Replacing \texttt{p6v11} by Model~A in their analysis further reduced the value to $F_b = 4.9 \times 10^{-11} \ \text{counts} \ \text{cm}^{-2} \ \text{s}^{-1}$, whereas other variations in their analysis (such as taking a $30^\circ$ radius ROI instead of their default choice of $10^\circ$) gave rise to larger values of $F_b \sim 1 \times 10^{-10} \ \text{counts} \ \text{cm}^{-2} \ \text{s}^{-1}$.
Model~O was used in an analysis by Ref.~\cite{Buschmann2020}, however, in that work the SCDs were subject to a sharp cutoff at lower fluxes as a partial attempt to mitigate Poisson and PS confusion, which makes any comparison to their results less meaningful.
\par More generally, an important difference between the SCD derived in the present work and those that have been obtained previously is that our SCD describes the full emission of the GCE. The results from earlier works were derived only for the PS contribution -- a separate model was included for the Poissonian contributions. Given the previously discussed inherent ambiguity between PS and diffuse contributions, results obtained through these different methods cannot be compared unambiguously. Accordingly, as a cross-check, we also perform a fit of the \emph{Fermi} map with \texttt{NPTFit} in our ROI, taking the same templates as in our NN training and omitting a Poissonian GCE ``DM'' template, such that faint GCE flux is expected to affect the lower end of the predicted GCE SCD, similarly to our NN-based approach. As in previous studies (e.g. Refs.~\cite{Lee2016, Leane2020, Leane2020a}), we model the SCD in \texttt{NPTFit} with a singly-broken power law. Our priors allow steep negative slopes up to $n_1 = 30$ for the GCE and disk SCDs above the flux break (see App.~\ref{sec:comparison_nptfit} for additional details).
Intriguingly, we find a best-fit estimate for the flux break of $F_b = 5.0 \times 10^{-11} \ \text{counts} \ \text{cm}^{-2} \ \text{s}^{-1}$ ($\bar{S} = 4.5$ expected counts), which is similar to the peak of the SCD favored by our NN (although \texttt{NPTFit} prefers a much narrower shape). On the other hand, when repeating the same \texttt{NPTFit} analysis with \texttt{p6v11} in place of Model~O, we obtain a much brighter SCD for the GCE with best-fit flux break $F_b = 1.5 \times 10^{-10} \ \text{counts} \ \text{cm}^{-2} \ \text{s}^{-1}$ ($\bar{S} = 13.5$ expected counts), consistent with previous \texttt{p6v11}-based studies. Thus, modeling the Galactic foregrounds with the Model~O instead of \texttt{p6v11} appears to shift the preferred GCE SCD to considerably fainter fluxes with \texttt{NPTFit}. Importantly, the faint peaks of the GCE SCDs obtained in our Model~O-based analyses with the NN and \texttt{NPTFit} are much more similar than the peaks arising from different \texttt{NPTFit} analyses that use different diffuse templates. We reiterate that Model~O has been found to give a considerably better fit to the \emph{Fermi} map than \texttt{p6v11} \cite{Buschmann2020}. 
\par It has already been noted in earlier studies that the choice of the diffuse template may bias the inferred SCD and even affect the preference for a Poissonian vs. PS-like GCE: within $10^\circ$, Ref.~\cite{Leane2020} found that model \texttt{p6v11} leads to an overwhelmingly large Bayes factor of $4 \times 10^{15}$ in favor of a PS-like GCE, whereas the purely \texttt{Galprop}-based~\cite{Strong:1998pw} Model F yields a Bayes factor of only $1$, indicating no preference for PS-like emission (and allowing separate template normalizations $A$ for the two hemispheres weakens the evidence to a Bayes factor $< 10$ even with \texttt{p6v11}). We study the impact of different sources of mismodeling on the predictions of $f^{\boldsymbol{\omega}}$ and $g^{\boldsymbol{\varpi}}$ in Sec.~\ref{sec:mismodeling_experiment}. 
\par Whilst we will address the question as to what constraints on the Poissonian GCE flux can be derived based on the estimated SCD in Sec.~\ref{sec:constraining_Poisson}, let us already comment on the results we obtained in \citetalias{List2020b} treating GCE PS (non-Poissonian) and GCE DM (Poissonian) as two separate templates: there, our NN found $(8.6 \pm 1.7)\%$ and $(0.3 \pm 1.2)\%$ flux of GCE DM and PSs, respectively. As we showed in Fig.~S4 in the Supplementary Material of \citetalias{List2020b}, confusion between DM and (very) dim PSs is common even when PSs make up the entire GCE (see Ref.~\cite{Chang2019} for an assessment of DM / PS misattribution with \texttt{NPTFit}). In light of the SCD predicted by $g^{\boldsymbol{\varpi}}$ assigning the bulk of the GCE flux to PSs with $< 5$ expected counts, the preference of our simpler NN in \citetalias{List2020b} for a Poissonian GCE would still be comprehensible even if the GCE were fully explained by PSs that follow this SCD without any Poissonian contribution.
\par For the disk PSs, $g^{\boldsymbol{\varpi}}$ prefers a brighter SCD framed by the 1-photon line and the 3FGL threshold on either side, which peaks at a flux of $F = 1.1 \times 10^{-10} \ \text{counts} \ \text{cm}^{-2} \ \text{s}^{-1}$ ($\bar{S} \sim 10$ expected counts). In view of the 3FGL mask excluding the known bright sources from our ROI, it is reassuring that the brightest PSs that our NN identifies lie just at the 3FGL threshold. A fainter SCD for the GCE PSs in comparison with the disk could possibly be attributed to the differing star formation histories in the Galactic Bulge and the Galactic Disk, causing the GCE PSs to be older and hence dimmer than their disk counterparts (e.g. Ref.~\cite{Crocker2017}).
\par The faint nature of the median ($\tau = 0.5$) SCD for the GCE as estimated by our NN would imply that a large number of PSs is required to explain the GCE flux, assuming there is no Poisson contribution e.g. from annihilating DM: in our masked ROI, integrating the median estimate for $dN / d (\log_{10} F)$ over the logarithmic flux $d (\log_{10} F)$ yields an expected number of $N \sim$ 10,100 GCE PSs, which translates to $N \sim$ 29,300 PSs in the entire sky when multiplying with $\int_{\text{sky}} T_\text{GCE} \, dA \ / \ \int_{\text{ROI}} T_\text{GCE} \, dA$, where $T_\text{GCE}$ denotes the generalized NFW-squared template for the GCE. For the quantile levels $\tau = 0.05$ and $0.95$, we obtain 10,300 and 189,500 GCE PSs in the sky, respectively. Our cross-check with \texttt{NPTFit} yields $N \sim$ 3,900 PSs in our ROI ($N \sim$ 11,200 in the entire sky) using Model~O (but only $N \sim 600$ in our ROI or 1,800 in the sky with \texttt{p6v11}). 
\par We emphasize that our NN analysis, as is the case for NPTF, is agnostic as to the physical origin of the GCE emission and does not take the energy spectrum of the photons into account. Although our results can therefore not directly be compared to the findings of MSP population studies, it is still interesting to discuss whether our estimates could be accommodated by an unresolved population of MSPs in the Galactic Center region. As early as 2005, before \emph{Fermi} launched, Ref.~\cite{Wang2005} suggested that $\gamma$-ray observations of the Galactic Center by \emph{EGRET}, which measured a spectrum with a break at several GeV, were consistent with thousands of unresolved MSPs in the region. More recent studies that make use of the \emph{Fermi} data have refined these findings. Ref.~\cite{Gonthier2018} estimated that 34,200 MSPs and 20,000 $-$ 50,200 MSPs at 68\% confidence can explain the GCE, respectively, while Ref.~\cite{Yuan2014} obtained the somewhat lower estimate of 10,000 $-$ 20,000 MSPs. Ref.~\cite{Ploeg2020} suggested a similar number of MSPs (specifically 17,900 $-$ 82,200 MSPs at 95\% confidence, see their Fig.~9), and that a less luminous population was expected in the bulge as compared to the disk (see their Fig.~6) -- although those authors did model the GCE as a boxy and a nuclear bulge rather than a NFW-squared template. Our NN results agree with the conclusion of Ref.~\cite{Hooper2016} that if MSPs make up the GCE, they must be fainter than the Galactic Disk population, although they found that 2,000 $-$ 13,750 bulge MSPs suffice. Recently, Ref.~\cite{Gautam2021} showed that MSPs formed by accretion-induced collapse (rather than through ``recycling'' of old neutron stars, e.g. Ref.~\cite{Radhakrishnan1982}) could explain both the GCE and the microwave haze from the inner Galaxy.
The SCD derived in that work (using the same ROI as herein) peaks at a flux of $F = 6 \times 10^{-12} \ \text{counts} \ \text{cm}^{-2} \ \text{s}^{-1}$, even below the 1-photon line, and corresponds to a population without any MSPs brighter than $F \geq 10^{-10} \ \text{counts} \ \text{cm}^{-2} \ \text{s}^{-1}$ (see their Fig.~5). In view of the uncertainties in our SCD estimate at the low flux end, such a population could be compatible with the results of our NN-based analysis. In contrast, earlier works located a sizable amount of the GCE flux just below the \emph{Fermi} detection threshold, implying that $\sim$~1,000 MSPs \cite{Gordon:2013vta} or several hundred PSs within a ROI of $10^\circ$ around the Galactic Center \cite{Lee2016} would be enough to explain the GCE. All these estimates must be interpreted with caution, however, as the exact numbers depend on the cutoff of the SCD at the low flux end. To account for the possibility that PSs make up only a fraction of the GCE flux, we integrate downwards over the flux bins until 50\% of the GCE flux is reached, starting at the bright flux end of our median estimate for the SCD. We find that a population of $\sim$~2,100 PSs in our ROI (6,300 PSs in the sky) brighter than $F = 1.4 \times 10^{-11} \ \text{counts} \ \text{cm}^{-2} \ \text{s}^{-1}$ (corresponding to $\bar{S} = 1.3$ expected counts per PS) could explain half of the excess emission. 

\section{Constraining the Poisson flux fraction}
\label{sec:constraining_Poisson}
The results in the previous section, in particular those shown in Fig.~\ref{fig:fermi_results}, represent our detailed findings for the nature of the GCE.  
Nevertheless, arguably the most important question related to the excess is whether the emission is consistent with DM annihilation, and a specific SCD does not immediately answer this question. Of course, in Sec.~\ref{sec:toy}, we showed that $g^{\boldsymbol{\varpi}}$ can be expected to produce an $F \, dN / d (\log_{10} F)$ peaked below the 1-photon line for a Poissonian input (see Fig.~\ref{fig:isotropic_poisson}), and given that this is what we expect for DM, these results would appear to weigh against a purely DM origin for the excess. In this section we will firm up this intuition and, in particular, introduce a summary statistic that can be used to shed light on the PS vs. Poissonian nature of the excess.

In doing so, we must account for the inherent degeneracy between Poisson and PS flux that has plagued previous results. If the GCE is truly Poissonian in nature, then we cannot exclude a PS origin. There will always remain an indistinguishable scenario where the flux arises from a large population of dim astrophysical sources, each of which produces far fewer than a single photon on average. In such an event, the PS hypothesis might be resolved by future measurements that push the 1-photon line to smaller fluxes, but the existing \emph{Fermi} data could not resolve the PS vs. DM debate. The inverse, however, is not true. If the GCE in fact has a PS origin, then as the sources become brighter, the dataset becomes less consistent with Poisson emission. In detail, it is possible to set an upper limit on the Poissonian fraction of the flux associated with a given template, which we denote by $\eta_P$. Using this exact logic, we will set a limit on $\eta_P$ for the GCE template emission. Doing so, we will find that for the analysis choices made in the present work, the GCE is consistent with an $\mathcal{O}(1)$ fraction arising from genuine non-Poissonian PS emission.

We can obtain a simple estimate of $\eta_P$ using the SCD determined by $g^{\boldsymbol{\varpi}}$ directly, and this is the first approach we will consider. However, we will find this approach is not sufficiently sensitive to obtain non-trivial constraints from the \emph{Fermi} map, and as such we will introduce an additional NN which will improve the sharpness. The Poisson flux fraction $\eta_P$ can also be determined using conventional likelihood based techniques (as we will outline below, with a detailed description provided in App.~\ref{sec:frequentist_llh_iso}), and we will validate our NN approach by benchmarking it against a frequentist computation of $\eta_P$ in a simple test scenario. Then, we will turn toward the realistic scenario. First, we will verify that the constraints we obtain for simulated (approximately) \emph{Fermi}-like maps are well calibrated, meaning that, for example, our 95\%-confidence constraint on $\eta_P$ lies above the true value for $\sim$~95\% of the maps. Then, we proceed to constrain the Poissonian component of the GCE in the \emph{Fermi} data.

\subsection{A simple estimate of $\eta_P$ from the SCD}
\label{sec:constraining_poisson_simple}
Let us recall the intuitive interpretation of the histogram labels: the relative cumulative histogram $\mathbf{U} = (U_j)_{j=1}^M$ expresses the fraction of flux coming from PSs \emph{at most} as bright as the value of $\log_{10} F$ associated with bin $j \in \{1, \ldots, M\}$. The fact that both $(U_j)_{j=1}^M$ and $\eta_P$ express flux fractions suggests that a simple estimator for the Poissonian flux fraction $\eta_P$ of a template can be directly obtained from the median estimate of $(U_j)_{j=1}^M$ provided by $g^{\boldsymbol{\varpi}}$. Since the PS / Poisson degeneracy decreases with increasing PS brightness, we can cut off the high-flux end of the SCD beyond a particular flux where we can be certain (at confidence level $\alpha \in [0, 1]$) that the entire flux located to the right of the cut-off is PS-like, and take the remaining flux fraction to the left of the cut-off (given by the \emph{cumulative} histogram evaluated at the cut-off) as an estimate of $\eta_P$. For a given value of $\alpha$, we define the cut-off such that the retained flux to the left is indeed greater than the true Poissonian flux fraction $\eta_P$ for $(100 \times \alpha)\%$ of the maps. In other words, the defining condition for the cut-off values is that the resulting constraints on $\eta_P$ are well-calibrated (with respect to the calibration dataset). 

In what follows, we will formulate this idea more precisely. Specifically, we determine a flux $\phi^*(\alpha)$ as a function of $\alpha \in [0, 1]$ such that interpolating the median relative cumulative $F \, dN / d (\log_{10} F)$, i.e. $\tilde{Q}^{\boldsymbol{\varpi}}(\mathbf{x} \,; 0.5)$, to this flux value can be expected to exceed $\eta_P$ with a probability of $\alpha$. Formally, this can be written as the following optimization problem: find $\phi^*= \phi^*(\alpha)$ such that for all confidence levels $\alpha \in [0, 1]$
\begin{equation}
    \left\langle\mathbb{I}\left[\eta_P(\mathbf{x}) \, \leq \, \tilde{Q}^{\boldsymbol{\varpi}}\left(\mathbf{x}; \, 0.5\right)\vert_{\phi^*(\alpha)}\right] \right\rangle_{\mathbf{x}} = \alpha,
    \label{eq:simple_FF_estimator}
\end{equation}
where $\eta_P(\mathbf{x})$ is the true Poissonian flux fraction of the template under consideration for map $\mathbf{x}$, and we write $\tilde{Q}^{\boldsymbol{\varpi}}\left(\mathbf{x}; \, 0.5\right)\vert_{\phi^*(\alpha)}$ for the piecewise linear interpolation of the predicted median cumulative histogram to the value of $\log_{10} F = \phi^*(\alpha)$. The sample average $\langle \cdot \rangle_{\mathbf{x}}$ is taken over a sufficiently large calibration dataset $\mathcal{X}_\text{cal}$.  Importantly, we emphasize that the definition of $\mathcal{X}_\text{cal}$ implicitly encodes the priors with respect to which the calibration property in Eq.~\eqref{eq:simple_FF_estimator} shall be satisfied: for example, if the set $\mathcal{X}_\text{cal}$ contains disproportionately many maps with a very large Poisson flux (i.e. $\eta_P \approx 1$), the calibration property requires large values of $\tilde{Q}^{\boldsymbol{\varpi}}\left(\mathbf{x}; \, 0.5\right)\vert_{\phi^*(\alpha)}$ in order for the inequality to hold true for $(100 \times \alpha)\%$ of the maps in $\mathcal{X}_\text{cal}$, giving rise to large values of $\phi^*(\alpha)$ in comparison with a calibration set $\mathcal{X}_\text{cal}$ that contains mainly PS-dominated maps. Throughout this section, we choose a non-informative prior for the Poissonian fraction $\eta_P$, implying that we generate the calibration dataset $\mathcal{X}_\text{cal}$ in such a way that $\eta_P$ is uniformly distributed in $[0, 1]$.
\par The simple estimator for the Poissonian flux fraction in Eq.~\eqref{eq:simple_FF_estimator} yields well-calibrated constraints by construction in that the true Poissonian flux fraction $\eta_P$ can be expected to fall $(100 \times \alpha)$\% of the times below the estimate $\tilde{Q}^{\boldsymbol{\varpi}}\left(\mathbf{x}; \, 0.5\right)\vert_{\phi^*(\alpha)}$ when drawing maps $\mathbf{x}$ from the distribution represented by the set $\mathcal{X}_\text{cal}$ that the estimator was calibrated on. However, as this estimator merely evaluates the estimated median histogram at a fixed value for each $\alpha$ without taking into account the shape of the histogram, the resulting constraints are quite weak: in fact, this estimator yields the trivial constraint $\tilde{\eta}_P = 100$\% at $\alpha = 95$\% confidence for the Poissonian GCE contribution in the \emph{Fermi} map when applied to the median GCE SCD predicted by our NN. The results of a benchmark test for this simple estimator are provided in App.~\ref{sec:systematic_constraints_simple}.

\subsection{Evaluating $\tilde{\eta}_P$ with an additional NN}
\label{sec:constraining_poisson_with_NN}
\par In order to obtain a more powerful estimator, we replace $\tilde{Q}^{\boldsymbol{\varpi}}\left(\mathbf{x}; \, 0.5\right)\vert_{\phi^*(\alpha)}$ by a function $\tilde{\Phi}$, which takes the entire median histogram and the confidence level $\alpha$ as inputs, i.e. $\tilde{\Phi} = \tilde{\Phi}(\tilde{q}^{\boldsymbol{\varpi}}\left(\mathbf{x}; \, 0.5); \, \alpha \right)$. Here, $\tilde{q}^{\boldsymbol{\varpi}}$ stands for the estimated (relative) density histogram, which is related to the cumulative histogram by $\tilde{Q}^{\boldsymbol{\varpi}}_j = \sum_{m=1}^j \tilde{q}^{\boldsymbol{\varpi}}_m$. 
This leads to the following modified optimization problem: find $\tilde{\Phi}$ such that for all $\alpha \in [0, 1]$
\begin{equation}
    \left\langle\mathbb{I}\left[\eta_P(\mathbf{x}) \, \leq \, \tilde{\Phi}(\tilde{q}^{\boldsymbol{\varpi}}\left(\mathbf{x}; \, 0.5); \, \alpha \right)\right] \right\rangle_{\mathbf{x}} = \alpha.
\label{eq:NN_FF_estimator}
\end{equation}
Note that Eq.~\eqref{eq:NN_FF_estimator} again requires the estimator $\tilde{\Phi}$ to be well-calibrated, but does not enforce it to be \emph{sharp}; for example, the simple estimator in Eq.~\eqref{eq:simple_FF_estimator} given by $\tilde{\Phi}\left(\tilde{q}^{\boldsymbol{\varpi}}(\mathbf{x}; \, 0.5); \, \alpha \right) = \tilde{Q}^{\boldsymbol{\varpi}}\left(\mathbf{x}; \, 0.5\right)\vert_{\phi^*(\alpha)}$ is a valid solution to Eq.~\eqref{eq:NN_FF_estimator}. Naturally, we are interested in finding a function $\tilde{\Phi}$ able to provide constraints on the Poissonian flux that are as tight as possible. Rather than making an explicit ansatz for $\tilde{\Phi}: (\tilde{q}^{\boldsymbol{\varpi}}\left(\mathbf{x}; \, 0.5); \, \alpha \right) \mapsto \tilde{\eta}_P(\mathbf{x}; \alpha) = \tilde{\Phi}(\tilde{q}^{\boldsymbol{\varpi}}\left(\mathbf{x}; \, 0.5); \, \alpha \right)$, we again resort to \emph{machine learning}: we take $\tilde{\Phi}$ to be a NN $h^{\boldsymbol{\nu}}$ with weights $\boldsymbol{\nu}$ and train it using the pinball loss function (see Eq.~\eqref{eq:pinball_loss}), where the confidence level $\alpha$ plays the role of the quantile level $\tau$ in this case. Now, the calibration dataset $\mathcal{X}_\text{cal}$ is given by the dataset used for the training of $h^{\boldsymbol{\nu}}$. Thus, the priors used for the training data generation implicitly set the priors with respect to which the calibration property in Eq.~\eqref{eq:NN_FF_estimator} will be encouraged during the NN training. 
\par In our experiments presented below, we take $h^{\boldsymbol{\nu}}$ to be a standard fully-connected NN with two hidden layers consisting of $256$ neurons each, which are followed by ReLU activation functions. For the output layer that yields the estimate of the Poissonian flux fraction $\tilde{\eta}_P$, we take a sigmoid activation function to enforce $\tilde{\eta}_P \in (0, 1)$. The training of $h^{\boldsymbol{\nu}}$ consists of 200 epochs, each batch contains 2,048 histograms, and we use an Adam optimizer \cite{Kingma2014} with initial learning rate $10^{-3}$ that exponentially decays to $10^{-4}$ by the end of the training. Just like we did for $\tau$ when training the NN $g^{\boldsymbol{\varpi}}$, we randomly draw an individual confidence level $\alpha \sim U([0, 1])$ for each histogram.
\par We expect the introduction of $h^{\boldsymbol{\nu}}$ will improve the sharpness of our estimator. However, before applying this method directly to the \emph{Fermi} map, we first benchmark its prediction against a frequentist limit obtained with an analytic likelihood function in a simple scenario where the likelihood approach can be reliably calculated.

\subsection{Benchmarking the NN estimator \texorpdfstring{$h^{\boldsymbol{\nu}}$}{$h^\nu$} in an isotropic example without a PSF}
\label{sec:constraining_poisson_iso_without_PSF}
\begin{figure}
\centering
  \noindent
   \resizebox{0.85\columnwidth}{!}{
    \includegraphics{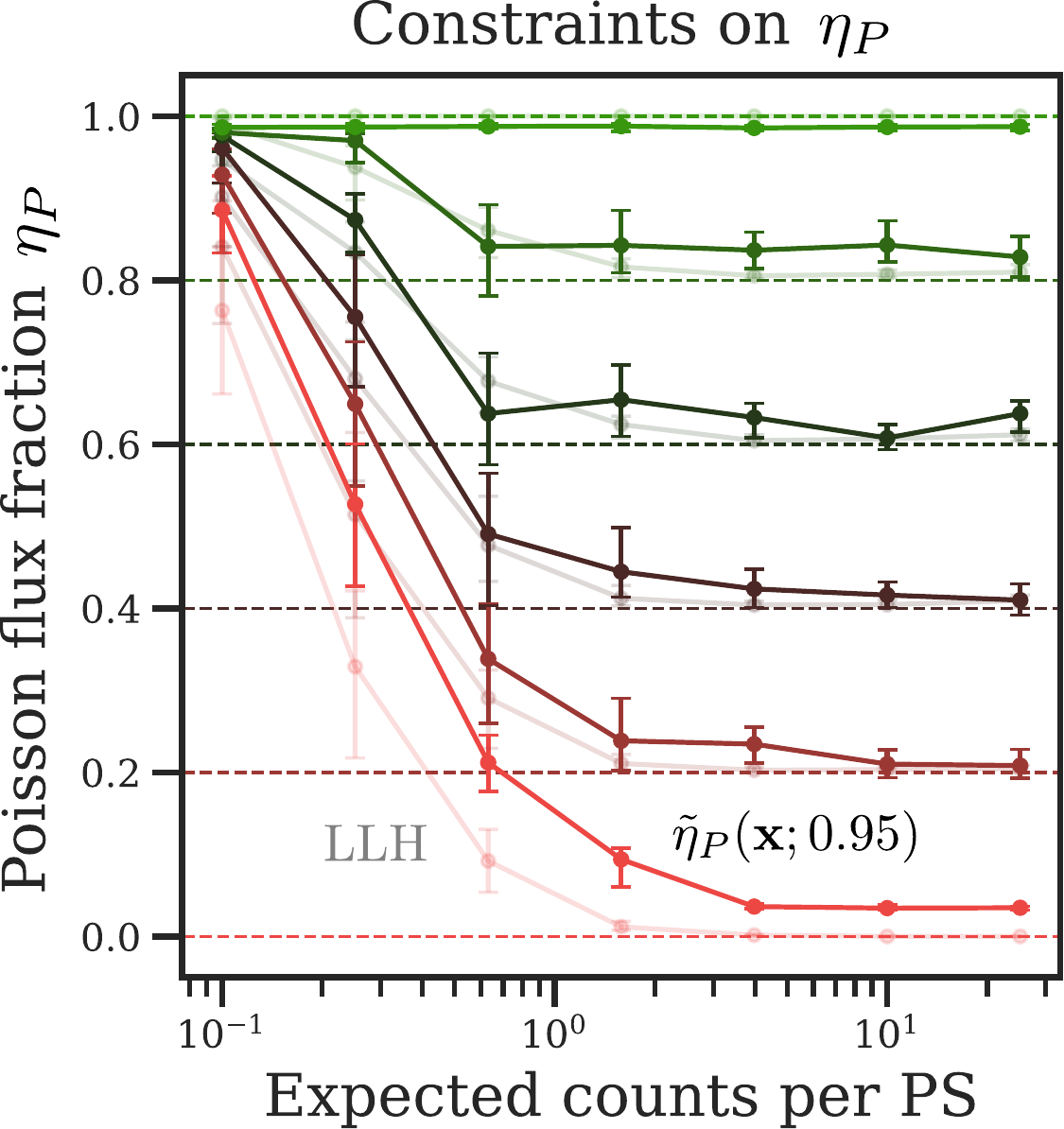}
    }
    \caption{Constraints on the Poissonian flux fraction $\eta_P$ for a single isotropically distributed PS population without a PSF. We consider 7 SCDs, given by Dirac delta distributions such that all the PSs of the population have the same expected number of counts per PS as indicated on the $x$-axis. We apply our method to maps with $\eta_P$ ranging from 0\% (bottom, red) to 100\% (top, green) in steps of 20\%. The bright lines show the constraints from $h^{\boldsymbol{\nu}}$ at 95\% confidence, while the faint lines in the background correspond to the 95\% frequentist limits based on the analytic likelihood (LLH). The errorbars indicate the 68\% scatter over 64 realizations for each combination of SCD and $\eta_P$. For PS populations as faint as $0.25$ expected counts per PS, the NN (the likelihood-based approach) can rule out more than half (a third) of the flux being Poissonian at 95\% confidence in the absence of Poissonian emission ($\eta_P = 0$). The NN constraints are not much weaker than their likelihood-based counterparts, particularly for $\eta_P > 0$.}
    \label{fig:constraints_iso_no_PSF}
\end{figure}
To validate our method for constraining the Poissonian flux fraction $\eta_P$ based on the SCD histogram predicted by $g^{\boldsymbol{\varpi}}$ using a second NN $h^{\boldsymbol{\nu}}$, we directly compare our results to those obtained by determining a frequentist one-sided 95\% upper limit on $\eta_P$ using an analytic likelihood approach. In particular, for a direct comparison we will quote the value of the NN-based estimate $\tilde{\eta}_P$ determined by $h^{\boldsymbol{\nu}}$ at $\alpha = 95\%$. To obtain frequentist limits for a given map $\mathbf{x}$, we consider the test statistic in terms of the logarithmic profile likelihoods
\begin{equation}
\begin{aligned}
    \text{TS}(\eta_P) = -2 \Big[ &\ln\left( p(\mathbf{x} \, | \, \eta_P, \hat{\boldsymbol{\theta}}(\eta_P)) \right) \\
    - &\ln\left( p(\mathbf{x} \, | \, \hat{\eta}_P, \hat{\boldsymbol{\theta}}(\hat{\eta}_P))\right) \Big],
\end{aligned}
\end{equation}
where $\hat{\eta}_P$ is the maximum likelihood estimate for the Poissonian  flux fraction and $\hat{\boldsymbol{\theta}}(\eta_P)$ denotes the remaining parameters describing the PS population that maximize the likelihood for a given Poisson flux fraction $\eta_P$ (namely the expected number of PSs and the total number of expected counts, see App.~\ref{sec:frequentist_llh_iso} for more details). From Wilks' theorem \cite{Wilks1938}, it follows that this test statistic is asymptotically $\chi^2$-distributed with one degree of freedom. Hence, we will report the frequentist one-sided upper $\alpha$-confidence limit as the value $\eta_P(\alpha)$ where the test statistic takes the value $\text{TS} = F^{-1}_{\chi^2_1}(2 \alpha - 1)$ for $\alpha > 0.5$ and $\eta_P(\alpha) > \hat{\eta}_P$ (where $F^{-1}_{\chi^2_1}$ denotes the quantile function of the $\chi^2_1$ distribution), e.g. $\text{TS} = 2.71$ for $\alpha = 95\%$ confidence.

The comparison is performed on a particularly simple example: we revisit the scenario of a single isotropically distributed PS population considered in Sec.~\ref{sec:toy}. However, we now consider the case \emph{without an instrumental PSF} that would introduce correlations between the pixels. As explained in Ref.~\cite{Collin:2021ufc}, existing methods to analytically compute the PS likelihood (in particular, the NPTF and CPG) rely on an approximate description of pixel-to-pixel correlations induced by the PSF (see Ref.~\cite{Collin:2021ufc} and also the discussion in Sec.~\ref{sec:cnns}), and so by assuming the direction of the incident photons is reconstructed exactly we can compute the true image likelihood exactly (and in fact in this limit the NPTF and CPG likelihoods reduce to the same form).
\par For the training of the NN $g^{\boldsymbol{\varpi}}$ (which predicts the SCD given a photon-count map), we take each count map to be the sum of \emph{two} individual maps stemming from two different isotropically distributed PS populations, as we did for the GCE flux in the realistic scenario in Sec.~\ref{sec:fermi}. This is because we intend to subsequently \emph{evaluate} the trained NN $g^{\boldsymbol{\varpi}}$ on mixed PS + Poisson maps to generate training data for the NN $h^{\boldsymbol{\nu}}$ whose task will then be to constrain the Poissonian flux component in the underlying map based on the SCD predicted by $g^{\boldsymbol{\varpi}}$, as described in Sec.~\ref{sec:constraining_poisson_with_NN}. Note that we do not include \emph{genuinely} Poissonian emission already in the training data for $g^{\boldsymbol{\varpi}}$ because there is no ``correct SCD'' for Poissonian flux that we could use as a label for the training of $g^{\boldsymbol{\varpi}}$. To ensure that the training dataset for $g^{\boldsymbol{\varpi}}$ includes maps so faint that they cannot be distinguished from Poisson emission at high confidence, not even with the analytic likelihood, we extend our prior range for the location parameter of the skew normal distributed SCDs in $\log_{10}(F)$-space from $[-1, 1.5]$ to $[-2, 1.5]$ (where $\text{counts} \ \text{cm}^{-2} \ \text{s}^{-1}$ is the reference unit); see App.~\ref{sec:NN_priors} for further details. We take a uniform exposure of $1 \ \text{cm}^{2} \ \text{s}$ again, implying that flux and counts have the same numerical values. We repeat the NN training of $g^{\boldsymbol{\varpi}}$ described in Sec.~\ref{sec:toy} for this case. In the next step, we generate 102,400 maps with 50,000 expected counts each, which will be used for creating the training and testing datasets for $h^{\boldsymbol{\nu}}$. The counts in each map are the sum of a Poissonian and a non-Poissonian PS template map, where the Poissonian flux fraction $\eta_P \sim U([0, 1])$ is randomly drawn between 0 and 1. Then, we evaluate the trained NN $g^{\boldsymbol{\varpi}}$ on these maps and use 4/5 of the predicted SCD histograms as the training data for $h^{\boldsymbol{\nu}}$, keeping the other 1/5 as an independent testing dataset. In the training of $h^{\boldsymbol{\nu}}$, the true label is given by the Poisson flux fraction $\eta_P$. We emphasize that for maps that contain flux from faint PSs, a fraction of the PS flux is indistinguishable from Poissonian flux (importantly, however, this flux is not accounted for by $\eta_P$). Since faint flux in training maps can be genuinely Poissonian, come from faint PSs, or consist of a mixture of both, $h^{\boldsymbol{\nu}}$ will not be able to derive tight constraints on $\eta_P$ in maps with a large faint flux component because overconfident predictions during the training are penalized by the pinball loss, which compares the $\alpha$-quantiles $\tilde{\eta}_P(\mathbf{x}; \alpha)$ estimated by $h^{\boldsymbol{\nu}}$ with the label $\eta_P(\mathbf{x})$. To ensure the physical degeneracy in these scenarios is reproduced in the prediction of $h^{\boldsymbol{\nu}}$, it is crucial that the training dataset contains maps with faint flux that is entirely Poissonian, which prevents $h^{\boldsymbol{\nu}}$ from speculating on a PS-like flux component whenever the SCD estimate produced by $g^{\boldsymbol{\varpi}}$ is so faint that it does not allow $h^{\boldsymbol{\nu}}$ to rule out a Poissonian origin.
\par To systematically assess the constraining power for varying PS brightness, we evaluate the trained NN $h^{\boldsymbol{\nu}}$ on estimated SCD histograms corresponding to maps whose counts are composed of a Poissonian contribution and a PS-like non-Poissonian contribution from a single homogeneous population of PSs with identical flux. As in the maps underlying the histograms used for the training of $h^{\boldsymbol{\nu}}$, the total flux in all these mixed PS + Poisson maps corresponds to 50,000 expected counts, resulting in $1.36 = $ 50,000 $/$ 36,868 expected counts in each pixel of our ROI with radius $25^\circ$. 
\par Figure~\ref{fig:constraints_iso_no_PSF} shows the 95\%-confidence constraints estimated by $h^{\boldsymbol{\nu}}$ as a function of the expected counts per PS. The different colors indicate the true Poissonian flux fraction $\eta_P$, from 0\% (red) to 100\% (green) in steps of 20\%. The constraints with the likelihood-based approach are given by the faint lines in the background. Interestingly, a substantial fraction of the PS flux can be distinguished from Poisson emission even for populations of PSs emitting on average $< 1$ count each. At the 1-photon line, the fraction of flux that the NN $h^{\boldsymbol{\nu}}$ cannot attribute to PSs at 95\% confidence is $< 20\%$ for $\eta_P = 0$. Although the constraints with the frequentist likelihood function based approach are sharper than their NN-based counterparts for small values of $\eta_P$, the difference in constraining power is rather modest, and our NN is able to provide tight constraints. We remark that while the likelihood-based constraints are directly inferred from the counts in each of the 36,868 pixels, $h^{\boldsymbol{\nu}}$ relies on only $M = 22$ histogram values as an input, which act as a ``summary statistic''.
\par The behavior of the constraints $\tilde{\eta}_P$ produced by $h^{\boldsymbol{\nu}}$ for bright populations with a large number of expected counts per PS reflects the necessity to comply with the calibration property in Eq.~\eqref{eq:NN_FF_estimator}: as the true Poissonian flux fraction $\eta_P$ is uniformly distributed over the training dataset $\mathcal{X}_\text{cal}$, $\sim5\%$ of the maps in $\mathcal{X}_\text{cal}$ have $\eta_P \geq 0.95$. A trivial estimator $\tilde{\eta}_P$ that completely ignores the input could therefore output the constant constraint $\tilde{\eta}_P(\mathbf{x}; 0.95) = 0.95$ and would be right for $\sim95\%$ of the histograms belonging to the maps in $\mathcal{X}_\text{cal}$, just as required by Eq.~\eqref{eq:NN_FF_estimator}. However, a more powerful estimator will realize that the conditional probability of $\eta_P \geq 0.95$ given a very faint (bright) SCD estimate as an input is greater (less) than $5\%$. The specific choice of the priors for the SCDs modulates the risk that the NN $h^{\boldsymbol{\nu}}$ can take by estimating a value $\tilde{\eta}_P(\mathbf{x}; 0.95)$ slightly below $1$ for very faint histograms (e.g. $\tilde{\eta}_P(\mathbf{x}; 0.95) = 0.986$ for $\eta_P = 1$ in Fig.~\ref{fig:constraints_iso_no_PSF}) while still being correct $\sim95\%$ of the times. Indeed, we confirmed that when using the lower limit $-1$ instead of $-2$ for the prior range of the SCD location parameter when generating $\mathcal{X}_\text{cal}$, which on average gives rise to brighter PS populations, the estimates $\tilde{\eta}_P(\mathbf{x}; 0.95)$ produced by $h^{\boldsymbol{\nu}}$ for $\eta_P = 1$ increase to $0.998$ owing to the higher probability for a very faint histogram to belong to a purely Poissonian map. A similar argument applies for $\eta_P < 1$: the estimates $\tilde{\eta}_P$ converge to those values that allow $h^{\boldsymbol{\nu}}$ to underestimate the true Poissonian fraction $\eta_P$ roughly $5\%$ of the time for $\alpha = 0.95$. This causes the NN to not exclude a small fraction $\tilde{\eta}_P(\mathbf{x}; 0.95) \approx 3\%$ of Poissonian flux being hidden in 100\% PS maps, even for relatively bright PS populations.

\subsection{Constraining a Poissonian GCE}
\label{sec:constraining_poisson_gce}
Now, we apply our validated approach for constraining the Poissonian flux fraction $\eta_P$ to the GCE template in the realistic scenario from Sec.~\ref{sec:fermi} where all the templates are present, $|b| \leq 2^\circ$ and known 3FGL sources are masked, and we take the non-uniform \emph{Fermi} exposure as well as the \emph{Fermi} PSF into account. (Again, we emphasize that when we include the PSF, existing likelihood-based approaches no longer fully describe the statistics of the map correctly.) We proceed similarly to the isotropic case in the previous section; however, the constraints provided by $h^{\boldsymbol{\nu}}$ are expected to be considerably weaker now in view of the increased difficulty of the problem. More specifically, the uncertainties in the SCD estimates are larger now (compare Figs.~\ref{fig:isotropic_results} and \ref{fig:simulated_hist_plot}, and also the sharpness plots in Fig.~\ref{fig:UQ_plot}), for which reason the true SCD might deviate more from the estimated median histogram $\tilde{q}^{\boldsymbol{\varpi}}\left(\mathbf{x}; \, 0.5\right)$, which serves as the input for the NN $h^{\boldsymbol{\nu}}$. Hence, $h^{\boldsymbol{\nu}}$ needs to produce weaker constraints in order to achieve calibration.

\subsubsection{Training \texorpdfstring{$h^{\boldsymbol{\nu}}$}{$h^\nu$}}
Next, we outline how the additional NN $h^{\boldsymbol{\nu}}$ is trained. Firstly, let us emphasize that we do not retrain the NN $g^{\boldsymbol{\varpi}}$ that generates the SCDs used for training the estimator $h^{\boldsymbol{\nu}}$. Rather, we will \emph{evaluate} $g^{\boldsymbol{\varpi}}$ (which, recall, has only been trained on maps with a PS-like GCE composed of two template maps as described in Sec.~\ref{sec:fermi}) on a dataset of maps with a mixed PS + (genuinely) Poissonian GCE. We will then take the SCD estimates produced by $g^{\boldsymbol{\varpi}}$ for these maps as the training dataset for $h^{\boldsymbol{\nu}}$, with the correct labels in the training of $h^{\boldsymbol{\nu}}$ given by the Poissonian GCE flux fractions $\eta_P$ of the maps underlying the input SCDs.
\par In detail, we generate 102,400 maps. We fix the expected flux fraction of each template to be the best-fit prediction of NN $f^{\boldsymbol{\omega}}$, and we use the median SCD as estimated by the NN $g^{\boldsymbol{\varpi}}$ for the disk PSs for all the maps (see Fig.~\ref{fig:fermi_results}). This is because we expect the uncertainty in the predicted \emph{Fermi} GCE SCD to outweigh the scatter in the GCE histogram predictions arising from small variations in the expected fluxes of the non-GCE templates. For the GCE itself, we allow for a wide range of possible compositions: we adopt a uniform prior for the fraction of the Poissonian GCE contribution $\eta_P$ in $[0\%, 100\%]$, and we draw the SCD parameters for the complementary GCE PS flux from our original priors that we already used to generate the $1.5 \times 10^6$ training and testing maps for NNs $f^{\boldsymbol{\omega}}$ and $g^{\boldsymbol{\varpi}}$, only adjusting the total expected flux in such a way that the expected total GCE flux (PS + Poisson) matches the best-fit estimate of $f^{\boldsymbol{\omega}}$. Thus, the GCE PSs in each of the 102,400 maps may range from nearly as faint as Poisson emission to above the 3FGL threshold, and they constitute 0 $-$ 100\% of the GCE flux with uniform probability. Since the GCE counts in each of the training maps for $f^{\boldsymbol{\omega}}$ and $g^{\boldsymbol{\varpi}}$ are the sum of \emph{two} independent GCE PS template maps, these NNs have been trained on maps that contain two PS populations, one of which is \emph{virtually} as faint as genuinely Poissonian emission, implying that this mixed Poissonian + PS-like GCE case does not require the NNs to extrapolate to an unknown region in the input space.
\par We then evaluate $g^{\boldsymbol{\varpi}}$ for each of these maps $\mathbf{x}$ to obtain the estimated median histograms $\tilde{q}^{\boldsymbol{\varpi}}\left(\mathbf{x}; \, 0.5\right)$. We randomly put aside 1/5 of the GCE histogram predictions for the 102,400 maps for validation and take the remaining 4/5 to be the training data for $h^{\boldsymbol{\nu}}$. We use the same NN architecture, hyperparameters, and training procedure for $h^{\boldsymbol{\nu}}$ as in the isotropic example without a PSF in Sec.~\ref{sec:constraining_poisson_iso_without_PSF}.

\subsubsection{Validation on simulated data}
\begin{figure}
\centering
  \noindent
   \resizebox{0.65\columnwidth}{!}{
    \includegraphics{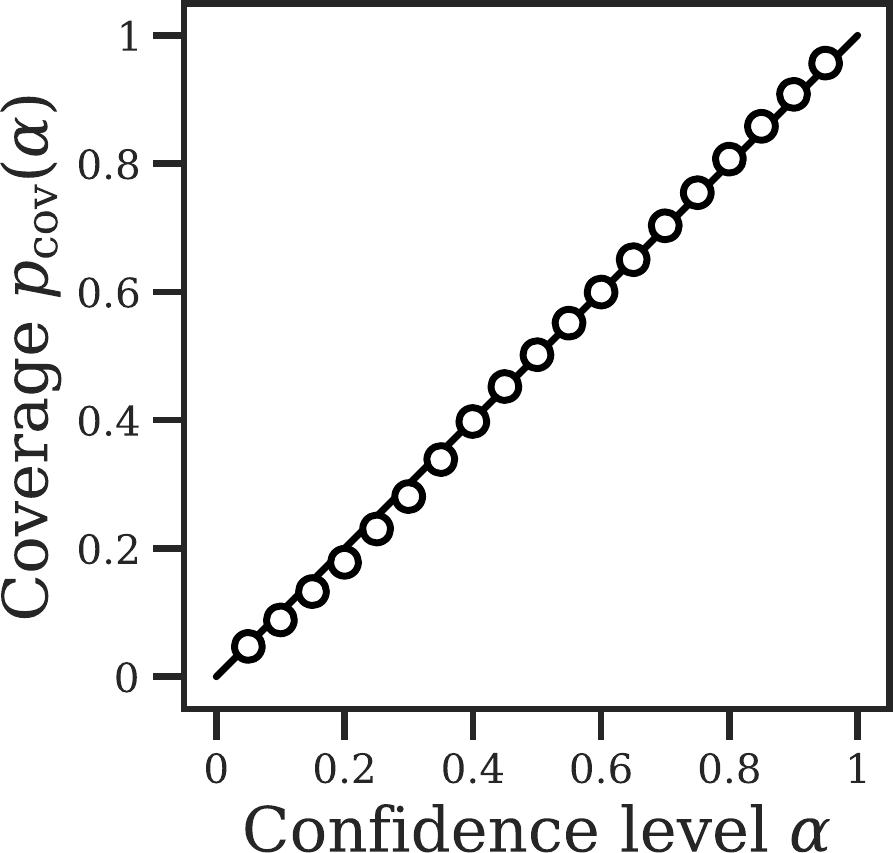}
    }
    \caption{Calibration plot for the estimated Poissonian GCE flux fraction $\tilde{\eta}_P(\mathbf{x}; \alpha) = \tilde{\Phi}(\tilde{q}^{\boldsymbol{\varpi}}\left(\mathbf{x}; \, 0.5); \, \alpha \right)$ produced by the NN $h^{\boldsymbol{\nu}}$ in the realistic scenario. The coverage on the $y$-axis is computed as the fraction of samples for which the estimate for a given confidence level $\alpha$ lies above the true value $\eta_P(\mathbf{x})$, i.e., by the left-hand side of Eq.~\eqref{eq:NN_FF_estimator}, where the sample average $\langle \cdot \rangle_{\mathbf{x}}$ is taken over the 20,480 histograms in the testing dataset.}
    \label{fig:calibration_constraining_Poisson}
\end{figure}
\begin{figure}
\centering
  \noindent
   \resizebox{1\columnwidth}{!}{
    \includegraphics{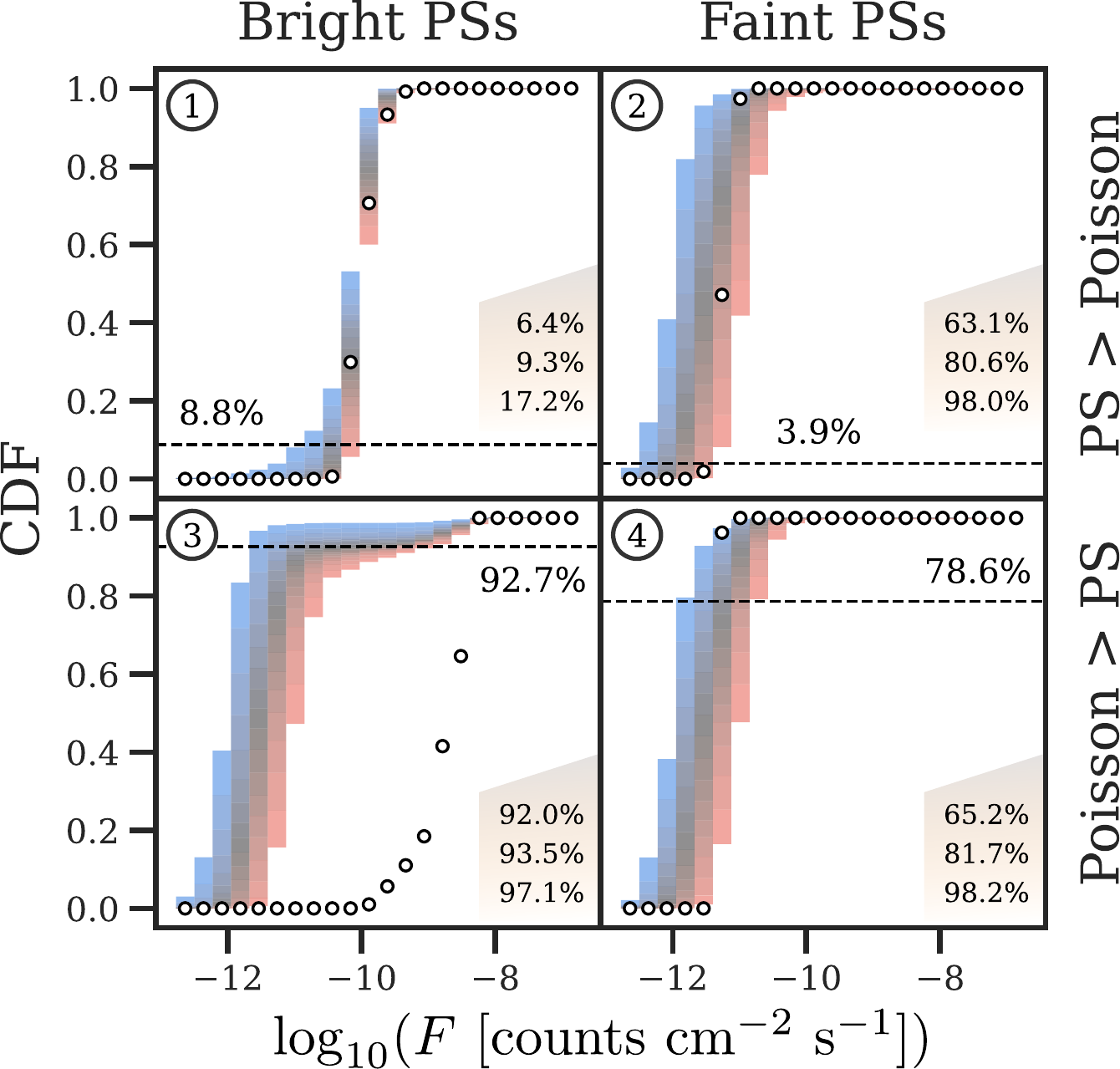}
    }
    \caption{Constraints on the Poissonian flux fraction $\eta_P$ for 4 simulated maps with a mixed PS + Poisson GCE from the testing dataset for $h^{\boldsymbol{\nu}}$ in the realistic scenario. The true Poissonian GCE flux fraction $\eta_P$ in each map is indicated by the horizontal dashed line, and its value is reported above or below.
    The white circles follow the normalized cumulative $F \, dN / d (\log_{10} F)$ that describes the GCE PS emission in each map. The colored regions show the estimated 5 $-$ 95\% quantiles produced by $g^{\boldsymbol{\varpi}}$, which agree with the true SCD of the GCE PSs for small $\eta_P$ and move to lower fluxes as $\eta_P$ increases.
    Maps 1 and 2 are PS-dominated, whereas the majority of the flux in maps 3 and 4 is Poissonian. The PS populations in maps 1 and 3 are relatively bright, while the PSs in maps 2 and 4 are faint and emit $\lesssim 1$ count per PS on average. Consequently, our NN $h^{\boldsymbol{\nu}}$ provides tight constraints $\tilde{\eta}_P$ only for maps 1 and 3, given by the three percentages on the right-hand side of each panel for confidence levels $\alpha = $ 50\%, 70\%, and 95\% (top to bottom). In contrast, the constraints for maps 2 and 4 are very similar, despite the big difference in the Poissonian GCE flux fraction $\eta_P$, and do not permit excluding a fully Poissonian GCE for either of the two.}
    \label{fig:selected_constraints}
\end{figure}
\begin{figure*}
\centering
  \noindent
   \resizebox{.8\textwidth}{!}{
    \includegraphics{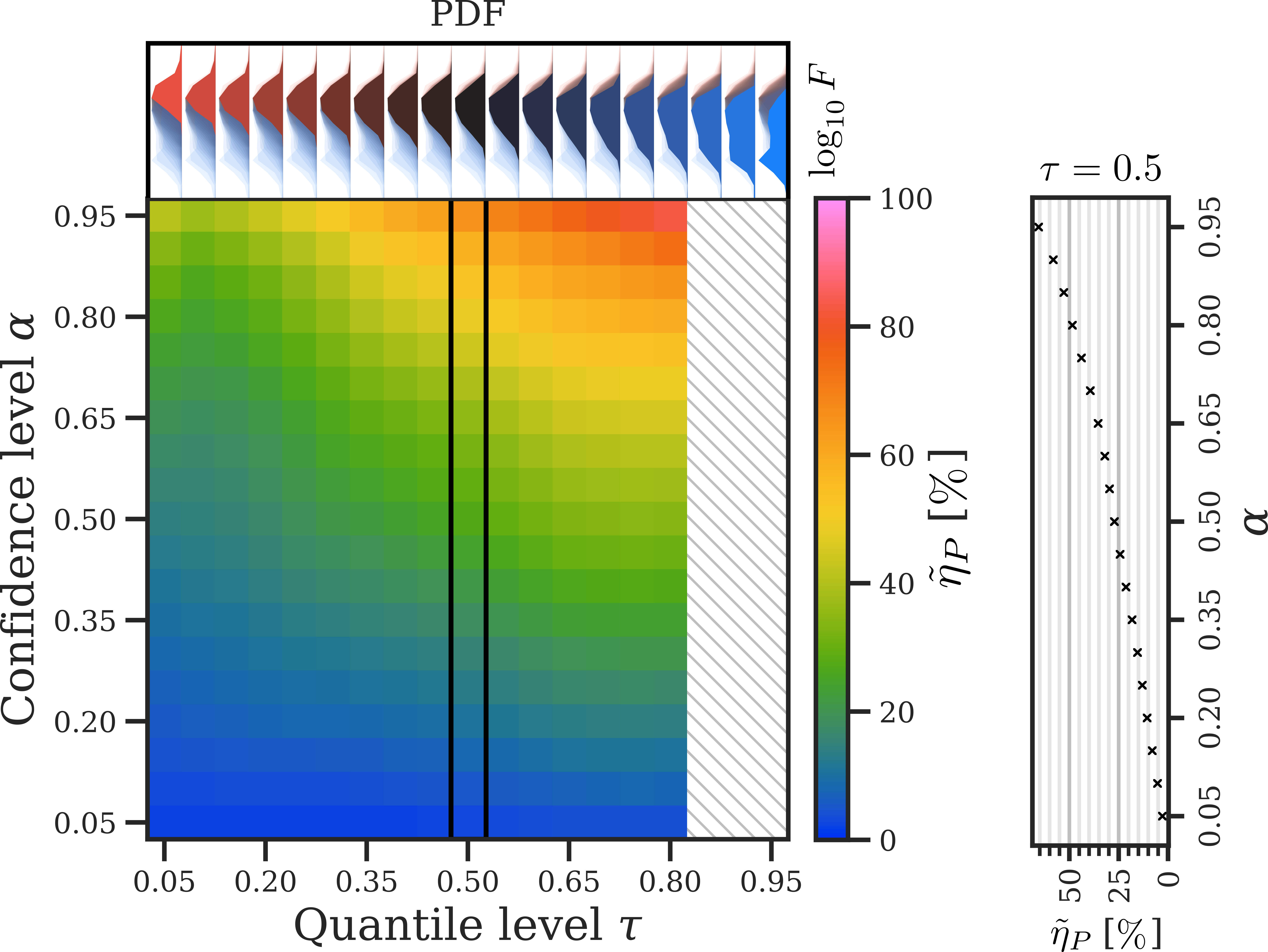}
    }
    \caption{Constraints on the Poissonian fraction $\eta_P$ of the GCE flux in the \emph{Fermi} map as estimated by the NN $h^{\boldsymbol{\nu}}$. The constraints are shown as a function of the quantile level $\tau$ for the SCD estimate from $g^{\boldsymbol{\varpi}}$ (columns) and confidence level $\alpha$ for the constraint (rows). The column for the median SCD histogram ($\tau = 0.5$) is surrounded by a box, and the corresponding constraints are shown in detail in the panel on the right. For orientation, the SCD estimate associated with each quantile level $\tau$ is highlighted in the panel above the constraints, with the SCDs for the other quantile levels plotted faintly in the background (see Fig.~\ref{fig:fermi_results} for a more explanatory plot of the \emph{Fermi} SCD estimates). For the median SCD estimate, we obtain the constraint $\tilde{\eta}_P = 65.6\%$ at $\alpha = $ 95\% confidence. For quantile levels $\tau \geq 0.85$, the histogram estimates lie outside the input space used for training $h^{\boldsymbol{\nu}}$, which is why we exclude this region from the plot (see main text).
    }
    \label{fig:fermi_constraints}
\end{figure*}
Figure~\ref{fig:calibration_constraining_Poisson} shows a calibration plot, where the coverage is computed as the sample average on the left-hand side in Eq.~\eqref{eq:NN_FF_estimator} over the 20,480 test samples. For small confidence levels $< 0.4$, the coverage lies slightly below the identity line, but the NN estimator $h^{\boldsymbol{\nu}} = \tilde{\Phi}$ is generally well calibrated, and the deviation from perfect calibration as defined in Eq.~\eqref{eq:NN_FF_estimator} is small. For example, the coverage at confidence level $\alpha = 0.95$ is given by $p_\text{cov}(0.95) = 0.957$. Because of the previously discussed degeneracy between faint PSs and Poisson emission, we expect $h^{\boldsymbol{\nu}}$ to provide tight constraints on the Poisson flux only for sufficiently bright PSs, as already seen in the isotropic example above. 
\par In Fig.~\ref{fig:selected_constraints}, we illustrate this behavior by considering four selected samples from our test dataset. The white circles trace the \emph{true} cumulative $F \, dN / d (\log_{10} F)$ of the GCE PS component, which characterizes the brightness of the GCE PSs contained in the respective map. The colored regions show the predicted quantiles provided by $g^{\boldsymbol{\varpi}}$ for each map which, in addition to the GCE PSs, contains a Poissonian GCE component. The estimated median distribution is the input for the NN $h^{\boldsymbol{\nu}}$. The true Poissonian flux contribution to the GCE $\eta_P$ is marked by horizontal dotted lines and stated above or below the lines. For Poissonian GCE flux fractions $\eta_P \approx 0$, the estimated SCD quantiles provided by $g^{\boldsymbol{\varpi}}$ coincide with the true SCD of the GCE PS component as expected (top panels). 
In the boxes on the right-hand side, we report the $\alpha = 50$\%, $70$\%, and $95$\% constraints produced by $h^{\boldsymbol{\nu}}$. The GCE in map 1 is dominated by relatively bright PSs, and the Poissonian flux only accounts for $\eta_P = 8.8$\% of the GCE. In this case, $h^{\boldsymbol{\nu}}$ is able to constrain $\eta_P$ to be less than $17.2$\% at $95$\% confidence. In contrast, the PSs in map 2, which constitute $96.1$\% of the GCE emission in the map, are not much brighter than Poissonian flux. Consequently, $h^{\boldsymbol{\nu}}$ cannot exclude that the GCE in the underlying map is almost entirely Poissonian. Maps 3 and 4 are dominated by Poissonian emission, with a small and moderate contribution of bright and faint PSs, respectively. This leads to a much narrower distribution of $\tilde{\eta}_P$ for map 3 ($\tilde{\eta}_P = 92.0$\% (97.1\%) at 50\% (95\%) confidence), whereas the constraints derived for map 4 are very similar to those for the faint PS-dominated map 2, reflecting the faint PS vs. Poisson degeneracy.

We remark that the case of two or more different GCE PS populations is not considered here (which would require training the NN $g^{\boldsymbol{\varpi}}$ on maps with $\geq 3$ PS populations because Poissonian flux is treated as a very faint PS population). This choice will impact the high confidence in the results for map 3 that the flux can be attributed to Poisson emission ($\tilde{\eta}_P$ is clustered near the true value). As the NN has not seen situations with more than two PS emission components, once it identifies the bright PS population, it can say confidently the remaining flux should be Poissonian. While this behavior will lead to more conservative constraints on $\eta_P$ in situations where, for example, the true distribution is a combination of Poisson emission and two separate PS populations, one dim and one bright, stronger constraints could be established in principle.

\subsubsection{Application to the Fermi map}
We now apply our approach for constraining $\eta_P$ to the \emph{Fermi} data. Recall that we have only used a single histogram for each map as the input to $h^{\boldsymbol{\nu}}$ during the training, namely the \emph{median} prediction $\tilde{q}^{\boldsymbol{\varpi}}(\mathbf{x}; \, 0.5)$; however, $g^{\boldsymbol{\varpi}}$ provides an estimate of the GCE SCD in the \emph{Fermi} map for \emph{any} quantile level $\tau \in (0, 1)$. As such, we can evaluate $h^{\boldsymbol{\nu}}$ individually for SCD histograms corresponding to different quantile levels in order to derive constraints on $\eta_P$ as a function $\tau$.\footnote{We also considered training $h^{\boldsymbol{\nu}}$ simultaneously on histograms for multiple quantile levels $\tau$, but this led to very similar constraints on $\eta_P$ for the \emph{Fermi} map in our experiments (less than 2\% difference for all confidence levels $\alpha$ as compared to only using the median prediction).}
\par Figure~\ref{fig:fermi_constraints} shows the estimated Poisson flux fraction $\tilde{\eta}_P$ as a function of the quantile level $\tau$ and the confidence level $\alpha$. The column $\tau = 0.5$ (surrounded by a box and also shown in detail on the right-hand side) is for the median histogram, and lower (higher) quantile levels correspond to brighter (fainter) SCDs. The density $F \, dN / d (\log_{10} F)$ histogram for each quantile level $\tau$ (that is, the input to $h^{\boldsymbol{\nu}}$) is illustrated in the panel above for orientation, and the color for each $\tau$ is the same as in the GCE panel in Fig.~\ref{fig:fermi_results}. For our median SCD, we obtain a constraint of $\tilde{\eta}_P = $ 65.6\% (39.4\%) at $\alpha = $ 95\% (70\%) confidence. At the bright end, $h^{\boldsymbol{\nu}}$ excludes a $>$~50\% Poissonian component of the GCE at 95\% confidence for $\tau \geq 0.25$, whereas for fainter GCE SCDs considered plausible by $g^{\boldsymbol{\varpi}}$, the 95\% constraint increases to $\tilde{\eta}_P = 83$\% for $\tau = 0.8$. For even higher quantile levels, the cumulative SCD histograms are fainter than 99\% of the histograms shown to $h^{\boldsymbol{\nu}}$ during its training. For this reason we consider the arising constraints unreliable, and therefore exclude this region from the plot. Specifically, we exclude values of $\tau$ for which the cumulative histogram for the GCE in the \emph{Fermi} map exceeds the 99\%-quantile value computed over the training maps by more than $0.1\%$ in at least one bin, which is only the case for $\tau \geq 0.85$ in the lowest three bins. The reason that the training dataset for $h^{\boldsymbol{\nu}}$ does not contain histograms with flux in the lowest few bins is that the uncertainties far below the 1-photon line are large, and the median histograms ($\tau = 0.5$), which is what we used for training $h^{\boldsymbol{\nu}}$, only start increasing at somewhat larger fluxes. For fluxes $F \gtrsim 10^{-12} \ \text{counts} \ \text{cm}^{-2} \ \text{s}^{-1}$, the cumulative histogram for the \emph{Fermi} map falls well within the range of the training data even for $\tau = 0.95$ (for example, compare the values of the $\tau = 0.95$ estimate for the \emph{Fermi} map in Fig.~\ref{fig:fermi_results} with the $\tau = 0.5$ estimate for simulated maps with a purely Poissonian GCE in Fig.~\ref{fig:comparison_w_best_fit}, which will be discussed below).

To summarize this section, the GCE identified by our NN-based framework in the \emph{Fermi} map in Sec.~\ref{sec:fermi} is faint enough that we cannot conclusively attribute the emission to either a population of unresolved PSs such as MSPs or alternatively to Poissonian emission as expected for DM annihilation. This is in disagreement with earlier NPTF-based analyses that found the GCE PS population to lie just below the 3FGL threshold \cite{Lee2016}, which would have allowed the method for constraining $\eta_P$ we introduced in this section to exclude a large contribution from a Poisson-dominated GCE at high confidence. Instead, the SCD we infer allows us to exclude a GCE that comprises of more than two-thirds Poisson emission (at 95\% confidence, for the median SCD estimate), still implying the excess cannot be entirely due to DM. We stress that the novel method we have developed herein, in addition to making use of a state-of-the-art (albeit imperfect) diffuse model, further passes the tests that previously called into question the PS interpretation of NPTF analyses, such as the recovery of artificially injected GCE flux from the \emph{Fermi} map and robustness against an unmodeled asymmetry in the GCE. We will demonstrate both of these points in the next section.

\section{Robustness of our findings}
\label{sec:robustness}
Whereas the statistical uncertainties of the flux fractions in analyses of the inner Galaxy are at the percent level -- both with the NPTF and with our NN-based framework -- it is the systematic uncertainties in the modeling that have thus far precluded a definitive resolution of the GCE origin. For instance, the Bayes factor for a PS-like GCE can vary by as much as 15 orders of magnitude depending on the diffuse foreground model used for the analysis (see Ref.~\cite{Leane2020}, Tab.~1).
\par In this section, we perform three experiments to assess the robustness of our findings. First, we compare our SCD estimate for the GCE in the \emph{Fermi} map and the resulting constraint on $\eta_P$ to the NN predictions for simulated maps whose GCE is entirely Poissonian, but which otherwise correspond to our best-fit parameters for the \emph{Fermi} map. Then, we carry out a mismodeling experiment where we apply our NN to simulated maps generated using alternate templates for the diffuse foregrounds, disk PSs, \emph{Fermi} bubbles, and the GCE itself. Lastly, we consider the recovery of artificially injected GCE flux from the real \emph{Fermi} data. The inability of the NPTF to correctly recover synthetic Poissonian GCE flux in this diagnostic test reported by Ref.~\cite{Leane2019a} called into question the NPTF-based evidence for a PS interpretation of the GCE by Ref.~\cite{Lee2016} (however, Ref.~\cite{Buschmann2020} demonstrated that this issue is resolved when using the improved diffuse Model~O instead of \texttt{p6v11}). While we showed in \citetalias{List2020b} that our NN was generally able to accurately determine the flux fractions of different templates, we found that the probability of GCE PS flux being confused with Poissonian GCE flux increased as the PSs became fainter, and faint GCE PS flux injected into the \emph{Fermi} map was frequently misattributed to the Poissonian template (see Figs.~S4 and S30 in \citetalias{List2020b}). Here, we demonstrate that our unified approach for the GCE (that attempts to disentangle the PS-like from the Poissonian component only at a later stage of the analysis) is able to accurately recover both Poissonian and PS-like GCE flux from the \emph{Fermi} map.

\subsection{Comparison with simulated best-fit maps}
\label{sec:fermi_vs_best_fit_MC}
\begin{figure*}
\centering
  \noindent
   \resizebox{.9\textwidth}{!}{
    \includegraphics{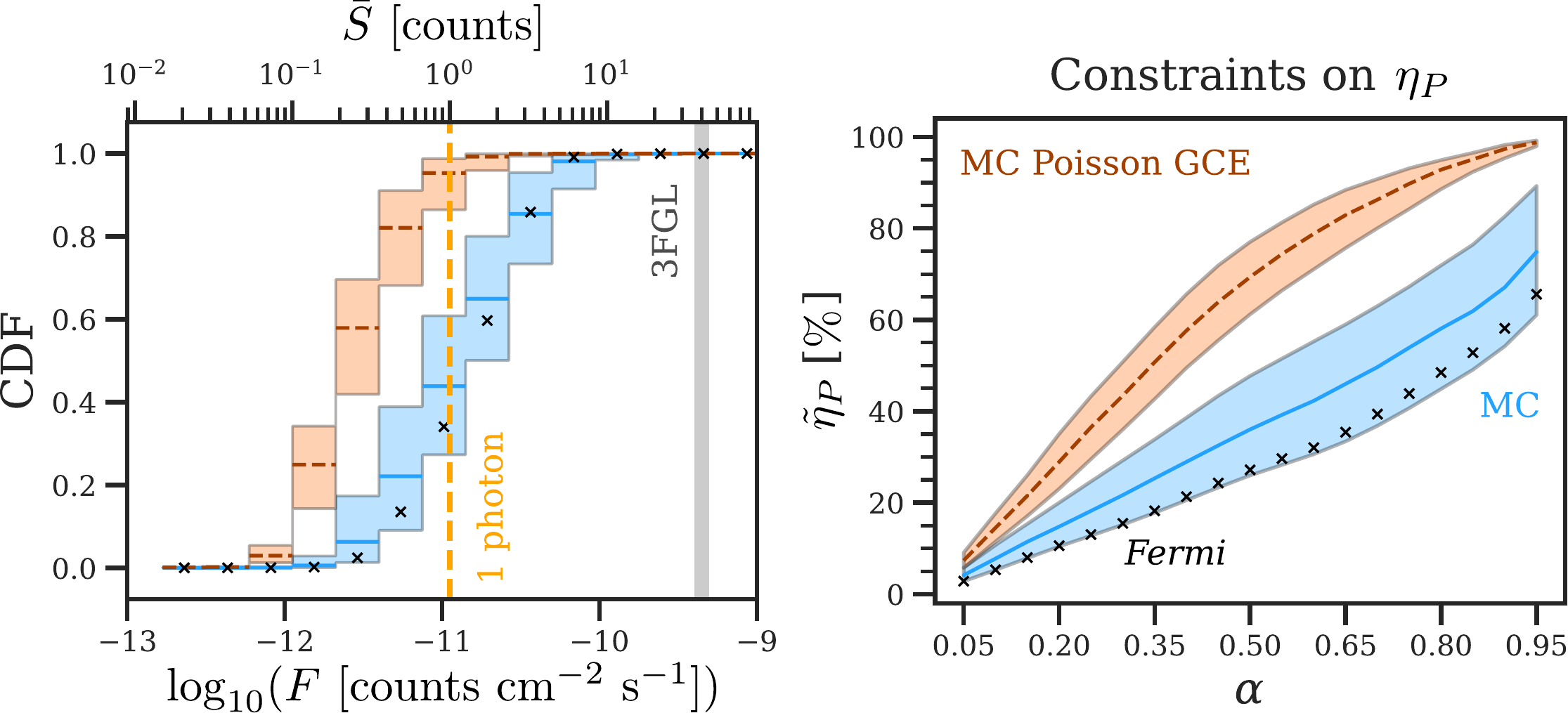}
    }
    \caption{Left: Predicted median ($\tau = 0.5$) cumulative SCD for the GCE in simulated MC maps and in the real \emph{Fermi} map (black crosses). The colored regions show the 68\% scatter over 1,024 MC realizations around the median (horizontal lines). The blue bands correspond to MC maps generated with all parameters set to the best-fit values determined from the \emph{Fermi} data, in particular with a PS-like GCE whose SCD is given by the \emph{Fermi} median prediction of $g^{\boldsymbol{\varpi}}$ shown in Fig.~\ref{fig:fermi_results}.
    For the maps represented by the orange regions, we use the same best-fit parameters for all the non-GCE templates, but we replace the PS-like GCE by an \emph{entirely Poissonian} GCE of the same total flux. In the Poissonian case, the sample median of the flux fraction located in flux bins at the 1-photon line or below is 95\%, but only 44\% for a PS-like GCE. Right: Constraints on the Poisson flux fraction $\eta_P$ derived from the median SCDs as a function of the confidence level $\alpha$. For the MC maps with a Poissonian GCE, the constraints reach $\sim100\%$ at $\alpha = 95\%$ confidence, while the sample median of the 95\%-confidence constraint for the \emph{Fermi} mock MC maps with a PS-like GCE is $74.8\%$. The median CDF estimate for the real \emph{Fermi} data is slightly brighter than the sample median of the MC maps, and the resulting constraints are therefore slightly stronger, but both SCD and constraints fall within the 68\% scatter over the MC realizations.}
    \label{fig:comparison_w_best_fit}
\end{figure*}
As a first robustness check, we compare our predicted SCD for the GCE in the \emph{Fermi} data with simulated best-fit maps. We generate 1,024 realizations corresponding to the best-fit flux fractions and median SCDs (for disk and GCE PSs) predicted by $f^{\boldsymbol{\omega}}$ and $g^{\boldsymbol{\varpi}}$ for the \emph{Fermi} map. Additionally, we simulate 1,024 maps with the same best-fit parameters, but with an entirely Poissonian GCE for comparison. Throughout this experiment, we only consider the median estimates for the SCD, i.e. $\tau = 0.5$. The left panel in Fig.~\ref{fig:comparison_w_best_fit} shows that for a 100\% Poissonian GCE in simulated maps, the median cumulative SCD reaches values close to one near the 1-photon line. In contrast, for the simulated maps with a PS-like GCE that follows the median SCD for the real data, the median SCD over the realizations locates roughly half the GCE in flux bins to the right of the 1-photon line. The median SCD in the real \emph{Fermi} data mostly lies somewhat below the sample median of the simulated best-fit maps, but falls within the 68\% scatter. The constraints on the Poisson flux fraction $\eta_P$ provided by $h^{\boldsymbol{\nu}}$ are plotted in the right panel, as a function of the confidence level $\alpha$. As the median SCD for the real \emph{Fermi} data is slightly brighter than the sample median of the simulated maps, the resulting constraints are slightly sharper, but well within the scatter over the simulated maps. For 97.4\% of the simulated maps with a Poissonian GCE, the 95\%-confidence constraint on $\eta_P$ exceeds 95\%, in comparison to the constraint $\tilde{\eta}_P = 65.6\%$ for the real \emph{Fermi} map, corroborating the preference for a PS-like GCE component over a purely Poissonian GCE.

\subsection{Mismodeling experiments for simulated maps}
\label{sec:mismodeling_experiment}
\begin{figure*}
\centering
  \noindent
   \resizebox{1\textwidth}{!}{
    \includegraphics{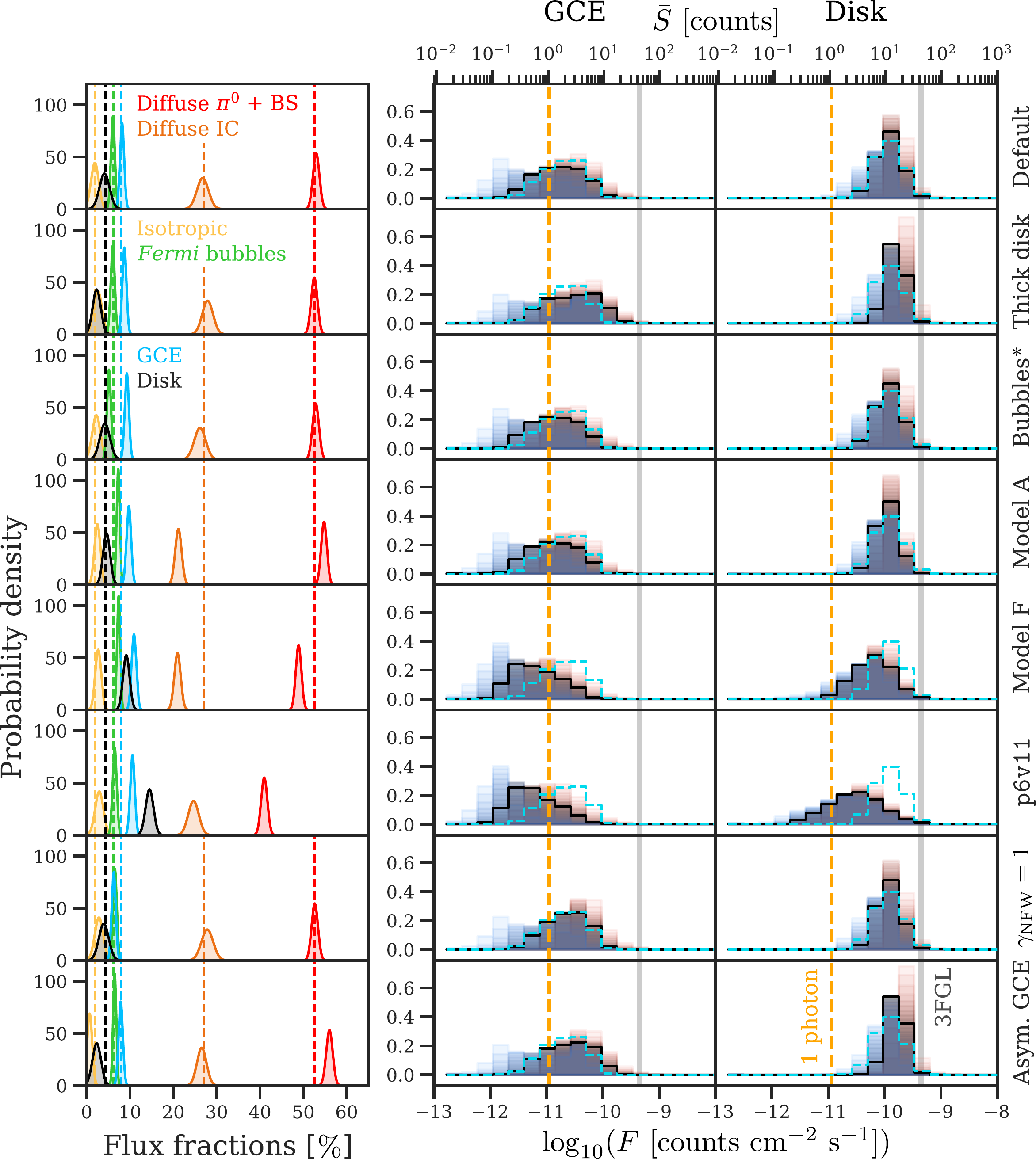}
    }
    \caption{Robustness of the NNs $f^{\boldsymbol{\omega}}$ and $g^{\boldsymbol{\varpi}}$ against mismodeling. The upper row shows the NN predictions (flux fractions and SCDs) for simulated \emph{Fermi} best-fit maps (median over 256 realizations) generated using the same templates as for the NN training. For the SCDs, we compute the median over the realizations for each quantile level $\tau$ of the cumulative histogram and plot the associated density histograms, where the colors again stand for quantile levels $\tau = 0.05 - 0.95$ (from red to blue).
    Each of the subsequent rows corresponds to a different mismodeling scenario, where one (or two in the case of diffuse mismodeling) template is replaced by an alternate template that describes the same physical process, as detailed in the text. Dashed lines mark the true flux fractions and SCDs. Whereas the NN predictions appear quite robust to
    varying the shape of the GCE, the \emph{Fermi} bubbles, and the disk, large deviations between the true and modeled diffuse model may lead to biases in the flux fractions and the SCDs (see the columns for \texttt{p6v11} and, to a somewhat lesser extent, Model~F).}
    \label{fig:mismodelling}
\end{figure*}
Since discrepancies between the templates and the true morphology of the $\gamma$-ray sources could bias the flux fractions and SCDs, or even lead to a spurious preference for a Poissonian or PS-like GCE in analyses of the \emph{Fermi} photon-count map, we study the sensitivity of our NN predictions to different sources of mismodeling in this section. We generate 256 \emph{Fermi} best-fit maps that correspond to the median flux fractions and SCDs estimated by $f^{\boldsymbol{\omega}}$ and $g^{\boldsymbol{\varpi}}$, respectively, using the same templates as for the NN training (just as in the experiment in Sec.~\ref{sec:fermi_vs_best_fit_MC}). These maps set the baseline for this example. The predicted flux fractions and SCDs (relative $F \, dN / d (\log_{10} F)$ density) are shown in the top row of Fig.~\ref{fig:mismodelling}, together with the correct labels (dashed lines). The \emph{cumulative} SCDs, which is what the NN $g^{\boldsymbol{\varpi}}$ is trained to optimize, as well as the resulting constraints on $\eta_P$ obtained from $h^{\boldsymbol{\nu}}$ are provided in Figs.~\ref{fig:mismodelling_cdf} and \ref{fig:mismodelling_eta} in App.~\ref{sec:mismodeling_cdfs}, respectively. In this case, where the templates perfectly match the data, the flux fractions are accurately recovered, and the estimated median SCDs are similar to the true histograms. 
\par Now, we consider different mismodeling scenarios by applying our NNs to maps in which a particular flux component was generated using a different template to that on which it was trained. We use 256 realizations for each scenario and take the same \emph{Fermi} best-fit flux fractions and SCDs as in the case without mismodeling. Thus, the results of this experiment display the bias arising from altering the ``truth'' (here represented by simulated \emph{Fermi} best-fit maps) while keeping our modeling fixed. The advantage of varying the truth rather than the templates used for the modeling is that it does not require retraining the NN for each scenario, which would be computationally expensive.
\par We consider the following cases, with the results shown in Fig.~\ref{fig:mismodelling}:
\begin{enumerate}
    \item \textbf{Default}: This represents the baseline case without any mismodeling. If our templates are a good model of the $\gamma$-ray sky in our ROI, the NN predictions should be close to the true values for the \emph{Fermi} map. However, note that even if our templates were a very poor description of the reality, the NN estimates for the simulated maps considered here should be similar to those for the real \emph{Fermi} map, simply because the simulated maps use the templates that the NNs were trained on, and the \emph{Fermi} best-fit parameters are the correct label. This is indeed the case: $f^{\boldsymbol{\omega}}$ correctly identifies the underlying flux fractions, and the median SCDs predicted by $g^{\boldsymbol{\varpi}}$ are similar to the truth. So, regardless of how well our templates describe the real \emph{Fermi} data, the simulated best-fit maps and the \emph{Fermi} map cause our NNs to produce (approximately) the same output. 
    
    \item \textbf{Thick disk}: For this case, we replace the thin disk template (scale height $z_s = 0.3 \ \text{kpc}$) by a thick disk template ($z_s = 1.0 \ \text{kpc}$). As a result, both the GCE and disk SCDs shift to slightly higher fluxes, while some of the disk PS flux is absorbed by the remaining templates, mostly the diffuse IC. Thus, if the thick disk were a better model for the real sky, but we use the thin disk template for the NN training, our NNs would be expected to underestimate the disk PS flux and to somewhat overestimate the PS brightness. 
    
    \item \textbf{Bubbles$^*$}: We use an alternate template for the \emph{Fermi} bubbles (where the star indicates the template is modified), which touches the Galactic Plane in the southern hemisphere (see the template delineated by the green lines in Fig.~S1 of \citetalias{List2020b}). The GCE template partially absorbs the unmodeled flux from the \emph{Fermi} bubbles, and the GCE SCD accordingly becomes slightly fainter.
    
    \item \textbf{Model~A}: Now, we turn to diffuse mismodeling. First, we replace the two template components of Model~O (pion decay + bremsstrahlung and IC) by their counterparts in Model~A. Among the models we consider, Model~A seems to be most similar to Model~O in that the effect on the SCDs is quite modest. The flux fractions of the diffuse $\pi^0 + \text{BS}$ template and the GCE are overestimated, whereas the diffuse IC flux is underestimated.
    
    \item \textbf{Model~F}: When replacing Model~O by Model~F, the NN $f^{\boldsymbol{\omega}}$ misinterprets a fraction of the diffuse flux to be disk and GCE flux. This also causes $g^{\boldsymbol{\varpi}}$ to predict fainter SCDs than the truth. Note that since we predict (and plot) the \emph{relative} SCDs, the probability mass under the histograms now corresponds to a larger total flux of the PS-like templates ($1.3 \times$ for the GCE and $2.2 \times$ for the disk). Since the NN mistakes a fraction of the diffuse flux for faint PSs, the SCDs start at a lower flux, well below the 1-photon line for the GCE template. The SCD cutoffs at the upper flux end are similar to the correct values, but the reduced density reflects the smaller \emph{relative} amount of bright PSs preferred by the NN due to the overestimated total PS flux.
    
    \item \texttt{\textbf{p6v11}}: The template \texttt{p6v11} provides a joint model for the diffuse flux from pion decay, bremsstrahlung, and IC scattering. Since it is the last official \emph{Fermi} model that does not include the \emph{Fermi} bubbles and other large-scale structures such as Loop 1, it is a popular choice for analyses of the inner Galaxy in which the \emph{Fermi} bubbles are modeled individually. However, it has been pointed out in previous studies that the hard IC component of \texttt{p6v11} may cause oversubtraction in the data \cite{Calore2015, Linden2016, Buschmann2020}. When we applied a NN trained using Model~O to simulated maps with diffuse flux described by \texttt{p6v11} in \citetalias{List2020b}, the flux ratio between the pions + bremsstrahlung and IC components was estimated to be $\sim1.4$ (see Fig.~S7 in \citetalias{List2020b}). However, both our NN and \texttt{NPTFit} favor a ratio close to $2$ (see Fig.~13) and hence a much smaller relative contribution of diffuse IC flux for the \emph{Fermi} map in our ROI, indicating a strong mismatch between \texttt{p6v11} and the preferred diffuse flux composition.
    Evaluating $f^{\boldsymbol{\omega}}$ and $g^{\boldsymbol{\varpi}}$ on simulated maps with \texttt{p6v11} flux (taken to be the sum of the best-fit \emph{Fermi} values for pion decay + bremsstrahlung and IC as determined by our NN trained on Model~O) therefore causes the NN predictions to strongly deviate from the truth: the total diffuse flux is underestimated by 14\%, and the faint disk PS flux is substantially overestimated.
    The bias that arises for the GCE SCD is very similar to the case of Model~F, and the mismatch with respect to the truth is exacerbated for the disk SCD, owing to the large fraction of diffuse flux that is misattributed to disk PSs.
    
    \item \textbf{$\gamma_\text{NFW} = 1.0$}: Here, we consider the robustness of our NN predictions against variations in the GCE morphology. We evaluate our NNs on maps with a GCE that follows an NFW-squared radial profile with $\gamma = 1.0$ instead of $\gamma = 1.2$. A small fraction of the GCE flux is absorbed by the other templates, which is unsurprising in view of $\gamma = 1.0$ modeling a less cuspy halo. The effect on the SCDs seems to be minor.
    
    \item \textbf{Asym. GCE}: Another test for the sensitivity with respect to the GCE morphology is to evaluate our NNs on maps with an asymmetric GCE template. This experiment is inspired by the findings of Refs.~\cite{Leane2020, Leane2020a} that identified a preference for a smooth asymmetric GCE in the \emph{Fermi} map with \texttt{NPTFit} in a ROI of radius $10^\circ$ when allowing the templates to float separately in the northern and southern hemisphere. We generate mock maps with an asymmetric GCE template defined as $T_\text{asym} = 2 \, T_\text{north} + T_\text{south}$ (where $T_\text{north}$ is the restriction of our default GCE template to the northern hemisphere, set to zero in the southern hemisphere, and conversely for $T_\text{south}$), yielding a north-to-south flux ratio of 2 for the GCE as found by the authors of Ref.~\cite{Leane2020} (using the diffuse model \texttt{p6v11}; see their Fig.~1), while we leave the \emph{total} GCE flux unchanged. Interestingly, the prediction for the GCE flux fraction is barely affected and the SCD for the GCE moves only very slightly to the right. Instead, the diffuse template modeling pion decay and bremsstrahlung, which is brighter in the northern hemisphere, absorbs some flux to account for the asymmetry. Also, the NN detects less faint disk PS emission, causing the disk flux fraction to decrease and the disk SCD to move to slightly brighter fluxes.
\end{enumerate}

In summary, the NN predictions are quite robust against modest deviations in the shape of the disk, the \emph{Fermi} bubbles, and the GCE, whereas strong diffuse mismodeling biases the estimated flux fractions and SCDs. With regard to the diffuse model, let us mention that the predicted SCD for the GCE shifts toward fainter fluxes when evaluating our Model~O-trained NNs on maps with diffuse flux described by Model~A, Model~F, or \texttt{p6v11}. Thus, if the diffuse flux in the \emph{Fermi} map deviated from Model~O toward any of the alternate diffuse models considered in this work, our NN prediction would be expected to overestimate the GCE flux at the faint end of the SCD, implying that in reality the flux fraction of the GCE would be somewhat smaller and the SCD brighter than our predictions, further increasing the tension with a 100\% DM explanation. The preference for a larger GCE flux when using Model~O as compared to \texttt{p6v11} has already been pointed out in Ref.~\cite[see Fig.~3]{Buschmann2020} and in \citetalias{List2020b} (see Tab.~S1). Our findings in Fig.~\ref{fig:mismodelling} also highlight that biases arising from mismodeling in inner Galaxy analyses depend on a complex interplay between the different flux components: for example, diffuse mismodeling does in fact not always lead to a spurious preference for PSs, but can also produce an overly faint SCD estimate, caused by the misattribution of diffuse flux to the GCE template. This can be contrasted with studies using \texttt{NPTFit} that have found diffuse mismodeling generates an artificial preference for brighter PSs, in particular see Ref.~\cite[Fig.~6]{Chang2019}. The discussion in Sec.~\ref{sec:cnns} about the different ways the two methods behave in the presence of mismodeling (shown for a simpler form of mismodeling in Fig.~\ref{fig:reshuffling_sketch}) suggests that this conceptual difference could also explain the different behavior observed for more complex mismodeling in a realistic setting such as considered here. Our CNN, which performs a macroscale assessment of the maps, appears to perceive the (Poissonian) diffuse flux misattributed to the GCE and disk as being fairly smooth in nature despite the mismodeling, causing the SCDs to rise at the low flux end. On the other hand, the NPTF as a microscale method is unaware of the spatial structure of the mismodeling and interprets the increased variance as an indication for PS-like emission. The difference between the SCDs inferred by NPTF and our NN approach could be a useful diagnostic for the presence of mismodeling. This point merits further exploration, although we do not pursue that here.

\subsection{Recovering artificially injected GCE flux from the \emph{Fermi} map}
\label{sec:injection}
\begin{figure}
\centering
  \noindent
   \resizebox{1\columnwidth}{!}{
    \includegraphics{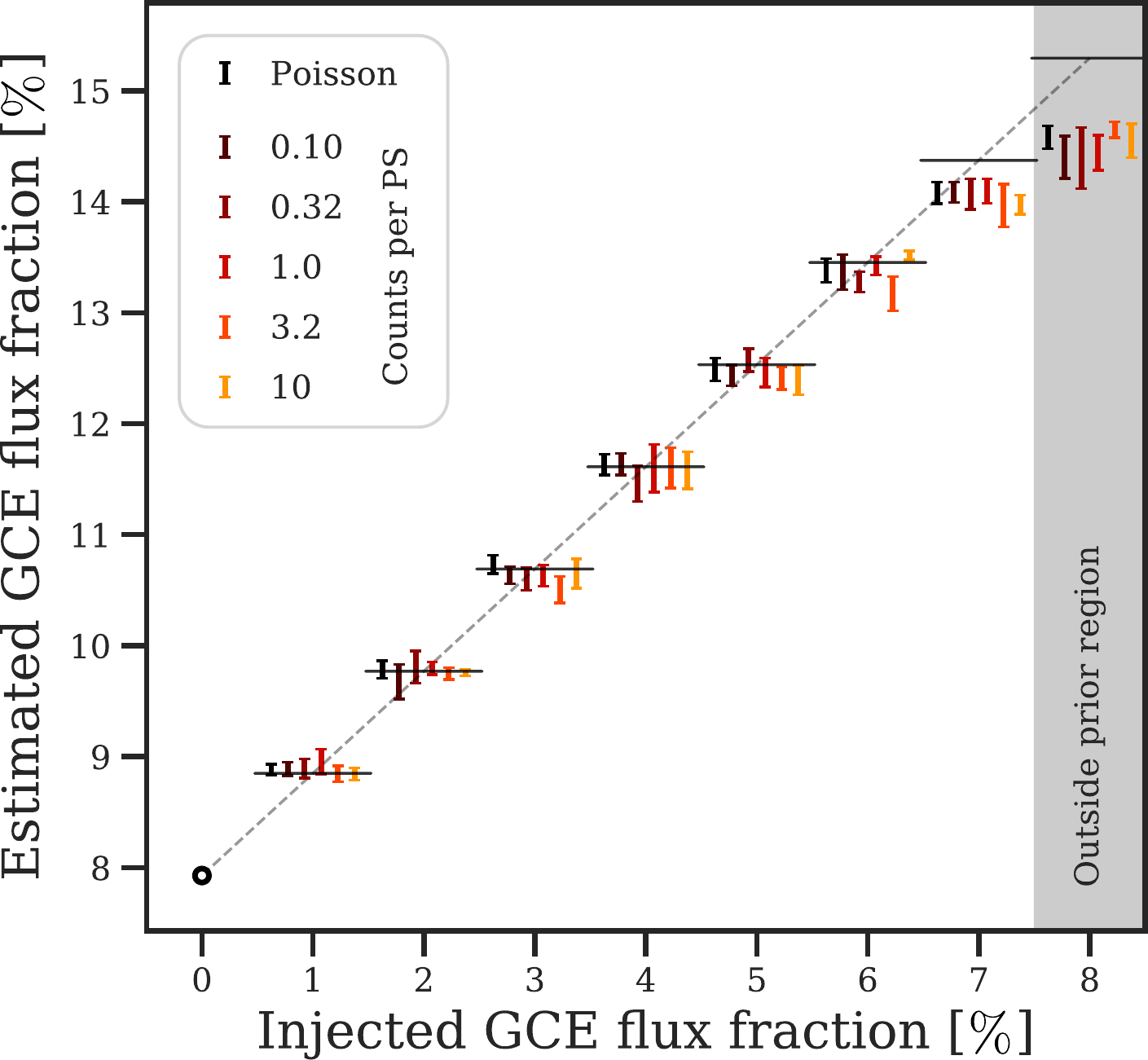}
    }
    \caption{Injected vs. estimated GCE flux fractions (post-injection) when artificially injecting GCE flux into the real \emph{Fermi} data. The first data point at 0\% injected flux corresponds to the GCE identified by our NN $f^{\boldsymbol{\omega}}$ in the original \emph{Fermi} map. We inject PS flux described by a Dirac delta $dN / dF$ located at 5 different fluxes (see the associated ``counts per PS'' in the legend), as well as Poissonian emission. The errorbars show the 68\% scatter around the median over 64 MC realizations in each case. The dashed diagonal line indicates the expected estimates given by the sum of the original GCE flux fraction and the artificially injected flux. For clarity, the 6 different cases are slightly offset horizontally around each injected flux fraction such that correct estimates lie on the horizontal lines. As long as the total (original + injected) GCE flux remains well within our prior limits, $f^{\boldsymbol{\omega}}$ accurately recovers the injected GCE flux irrespective of its nature (Poissonian / faint PSs / moderately bright PSs).}
    \label{fig:injection_FF}
\end{figure}
First considered in Ref.~\cite{Leane2019a}, the recovery of synthetic GCE flux injected into the \emph{Fermi} map is a powerful test for confirming that the results for the GCE are physical, rather than a spurious artifact resulting from oversubtraction of a GCE component or cross-talk between the GCE and non-GCE templates. In their analysis with the diffuse model \texttt{p6v11}, the authors of that work reported that even when injecting a Poissonian GCE above the \emph{Fermi} GCE level, NPTF incorrectly attributed the synthetic Poissonian GCE flux to the GCE PS template. Also, they demonstrated that the NPTF preferred an (unphysical) negative normalization for the Poissonian GCE template when allowed by the priors. More recently, Ref.~\cite{Buschmann2020} showed that replacing \texttt{p6v11} by Model~O, or alternatively applying a spherical-harmonic marginalization procedure, leads to correctly recovered flux fractions with the NPTF. 
\par In \citetalias{List2020b} (Sec.~S10, see in particular Fig.~S30), we considered the injection of both Poissonian and PS-like GCE flux into the \emph{Fermi} map, for different template choices. While we found Poissonian GCE flux to be recovered by our NN roughly as expected, (moderately) dim synthetic PS emission was frequently misattributed to the Poissonian template. In this section, we demonstrate that our novel unified approach for the Poissonian and PS GCE components enables the accurate recovery of injected GCE flux from the \emph{Fermi} map, be it Poissonian or PS-like. Importantly, the NN $f^{\boldsymbol{\omega}}$ now only needs to identify injected GCE flux as such, without distinguishing between Poisson / PS flux, and the NN $g^{\boldsymbol{\varpi}}$ assesses the brightness of the injected GCE flux while still not making a statement as to whether the injected flux is genuinely Poissonian or PS-like (rather, we address this question separately as explained in Sec.~\ref{sec:constraining_Poisson}). We consider the injection of Poissonian GCE emission, as well as GCE PSs described by a Dirac delta $dN / dF$ at fluxes $F = $ 0.11, 0.35, 1.1, 3.5, and $11 \times 10^{-11} \ \text{counts} \ \text{cm}^{-2} \ \text{s}^{-1}$, corresponding to 0.10, 0.32, 1.0, 3.2, and 10 expected counts per PS. Thus, the injected PSs span a flux range from far below the 1-photon line to the brightest GCE PSs identified by our NN in the \emph{Fermi} map.
\par Figure~\ref{fig:injection_FF} shows the injected vs. estimated GCE flux fraction (post-injection) for these 6 cases. The NN $f^{\boldsymbol{\omega}}$ accurately recovers the synthetic GCE flux in each case. For total GCE flux fractions (original + injected) close to or above the maximum GCE flux fraction contained in the training maps (as determined by our priors on the template normalizations and SCD parameters), the GCE flux fractions are slightly underestimated. Note that since the GCE counts in the training maps are composed of two GCE template maps, there are few training maps with very small (large) GCE flux fractions $\sim0$\% ($\sim15$\%), as both GCE template maps need to have a very small (large) flux for this to occur (see also Fig.~\ref{fig:flux_fraction_errors} for a typical sample of flux fractions for each template). However, this is not a flaw in our methodology, but rather the result of our narrow priors around the expected \emph{Fermi} values used for the generation of the training maps, which does not permit the analysis of maps whose composition deviates considerably from the real \emph{Fermi} data without retraining our NNs.
\par In App.~\ref{sec:injection_scds}, we show how the SCD predictions are affected by the injected GCE flux, depending on its Poissonian / PS-like nature. Also, we discuss how the constraints on the Poisson flux fraction $\eta_P$ derived by $h^{\boldsymbol{\nu}}$ from the SCDs vary as a function of the injected GCE flux in each case. In short, we find that injected flux from sources brighter (fainter) than the peak of our median SCD at $F = (3 - 4) \times 10^{-11} \ \text{counts} \ \text{cm}^{-2} \ \text{s}^{-1}$ shifts the SCD predictions to higher (lower) fluxes. In particular, we identify $F \sim 3.5 \times 10^{-11} \ \text{counts} \ \text{cm}^{-2} \ \text{s}^{-1}$ as the ``characteristic'' brightness of the GCE in the \emph{Fermi} map as judged by our NNs, which leaves the constraints on $\eta_P$ approximately unaffected.

\section{Conclusions}
\label{sec:conclusion}
In this paper, we have presented a two-step framework for a NN-based analysis of the $\gamma$-ray photon counts from the inner Galaxy. In the first step, we utilize a trained NN $f^{\boldsymbol{\omega}}$ as a template fitting tool, which yields the flux fraction of each spatial template and the associated uncertainty.
Then, we introduced a second NN $g^{\boldsymbol{\varpi}}$ to predict the SCDs of the (potentially) PS-like templates by means of the \emph{Earth Mover's Pinball Loss} that expresses the distribution over possible histograms in terms of quantiles. At this second stage, we harness the estimated flux fractions from $f^{\boldsymbol{\omega}}$ to compute a residual map, which is fed to $g^{\boldsymbol{\varpi}}$ as an additional input channel.
After validating our framework for a single isotropically distributed PS population and for \emph{Fermi} mock maps, we presented our findings for the real \emph{Fermi} map. Our NN identifies a GCE in the data, which accounts for $(7.9 \pm 0.5)$\% of the flux in our ROI. As to the SCD, we find a faint GCE that would require $\mathcal{O}(10^4)$ PSs to explain the entire GCE flux (and at least $\mathcal{O}(10^3)$ PSs to explain the brightest half of it), given that our NN $g^{\boldsymbol{\varpi}}$ assigns almost all of the flux to PSs that emit $< 10$ counts each. Our median estimate of in total 29,300 GCE PSs is broadly consistent with population studies of MSPs in the Galactic Center, for instance the 17,900 $-$ 82,200 predicted at 95\% confidence by Ref.~\cite{Ploeg2020}. Nonetheless, our results do stand in contrast with the earlier analyses (e.g. Ref.~\cite{Lee2016} which used NPTF) that suggested many of the GCE PSs lie just below the \emph{Fermi} detection threshold, which would require only several hundred PSs to explain the GCE within a radius of $10^\circ$ around the Galactic Center (the region where the presence of an excess has been firmly established, e.g. Ref.~\cite{Daylan2016}).
Uncertainties in the diffuse model play a key role here: using Model~O rather than \texttt{p6v11} as in Ref.~\cite{Lee2016}, we find a population of sources fainter than $10$ expected counts per PS with \texttt{NPTFit} as well, albeit described by a much narrower SCD that locates nearly no flux below the 1-photon line (see App.~\ref{sec:comparison_nptfit}). Whilst NPTF analyses may underestimate the power at the faint end of the SCD, especially in the presence of an instrument PSF and diffuse foregrounds \cite[Sec.~III]{Chang2019}, and steeply peaked SCDs have been found to also occur as artifacts of mismodeling \cite[Sec.~VII~B]{Leane2020a}, we note that the wider SCD that our NN $g^{\boldsymbol{\varpi}}$ prefers for the GCE is subject to systematic modeling uncertainties as well (see Sec.~\ref{sec:mismodeling_experiment}). Also, for faint PS flux that follows a narrow $dN / dF$, the uncertainty regions in the SCD predicted by $g^{\boldsymbol{\varpi}}$ may extend to neighboring bins, possibly overestimating the width of the SCD (see also App.~\ref{sec:injection_scds}).
\par Finally, we have introduced a NN-based approach for constraining the Poissonian GCE component based on the estimated SCD histogram. We have shown that for an isotropic PS population in the absence of a PSF, our NN estimator $h^{\boldsymbol{\nu}}$ yields tight constraints on the Poissonian flux fraction $\eta_P$. For example, for a population with only $0.6$ expected counts per PS, our approach allows distinguishing $\sim$~80\% of the flux from Poissonian emission at 95\% confidence, in comparison to $\sim$~91\% of the flux using the analytic likelihood, which can be exactly computed in this simple case without a PSF. When applying our approach to the real \emph{Fermi} map, the preference of $g^{\boldsymbol{\varpi}}$ for a faint SCD prevents the NN $h^{\boldsymbol{\nu}}$ from excluding a Poisson-dominated GCE at high confidence; still, for the median SCD, we obtain the constraint $\tilde{\eta}_P = 66\%$ at 95\% confidence, suggesting that an $\mathcal{O}(1)$ fraction of the GCE is due to point-like structure, which may be astrophysical sources.
\par As pointed out in much of the recent work on the GCE (e.g. Refs.~\cite{Chang2019, Leane2019a, Leane2020, Leane2020a, Buschmann2020}), the results of any GCE analysis must be interpreted with caution due to potential biases caused by mismodeling. Although modeling uncertainties -- most importantly of the Galactic foregrounds -- and, as shown in this work, the inherent degeneracy between faint PSs and Poisson emission, currently do not permit us to give a definitive answer as to whether flux from DM annihilation is present in the \emph{Fermi} map, we have demonstrated that our approach is robust against various sources of mismodeling such as a north-south asymmetry of the GCE, and is able to accurately recover artificial Poissonian and PS-like GCE flux from the \emph{Fermi} map. Let us highlight again that while mismodeling may erroneously ``flip the switch'' between DM and PSs in existing approaches that include a separate model for the two components, our unified approach entirely abandons the concept of such a switch in view of the Poissonian vs. faint PS degeneracy and instead naturally includes Poissonian emission at the low flux end of the SCD where the discriminatory power of our NN is exhausted, implying that increasing mismodeling causes an incremental shift of the prediction rather than a sudden change in the DM vs. PS preference. Whilst the recently developed Model~O, which we have used herein, provides a much better fit to the \emph{Fermi} data than diffuse models used in earlier analyses such as \texttt{p6v11}, it does not describe the data at the level of Poisson noise at energies $\lesssim 4$ GeV either (Ref.~\cite{Buschmann2020}, Fig.~17). Thus, further progress with regard to the diffuse template has the potential to considerably reduce systematic uncertainties. Moreover, next generation radio telescopes (first and foremost the Square Kilometre Array) are expected to detect many currently unresolved MSPs belonging to the putative population in the Galactic Bulge \cite{Macquart2015, Calore2016}. Another interesting approach at radio frequencies is the search for synchrotron radiation arising from DM annihilation, which can provide stringent constraints on WIMP mass and annihilation cross-section, as recently derived in Ref.~\cite{Regis2021}. At the same time, the development and improvement of analysis methods for $\gamma$-ray maps continue: as elaborated in Sec.~\ref{sec:cnns}, different methods exhibit different behavior in the presence of mismodeling. As such, a more complete and robust picture of the $\gamma$-ray emission from the inner Galaxy can be obtained by bundling multiple approaches, with discrepancies between the results providing valuable clues to possible shortcomings in the modeling.
\par With this work, we build on our deep learning-based framework in \citetalias{List2020b}, further showing (1) that NNs are able to recover the SCD of PS populations from photon-count maps and (2) how the SCD estimates can be exploited to constrain the Poissonian flux fraction $\eta_P$ using a separate NN. Regarding extensions of our work, one potential avenue is to incorporate information about the energy of the photon counts into our framework. Furthermore, equipping the templates with additional degrees of freedom enables a more flexible modeling and, in turn, more robust results. In this spirit, Ref.~\cite{Mishra-Sharma2020} showed that machine learning techniques such as Gaussian processes and normalizing flows yield promising results. Whilst we do not model any DM substructure in this work in line with previous NPTF-based analyses, it would be interesting to study the effect of DM subhalos, which can cause deviations of DM annihilation from Poisson emission, making the signal appear more PS-like (see e.g. Refs.~\cite{Somalwar2020, Runburg2021}). Finally, deep learning-based analyses have a great potential for shedding light on other regions of the sky, e.g. the $\gamma$-ray excess recently identified in M31 \cite{Karwin2019}, for which DM annihilation has also been proposed as a possible explanation \cite{Burns2020}.

\begin{acknowledgments}
We thank G. Collin, S. Mishra-Sharma, D. Shih, and T. Slatyer for comments on a draft version of this work.
FL thanks I. Bhat for fruitful discussions at an earlier stage of this project. 
NLR benefited from discussions with G. Collin and S. Mishra-Sharma related to the importance of the degeneracy between Poisson emission and dim point sources.
We also thank the anonymous referee for their feedback, which improved the quality of this work.
The authors acknowledge the National Computational Infrastructure (NCI), which is supported by the Australian Government, for providing services and computational resources on the supercomputer Gadi that have contributed to the research results reported within this paper.
The authors further acknowledge the technical assistance provided by the Sydney Informatics Hub, a Core Research Facility of the University of Sydney, and the generous allocation of resources through the Computational Grand Challenge program.
Our work also made use of resources provided by the National Energy Research Scientific Computing Center, a US Department of Energy Office of Science User Facility supported by Contract No. DE-AC02-05CH11231.
The authors thank the \emph{Fermi} collaboration for making the data publicly available.
FL is supported by the University of Sydney International Scholarship (USydIS).
NLR is supported in part by the Miller Institute for Basic Research in Science at the University of California, Berkeley, and thanks the University of Melbourne for their hospitality while this work was being completed.
\par \emph{Software}: \texttt{matplotlib} \cite{mpl}, \texttt{seaborn} \cite{seaborn}, \texttt{numpy} \cite{npy}, \texttt{scipy} \cite{scipy}, \texttt{numba} \cite{numba}, \texttt{healpy} \cite{healpy}, \texttt{Tensorflow} \cite{Abadi2016}, \texttt{Keras} \cite{keras}, \texttt{ray} \cite{ray}, \texttt{NPTFit} \cite{Mishra-Sharma2017}, \texttt{NPTFit-Sim} \cite{NPTFit-Sim}, \texttt{iminuit} \cite{iminuit}, \texttt{dill} \cite{dill}, \texttt{cloudpickle},\footnote{\href{https://github.com/cloudpipe/cloudpickle}{https://github.com/cloudpipe/cloudpickle}} \texttt{colorcet}.\footnote{\href{https://github.com/holoviz/colorcet}{https://github.com/holoviz/colorcet}} Also, we used the \texttt{arXiv} preprint repository and the free software \texttt{Inkscape}.\footnote{\href{https://inkscape.org/}{https://inkscape.org/}}
\end{acknowledgments}


\begin{thebibliography}{119}%
\makeatletter
\providecommand \@ifxundefined [1]{%
 \@ifx{#1\undefined}
}%
\providecommand \@ifnum [1]{%
 \ifnum #1\expandafter \@firstoftwo
 \else \expandafter \@secondoftwo
 \fi
}%
\providecommand \@ifx [1]{%
 \ifx #1\expandafter \@firstoftwo
 \else \expandafter \@secondoftwo
 \fi
}%
\providecommand \natexlab [1]{#1}%
\providecommand \enquote  [1]{``#1''}%
\providecommand \bibnamefont  [1]{#1}%
\providecommand \bibfnamefont [1]{#1}%
\providecommand \citenamefont [1]{#1}%
\providecommand \href@noop [0]{\@secondoftwo}%
\providecommand \href [0]{\begingroup \@sanitize@url \@href}%
\providecommand \@href[1]{\@@startlink{#1}\@@href}%
\providecommand \@@href[1]{\endgroup#1\@@endlink}%
\providecommand \@sanitize@url [0]{\catcode `\\12\catcode `\$12\catcode
  `\&12\catcode `\#12\catcode `\^12\catcode `\_12\catcode `\%12\relax}%
\providecommand \@@startlink[1]{}%
\providecommand \@@endlink[0]{}%
\providecommand \url  [0]{\begingroup\@sanitize@url \@url }%
\providecommand \@url [1]{\endgroup\@href {#1}{\urlprefix }}%
\providecommand \urlprefix  [0]{URL }%
\providecommand \Eprint [0]{\href }%
\providecommand \doibase [0]{https://doi.org/}%
\providecommand \selectlanguage [0]{\@gobble}%
\providecommand \bibinfo  [0]{\@secondoftwo}%
\providecommand \bibfield  [0]{\@secondoftwo}%
\providecommand \translation [1]{[#1]}%
\providecommand \BibitemOpen [0]{}%
\providecommand \bibitemStop [0]{}%
\providecommand \bibitemNoStop [0]{.\EOS\space}%
\providecommand \EOS [0]{\spacefactor3000\relax}%
\providecommand \BibitemShut  [1]{\csname bibitem#1\endcsname}%
\let\auto@bib@innerbib\@empty
\bibitem [{\citenamefont {Bertone}\ \emph {et~al.}(2005)\citenamefont
  {Bertone}, \citenamefont {Hooper},\ and\ \citenamefont {Silk}}]{Bertone2005}%
  \BibitemOpen
  \bibfield  {author} {\bibinfo {author} {\bibfnamefont {G.}~\bibnamefont
  {Bertone}}, \bibinfo {author} {\bibfnamefont {D.}~\bibnamefont {Hooper}},\
  and\ \bibinfo {author} {\bibfnamefont {J.}~\bibnamefont {Silk}},\ }\href
  {https://doi.org/10.1016/j.physrep.2004.08.031} {\bibfield  {journal}
  {\bibinfo  {journal} {Phys. Rep.}\ }\textbf {\bibinfo {volume} {405}},\
  \bibinfo {pages} {279} (\bibinfo {year} {2005})},\ \Eprint
  {https://arxiv.org/abs/hep-ph/0404175} {arXiv:hep-ph/0404175} \BibitemShut
  {NoStop}%
\bibitem [{\citenamefont {Aghanim}\ \emph {et~al.}(2020)\citenamefont
  {Aghanim}, \citenamefont {Akrami}, \citenamefont {Ashdown}, \citenamefont
  {Aumont}, \citenamefont {Baccigalupi}, \citenamefont {Ballardini},
  \citenamefont {Banday}, \citenamefont {Barreiro} \emph
  {et~al.}}]{PlanckCollaboration2018}%
  \BibitemOpen
  \bibfield  {author} {\bibinfo {author} {\bibfnamefont {N.}~\bibnamefont
  {Aghanim}}, \bibinfo {author} {\bibfnamefont {Y.}~\bibnamefont {Akrami}},
  \bibinfo {author} {\bibfnamefont {M.}~\bibnamefont {Ashdown}}, \bibinfo
  {author} {\bibfnamefont {J.}~\bibnamefont {Aumont}}, \bibinfo {author}
  {\bibfnamefont {C.}~\bibnamefont {Baccigalupi}}, \bibinfo {author}
  {\bibfnamefont {M.}~\bibnamefont {Ballardini}}, \bibinfo {author}
  {\bibfnamefont {A.~J.}\ \bibnamefont {Banday}}, \bibinfo {author}
  {\bibfnamefont {R.~B.}\ \bibnamefont {Barreiro}}, \emph {et~al.},\ }\href
  {https://doi.org/10.1051/0004-6361/201833910} {\bibfield  {journal} {\bibinfo
   {journal} {A\&A}\ }\textbf {\bibinfo {volume} {641}},\ \bibinfo {pages} {A6}
  (\bibinfo {year} {2020})},\ \Eprint {https://arxiv.org/abs/1807.06209}
  {arXiv:1807.06209} \BibitemShut {NoStop}%
\bibitem [{\citenamefont {Murgia}(2020)}]{Murgia:2020dzu}%
  \BibitemOpen
  \bibfield  {author} {\bibinfo {author} {\bibfnamefont {S.}~\bibnamefont
  {Murgia}},\ }\href {https://doi.org/10.1146/annurev-nucl-101916-123029}
  {\bibfield  {journal} {\bibinfo  {journal} {Ann. Rev. Nucl. Part. Sci.}\
  }\textbf {\bibinfo {volume} {70}},\ \bibinfo {pages} {455} (\bibinfo {year}
  {2020})}\BibitemShut {NoStop}%
\bibitem [{\citenamefont {Calore}\ \emph {et~al.}(2015)\citenamefont {Calore},
  \citenamefont {Cholis},\ and\ \citenamefont {Weniger}}]{Calore2015}%
  \BibitemOpen
  \bibfield  {author} {\bibinfo {author} {\bibfnamefont {F.}~\bibnamefont
  {Calore}}, \bibinfo {author} {\bibfnamefont {I.}~\bibnamefont {Cholis}},\
  and\ \bibinfo {author} {\bibfnamefont {C.}~\bibnamefont {Weniger}},\ }\href
  {https://doi.org/10.1088/1475-7516/2015/03/038} {\bibfield  {journal}
  {\bibinfo  {journal} {JCAP}\ }\textbf {\bibinfo {volume} {2015}}\bibfield
  {number} {\bibinfo  {number} { (03)},\ \bibinfo {pages} {038}},\ }\Eprint
  {https://arxiv.org/abs/1409.0042} {arXiv:1409.0042} \BibitemShut {NoStop}%
\bibitem [{\citenamefont {Daylan}\ \emph {et~al.}(2016)\citenamefont {Daylan},
  \citenamefont {Finkbeiner}, \citenamefont {Hooper}, \citenamefont {Linden},
  \citenamefont {Portillo}, \citenamefont {Rodd},\ and\ \citenamefont
  {Slatyer}}]{Daylan2016}%
  \BibitemOpen
  \bibfield  {author} {\bibinfo {author} {\bibfnamefont {T.}~\bibnamefont
  {Daylan}}, \bibinfo {author} {\bibfnamefont {D.~P.}\ \bibnamefont
  {Finkbeiner}}, \bibinfo {author} {\bibfnamefont {D.}~\bibnamefont {Hooper}},
  \bibinfo {author} {\bibfnamefont {T.}~\bibnamefont {Linden}}, \bibinfo
  {author} {\bibfnamefont {S.~K.}\ \bibnamefont {Portillo}}, \bibinfo {author}
  {\bibfnamefont {N.~L.}\ \bibnamefont {Rodd}},\ and\ \bibinfo {author}
  {\bibfnamefont {T.~R.}\ \bibnamefont {Slatyer}},\ }\href
  {https://doi.org/10.1016/j.dark.2015.12.005} {\bibfield  {journal} {\bibinfo
  {journal} {Phys. Dark Univ.}\ }\textbf {\bibinfo {volume} {12}},\ \bibinfo
  {pages} {1} (\bibinfo {year} {2016})},\ \Eprint
  {https://arxiv.org/abs/1402.6703} {arXiv:1402.6703} \BibitemShut {NoStop}%
\bibitem [{\citenamefont {Goodenough}\ and\ \citenamefont
  {Hooper}(2009)}]{Goodenough2009}%
  \BibitemOpen
  \bibfield  {author} {\bibinfo {author} {\bibfnamefont {L.}~\bibnamefont
  {Goodenough}}\ and\ \bibinfo {author} {\bibfnamefont {D.}~\bibnamefont
  {Hooper}},\ }\href {http://arxiv.org/abs/0910.2998} {\bibfield  {journal}
  {\bibinfo  {journal} {preprint (arXiv:0910.2998)}\ } (\bibinfo {year}
  {2009})}\BibitemShut {NoStop}%
\bibitem [{\citenamefont {Hooper}\ and\ \citenamefont
  {Goodenough}(2011)}]{Hooper:2010mq}%
  \BibitemOpen
  \bibfield  {author} {\bibinfo {author} {\bibfnamefont {D.}~\bibnamefont
  {Hooper}}\ and\ \bibinfo {author} {\bibfnamefont {L.}~\bibnamefont
  {Goodenough}},\ }\href {https://doi.org/10.1016/j.physletb.2011.02.029}
  {\bibfield  {journal} {\bibinfo  {journal} {Phys. Lett. B}\ }\textbf
  {\bibinfo {volume} {697}},\ \bibinfo {pages} {412} (\bibinfo {year}
  {2011})},\ \Eprint {https://arxiv.org/abs/1010.2752} {arXiv:1010.2752}
  \BibitemShut {NoStop}%
\bibitem [{\citenamefont {Hooper}\ and\ \citenamefont
  {Linden}(2011)}]{Hooper:2011ti}%
  \BibitemOpen
  \bibfield  {author} {\bibinfo {author} {\bibfnamefont {D.}~\bibnamefont
  {Hooper}}\ and\ \bibinfo {author} {\bibfnamefont {T.}~\bibnamefont
  {Linden}},\ }\href {https://doi.org/10.1103/PhysRevD.84.123005} {\bibfield
  {journal} {\bibinfo  {journal} {Phys. Rev. D}\ }\textbf {\bibinfo {volume}
  {84}},\ \bibinfo {pages} {123005} (\bibinfo {year} {2011})},\ \Eprint
  {https://arxiv.org/abs/1110.0006} {arXiv:1110.0006} \BibitemShut {NoStop}%
\bibitem [{\citenamefont {Abazajian}\ and\ \citenamefont
  {Kaplinghat}(2012)}]{Abazajian:2012pn}%
  \BibitemOpen
  \bibfield  {author} {\bibinfo {author} {\bibfnamefont {K.~N.}\ \bibnamefont
  {Abazajian}}\ and\ \bibinfo {author} {\bibfnamefont {M.}~\bibnamefont
  {Kaplinghat}},\ }\href {https://doi.org/10.1103/PhysRevD.86.083511}
  {\bibfield  {journal} {\bibinfo  {journal} {Phys. Rev. D}\ }\textbf {\bibinfo
  {volume} {86}},\ \bibinfo {pages} {083511} (\bibinfo {year} {2012})},\
  \bibinfo {note} {[Erratum: Phys. Rev. D 87, 129902 (2013)]},\ \Eprint
  {https://arxiv.org/abs/1207.6047} {arXiv:1207.6047} \BibitemShut {NoStop}%
\bibitem [{\citenamefont {Hooper}\ and\ \citenamefont
  {Slatyer}(2013)}]{Hooper:2013rwa}%
  \BibitemOpen
  \bibfield  {author} {\bibinfo {author} {\bibfnamefont {D.}~\bibnamefont
  {Hooper}}\ and\ \bibinfo {author} {\bibfnamefont {T.~R.}\ \bibnamefont
  {Slatyer}},\ }\href {https://doi.org/10.1016/j.dark.2013.06.003} {\bibfield
  {journal} {\bibinfo  {journal} {Phys. Dark Univ.}\ }\textbf {\bibinfo
  {volume} {2}},\ \bibinfo {pages} {118} (\bibinfo {year} {2013})},\ \Eprint
  {https://arxiv.org/abs/1302.6589} {arXiv:1302.6589} \BibitemShut {NoStop}%
\bibitem [{\citenamefont {Gordon}\ and\ \citenamefont
  {Macias}(2013)}]{Gordon:2013vta}%
  \BibitemOpen
  \bibfield  {author} {\bibinfo {author} {\bibfnamefont {C.}~\bibnamefont
  {Gordon}}\ and\ \bibinfo {author} {\bibfnamefont {O.}~\bibnamefont
  {Macias}},\ }\href {https://doi.org/10.1103/PhysRevD.88.083521} {\bibfield
  {journal} {\bibinfo  {journal} {Phys. Rev. D}\ }\textbf {\bibinfo {volume}
  {88}},\ \bibinfo {pages} {083521} (\bibinfo {year} {2013})},\ \Eprint
  {https://arxiv.org/abs/1306.5725} {arXiv:1306.5725} \BibitemShut {NoStop}%
\bibitem [{\citenamefont {Abazajian}\ \emph {et~al.}(2014)\citenamefont
  {Abazajian}, \citenamefont {Canac}, \citenamefont {Horiuchi},\ and\
  \citenamefont {Kaplinghat}}]{Abazajian2014}%
  \BibitemOpen
  \bibfield  {author} {\bibinfo {author} {\bibfnamefont {K.~N.}\ \bibnamefont
  {Abazajian}}, \bibinfo {author} {\bibfnamefont {N.}~\bibnamefont {Canac}},
  \bibinfo {author} {\bibfnamefont {S.}~\bibnamefont {Horiuchi}},\ and\
  \bibinfo {author} {\bibfnamefont {M.}~\bibnamefont {Kaplinghat}},\ }\href
  {https://doi.org/10.1103/PhysRevD.90.023526} {\bibfield  {journal} {\bibinfo
  {journal} {Phys. Rev. D}\ }\textbf {\bibinfo {volume} {90}},\ \bibinfo
  {pages} {023526} (\bibinfo {year} {2014})},\ \Eprint
  {https://arxiv.org/abs/1402.4090} {arXiv:1402.4090} \BibitemShut {NoStop}%
\bibitem [{\citenamefont {Abazajian}\ \emph {et~al.}(2015)\citenamefont
  {Abazajian}, \citenamefont {Canac}, \citenamefont {Horiuchi}, \citenamefont
  {Kaplinghat},\ and\ \citenamefont {Kwa}}]{Abazajian:2014hsa}%
  \BibitemOpen
  \bibfield  {author} {\bibinfo {author} {\bibfnamefont {K.~N.}\ \bibnamefont
  {Abazajian}}, \bibinfo {author} {\bibfnamefont {N.}~\bibnamefont {Canac}},
  \bibinfo {author} {\bibfnamefont {S.}~\bibnamefont {Horiuchi}}, \bibinfo
  {author} {\bibfnamefont {M.}~\bibnamefont {Kaplinghat}},\ and\ \bibinfo
  {author} {\bibfnamefont {A.}~\bibnamefont {Kwa}},\ }\href
  {https://doi.org/10.1088/1475-7516/2015/07/013} {\bibfield  {journal}
  {\bibinfo  {journal} {JCAP}\ }\textbf {\bibinfo {volume} {2015}}\bibfield
  {number} {\bibinfo  {number} { (07)},\ \bibinfo {pages} {013}},\ }\Eprint
  {https://arxiv.org/abs/1410.6168} {arXiv:1410.6168} \BibitemShut {NoStop}%
\bibitem [{\citenamefont {Ajello}\ \emph {et~al.}(2016)\citenamefont {Ajello},
  \citenamefont {Albert}, \citenamefont {Atwood}, \citenamefont {Barbiellini},
  \citenamefont {Bastieri}, \citenamefont {Bechtol}, \citenamefont
  {Bellazzini}, \citenamefont {Bissaldi} \emph {et~al.}}]{Ajello2016}%
  \BibitemOpen
  \bibfield  {author} {\bibinfo {author} {\bibfnamefont {M.}~\bibnamefont
  {Ajello}}, \bibinfo {author} {\bibfnamefont {A.}~\bibnamefont {Albert}},
  \bibinfo {author} {\bibfnamefont {W.~B.}\ \bibnamefont {Atwood}}, \bibinfo
  {author} {\bibfnamefont {G.}~\bibnamefont {Barbiellini}}, \bibinfo {author}
  {\bibfnamefont {D.}~\bibnamefont {Bastieri}}, \bibinfo {author}
  {\bibfnamefont {K.}~\bibnamefont {Bechtol}}, \bibinfo {author} {\bibfnamefont
  {R.}~\bibnamefont {Bellazzini}}, \bibinfo {author} {\bibfnamefont
  {E.}~\bibnamefont {Bissaldi}}, \emph {et~al.},\ }\href
  {https://doi.org/10.3847/0004-637X/819/1/44} {\bibfield  {journal} {\bibinfo
  {journal} {ApJ}\ }\textbf {\bibinfo {volume} {819}},\ \bibinfo {pages} {44}
  (\bibinfo {year} {2016})},\ \Eprint {https://arxiv.org/abs/1511.02938}
  {arXiv:1511.02938} \BibitemShut {NoStop}%
\bibitem [{\citenamefont {Linden}\ \emph {et~al.}(2016)\citenamefont {Linden},
  \citenamefont {Rodd}, \citenamefont {Safdi},\ and\ \citenamefont
  {Slatyer}}]{Linden2016}%
  \BibitemOpen
  \bibfield  {author} {\bibinfo {author} {\bibfnamefont {T.}~\bibnamefont
  {Linden}}, \bibinfo {author} {\bibfnamefont {N.~L.}\ \bibnamefont {Rodd}},
  \bibinfo {author} {\bibfnamefont {B.~R.}\ \bibnamefont {Safdi}},\ and\
  \bibinfo {author} {\bibfnamefont {T.~R.}\ \bibnamefont {Slatyer}},\ }\href
  {https://doi.org/10.1103/PhysRevD.94.103013} {\bibfield  {journal} {\bibinfo
  {journal} {Phys. Rev. D}\ }\textbf {\bibinfo {volume} {94}},\ \bibinfo
  {pages} {103013} (\bibinfo {year} {2016})},\ \Eprint
  {https://arxiv.org/abs/1604.01026} {arXiv:1604.01026} \BibitemShut {NoStop}%
\bibitem [{\citenamefont {Macias}\ \emph {et~al.}(2018)\citenamefont {Macias},
  \citenamefont {Gordon}, \citenamefont {Crocker}, \citenamefont {Coleman},
  \citenamefont {Paterson}, \citenamefont {Horiuchi},\ and\ \citenamefont
  {Pohl}}]{Macias:2016nev}%
  \BibitemOpen
  \bibfield  {author} {\bibinfo {author} {\bibfnamefont {O.}~\bibnamefont
  {Macias}}, \bibinfo {author} {\bibfnamefont {C.}~\bibnamefont {Gordon}},
  \bibinfo {author} {\bibfnamefont {R.~M.}\ \bibnamefont {Crocker}}, \bibinfo
  {author} {\bibfnamefont {B.}~\bibnamefont {Coleman}}, \bibinfo {author}
  {\bibfnamefont {D.}~\bibnamefont {Paterson}}, \bibinfo {author}
  {\bibfnamefont {S.}~\bibnamefont {Horiuchi}},\ and\ \bibinfo {author}
  {\bibfnamefont {M.}~\bibnamefont {Pohl}},\ }\href
  {https://doi.org/10.1038/s41550-018-0414-3} {\bibfield  {journal} {\bibinfo
  {journal} {Nat. Astron.}\ }\textbf {\bibinfo {volume} {2}},\ \bibinfo {pages}
  {387} (\bibinfo {year} {2018})},\ \Eprint {https://arxiv.org/abs/1611.06644}
  {arXiv:1611.06644} \BibitemShut {NoStop}%
\bibitem [{\citenamefont {Clark}\ \emph {et~al.}(2018)\citenamefont {Clark},
  \citenamefont {Scott}, \citenamefont {Trotta},\ and\ \citenamefont
  {Lewis}}]{Clark2018}%
  \BibitemOpen
  \bibfield  {author} {\bibinfo {author} {\bibfnamefont {H.~A.}\ \bibnamefont
  {Clark}}, \bibinfo {author} {\bibfnamefont {P.}~\bibnamefont {Scott}},
  \bibinfo {author} {\bibfnamefont {R.}~\bibnamefont {Trotta}},\ and\ \bibinfo
  {author} {\bibfnamefont {G.~F.}\ \bibnamefont {Lewis}},\ }\href
  {https://doi.org/10.1088/1475-7516/2018/07/060} {\bibfield  {journal}
  {\bibinfo  {journal} {JCAP}\ }\textbf {\bibinfo {volume} {2018}}\bibfield
  {number} {\bibinfo  {number} { (07)},\ \bibinfo {pages} {060}},\ }\Eprint
  {https://arxiv.org/abs/1612.01539} {arXiv:1612.01539} \BibitemShut {NoStop}%
\bibitem [{\citenamefont {Mirabal}(2013)}]{Mirabal2013}%
  \BibitemOpen
  \bibfield  {author} {\bibinfo {author} {\bibfnamefont {N.}~\bibnamefont
  {Mirabal}},\ }\href {https://doi.org/10.1093/mnras/stt1740} {\bibfield
  {journal} {\bibinfo  {journal} {MNRAS}\ }\textbf {\bibinfo {volume} {436}},\
  \bibinfo {pages} {2461} (\bibinfo {year} {2013})},\ \Eprint
  {https://arxiv.org/abs/1309.3428} {arXiv:1309.3428} \BibitemShut {NoStop}%
\bibitem [{\citenamefont {Petrovi{\'c}}\ \emph {et~al.}(2015)\citenamefont
  {Petrovi{\'c}}, \citenamefont {Serpico},\ and\ \citenamefont
  {Zaharijas}}]{Petrovic:2014xra}%
  \BibitemOpen
  \bibfield  {author} {\bibinfo {author} {\bibfnamefont {J.}~\bibnamefont
  {Petrovi{\'c}}}, \bibinfo {author} {\bibfnamefont {P.~D.}\ \bibnamefont
  {Serpico}},\ and\ \bibinfo {author} {\bibfnamefont {G.}~\bibnamefont
  {Zaharijas}},\ }\href {https://doi.org/10.1088/1475-7516/2015/02/023}
  {\bibfield  {journal} {\bibinfo  {journal} {JCAP}\ }\textbf {\bibinfo
  {volume} {2015}}\bibfield  {number} {\bibinfo  {number} { (02)},\ \bibinfo
  {pages} {023}},\ }\Eprint {https://arxiv.org/abs/1411.2980} {arXiv:1411.2980}
  \BibitemShut {NoStop}%
\bibitem [{\citenamefont {Yuan}\ and\ \citenamefont
  {Ioka}(2015)}]{Yuan:2014yda}%
  \BibitemOpen
  \bibfield  {author} {\bibinfo {author} {\bibfnamefont {Q.}~\bibnamefont
  {Yuan}}\ and\ \bibinfo {author} {\bibfnamefont {K.}~\bibnamefont {Ioka}},\
  }\href {https://doi.org/10.1088/0004-637X/802/2/124} {\bibfield  {journal}
  {\bibinfo  {journal} {ApJ}\ }\textbf {\bibinfo {volume} {802}},\ \bibinfo
  {pages} {124} (\bibinfo {year} {2015})},\ \Eprint
  {https://arxiv.org/abs/1411.4363} {arXiv:1411.4363} \BibitemShut {NoStop}%
\bibitem [{\citenamefont {Brandt}\ and\ \citenamefont
  {Kocsis}(2015)}]{Brandt:2015ula}%
  \BibitemOpen
  \bibfield  {author} {\bibinfo {author} {\bibfnamefont {T.~D.}\ \bibnamefont
  {Brandt}}\ and\ \bibinfo {author} {\bibfnamefont {B.}~\bibnamefont
  {Kocsis}},\ }\href {https://doi.org/10.1088/0004-637X/812/1/15} {\bibfield
  {journal} {\bibinfo  {journal} {ApJ}\ }\textbf {\bibinfo {volume} {812}},\
  \bibinfo {pages} {15} (\bibinfo {year} {2015})},\ \Eprint
  {https://arxiv.org/abs/1507.05616} {arXiv:1507.05616} \BibitemShut {NoStop}%
\bibitem [{\citenamefont {O'Leary}\ \emph {et~al.}(2015)\citenamefont
  {O'Leary}, \citenamefont {Kistler}, \citenamefont {Kerr},\ and\ \citenamefont
  {Dexter}}]{OLeary2015}%
  \BibitemOpen
  \bibfield  {author} {\bibinfo {author} {\bibfnamefont {R.~M.}\ \bibnamefont
  {O'Leary}}, \bibinfo {author} {\bibfnamefont {M.~D.}\ \bibnamefont
  {Kistler}}, \bibinfo {author} {\bibfnamefont {M.}~\bibnamefont {Kerr}},\ and\
  \bibinfo {author} {\bibfnamefont {J.}~\bibnamefont {Dexter}},\ }\href
  {http://arxiv.org/abs/1504.02477} {\bibfield  {journal} {\bibinfo  {journal}
  {preprint (arXiv:1504.02477)}\ } (\bibinfo {year} {2015})}\BibitemShut
  {NoStop}%
\bibitem [{\citenamefont {Carlson}\ and\ \citenamefont
  {Profumo}(2014)}]{Carlson2014}%
  \BibitemOpen
  \bibfield  {author} {\bibinfo {author} {\bibfnamefont {E.}~\bibnamefont
  {Carlson}}\ and\ \bibinfo {author} {\bibfnamefont {S.}~\bibnamefont
  {Profumo}},\ }\href {https://doi.org/10.1103/PhysRevD.90.023015} {\bibfield
  {journal} {\bibinfo  {journal} {Phys. Rev. D}\ }\textbf {\bibinfo {volume}
  {90}},\ \bibinfo {pages} {023015} (\bibinfo {year} {2014})},\ \Eprint
  {https://arxiv.org/abs/1405.7685} {arXiv:1405.7685} \BibitemShut {NoStop}%
\bibitem [{\citenamefont {Petrovi{\'{c}}}\ \emph {et~al.}(2014)\citenamefont
  {Petrovi{\'{c}}}, \citenamefont {Serpico},\ and\ \citenamefont
  {Zaharija{\v{s}}}}]{Petrovic2014}%
  \BibitemOpen
  \bibfield  {author} {\bibinfo {author} {\bibfnamefont {J.}~\bibnamefont
  {Petrovi{\'{c}}}}, \bibinfo {author} {\bibfnamefont {P.~D.}\ \bibnamefont
  {Serpico}},\ and\ \bibinfo {author} {\bibfnamefont {G.}~\bibnamefont
  {Zaharija{\v{s}}}},\ }\href {https://doi.org/10.1088/1475-7516/2014/10/052}
  {\bibfield  {journal} {\bibinfo  {journal} {JCAP}\ }\textbf {\bibinfo
  {volume} {2014}}\bibfield  {number} {\bibinfo  {number} { (10)},\ \bibinfo
  {pages} {052}},\ }\Eprint {https://arxiv.org/abs/1405.7928} {arXiv:1405.7928}
  \BibitemShut {NoStop}%
\bibitem [{\citenamefont {Cholis}\ \emph {et~al.}(2015)\citenamefont {Cholis},
  \citenamefont {Evoli}, \citenamefont {Calore}, \citenamefont {Linden},
  \citenamefont {Weniger},\ and\ \citenamefont {Hooper}}]{Cholis2015}%
  \BibitemOpen
  \bibfield  {author} {\bibinfo {author} {\bibfnamefont {I.}~\bibnamefont
  {Cholis}}, \bibinfo {author} {\bibfnamefont {C.}~\bibnamefont {Evoli}},
  \bibinfo {author} {\bibfnamefont {F.}~\bibnamefont {Calore}}, \bibinfo
  {author} {\bibfnamefont {T.}~\bibnamefont {Linden}}, \bibinfo {author}
  {\bibfnamefont {C.}~\bibnamefont {Weniger}},\ and\ \bibinfo {author}
  {\bibfnamefont {D.}~\bibnamefont {Hooper}},\ }\href
  {https://doi.org/10.1088/1475-7516/2015/12/005} {\bibfield  {journal}
  {\bibinfo  {journal} {JCAP}\ }\textbf {\bibinfo {volume} {2015}}\bibfield
  {number} {\bibinfo  {number} { (12)}},\ }\Eprint
  {https://arxiv.org/abs/1506.05119} {arXiv:1506.05119} \BibitemShut {NoStop}%
\bibitem [{\citenamefont {Ploeg}\ \emph {et~al.}(2017)\citenamefont {Ploeg},
  \citenamefont {Gordon}, \citenamefont {Crocker},\ and\ \citenamefont
  {Macias}}]{Ploeg2017}%
  \BibitemOpen
  \bibfield  {author} {\bibinfo {author} {\bibfnamefont {H.}~\bibnamefont
  {Ploeg}}, \bibinfo {author} {\bibfnamefont {C.}~\bibnamefont {Gordon}},
  \bibinfo {author} {\bibfnamefont {R.}~\bibnamefont {Crocker}},\ and\ \bibinfo
  {author} {\bibfnamefont {O.}~\bibnamefont {Macias}},\ }\href
  {https://doi.org/10.1088/1475-7516/2017/08/015} {\bibfield  {journal}
  {\bibinfo  {journal} {JCAP}\ }\textbf {\bibinfo {volume} {2017}}\bibfield
  {number} {\bibinfo  {number} { (08)},\ \bibinfo {pages} {015}},\ }\Eprint
  {https://arxiv.org/abs/1705.00806} {arXiv:1705.00806} \BibitemShut {NoStop}%
\bibitem [{\citenamefont {Bartels}\ \emph {et~al.}(2018)\citenamefont
  {Bartels}, \citenamefont {Storm}, \citenamefont {Weniger},\ and\
  \citenamefont {Calore}}]{Bartels:2017vsx}%
  \BibitemOpen
  \bibfield  {author} {\bibinfo {author} {\bibfnamefont {R.}~\bibnamefont
  {Bartels}}, \bibinfo {author} {\bibfnamefont {E.}~\bibnamefont {Storm}},
  \bibinfo {author} {\bibfnamefont {C.}~\bibnamefont {Weniger}},\ and\ \bibinfo
  {author} {\bibfnamefont {F.}~\bibnamefont {Calore}},\ }\href
  {https://doi.org/10.1038/s41550-018-0531-z} {\bibfield  {journal} {\bibinfo
  {journal} {Nat. Astron.}\ }\textbf {\bibinfo {volume} {2}},\ \bibinfo {pages}
  {819} (\bibinfo {year} {2018})},\ \Eprint {https://arxiv.org/abs/1711.04778}
  {arXiv:1711.04778} \BibitemShut {NoStop}%
\bibitem [{\citenamefont {Macias}\ \emph {et~al.}(2019)\citenamefont {Macias},
  \citenamefont {Horiuchi}, \citenamefont {Kaplinghat}, \citenamefont {Gordon},
  \citenamefont {Crocker},\ and\ \citenamefont {Nataf}}]{Macias:2019omb}%
  \BibitemOpen
  \bibfield  {author} {\bibinfo {author} {\bibfnamefont {O.}~\bibnamefont
  {Macias}}, \bibinfo {author} {\bibfnamefont {S.}~\bibnamefont {Horiuchi}},
  \bibinfo {author} {\bibfnamefont {M.}~\bibnamefont {Kaplinghat}}, \bibinfo
  {author} {\bibfnamefont {C.}~\bibnamefont {Gordon}}, \bibinfo {author}
  {\bibfnamefont {R.~M.}\ \bibnamefont {Crocker}},\ and\ \bibinfo {author}
  {\bibfnamefont {D.~M.}\ \bibnamefont {Nataf}},\ }\href
  {https://doi.org/10.1088/1475-7516/2019/09/042} {\bibfield  {journal}
  {\bibinfo  {journal} {JCAP}\ }\textbf {\bibinfo {volume} {2019}}\bibfield
  {number} {\bibinfo  {number} { (09)},\ \bibinfo {pages} {042}},\ }\Eprint
  {https://arxiv.org/abs/1901.03822} {arXiv:1901.03822} \BibitemShut {NoStop}%
\bibitem [{\citenamefont {Abazajian}\ \emph {et~al.}(2020)\citenamefont
  {Abazajian}, \citenamefont {Horiuchi}, \citenamefont {Kaplinghat},
  \citenamefont {Keeley},\ and\ \citenamefont {Macias}}]{Abazajian2020}%
  \BibitemOpen
  \bibfield  {author} {\bibinfo {author} {\bibfnamefont {K.~N.}\ \bibnamefont
  {Abazajian}}, \bibinfo {author} {\bibfnamefont {S.}~\bibnamefont {Horiuchi}},
  \bibinfo {author} {\bibfnamefont {M.}~\bibnamefont {Kaplinghat}}, \bibinfo
  {author} {\bibfnamefont {R.~E.}\ \bibnamefont {Keeley}},\ and\ \bibinfo
  {author} {\bibfnamefont {O.}~\bibnamefont {Macias}},\ }\href
  {https://doi.org/10.1103/PhysRevD.102.043012} {\bibfield  {journal} {\bibinfo
   {journal} {Phys. Rev. D}\ }\textbf {\bibinfo {volume} {102}},\ \bibinfo
  {pages} {043012} (\bibinfo {year} {2020})},\ \Eprint
  {https://arxiv.org/abs/2003.10416} {arXiv:2003.10416} \BibitemShut {NoStop}%
\bibitem [{\citenamefont {Calore}\ \emph {et~al.}(2021)\citenamefont {Calore},
  \citenamefont {Donato},\ and\ \citenamefont {Manconi}}]{Calore2021}%
  \BibitemOpen
  \bibfield  {author} {\bibinfo {author} {\bibfnamefont {F.}~\bibnamefont
  {Calore}}, \bibinfo {author} {\bibfnamefont {F.}~\bibnamefont {Donato}},\
  and\ \bibinfo {author} {\bibfnamefont {S.}~\bibnamefont {Manconi}},\ }\href
  {http://arxiv.org/abs/2102.12497} {\bibfield  {journal} {\bibinfo  {journal}
  {preprint (arXiv:2102.12497)}\ } (\bibinfo {year} {2021})}\BibitemShut
  {NoStop}%
\bibitem [{\citenamefont {Di~Mauro}(2020)}]{DiMauro:2020rcr}%
  \BibitemOpen
  \bibfield  {author} {\bibinfo {author} {\bibfnamefont {M.}~\bibnamefont
  {Di~Mauro}},\ }\href {https://doi.org/10.1103/PhysRevD.102.103013} {\bibfield
   {journal} {\bibinfo  {journal} {Phys. Rev. D}\ }\textbf {\bibinfo {volume}
  {102}},\ \bibinfo {pages} {103013} (\bibinfo {year} {2020})},\ \Eprint
  {https://arxiv.org/abs/2010.02231} {arXiv:2010.02231} \BibitemShut {NoStop}%
\bibitem [{\citenamefont {{Di Mauro}}(2021)}]{DiMauro2021}%
  \BibitemOpen
  \bibfield  {author} {\bibinfo {author} {\bibfnamefont {M.}~\bibnamefont {{Di
  Mauro}}},\ }\href {https://doi.org/10.1103/PhysRevD.103.063029} {\bibfield
  {journal} {\bibinfo  {journal} {Phys. Rev. D}\ }\textbf {\bibinfo {volume}
  {103}},\ \bibinfo {pages} {063029} (\bibinfo {year} {2021})},\ \Eprint
  {https://arxiv.org/abs/2101.04694} {arXiv:2101.04694} \BibitemShut {NoStop}%
\bibitem [{\citenamefont {Lee}\ \emph {et~al.}(2016)\citenamefont {Lee},
  \citenamefont {Lisanti}, \citenamefont {Safdi}, \citenamefont {Slatyer},\
  and\ \citenamefont {Xue}}]{Lee2016}%
  \BibitemOpen
  \bibfield  {author} {\bibinfo {author} {\bibfnamefont {S.~K.}\ \bibnamefont
  {Lee}}, \bibinfo {author} {\bibfnamefont {M.}~\bibnamefont {Lisanti}},
  \bibinfo {author} {\bibfnamefont {B.~R.}\ \bibnamefont {Safdi}}, \bibinfo
  {author} {\bibfnamefont {T.~R.}\ \bibnamefont {Slatyer}},\ and\ \bibinfo
  {author} {\bibfnamefont {W.}~\bibnamefont {Xue}},\ }\href
  {https://doi.org/10.1103/PhysRevLett.116.051103} {\bibfield  {journal}
  {\bibinfo  {journal} {Phys. Rev. Lett.}\ }\textbf {\bibinfo {volume} {116}},\
  \bibinfo {pages} {051103} (\bibinfo {year} {2016})},\ \Eprint
  {https://arxiv.org/abs/1506.05124} {arXiv:1506.05124} \BibitemShut {NoStop}%
\bibitem [{\citenamefont {Mishra-Sharma}\ \emph {et~al.}(2017)\citenamefont
  {Mishra-Sharma}, \citenamefont {Rodd},\ and\ \citenamefont
  {Safdi}}]{Mishra-Sharma2017}%
  \BibitemOpen
  \bibfield  {author} {\bibinfo {author} {\bibfnamefont {S.}~\bibnamefont
  {Mishra-Sharma}}, \bibinfo {author} {\bibfnamefont {N.~L.}\ \bibnamefont
  {Rodd}},\ and\ \bibinfo {author} {\bibfnamefont {B.~R.}\ \bibnamefont
  {Safdi}},\ }\href {https://doi.org/10.3847/1538-3881/aa6d5f} {\bibfield
  {journal} {\bibinfo  {journal} {The Astronomical Journal}\ }\textbf {\bibinfo
  {volume} {153}},\ \bibinfo {pages} {253} (\bibinfo {year} {2017})},\ \Eprint
  {https://arxiv.org/abs/1612.03173} {arXiv:1612.03173} \BibitemShut {NoStop}%
\bibitem [{\citenamefont {Malyshev}\ and\ \citenamefont
  {Hogg}(2011)}]{Malyshev:2011zi}%
  \BibitemOpen
  \bibfield  {author} {\bibinfo {author} {\bibfnamefont {D.}~\bibnamefont
  {Malyshev}}\ and\ \bibinfo {author} {\bibfnamefont {D.~W.}\ \bibnamefont
  {Hogg}},\ }\href {https://doi.org/10.1088/0004-637X/738/2/181} {\bibfield
  {journal} {\bibinfo  {journal} {ApJ}\ }\textbf {\bibinfo {volume} {738}},\
  \bibinfo {pages} {181} (\bibinfo {year} {2011})},\ \Eprint
  {https://arxiv.org/abs/1104.0010} {arXiv:1104.0010} \BibitemShut {NoStop}%
\bibitem [{\citenamefont {Collin}\ \emph {et~al.}(2021)\citenamefont {Collin},
  \citenamefont {Rodd}, \citenamefont {Erjavec},\ and\ \citenamefont
  {Perez}}]{Collin:2021ufc}%
  \BibitemOpen
  \bibfield  {author} {\bibinfo {author} {\bibfnamefont {G.~H.}\ \bibnamefont
  {Collin}}, \bibinfo {author} {\bibfnamefont {N.~L.}\ \bibnamefont {Rodd}},
  \bibinfo {author} {\bibfnamefont {T.}~\bibnamefont {Erjavec}},\ and\ \bibinfo
  {author} {\bibfnamefont {K.}~\bibnamefont {Perez}},\ }\href
  {http://arxiv.org/abs/2104.04529} {\bibfield  {journal} {\bibinfo  {journal}
  {preprint (arXiv:2104.04529)}\ } (\bibinfo {year} {2021})}\BibitemShut
  {NoStop}%
\bibitem [{\citenamefont {Lee}\ \emph {et~al.}(2015)\citenamefont {Lee},
  \citenamefont {Lisanti},\ and\ \citenamefont {Safdi}}]{Lee:2014mza}%
  \BibitemOpen
  \bibfield  {author} {\bibinfo {author} {\bibfnamefont {S.~K.}\ \bibnamefont
  {Lee}}, \bibinfo {author} {\bibfnamefont {M.}~\bibnamefont {Lisanti}},\ and\
  \bibinfo {author} {\bibfnamefont {B.~R.}\ \bibnamefont {Safdi}},\ }\href
  {https://doi.org/10.1088/1475-7516/2015/05/056} {\bibfield  {journal}
  {\bibinfo  {journal} {JCAP}\ }\textbf {\bibinfo {volume} {2015}}\bibfield
  {number} {\bibinfo  {number} { (05)},\ \bibinfo {pages} {056}},\ }\Eprint
  {https://arxiv.org/abs/1412.6099} {arXiv:1412.6099} \BibitemShut {NoStop}%
\bibitem [{\citenamefont {Bartels}\ \emph {et~al.}(2016)\citenamefont
  {Bartels}, \citenamefont {Krishnamurthy},\ and\ \citenamefont
  {Weniger}}]{Bartels2016}%
  \BibitemOpen
  \bibfield  {author} {\bibinfo {author} {\bibfnamefont {R.}~\bibnamefont
  {Bartels}}, \bibinfo {author} {\bibfnamefont {S.}~\bibnamefont
  {Krishnamurthy}},\ and\ \bibinfo {author} {\bibfnamefont {C.}~\bibnamefont
  {Weniger}},\ }\href {https://doi.org/10.1103/PhysRevLett.116.051102}
  {\bibfield  {journal} {\bibinfo  {journal} {Phys. Rev. Lett.}\ }\textbf
  {\bibinfo {volume} {116}},\ \bibinfo {pages} {051102} (\bibinfo {year}
  {2016})},\ \Eprint {https://arxiv.org/abs/1506.05104} {arXiv:1506.05104}
  \BibitemShut {NoStop}%
\bibitem [{\citenamefont {Zhong}\ \emph {et~al.}(2020)\citenamefont {Zhong},
  \citenamefont {McDermott}, \citenamefont {Cholis},\ and\ \citenamefont
  {Fox}}]{Zhong:2019ycb}%
  \BibitemOpen
  \bibfield  {author} {\bibinfo {author} {\bibfnamefont {Y.~M.}\ \bibnamefont
  {Zhong}}, \bibinfo {author} {\bibfnamefont {S.~D.}\ \bibnamefont
  {McDermott}}, \bibinfo {author} {\bibfnamefont {I.}~\bibnamefont {Cholis}},\
  and\ \bibinfo {author} {\bibfnamefont {P.~J.}\ \bibnamefont {Fox}},\ }\href
  {https://doi.org/10.1103/PhysRevLett.124.231103} {\bibfield  {journal}
  {\bibinfo  {journal} {Phys. Rev. Lett.}\ }\textbf {\bibinfo {volume} {124}},\
  \bibinfo {pages} {231103} (\bibinfo {year} {2020})},\ \Eprint
  {https://arxiv.org/abs/1911.12369} {arXiv:1911.12369} \BibitemShut {NoStop}%
\bibitem [{\citenamefont {Abdollahi}\ \emph {et~al.}(2020)\citenamefont
  {Abdollahi}, \citenamefont {Acero}, \citenamefont {Ackermann}, \citenamefont
  {Ajello}, \citenamefont {Atwood}, \citenamefont {Axelsson}, \citenamefont
  {Baldini}, \citenamefont {Ballet} \emph {et~al.}}]{Abdollahi2020}%
  \BibitemOpen
  \bibfield  {author} {\bibinfo {author} {\bibfnamefont {S.}~\bibnamefont
  {Abdollahi}}, \bibinfo {author} {\bibfnamefont {F.}~\bibnamefont {Acero}},
  \bibinfo {author} {\bibfnamefont {M.}~\bibnamefont {Ackermann}}, \bibinfo
  {author} {\bibfnamefont {M.}~\bibnamefont {Ajello}}, \bibinfo {author}
  {\bibfnamefont {W.~B.}\ \bibnamefont {Atwood}}, \bibinfo {author}
  {\bibfnamefont {M.}~\bibnamefont {Axelsson}}, \bibinfo {author}
  {\bibfnamefont {L.}~\bibnamefont {Baldini}}, \bibinfo {author} {\bibfnamefont
  {J.}~\bibnamefont {Ballet}}, \emph {et~al.},\ }\href
  {https://doi.org/10.3847/1538-4365/ab6bcb} {\bibfield  {journal} {\bibinfo
  {journal} {ApJ Supplement Series}\ }\textbf {\bibinfo {volume} {247}},\
  \bibinfo {pages} {33} (\bibinfo {year} {2020})},\ \Eprint
  {https://arxiv.org/abs/1902.10045} {arXiv:1902.10045} \BibitemShut {NoStop}%
\bibitem [{\citenamefont {Leane}\ and\ \citenamefont
  {Slatyer}(2019)}]{Leane2019a}%
  \BibitemOpen
  \bibfield  {author} {\bibinfo {author} {\bibfnamefont {R.~K.}\ \bibnamefont
  {Leane}}\ and\ \bibinfo {author} {\bibfnamefont {T.~R.}\ \bibnamefont
  {Slatyer}},\ }\href {https://doi.org/10.1103/PhysRevLett.123.241101}
  {\bibfield  {journal} {\bibinfo  {journal} {Phys. Rev. Lett.}\ }\textbf
  {\bibinfo {volume} {123}},\ \bibinfo {pages} {241101} (\bibinfo {year}
  {2019})},\ \Eprint {https://arxiv.org/abs/1904.08430} {arXiv:1904.08430}
  \BibitemShut {NoStop}%
\bibitem [{\citenamefont {Buschmann}\ \emph {et~al.}(2020)\citenamefont
  {Buschmann}, \citenamefont {Rodd}, \citenamefont {Safdi}, \citenamefont
  {Chang}, \citenamefont {Mishra-Sharma}, \citenamefont {Lisanti},\ and\
  \citenamefont {Macias}}]{Buschmann2020}%
  \BibitemOpen
  \bibfield  {author} {\bibinfo {author} {\bibfnamefont {M.}~\bibnamefont
  {Buschmann}}, \bibinfo {author} {\bibfnamefont {N.~L.}\ \bibnamefont {Rodd}},
  \bibinfo {author} {\bibfnamefont {B.~R.}\ \bibnamefont {Safdi}}, \bibinfo
  {author} {\bibfnamefont {L.~J.}\ \bibnamefont {Chang}}, \bibinfo {author}
  {\bibfnamefont {S.}~\bibnamefont {Mishra-Sharma}}, \bibinfo {author}
  {\bibfnamefont {M.}~\bibnamefont {Lisanti}},\ and\ \bibinfo {author}
  {\bibfnamefont {O.}~\bibnamefont {Macias}},\ }\href
  {https://doi.org/10.1103/PhysRevD.102.023023} {\bibfield  {journal} {\bibinfo
   {journal} {Phys. Rev. D}\ }\textbf {\bibinfo {volume} {102}},\ \bibinfo
  {pages} {023023} (\bibinfo {year} {2020})},\ \Eprint
  {https://arxiv.org/abs/2002.12373} {arXiv:2002.12373} \BibitemShut {NoStop}%
\bibitem [{\citenamefont {Leane}\ and\ \citenamefont
  {Slatyer}(2020{\natexlab{a}})}]{Leane2020}%
  \BibitemOpen
  \bibfield  {author} {\bibinfo {author} {\bibfnamefont {R.~K.}\ \bibnamefont
  {Leane}}\ and\ \bibinfo {author} {\bibfnamefont {T.~R.}\ \bibnamefont
  {Slatyer}},\ }\href {https://doi.org/10.1103/PhysRevLett.125.121105}
  {\bibfield  {journal} {\bibinfo  {journal} {Phys. Rev. Lett.}\ }\textbf
  {\bibinfo {volume} {125}},\ \bibinfo {pages} {121105} (\bibinfo {year}
  {2020}{\natexlab{a}})},\ \Eprint {https://arxiv.org/abs/2002.12370}
  {arXiv:2002.12370} \BibitemShut {NoStop}%
\bibitem [{\citenamefont {Leane}\ and\ \citenamefont
  {Slatyer}(2020{\natexlab{b}})}]{Leane2020a}%
  \BibitemOpen
  \bibfield  {author} {\bibinfo {author} {\bibfnamefont {R.~K.}\ \bibnamefont
  {Leane}}\ and\ \bibinfo {author} {\bibfnamefont {T.~R.}\ \bibnamefont
  {Slatyer}},\ }\href {https://link.aps.org/doi/10.1103/PhysRevD.102.063019}
  {\bibfield  {journal} {\bibinfo  {journal} {Phys. Rev. D}\ }\textbf {\bibinfo
  {volume} {102}},\ \bibinfo {pages} {063019} (\bibinfo {year}
  {2020}{\natexlab{b}})},\ \Eprint {https://arxiv.org/abs/2002.12371}
  {arXiv:2002.12371} \BibitemShut {NoStop}%
\bibitem [{\citenamefont {Storm}\ \emph {et~al.}(2017)\citenamefont {Storm},
  \citenamefont {Weniger},\ and\ \citenamefont {Calore}}]{Storm2017}%
  \BibitemOpen
  \bibfield  {author} {\bibinfo {author} {\bibfnamefont {E.}~\bibnamefont
  {Storm}}, \bibinfo {author} {\bibfnamefont {C.}~\bibnamefont {Weniger}},\
  and\ \bibinfo {author} {\bibfnamefont {F.}~\bibnamefont {Calore}},\ }\href
  {https://doi.org/10.1088/1475-7516/2017/08/022} {\bibfield  {journal}
  {\bibinfo  {journal} {JCAP}\ }\textbf {\bibinfo {volume} {2017}}\bibfield
  {number} {\bibinfo  {number} { (08)},\ \bibinfo {pages} {022}},\ }\Eprint
  {https://arxiv.org/abs/1705.04065} {arXiv:1705.04065} \BibitemShut {NoStop}%
\bibitem [{\citenamefont {Mishra-Sharma}\ and\ \citenamefont
  {Cranmer}(2020)}]{Mishra-Sharma2020}%
  \BibitemOpen
  \bibfield  {author} {\bibinfo {author} {\bibfnamefont {S.}~\bibnamefont
  {Mishra-Sharma}}\ and\ \bibinfo {author} {\bibfnamefont {K.}~\bibnamefont
  {Cranmer}},\ }\href {http://arxiv.org/abs/2010.10450} {\bibfield  {journal}
  {\bibinfo  {journal} {preprint (arXiv:2010.10450)}\ } (\bibinfo {year}
  {2020})}\BibitemShut {NoStop}%
\bibitem [{\citenamefont {Caron}\ \emph {et~al.}(2018)\citenamefont {Caron},
  \citenamefont {G{\'{o}}mez-Vargas}, \citenamefont {Hendriks},\ and\
  \citenamefont {de~Austri}}]{Caron2018}%
  \BibitemOpen
  \bibfield  {author} {\bibinfo {author} {\bibfnamefont {S.}~\bibnamefont
  {Caron}}, \bibinfo {author} {\bibfnamefont {G.~A.}\ \bibnamefont
  {G{\'{o}}mez-Vargas}}, \bibinfo {author} {\bibfnamefont {L.}~\bibnamefont
  {Hendriks}},\ and\ \bibinfo {author} {\bibfnamefont {R.~R.}\ \bibnamefont
  {de~Austri}},\ }\href {https://doi.org/10.1088/1475-7516/2018/05/058}
  {\bibfield  {journal} {\bibinfo  {journal} {JCAP}\ }\textbf {\bibinfo
  {volume} {2018}}\bibfield  {number} {\bibinfo  {number} { (05)},\ \bibinfo
  {pages} {058}},\ }\Eprint {https://arxiv.org/abs/1708.06706}
  {arXiv:1708.06706} \BibitemShut {NoStop}%
\bibitem [{\citenamefont {List}\ \emph {et~al.}(2020)\citenamefont {List},
  \citenamefont {Rodd}, \citenamefont {Lewis},\ and\ \citenamefont
  {Bhat}}]{List2020b}%
  \BibitemOpen
  \bibfield  {author} {\bibinfo {author} {\bibfnamefont {F.}~\bibnamefont
  {List}}, \bibinfo {author} {\bibfnamefont {N.~L.}\ \bibnamefont {Rodd}},
  \bibinfo {author} {\bibfnamefont {G.~F.}\ \bibnamefont {Lewis}},\ and\
  \bibinfo {author} {\bibfnamefont {I.}~\bibnamefont {Bhat}},\ }\href
  {https://doi.org/10.1103/PhysRevLett.125.241102} {\bibfield  {journal}
  {\bibinfo  {journal} {Phys. Rev. Lett.}\ }\textbf {\bibinfo {volume} {125}},\
  \bibinfo {pages} {241102} (\bibinfo {year} {2020})},\ \Eprint
  {https://arxiv.org/abs/2006.12504} {arXiv:2006.12504} \BibitemShut {NoStop}%
\bibitem [{\citenamefont {Chang}\ \emph {et~al.}(2020)\citenamefont {Chang},
  \citenamefont {Mishra-Sharma}, \citenamefont {Lisanti}, \citenamefont
  {Buschmann}, \citenamefont {Rodd},\ and\ \citenamefont {Safdi}}]{Chang2019}%
  \BibitemOpen
  \bibfield  {author} {\bibinfo {author} {\bibfnamefont {L.~J.}\ \bibnamefont
  {Chang}}, \bibinfo {author} {\bibfnamefont {S.}~\bibnamefont
  {Mishra-Sharma}}, \bibinfo {author} {\bibfnamefont {M.}~\bibnamefont
  {Lisanti}}, \bibinfo {author} {\bibfnamefont {M.}~\bibnamefont {Buschmann}},
  \bibinfo {author} {\bibfnamefont {N.~L.}\ \bibnamefont {Rodd}},\ and\
  \bibinfo {author} {\bibfnamefont {B.~R.}\ \bibnamefont {Safdi}},\ }\href
  {https://doi.org/10.1103/PhysRevD.101.023014} {\bibfield  {journal} {\bibinfo
   {journal} {Phys. Rev. D}\ }\textbf {\bibinfo {volume} {101}},\ \bibinfo
  {pages} {023014} (\bibinfo {year} {2020})},\ \Eprint
  {https://arxiv.org/abs/1908.10874} {arXiv:1908.10874} \BibitemShut {NoStop}%
\bibitem [{\citenamefont {Acero}\ \emph {et~al.}(2015)\citenamefont {Acero},
  \citenamefont {Ackermann}, \citenamefont {Ajello}, \citenamefont {Albert},
  \citenamefont {Atwood}, \citenamefont {Axelsson}, \citenamefont {Baldini},
  \citenamefont {Ballet} \emph {et~al.}}]{Acero2015}%
  \BibitemOpen
  \bibfield  {author} {\bibinfo {author} {\bibfnamefont {F.}~\bibnamefont
  {Acero}}, \bibinfo {author} {\bibfnamefont {M.}~\bibnamefont {Ackermann}},
  \bibinfo {author} {\bibfnamefont {M.}~\bibnamefont {Ajello}}, \bibinfo
  {author} {\bibfnamefont {A.}~\bibnamefont {Albert}}, \bibinfo {author}
  {\bibfnamefont {W.~B.}\ \bibnamefont {Atwood}}, \bibinfo {author}
  {\bibfnamefont {M.}~\bibnamefont {Axelsson}}, \bibinfo {author}
  {\bibfnamefont {L.}~\bibnamefont {Baldini}}, \bibinfo {author} {\bibfnamefont
  {J.}~\bibnamefont {Ballet}}, \emph {et~al.},\ }\href
  {https://doi.org/10.1088/0067-0049/218/2/23} {\bibfield  {journal} {\bibinfo
  {journal} {ApJ Supplement Series}\ }\textbf {\bibinfo {volume} {218}},\
  \bibinfo {pages} {23} (\bibinfo {year} {2015})},\ \Eprint
  {https://arxiv.org/abs/1501.02003} {arXiv:1501.02003} \BibitemShut {NoStop}%
\bibitem [{\citenamefont {Lecun}\ \emph {et~al.}(1998)\citenamefont {Lecun},
  \citenamefont {Bottou}, \citenamefont {Bengio},\ and\ \citenamefont
  {Haffner}}]{Lecun1998}%
  \BibitemOpen
  \bibfield  {author} {\bibinfo {author} {\bibfnamefont {Y.}~\bibnamefont
  {Lecun}}, \bibinfo {author} {\bibfnamefont {L.}~\bibnamefont {Bottou}},
  \bibinfo {author} {\bibfnamefont {Y.}~\bibnamefont {Bengio}},\ and\ \bibinfo
  {author} {\bibfnamefont {P.}~\bibnamefont {Haffner}},\ }\href
  {https://doi.org/10.1109/5.726791} {\bibfield  {journal} {\bibinfo  {journal}
  {Proceedings of the IEEE}\ }\textbf {\bibinfo {volume} {86}},\ \bibinfo
  {pages} {2278} (\bibinfo {year} {1998})}\BibitemShut {NoStop}%
\bibitem [{\citenamefont {Perraudin}\ \emph {et~al.}(2019)\citenamefont
  {Perraudin}, \citenamefont {Defferrard}, \citenamefont {Kacprzak},\ and\
  \citenamefont {Sgier}}]{Perraudin2019a}%
  \BibitemOpen
  \bibfield  {author} {\bibinfo {author} {\bibfnamefont {N.}~\bibnamefont
  {Perraudin}}, \bibinfo {author} {\bibfnamefont {M.}~\bibnamefont
  {Defferrard}}, \bibinfo {author} {\bibfnamefont {T.}~\bibnamefont
  {Kacprzak}},\ and\ \bibinfo {author} {\bibfnamefont {R.}~\bibnamefont
  {Sgier}},\ }\href {https://doi.org/10.1016/j.ascom.2019.03.004} {\bibfield
  {journal} {\bibinfo  {journal} {Astronomy and Computing}\ }\textbf {\bibinfo
  {volume} {27}},\ \bibinfo {pages} {130} (\bibinfo {year} {2019})},\ \Eprint
  {https://arxiv.org/abs/1810.12186} {arXiv:1810.12186} \BibitemShut {NoStop}%
\bibitem [{\citenamefont {Defferrard}\ \emph {et~al.}(2020)\citenamefont
  {Defferrard}, \citenamefont {Milani}, \citenamefont {Gusset},\ and\
  \citenamefont {Perraudin}}]{Defferrard2020}%
  \BibitemOpen
  \bibfield  {author} {\bibinfo {author} {\bibfnamefont {M.}~\bibnamefont
  {Defferrard}}, \bibinfo {author} {\bibfnamefont {M.}~\bibnamefont {Milani}},
  \bibinfo {author} {\bibfnamefont {F.}~\bibnamefont {Gusset}},\ and\ \bibinfo
  {author} {\bibfnamefont {N.}~\bibnamefont {Perraudin}},\ }in\ \href
  {https://openreview.net/forum?id=B1e3OlStPB} {\emph {\bibinfo {booktitle}
  {8th International Conference on Learning Representations, ICLR 2020}}}\
  (\bibinfo {year} {2020})\BibitemShut {NoStop}%
\bibitem [{\citenamefont {Smith}\ \emph {et~al.}(2018)\citenamefont {Smith},
  \citenamefont {Kindermans}, \citenamefont {Ying},\ and\ \citenamefont
  {Le}}]{Smith2018}%
  \BibitemOpen
  \bibfield  {author} {\bibinfo {author} {\bibfnamefont {S.~L.}\ \bibnamefont
  {Smith}}, \bibinfo {author} {\bibfnamefont {P.~J.}\ \bibnamefont
  {Kindermans}}, \bibinfo {author} {\bibfnamefont {C.}~\bibnamefont {Ying}},\
  and\ \bibinfo {author} {\bibfnamefont {Q.~V.}\ \bibnamefont {Le}},\
  }\href@noop {} {\bibfield  {journal} {\bibinfo  {journal} {6th International
  Conference on Learning Representations, ICLR 2018}\ } (\bibinfo {year}
  {2018})},\ \Eprint {https://arxiv.org/abs/1711.00489} {arXiv:1711.00489}
  \BibitemShut {NoStop}%
\bibitem [{\citenamefont {Gorski}\ \emph {et~al.}(2005)\citenamefont {Gorski},
  \citenamefont {Hivon}, \citenamefont {Banday}, \citenamefont {Wandelt},
  \citenamefont {Hansen}, \citenamefont {Reinecke},\ and\ \citenamefont
  {Bartelmann}}]{Gorski2005}%
  \BibitemOpen
  \bibfield  {author} {\bibinfo {author} {\bibfnamefont {K.~M.}\ \bibnamefont
  {Gorski}}, \bibinfo {author} {\bibfnamefont {E.}~\bibnamefont {Hivon}},
  \bibinfo {author} {\bibfnamefont {A.~J.}\ \bibnamefont {Banday}}, \bibinfo
  {author} {\bibfnamefont {B.~D.}\ \bibnamefont {Wandelt}}, \bibinfo {author}
  {\bibfnamefont {F.~K.}\ \bibnamefont {Hansen}}, \bibinfo {author}
  {\bibfnamefont {M.}~\bibnamefont {Reinecke}},\ and\ \bibinfo {author}
  {\bibfnamefont {M.}~\bibnamefont {Bartelmann}},\ }\href
  {https://doi.org/10.1086/427976} {\bibfield  {journal} {\bibinfo  {journal}
  {ApJ}\ }\textbf {\bibinfo {volume} {622}},\ \bibinfo {pages} {759} (\bibinfo
  {year} {2005})},\ \Eprint {https://arxiv.org/abs/astro-ph/0409513}
  {arXiv:astro-ph/0409513} \BibitemShut {NoStop}%
\bibitem [{\citenamefont {Ioffe}\ and\ \citenamefont
  {Szegedy}(2015)}]{Ioffe2015}%
  \BibitemOpen
  \bibfield  {author} {\bibinfo {author} {\bibfnamefont {S.}~\bibnamefont
  {Ioffe}}\ and\ \bibinfo {author} {\bibfnamefont {C.}~\bibnamefont
  {Szegedy}},\ }\href@noop {} {\bibfield  {journal} {\bibinfo  {journal} {32nd
  International Conference on Machine Learning, ICML 2015}\ ,\ \bibinfo {pages}
  {448}} (\bibinfo {year} {2015})},\ \Eprint {https://arxiv.org/abs/1502.03167}
  {arXiv:1502.03167} \BibitemShut {NoStop}%
\bibitem [{\citenamefont {Ulyanov}\ \emph {et~al.}(2016)\citenamefont
  {Ulyanov}, \citenamefont {Vedaldi},\ and\ \citenamefont
  {Lempitsky}}]{Ulyanov2016}%
  \BibitemOpen
  \bibfield  {author} {\bibinfo {author} {\bibfnamefont {D.}~\bibnamefont
  {Ulyanov}}, \bibinfo {author} {\bibfnamefont {A.}~\bibnamefont {Vedaldi}},\
  and\ \bibinfo {author} {\bibfnamefont {V.}~\bibnamefont {Lempitsky}},\ }\href
  {http://arxiv.org/abs/1607.08022} {\bibfield  {journal} {\bibinfo  {journal}
  {preprint (arXiv:1607.08022)}\ } (\bibinfo {year} {2016})}\BibitemShut
  {NoStop}%
\bibitem [{\citenamefont {McDermott}\ \emph {et~al.}(2016)\citenamefont
  {McDermott}, \citenamefont {Fox}, \citenamefont {Cholis},\ and\ \citenamefont
  {Lee}}]{McDermott:2015ydv}%
  \BibitemOpen
  \bibfield  {author} {\bibinfo {author} {\bibfnamefont {S.~D.}\ \bibnamefont
  {McDermott}}, \bibinfo {author} {\bibfnamefont {P.~J.}\ \bibnamefont {Fox}},
  \bibinfo {author} {\bibfnamefont {I.}~\bibnamefont {Cholis}},\ and\ \bibinfo
  {author} {\bibfnamefont {S.~K.}\ \bibnamefont {Lee}},\ }\href
  {https://doi.org/10.1088/1475-7516/2016/07/045} {\bibfield  {journal}
  {\bibinfo  {journal} {JCAP}\ }\textbf {\bibinfo {volume} {2016}}\bibfield
  {number} {\bibinfo  {number} { (07)},\ \bibinfo {pages} {045}},\ }\Eprint
  {https://arxiv.org/abs/1512.00012} {arXiv:1512.00012} \BibitemShut {NoStop}%
\bibitem [{\citenamefont {Balaji}\ \emph {et~al.}(2018)\citenamefont {Balaji},
  \citenamefont {Cholis}, \citenamefont {Fox},\ and\ \citenamefont
  {McDermott}}]{Balaji:2018rwz}%
  \BibitemOpen
  \bibfield  {author} {\bibinfo {author} {\bibfnamefont {B.}~\bibnamefont
  {Balaji}}, \bibinfo {author} {\bibfnamefont {I.}~\bibnamefont {Cholis}},
  \bibinfo {author} {\bibfnamefont {P.~J.}\ \bibnamefont {Fox}},\ and\ \bibinfo
  {author} {\bibfnamefont {S.~D.}\ \bibnamefont {McDermott}},\ }\href
  {https://doi.org/10.1103/PhysRevD.98.043009} {\bibfield  {journal} {\bibinfo
  {journal} {Phys. Rev. D}\ }\textbf {\bibinfo {volume} {98}},\ \bibinfo
  {pages} {043009} (\bibinfo {year} {2018})},\ \Eprint
  {https://arxiv.org/abs/1803.01952} {arXiv:1803.01952} \BibitemShut {NoStop}%
\bibitem [{\citenamefont {Brewer}\ \emph {et~al.}(2013)\citenamefont {Brewer},
  \citenamefont {Foreman-Mackey},\ and\ \citenamefont {Hogg}}]{Brewer2013}%
  \BibitemOpen
  \bibfield  {author} {\bibinfo {author} {\bibfnamefont {B.~J.}\ \bibnamefont
  {Brewer}}, \bibinfo {author} {\bibfnamefont {D.}~\bibnamefont
  {Foreman-Mackey}},\ and\ \bibinfo {author} {\bibfnamefont {D.~W.}\
  \bibnamefont {Hogg}},\ }\href {https://doi.org/10.1088/0004-6256/146/1/7}
  {\bibfield  {journal} {\bibinfo  {journal} {The Astronomical Journal}\
  }\textbf {\bibinfo {volume} {146}},\ \bibinfo {pages} {7} (\bibinfo {year}
  {2013})},\ \Eprint {https://arxiv.org/abs/1211.5805} {arXiv:1211.5805}
  \BibitemShut {NoStop}%
\bibitem [{\citenamefont {Daylan}\ \emph {et~al.}(2017)\citenamefont {Daylan},
  \citenamefont {Portillo},\ and\ \citenamefont {Finkbeiner}}]{Daylan2017}%
  \BibitemOpen
  \bibfield  {author} {\bibinfo {author} {\bibfnamefont {T.}~\bibnamefont
  {Daylan}}, \bibinfo {author} {\bibfnamefont {S.~K.~N.}\ \bibnamefont
  {Portillo}},\ and\ \bibinfo {author} {\bibfnamefont {D.~P.}\ \bibnamefont
  {Finkbeiner}},\ }\href {https://doi.org/10.3847/1538-4357/aa679e} {\bibfield
  {journal} {\bibinfo  {journal} {ApJ}\ }\textbf {\bibinfo {volume} {839}},\
  \bibinfo {pages} {4} (\bibinfo {year} {2017})},\ \Eprint
  {https://arxiv.org/abs/1607.04637} {arXiv:1607.04637} \BibitemShut {NoStop}%
\bibitem [{\citenamefont {Portillo}\ \emph {et~al.}(2017)\citenamefont
  {Portillo}, \citenamefont {Lee}, \citenamefont {Daylan},\ and\ \citenamefont
  {Finkbeiner}}]{Portillo2017}%
  \BibitemOpen
  \bibfield  {author} {\bibinfo {author} {\bibfnamefont {S.~K.~N.}\
  \bibnamefont {Portillo}}, \bibinfo {author} {\bibfnamefont {B.~C.~G.}\
  \bibnamefont {Lee}}, \bibinfo {author} {\bibfnamefont {T.}~\bibnamefont
  {Daylan}},\ and\ \bibinfo {author} {\bibfnamefont {D.~P.}\ \bibnamefont
  {Finkbeiner}},\ }\href {https://doi.org/10.3847/1538-3881/aa8565} {\bibfield
  {journal} {\bibinfo  {journal} {The Astronomical Journal}\ }\textbf {\bibinfo
  {volume} {154}},\ \bibinfo {pages} {132} (\bibinfo {year} {2017})},\ \Eprint
  {https://arxiv.org/abs/1703.01303} {arXiv:1703.01303} \BibitemShut {NoStop}%
\bibitem [{\citenamefont {Ronneberger}\ \emph {et~al.}(2015)\citenamefont
  {Ronneberger}, \citenamefont {Fischer},\ and\ \citenamefont
  {Brox}}]{Ronneberger2015}%
  \BibitemOpen
  \bibfield  {author} {\bibinfo {author} {\bibfnamefont {O.}~\bibnamefont
  {Ronneberger}}, \bibinfo {author} {\bibfnamefont {P.}~\bibnamefont
  {Fischer}},\ and\ \bibinfo {author} {\bibfnamefont {T.}~\bibnamefont
  {Brox}},\ }\href@noop {} {\bibfield  {journal} {\bibinfo  {journal}
  {International Conference on Medical image computing and computer-assisted
  intervention}\ ,\ \bibinfo {pages} {234}} (\bibinfo {year} {2015})},\ \Eprint
  {https://arxiv.org/abs/1505.04597} {arXiv:1505.04597} \BibitemShut {NoStop}%
\bibitem [{\citenamefont {Caron}\ \emph {et~al.}(2021)\citenamefont {Caron},
  \citenamefont {Dijkstra}, \citenamefont {Eckner}, \citenamefont {Hendriks},
  \citenamefont {J{\'{o}}hannesson}, \citenamefont {Panes}, \citenamefont
  {de~Austri},\ and\ \citenamefont {Zaharijas}}]{Caron2021}%
  \BibitemOpen
  \bibfield  {author} {\bibinfo {author} {\bibfnamefont {S.}~\bibnamefont
  {Caron}}, \bibinfo {author} {\bibfnamefont {K.}~\bibnamefont {Dijkstra}},
  \bibinfo {author} {\bibfnamefont {C.}~\bibnamefont {Eckner}}, \bibinfo
  {author} {\bibfnamefont {L.}~\bibnamefont {Hendriks}}, \bibinfo {author}
  {\bibfnamefont {G.}~\bibnamefont {J{\'{o}}hannesson}}, \bibinfo {author}
  {\bibfnamefont {B.}~\bibnamefont {Panes}}, \bibinfo {author} {\bibfnamefont
  {R.~R.}\ \bibnamefont {de~Austri}},\ and\ \bibinfo {author} {\bibfnamefont
  {G.}~\bibnamefont {Zaharijas}},\ }\href {http://arxiv.org/abs/2103.11068}
  {\bibfield  {journal} {\bibinfo  {journal} {preprint (arXiv:2103.11068)}\ }
  (\bibinfo {year} {2021})}\BibitemShut {NoStop}%
\bibitem [{\citenamefont {Jeffrey}\ \emph {et~al.}(2020)\citenamefont
  {Jeffrey}, \citenamefont {Alsing},\ and\ \citenamefont
  {Lanusse}}]{Jeffrey2020}%
  \BibitemOpen
  \bibfield  {author} {\bibinfo {author} {\bibfnamefont {N.}~\bibnamefont
  {Jeffrey}}, \bibinfo {author} {\bibfnamefont {J.}~\bibnamefont {Alsing}},\
  and\ \bibinfo {author} {\bibfnamefont {F.}~\bibnamefont {Lanusse}},\ }\href
  {https://doi.org/10.1093/mnras/staa3594} {\bibfield  {journal} {\bibinfo
  {journal} {MNRAS}\ }\textbf {\bibinfo {volume} {501}},\ \bibinfo {pages}
  {954} (\bibinfo {year} {2020})},\ \Eprint {https://arxiv.org/abs/2009.08459}
  {arXiv:2009.08459} \BibitemShut {NoStop}%
\bibitem [{\citenamefont {Hort{\'{u}}a}\ \emph {et~al.}(2020)\citenamefont
  {Hort{\'{u}}a}, \citenamefont {Malago},\ and\ \citenamefont
  {Volpi}}]{Hortua2020a}%
  \BibitemOpen
  \bibfield  {author} {\bibinfo {author} {\bibfnamefont {H.~J.}\ \bibnamefont
  {Hort{\'{u}}a}}, \bibinfo {author} {\bibfnamefont {L.}~\bibnamefont
  {Malago}},\ and\ \bibinfo {author} {\bibfnamefont {R.}~\bibnamefont
  {Volpi}},\ }\href
  {https://iopscience.iop.org/article/10.1088/2632-2153/aba6f1} {\bibfield
  {journal} {\bibinfo  {journal} {Mach. Learn.: Sci. Technol.}\ }\textbf
  {\bibinfo {volume} {1}},\ \bibinfo {pages} {035014} (\bibinfo {year}
  {2020})},\ \Eprint {https://arxiv.org/abs/2005.07694} {arXiv:2005.07694}
  \BibitemShut {NoStop}%
\bibitem [{\citenamefont {Trippe}\ and\ \citenamefont
  {Turner}(2018)}]{Trippe2018}%
  \BibitemOpen
  \bibfield  {author} {\bibinfo {author} {\bibfnamefont {B.~L.}\ \bibnamefont
  {Trippe}}\ and\ \bibinfo {author} {\bibfnamefont {R.~E.}\ \bibnamefont
  {Turner}},\ }\href {http://arxiv.org/abs/1802.04908} {\bibfield  {journal}
  {\bibinfo  {journal} {preprint (arXiv:1802.04908)}\ } (\bibinfo {year}
  {2018})}\BibitemShut {NoStop}%
\bibitem [{\citenamefont {Kobyzev}\ \emph {et~al.}(2019)\citenamefont
  {Kobyzev}, \citenamefont {Prince},\ and\ \citenamefont
  {Brubaker}}]{Kobyzev2019}%
  \BibitemOpen
  \bibfield  {author} {\bibinfo {author} {\bibfnamefont {I.}~\bibnamefont
  {Kobyzev}}, \bibinfo {author} {\bibfnamefont {S.~J.~D.}\ \bibnamefont
  {Prince}},\ and\ \bibinfo {author} {\bibfnamefont {M.~A.}\ \bibnamefont
  {Brubaker}},\ }\href@noop {} {\bibfield  {journal} {\bibinfo  {journal}
  {preprint (arXiv:1908.09257)}\ } (\bibinfo {year} {2019})},\ \Eprint
  {https://arxiv.org/abs/1908.09257} {arXiv:1908.09257} \BibitemShut {NoStop}%
\bibitem [{\citenamefont {Papamakarios}\ \emph {et~al.}(2021)\citenamefont
  {Papamakarios}, \citenamefont {Nalisnick}, \citenamefont {Rezende},
  \citenamefont {Mohamed},\ and\ \citenamefont
  {Lakshminarayanan}}]{Papamakarios2021}%
  \BibitemOpen
  \bibfield  {author} {\bibinfo {author} {\bibfnamefont {G.}~\bibnamefont
  {Papamakarios}}, \bibinfo {author} {\bibfnamefont {E.}~\bibnamefont
  {Nalisnick}}, \bibinfo {author} {\bibfnamefont {D.~J.}\ \bibnamefont
  {Rezende}}, \bibinfo {author} {\bibfnamefont {S.}~\bibnamefont {Mohamed}},\
  and\ \bibinfo {author} {\bibfnamefont {B.}~\bibnamefont {Lakshminarayanan}},\
  }\href@noop {} {\bibfield  {journal} {\bibinfo  {journal} {Journal of Machine
  Learning Research}\ }\textbf {\bibinfo {volume} {22}},\ \bibinfo {pages} {1}
  (\bibinfo {year} {2021})},\ \Eprint {https://arxiv.org/abs/1912.02762}
  {arXiv:1912.02762} \BibitemShut {NoStop}%
\bibitem [{\citenamefont {Miller}\ \emph {et~al.}(2020)\citenamefont {Miller},
  \citenamefont {Cole}, \citenamefont {Louppe},\ and\ \citenamefont
  {Weniger}}]{Miller2020}%
  \BibitemOpen
  \bibfield  {author} {\bibinfo {author} {\bibfnamefont {B.~K.}\ \bibnamefont
  {Miller}}, \bibinfo {author} {\bibfnamefont {A.}~\bibnamefont {Cole}},
  \bibinfo {author} {\bibfnamefont {G.}~\bibnamefont {Louppe}},\ and\ \bibinfo
  {author} {\bibfnamefont {C.}~\bibnamefont {Weniger}},\ }\href
  {http://arxiv.org/abs/2011.13951} {\bibfield  {journal} {\bibinfo  {journal}
  {preprint (arXiv:2011.13951)}\ } (\bibinfo {year} {2020})}\BibitemShut
  {NoStop}%
\bibitem [{\citenamefont {List}(2021)}]{List2021}%
  \BibitemOpen
  \bibfield  {author} {\bibinfo {author} {\bibfnamefont {F.}~\bibnamefont
  {List}},\ }\href@noop {} {\bibfield  {journal} {\bibinfo  {journal}
  {Proceedings of the 38th International Conference on Machine Learning}\ }
  (\bibinfo {year} {2021})},\ \Eprint {https://arxiv.org/abs/2106.02051}
  {arXiv:2106.02051} \BibitemShut {NoStop}%
\bibitem [{\citenamefont {Ramdas}\ \emph {et~al.}(2017)\citenamefont {Ramdas},
  \citenamefont {Trillos},\ and\ \citenamefont {Cuturi}}]{Ramdas2017}%
  \BibitemOpen
  \bibfield  {author} {\bibinfo {author} {\bibfnamefont {A.}~\bibnamefont
  {Ramdas}}, \bibinfo {author} {\bibfnamefont {N.}~\bibnamefont {Trillos}},\
  and\ \bibinfo {author} {\bibfnamefont {M.}~\bibnamefont {Cuturi}},\ }\href
  {https://doi.org/10.3390/e19020047} {\bibfield  {journal} {\bibinfo
  {journal} {Entropy}\ }\textbf {\bibinfo {volume} {19}},\ \bibinfo {pages}
  {47} (\bibinfo {year} {2017})},\ \Eprint {https://arxiv.org/abs/1509.02237}
  {arXiv:1509.02237} \BibitemShut {NoStop}%
\bibitem [{\citenamefont {Rubner}\ \emph {et~al.}(2000)\citenamefont {Rubner},
  \citenamefont {Tomasi},\ and\ \citenamefont {Guibas}}]{Rubner2000}%
  \BibitemOpen
  \bibfield  {author} {\bibinfo {author} {\bibfnamefont {Y.}~\bibnamefont
  {Rubner}}, \bibinfo {author} {\bibfnamefont {C.}~\bibnamefont {Tomasi}},\
  and\ \bibinfo {author} {\bibfnamefont {L.~J.}\ \bibnamefont {Guibas}},\
  }\href {https://doi.org/https://doi.org/10.1023/A:1026543900054} {\bibfield
  {journal} {\bibinfo  {journal} {International Journal of Computer Vision}\
  }\textbf {\bibinfo {volume} {40}},\ \bibinfo {pages} {99} (\bibinfo {year}
  {2000})}\BibitemShut {NoStop}%
\bibitem [{\citenamefont {Arjovsky}\ \emph {et~al.}(2017)\citenamefont
  {Arjovsky}, \citenamefont {Chintala},\ and\ \citenamefont
  {Bottou}}]{Arjovsky2017}%
  \BibitemOpen
  \bibfield  {author} {\bibinfo {author} {\bibfnamefont {M.}~\bibnamefont
  {Arjovsky}}, \bibinfo {author} {\bibfnamefont {S.}~\bibnamefont {Chintala}},\
  and\ \bibinfo {author} {\bibfnamefont {L.}~\bibnamefont {Bottou}},\
  }\href@noop {} {\bibfield  {journal} {\bibinfo  {journal} {International
  Conference on Machine Learning}\ ,\ \bibinfo {pages} {214}} (\bibinfo {year}
  {2017})},\ \Eprint {https://arxiv.org/abs/1701.07875} {arXiv:1701.07875}
  \BibitemShut {NoStop}%
\bibitem [{\citenamefont {Hou}\ \emph {et~al.}(2016)\citenamefont {Hou},
  \citenamefont {Yu},\ and\ \citenamefont {Samaras}}]{Hou2016}%
  \BibitemOpen
  \bibfield  {author} {\bibinfo {author} {\bibfnamefont {L.}~\bibnamefont
  {Hou}}, \bibinfo {author} {\bibfnamefont {C.-P.}\ \bibnamefont {Yu}},\ and\
  \bibinfo {author} {\bibfnamefont {D.}~\bibnamefont {Samaras}},\ }\href
  {http://arxiv.org/abs/1611.05916} {\bibfield  {journal} {\bibinfo  {journal}
  {preprint (arXiv:1611.05916)}\ } (\bibinfo {year} {2016})}\BibitemShut
  {NoStop}%
\bibitem [{\citenamefont {Koenker}\ and\ \citenamefont
  {Bassett}(1978)}]{Koenker1978}%
  \BibitemOpen
  \bibfield  {author} {\bibinfo {author} {\bibfnamefont {R.}~\bibnamefont
  {Koenker}}\ and\ \bibinfo {author} {\bibfnamefont {G.}~\bibnamefont
  {Bassett}},\ }\href@noop {} {\bibfield  {journal} {\bibinfo  {journal}
  {Econometrica}\ }\textbf {\bibinfo {volume} {46}},\ \bibinfo {pages} {33}
  (\bibinfo {year} {1978})}\BibitemShut {NoStop}%
\bibitem [{\citenamefont {Koenker}\ and\ \citenamefont
  {Hallock}(2001)}]{Koenker2001}%
  \BibitemOpen
  \bibfield  {author} {\bibinfo {author} {\bibfnamefont {R.}~\bibnamefont
  {Koenker}}\ and\ \bibinfo {author} {\bibfnamefont {K.~F.}\ \bibnamefont
  {Hallock}},\ }\href {https://doi.org/10.2307/j.ctvcm4j72.14} {\bibfield
  {journal} {\bibinfo  {journal} {Journal of Economic Perspectives}\ }\textbf
  {\bibinfo {volume} {15}},\ \bibinfo {pages} {143} (\bibinfo {year}
  {2001})}\BibitemShut {NoStop}%
\bibitem [{\citenamefont {Fox}\ and\ \citenamefont {Rubin}(1964)}]{Fox1964}%
  \BibitemOpen
  \bibfield  {author} {\bibinfo {author} {\bibfnamefont {M.}~\bibnamefont
  {Fox}}\ and\ \bibinfo {author} {\bibfnamefont {H.}~\bibnamefont {Rubin}},\
  }\href {https://doi.org/10.1214/aoms/1177700518} {\bibfield  {journal}
  {\bibinfo  {journal} {The Annals of Mathematical Statistics}\ }\textbf
  {\bibinfo {volume} {35}},\ \bibinfo {pages} {1019} (\bibinfo {year}
  {1964})}\BibitemShut {NoStop}%
\bibitem [{\citenamefont {Ferguson}(2014)}]{ferguson2014mathematical}%
  \BibitemOpen
  \bibfield  {author} {\bibinfo {author} {\bibfnamefont {T.~S.}\ \bibnamefont
  {Ferguson}},\ }\href {https://doi.org/10.1016/C2013-0-07705-5} {\emph
  {\bibinfo {title} {{Mathematical Statistics}}}},\ Vol.~\bibinfo {volume} {1}\
  (\bibinfo  {publisher} {Academic Press},\ \bibinfo {year} {2014})\BibitemShut
  {NoStop}%
\bibitem [{\citenamefont {Tagasovska}\ and\ \citenamefont
  {Lopez-Paz}(2018)}]{Tagasovska2018}%
  \BibitemOpen
  \bibfield  {author} {\bibinfo {author} {\bibfnamefont {N.}~\bibnamefont
  {Tagasovska}}\ and\ \bibinfo {author} {\bibfnamefont {D.}~\bibnamefont
  {Lopez-Paz}},\ }\href {http://arxiv.org/abs/1811.00908} {\bibfield  {journal}
  {\bibinfo  {journal} {preprint (arXiv:1811.00908)}\ } (\bibinfo {year}
  {2018})}\BibitemShut {NoStop}%
\bibitem [{\citenamefont {Shridhar}\ \emph {et~al.}(2018)\citenamefont
  {Shridhar}, \citenamefont {Laumann},\ and\ \citenamefont
  {Liwicki}}]{Shridhar2018}%
  \BibitemOpen
  \bibfield  {author} {\bibinfo {author} {\bibfnamefont {K.}~\bibnamefont
  {Shridhar}}, \bibinfo {author} {\bibfnamefont {F.}~\bibnamefont {Laumann}},\
  and\ \bibinfo {author} {\bibfnamefont {M.}~\bibnamefont {Liwicki}},\ }\href
  {http://arxiv.org/abs/1806.05978} {\bibfield  {journal} {\bibinfo  {journal}
  {preprint (arXiv:1806.05978)}\ } (\bibinfo {year} {2018})}\BibitemShut
  {NoStop}%
\bibitem [{\citenamefont {Kingma}\ and\ \citenamefont {Ba}(2014)}]{Kingma2014}%
  \BibitemOpen
  \bibfield  {author} {\bibinfo {author} {\bibfnamefont {D.~P.}\ \bibnamefont
  {Kingma}}\ and\ \bibinfo {author} {\bibfnamefont {J.}~\bibnamefont {Ba}},\
  }\href {http://arxiv.org/abs/1412.6980} {\bibfield  {journal} {\bibinfo
  {journal} {preprint (arXiv:1412.6980)}\ } (\bibinfo {year}
  {2014})}\BibitemShut {NoStop}%
\bibitem [{\citenamefont {Kuleshov}\ \emph {et~al.}(2018)\citenamefont
  {Kuleshov}, \citenamefont {Fenner},\ and\ \citenamefont
  {Ermon}}]{Kuleshov2018}%
  \BibitemOpen
  \bibfield  {author} {\bibinfo {author} {\bibfnamefont {V.}~\bibnamefont
  {Kuleshov}}, \bibinfo {author} {\bibfnamefont {N.}~\bibnamefont {Fenner}},\
  and\ \bibinfo {author} {\bibfnamefont {S.}~\bibnamefont {Ermon}},\
  }\href@noop {} {\bibfield  {journal} {\bibinfo  {journal} {International
  Conference on Machine Learning}\ ,\ \bibinfo {pages} {2796}} (\bibinfo {year}
  {2018})},\ \Eprint {https://arxiv.org/abs/1807.00263} {arXiv:1807.00263}
  \BibitemShut {NoStop}%
\bibitem [{\citenamefont {He}\ \emph {et~al.}(2016)\citenamefont {He},
  \citenamefont {Zhang}, \citenamefont {Ren},\ and\ \citenamefont
  {Sun}}]{He2016}%
  \BibitemOpen
  \bibfield  {author} {\bibinfo {author} {\bibfnamefont {K.}~\bibnamefont
  {He}}, \bibinfo {author} {\bibfnamefont {X.}~\bibnamefont {Zhang}}, \bibinfo
  {author} {\bibfnamefont {S.}~\bibnamefont {Ren}},\ and\ \bibinfo {author}
  {\bibfnamefont {J.}~\bibnamefont {Sun}},\ }\href
  {https://doi.org/10.1109/CVPR.2016.90} {\bibfield  {journal} {\bibinfo
  {journal} {Proceedings of the IEEE Conference on Computer Vision and Pattern
  Recognition}\ ,\ \bibinfo {pages} {770}} (\bibinfo {year} {2016})},\ \Eprint
  {https://arxiv.org/abs/1512.03385} {arXiv:1512.03385} \BibitemShut {NoStop}%
\bibitem [{\citenamefont {Su}\ \emph {et~al.}(2010)\citenamefont {Su},
  \citenamefont {Slatyer},\ and\ \citenamefont {Finkbeiner}}]{Su2010}%
  \BibitemOpen
  \bibfield  {author} {\bibinfo {author} {\bibfnamefont {M.}~\bibnamefont
  {Su}}, \bibinfo {author} {\bibfnamefont {T.~R.}\ \bibnamefont {Slatyer}},\
  and\ \bibinfo {author} {\bibfnamefont {D.~P.}\ \bibnamefont {Finkbeiner}},\
  }\href {https://doi.org/10.1088/0004-637X/724/2/1044} {\bibfield  {journal}
  {\bibinfo  {journal} {ApJ}\ }\textbf {\bibinfo {volume} {724}},\ \bibinfo
  {pages} {1044} (\bibinfo {year} {2010})},\ \Eprint
  {https://arxiv.org/abs/1005.5480} {arXiv:1005.5480} \BibitemShut {NoStop}%
\bibitem [{\citenamefont {Navarro}\ \emph {et~al.}(1997)\citenamefont
  {Navarro}, \citenamefont {Frenk},\ and\ \citenamefont {White}}]{Navarro1997}%
  \BibitemOpen
  \bibfield  {author} {\bibinfo {author} {\bibfnamefont {J.~F.}\ \bibnamefont
  {Navarro}}, \bibinfo {author} {\bibfnamefont {C.~S.}\ \bibnamefont {Frenk}},\
  and\ \bibinfo {author} {\bibfnamefont {S.~D.~M.}\ \bibnamefont {White}},\
  }\href {https://doi.org/10.1086/304888} {\bibfield  {journal} {\bibinfo
  {journal} {ApJ}\ }\textbf {\bibinfo {volume} {490}},\ \bibinfo {pages} {493}
  (\bibinfo {year} {1997})}\BibitemShut {NoStop}%
\bibitem [{\citenamefont {Strong}\ and\ \citenamefont
  {Moskalenko}(1998)}]{Strong:1998pw}%
  \BibitemOpen
  \bibfield  {author} {\bibinfo {author} {\bibfnamefont {A.~W.}\ \bibnamefont
  {Strong}}\ and\ \bibinfo {author} {\bibfnamefont {I.~V.}\ \bibnamefont
  {Moskalenko}},\ }\href {https://doi.org/10.1086/306470} {\bibfield  {journal}
  {\bibinfo  {journal} {ApJ}\ }\textbf {\bibinfo {volume} {509}},\ \bibinfo
  {pages} {212} (\bibinfo {year} {1998})},\ \Eprint
  {https://arxiv.org/abs/astro-ph/9807150} {arXiv:astro-ph/9807150}
  \BibitemShut {NoStop}%
\bibitem [{\citenamefont {Crocker}\ \emph {et~al.}(2017)\citenamefont
  {Crocker}, \citenamefont {Ruiter}, \citenamefont {Seitenzahl}, \citenamefont
  {Panther}, \citenamefont {Sim}, \citenamefont {Baumgardt}, \citenamefont
  {M{\"{o}}ller}, \citenamefont {Nataf}, \citenamefont {Ferrario},
  \citenamefont {Eldridge}, \citenamefont {White}, \citenamefont {Tucker},\
  and\ \citenamefont {Aharonian}}]{Crocker2017}%
  \BibitemOpen
  \bibfield  {author} {\bibinfo {author} {\bibfnamefont {R.~M.}\ \bibnamefont
  {Crocker}}, \bibinfo {author} {\bibfnamefont {A.~J.}\ \bibnamefont {Ruiter}},
  \bibinfo {author} {\bibfnamefont {I.~R.}\ \bibnamefont {Seitenzahl}},
  \bibinfo {author} {\bibfnamefont {F.~H.}\ \bibnamefont {Panther}}, \bibinfo
  {author} {\bibfnamefont {S.}~\bibnamefont {Sim}}, \bibinfo {author}
  {\bibfnamefont {H.}~\bibnamefont {Baumgardt}}, \bibinfo {author}
  {\bibfnamefont {A.}~\bibnamefont {M{\"{o}}ller}}, \bibinfo {author}
  {\bibfnamefont {D.~M.}\ \bibnamefont {Nataf}}, \bibinfo {author}
  {\bibfnamefont {L.}~\bibnamefont {Ferrario}}, \bibinfo {author}
  {\bibfnamefont {J.~J.}\ \bibnamefont {Eldridge}}, \bibinfo {author}
  {\bibfnamefont {M.}~\bibnamefont {White}}, \bibinfo {author} {\bibfnamefont
  {B.~E.}\ \bibnamefont {Tucker}},\ and\ \bibinfo {author} {\bibfnamefont
  {F.}~\bibnamefont {Aharonian}},\ }\href
  {https://doi.org/10.1038/s41550-017-0135} {\bibfield  {journal} {\bibinfo
  {journal} {Nat. Astronomy}\ }\textbf {\bibinfo {volume} {1}},\ \bibinfo
  {pages} {0135} (\bibinfo {year} {2017})},\ \Eprint
  {https://arxiv.org/abs/1607.03495} {arXiv:1607.03495} \BibitemShut {NoStop}%
\bibitem [{\citenamefont {Wang}\ \emph {et~al.}(2005)\citenamefont {Wang},
  \citenamefont {Jiang},\ and\ \citenamefont {Cheng}}]{Wang2005}%
  \BibitemOpen
  \bibfield  {author} {\bibinfo {author} {\bibfnamefont {W.}~\bibnamefont
  {Wang}}, \bibinfo {author} {\bibfnamefont {Z.~J.}\ \bibnamefont {Jiang}},\
  and\ \bibinfo {author} {\bibfnamefont {K.~S.}\ \bibnamefont {Cheng}},\ }\href
  {https://doi.org/10.1111/j.1365-2966.2005.08816.x} {\bibfield  {journal}
  {\bibinfo  {journal} {MNRAS}\ }\textbf {\bibinfo {volume} {358}},\ \bibinfo
  {pages} {263} (\bibinfo {year} {2005})}\BibitemShut {NoStop}%
\bibitem [{\citenamefont {Gonthier}\ \emph {et~al.}(2018)\citenamefont
  {Gonthier}, \citenamefont {Harding}, \citenamefont {Ferrara}, \citenamefont
  {Frederick}, \citenamefont {Mohr},\ and\ \citenamefont {Koh}}]{Gonthier2018}%
  \BibitemOpen
  \bibfield  {author} {\bibinfo {author} {\bibfnamefont {P.~L.}\ \bibnamefont
  {Gonthier}}, \bibinfo {author} {\bibfnamefont {A.~K.}\ \bibnamefont
  {Harding}}, \bibinfo {author} {\bibfnamefont {E.~C.}\ \bibnamefont
  {Ferrara}}, \bibinfo {author} {\bibfnamefont {S.~E.}\ \bibnamefont
  {Frederick}}, \bibinfo {author} {\bibfnamefont {V.~E.}\ \bibnamefont
  {Mohr}},\ and\ \bibinfo {author} {\bibfnamefont {Y.-M.}\ \bibnamefont
  {Koh}},\ }\href {https://doi.org/10.3847/1538-4357/aad08d} {\bibfield
  {journal} {\bibinfo  {journal} {ApJ}\ }\textbf {\bibinfo {volume} {863}},\
  \bibinfo {pages} {199} (\bibinfo {year} {2018})},\ \Eprint
  {https://arxiv.org/abs/1806.11215} {arXiv:1806.11215} \BibitemShut {NoStop}%
\bibitem [{\citenamefont {Yuan}\ and\ \citenamefont {Zhang}(2014)}]{Yuan2014}%
  \BibitemOpen
  \bibfield  {author} {\bibinfo {author} {\bibfnamefont {Q.}~\bibnamefont
  {Yuan}}\ and\ \bibinfo {author} {\bibfnamefont {B.}~\bibnamefont {Zhang}},\
  }\href {https://doi.org/10.1016/j.jheap.2014.06.001} {\bibfield  {journal}
  {\bibinfo  {journal} {Journal of High Energy Astrophysics}\ }\textbf
  {\bibinfo {volume} {3}},\ \bibinfo {pages} {1} (\bibinfo {year} {2014})},\
  \Eprint {https://arxiv.org/abs/1404.2318} {arXiv:1404.2318} \BibitemShut
  {NoStop}%
\bibitem [{\citenamefont {Ploeg}\ \emph {et~al.}(2020)\citenamefont {Ploeg},
  \citenamefont {Gordon}, \citenamefont {Crocker},\ and\ \citenamefont
  {Macias}}]{Ploeg2020}%
  \BibitemOpen
  \bibfield  {author} {\bibinfo {author} {\bibfnamefont {H.}~\bibnamefont
  {Ploeg}}, \bibinfo {author} {\bibfnamefont {C.}~\bibnamefont {Gordon}},
  \bibinfo {author} {\bibfnamefont {R.}~\bibnamefont {Crocker}},\ and\ \bibinfo
  {author} {\bibfnamefont {O.}~\bibnamefont {Macias}},\ }\href
  {https://doi.org/10.1088/1475-7516/2020/12/035} {\bibfield  {journal}
  {\bibinfo  {journal} {JCAP}\ }\textbf {\bibinfo {volume} {2020}}\bibfield
  {number} {\bibinfo  {number} { (12)}},\ }\Eprint
  {https://arxiv.org/abs/2008.10821} {arXiv:2008.10821} \BibitemShut {NoStop}%
\bibitem [{\citenamefont {Hooper}\ and\ \citenamefont
  {Mohlabeng}(2016)}]{Hooper2016}%
  \BibitemOpen
  \bibfield  {author} {\bibinfo {author} {\bibfnamefont {D.}~\bibnamefont
  {Hooper}}\ and\ \bibinfo {author} {\bibfnamefont {G.}~\bibnamefont
  {Mohlabeng}},\ }\href {https://doi.org/10.1088/1475-7516/2016/03/049}
  {\bibfield  {journal} {\bibinfo  {journal} {JCAP}\ }\textbf {\bibinfo
  {volume} {2016}}\bibfield  {number} {\bibinfo  {number} { (03)},\ \bibinfo
  {pages} {049}},\ }\Eprint {https://arxiv.org/abs/1512.04966}
  {arXiv:1512.04966} \BibitemShut {NoStop}%
\bibitem [{\citenamefont {Gautam}\ \emph {et~al.}(2021)\citenamefont {Gautam},
  \citenamefont {Crocker}, \citenamefont {Ferrario}, \citenamefont {Ruiter},
  \citenamefont {Ploeg}, \citenamefont {Gordon},\ and\ \citenamefont
  {Macias}}]{Gautam2021}%
  \BibitemOpen
  \bibfield  {author} {\bibinfo {author} {\bibfnamefont {A.}~\bibnamefont
  {Gautam}}, \bibinfo {author} {\bibfnamefont {R.~M.}\ \bibnamefont {Crocker}},
  \bibinfo {author} {\bibfnamefont {L.}~\bibnamefont {Ferrario}}, \bibinfo
  {author} {\bibfnamefont {A.~J.}\ \bibnamefont {Ruiter}}, \bibinfo {author}
  {\bibfnamefont {H.}~\bibnamefont {Ploeg}}, \bibinfo {author} {\bibfnamefont
  {C.}~\bibnamefont {Gordon}},\ and\ \bibinfo {author} {\bibfnamefont
  {O.}~\bibnamefont {Macias}},\ }\href {http://arxiv.org/abs/2106.00222}
  {\bibfield  {journal} {\bibinfo  {journal} {preprint (arXiv:2106.00222)}\ }
  (\bibinfo {year} {2021})}\BibitemShut {NoStop}%
\bibitem [{\citenamefont {Radhakrishnan}\ and\ \citenamefont
  {Srinivasan}(1982)}]{Radhakrishnan1982}%
  \BibitemOpen
  \bibfield  {author} {\bibinfo {author} {\bibfnamefont {V.}~\bibnamefont
  {Radhakrishnan}}\ and\ \bibinfo {author} {\bibfnamefont {G.}~\bibnamefont
  {Srinivasan}},\ }\href {https://www.jstor.org/stable/24087904} {\bibfield
  {journal} {\bibinfo  {journal} {Current Science}\ }\textbf {\bibinfo {volume}
  {51}},\ \bibinfo {pages} {1096} (\bibinfo {year} {1982})}\BibitemShut
  {NoStop}%
\bibitem [{\citenamefont {Wilks}(1938)}]{Wilks1938}%
  \BibitemOpen
  \bibfield  {author} {\bibinfo {author} {\bibfnamefont {S.~S.}\ \bibnamefont
  {Wilks}},\ }\href {https://doi.org/10.1214/aoms/1177732360} {\bibfield
  {journal} {\bibinfo  {journal} {The Annals of Mathematical Statistics}\
  }\textbf {\bibinfo {volume} {9}},\ \bibinfo {pages} {60} (\bibinfo {year}
  {1938})}\BibitemShut {NoStop}%
\bibitem [{\citenamefont {Macquart}\ and\ \citenamefont
  {Kanekar}(2015)}]{Macquart2015}%
  \BibitemOpen
  \bibfield  {author} {\bibinfo {author} {\bibfnamefont {J.~P.}\ \bibnamefont
  {Macquart}}\ and\ \bibinfo {author} {\bibfnamefont {N.}~\bibnamefont
  {Kanekar}},\ }\href {https://doi.org/10.1088/0004-637X/805/2/172} {\bibfield
  {journal} {\bibinfo  {journal} {ApJ}\ }\textbf {\bibinfo {volume} {805}},\
  \bibinfo {pages} {172} (\bibinfo {year} {2015})},\ \Eprint
  {https://arxiv.org/abs/1504.02492} {arXiv:1504.02492} \BibitemShut {NoStop}%
\bibitem [{\citenamefont {Calore}\ \emph {et~al.}(2016)\citenamefont {Calore},
  \citenamefont {Mauro}, \citenamefont {Donato}, \citenamefont {Hessels},\ and\
  \citenamefont {Weniger}}]{Calore2016}%
  \BibitemOpen
  \bibfield  {author} {\bibinfo {author} {\bibfnamefont {F.}~\bibnamefont
  {Calore}}, \bibinfo {author} {\bibfnamefont {M.~D.}\ \bibnamefont {Mauro}},
  \bibinfo {author} {\bibfnamefont {F.}~\bibnamefont {Donato}}, \bibinfo
  {author} {\bibfnamefont {J.~W.~T.}\ \bibnamefont {Hessels}},\ and\ \bibinfo
  {author} {\bibfnamefont {C.}~\bibnamefont {Weniger}},\ }\href
  {https://doi.org/10.3847/0004-637x/827/2/143} {\bibfield  {journal} {\bibinfo
   {journal} {ApJ}\ }\textbf {\bibinfo {volume} {827}},\ \bibinfo {pages} {143}
  (\bibinfo {year} {2016})},\ \Eprint {https://arxiv.org/abs/1512.06825}
  {arXiv:1512.06825} \BibitemShut {NoStop}%
\bibitem [{\citenamefont {Regis}\ \emph {et~al.}(2021)\citenamefont {Regis},
  \citenamefont {Reynoso-Cordova}, \citenamefont {Filipovi{\'{c}}},
  \citenamefont {Br{\"{u}}ggen}, \citenamefont {Carretti}, \citenamefont
  {Collier}, \citenamefont {Hopkins}, \citenamefont {Lenc}, \citenamefont
  {Maio}, \citenamefont {Marvil}, \citenamefont {Norris},\ and\ \citenamefont
  {Vernstrom}}]{Regis2021}%
  \BibitemOpen
  \bibfield  {author} {\bibinfo {author} {\bibfnamefont {M.}~\bibnamefont
  {Regis}}, \bibinfo {author} {\bibfnamefont {J.}~\bibnamefont
  {Reynoso-Cordova}}, \bibinfo {author} {\bibfnamefont {M.~D.}\ \bibnamefont
  {Filipovi{\'{c}}}}, \bibinfo {author} {\bibfnamefont {M.}~\bibnamefont
  {Br{\"{u}}ggen}}, \bibinfo {author} {\bibfnamefont {E.}~\bibnamefont
  {Carretti}}, \bibinfo {author} {\bibfnamefont {J.}~\bibnamefont {Collier}},
  \bibinfo {author} {\bibfnamefont {A.~M.}\ \bibnamefont {Hopkins}}, \bibinfo
  {author} {\bibfnamefont {E.}~\bibnamefont {Lenc}}, \bibinfo {author}
  {\bibfnamefont {U.}~\bibnamefont {Maio}}, \bibinfo {author} {\bibfnamefont
  {J.~R.}\ \bibnamefont {Marvil}}, \bibinfo {author} {\bibfnamefont {R.~P.}\
  \bibnamefont {Norris}},\ and\ \bibinfo {author} {\bibfnamefont
  {T.}~\bibnamefont {Vernstrom}},\ }\href {http://arxiv.org/abs/2106.08025}
  {\bibfield  {journal} {\bibinfo  {journal} {preprint (arXiv:2106.08025)}\ }
  (\bibinfo {year} {2021})}\BibitemShut {NoStop}%
\bibitem [{\citenamefont {Somalwar}\ \emph {et~al.}(2021)\citenamefont
  {Somalwar}, \citenamefont {Chang}, \citenamefont {Mishra-Sharma},\ and\
  \citenamefont {Lisanti}}]{Somalwar2020}%
  \BibitemOpen
  \bibfield  {author} {\bibinfo {author} {\bibfnamefont {J.~J.}\ \bibnamefont
  {Somalwar}}, \bibinfo {author} {\bibfnamefont {L.~J.}\ \bibnamefont {Chang}},
  \bibinfo {author} {\bibfnamefont {S.}~\bibnamefont {Mishra-Sharma}},\ and\
  \bibinfo {author} {\bibfnamefont {M.}~\bibnamefont {Lisanti}},\ }\href
  {https://doi.org/10.3847/1538-4357/abc87d} {\bibfield  {journal} {\bibinfo
  {journal} {ApJ}\ }\textbf {\bibinfo {volume} {906}},\ \bibinfo {pages} {57}
  (\bibinfo {year} {2021})},\ \Eprint {https://arxiv.org/abs/2009.00021}
  {arXiv:2009.00021} \BibitemShut {NoStop}%
\bibitem [{\citenamefont {Runburg}\ \emph {et~al.}(2021)\citenamefont
  {Runburg}, \citenamefont {Baxter},\ and\ \citenamefont
  {Kumar}}]{Runburg2021}%
  \BibitemOpen
  \bibfield  {author} {\bibinfo {author} {\bibfnamefont {J.}~\bibnamefont
  {Runburg}}, \bibinfo {author} {\bibfnamefont {E.~J.}\ \bibnamefont
  {Baxter}},\ and\ \bibinfo {author} {\bibfnamefont {J.}~\bibnamefont
  {Kumar}},\ }\href {http://arxiv.org/abs/2106.10399} {\bibfield  {journal}
  {\bibinfo  {journal} {preprint (arXiv:2106.10399)}\ } (\bibinfo {year}
  {2021})}\BibitemShut {NoStop}%
\bibitem [{\citenamefont {Karwin}\ \emph {et~al.}(2019)\citenamefont {Karwin},
  \citenamefont {Murgia}, \citenamefont {Campbell},\ and\ \citenamefont
  {Moskalenko}}]{Karwin2019}%
  \BibitemOpen
  \bibfield  {author} {\bibinfo {author} {\bibfnamefont {C.~M.}\ \bibnamefont
  {Karwin}}, \bibinfo {author} {\bibfnamefont {S.}~\bibnamefont {Murgia}},
  \bibinfo {author} {\bibfnamefont {S.}~\bibnamefont {Campbell}},\ and\
  \bibinfo {author} {\bibfnamefont {I.~V.}\ \bibnamefont {Moskalenko}},\ }\href
  {https://doi.org/10.3847/1538-4357/ab2880} {\bibfield  {journal} {\bibinfo
  {journal} {ApJ}\ }\textbf {\bibinfo {volume} {880}},\ \bibinfo {pages} {95}
  (\bibinfo {year} {2019})},\ \Eprint {https://arxiv.org/abs/1903.10533}
  {arXiv:1903.10533} \BibitemShut {NoStop}%
\bibitem [{\citenamefont {Burns}\ \emph {et~al.}(2021)\citenamefont {Burns},
  \citenamefont {Fieg}, \citenamefont {Rajaraman},\ and\ \citenamefont
  {Karwin}}]{Burns2020}%
  \BibitemOpen
  \bibfield  {author} {\bibinfo {author} {\bibfnamefont {A.~K.}\ \bibnamefont
  {Burns}}, \bibinfo {author} {\bibfnamefont {M.}~\bibnamefont {Fieg}},
  \bibinfo {author} {\bibfnamefont {A.}~\bibnamefont {Rajaraman}},\ and\
  \bibinfo {author} {\bibfnamefont {C.~M.}\ \bibnamefont {Karwin}},\ }\href
  {http://dx.doi.org/10.1103/PhysRevD.103.063023} {\bibfield  {journal}
  {\bibinfo  {journal} {Phys. Rev. D}\ }\textbf {\bibinfo {volume} {103}},\
  \bibinfo {pages} {063023} (\bibinfo {year} {2021})},\ \Eprint
  {https://arxiv.org/abs/2010.11650} {arXiv:2010.11650} \BibitemShut {NoStop}%
\bibitem [{\citenamefont {Hunter}(2007)}]{mpl}%
  \BibitemOpen
  \bibfield  {author} {\bibinfo {author} {\bibfnamefont {J.~D.}\ \bibnamefont
  {Hunter}},\ }\href {https://doi.org/10.1109/MCSE.2007.55} {\bibfield
  {journal} {\bibinfo  {journal} {Computing in Science \& Engineering}\
  }\textbf {\bibinfo {volume} {9}},\ \bibinfo {pages} {90} (\bibinfo {year}
  {2007})}\BibitemShut {NoStop}%
\bibitem [{\citenamefont {Waskom}\ \emph {et~al.}(2017)\citenamefont {Waskom}
  \emph {et~al.}}]{seaborn}%
  \BibitemOpen
  \bibfield  {author} {\bibinfo {author} {\bibfnamefont {M.}~\bibnamefont
  {Waskom}} \emph {et~al.},\ }\href {https://doi.org/10.5281/zenodo.883859}
  {\bibinfo {title} {mwaskom/seaborn: v0.8.1 (september 2017)}} (\bibinfo
  {year} {2017})\BibitemShut {NoStop}%
\bibitem [{\citenamefont {Oliphant}(2006)}]{npy}%
  \BibitemOpen
  \bibfield  {author} {\bibinfo {author} {\bibfnamefont {T.~E.}\ \bibnamefont
  {Oliphant}},\ }\href@noop {} {\emph {\bibinfo {title} {A guide to NumPy}}},\
  Vol.~\bibinfo {volume} {1}\ (\bibinfo  {publisher} {Trelgol Publishing USA},\
  \bibinfo {year} {2006})\BibitemShut {NoStop}%
\bibitem [{\citenamefont {{Virtanen}}\ \emph {et~al.}(2020)\citenamefont
  {{Virtanen}} \emph {et~al.}}]{scipy}%
  \BibitemOpen
  \bibfield  {author} {\bibinfo {author} {\bibfnamefont {P.}~\bibnamefont
  {{Virtanen}}} \emph {et~al.},\ }\href
  {https://doi.org/https://doi.org/10.1038/s41592-019-0686-2} {\bibfield
  {journal} {\bibinfo  {journal} {Nat. Methods}\ }\textbf {\bibinfo {volume}
  {17}},\ \bibinfo {pages} {261} (\bibinfo {year} {2020})}\BibitemShut
  {NoStop}%
\bibitem [{\citenamefont {Lam}\ \emph {et~al.}(2015)\citenamefont {Lam},
  \citenamefont {Pitrou},\ and\ \citenamefont {Seibert}}]{numba}%
  \BibitemOpen
  \bibfield  {author} {\bibinfo {author} {\bibfnamefont {S.~K.}\ \bibnamefont
  {Lam}}, \bibinfo {author} {\bibfnamefont {A.}~\bibnamefont {Pitrou}},\ and\
  \bibinfo {author} {\bibfnamefont {S.}~\bibnamefont {Seibert}},\ }\href
  {https://doi.org/10.1145/2833157.2833162} {\bibfield  {journal} {\bibinfo
  {journal} {Proceedings of the Second Workshop on the LLVM Compiler
  Infrastructure in HPC}\ } (\bibinfo {year} {2015})}\BibitemShut {NoStop}%
\bibitem [{\citenamefont {Zonca}\ \emph {et~al.}(2019)\citenamefont {Zonca},
  \citenamefont {Singer}, \citenamefont {Lenz}, \citenamefont {Reinecke},
  \citenamefont {Rosset}, \citenamefont {Hivon},\ and\ \citenamefont
  {Gorski}}]{healpy}%
  \BibitemOpen
  \bibfield  {author} {\bibinfo {author} {\bibfnamefont {A.}~\bibnamefont
  {Zonca}}, \bibinfo {author} {\bibfnamefont {L.}~\bibnamefont {Singer}},
  \bibinfo {author} {\bibfnamefont {D.}~\bibnamefont {Lenz}}, \bibinfo {author}
  {\bibfnamefont {M.}~\bibnamefont {Reinecke}}, \bibinfo {author}
  {\bibfnamefont {C.}~\bibnamefont {Rosset}}, \bibinfo {author} {\bibfnamefont
  {E.}~\bibnamefont {Hivon}},\ and\ \bibinfo {author} {\bibfnamefont
  {K.}~\bibnamefont {Gorski}},\ }\href {https://doi.org/10.21105/joss.01298}
  {\bibfield  {journal} {\bibinfo  {journal} {Journal of Open Source Software}\
  }\textbf {\bibinfo {volume} {4}},\ \bibinfo {pages} {1298} (\bibinfo {year}
  {2019})}\BibitemShut {NoStop}%
\bibitem [{\citenamefont {Abadi}\ \emph {et~al.}(2016)\citenamefont {Abadi},
  \citenamefont {Agarwal}, \citenamefont {Barham}, \citenamefont {Brevdo},
  \citenamefont {Chen}, \citenamefont {Citro}, \citenamefont {Corrado},
  \citenamefont {Davis} \emph {et~al.}}]{Abadi2016}%
  \BibitemOpen
  \bibfield  {author} {\bibinfo {author} {\bibfnamefont {M.}~\bibnamefont
  {Abadi}}, \bibinfo {author} {\bibfnamefont {A.}~\bibnamefont {Agarwal}},
  \bibinfo {author} {\bibfnamefont {P.}~\bibnamefont {Barham}}, \bibinfo
  {author} {\bibfnamefont {E.}~\bibnamefont {Brevdo}}, \bibinfo {author}
  {\bibfnamefont {Z.}~\bibnamefont {Chen}}, \bibinfo {author} {\bibfnamefont
  {C.}~\bibnamefont {Citro}}, \bibinfo {author} {\bibfnamefont {G.~S.}\
  \bibnamefont {Corrado}}, \bibinfo {author} {\bibfnamefont {A.}~\bibnamefont
  {Davis}}, \emph {et~al.},\ }\href {http://arxiv.org/abs/1603.04467}
  {\bibfield  {journal} {\bibinfo  {journal} {preprint (arXiv:1603.04467)}\ }
  (\bibinfo {year} {2016})}\BibitemShut {NoStop}%
\bibitem [{\citenamefont {Chollet}\ \emph {et~al.}(2015)\citenamefont {Chollet}
  \emph {et~al.}}]{keras}%
  \BibitemOpen
  \bibfield  {author} {\bibinfo {author} {\bibfnamefont {F.}~\bibnamefont
  {Chollet}} \emph {et~al.},\ }\href@noop {} {\bibinfo {title} {Keras}},\
  \bibinfo {howpublished} {\url{https://keras.io}} (\bibinfo {year}
  {2015})\BibitemShut {NoStop}%
\bibitem [{\citenamefont {Moritz}\ \emph {et~al.}(2017)\citenamefont {Moritz},
  \citenamefont {Nishihara}, \citenamefont {Wang}, \citenamefont {Tumanov},
  \citenamefont {Liaw}, \citenamefont {Liang}, \citenamefont {Paul},
  \citenamefont {Jordan},\ and\ \citenamefont {Stoica}}]{ray}%
  \BibitemOpen
  \bibfield  {author} {\bibinfo {author} {\bibfnamefont {P.}~\bibnamefont
  {Moritz}}, \bibinfo {author} {\bibfnamefont {R.}~\bibnamefont {Nishihara}},
  \bibinfo {author} {\bibfnamefont {S.}~\bibnamefont {Wang}}, \bibinfo {author}
  {\bibfnamefont {A.}~\bibnamefont {Tumanov}}, \bibinfo {author} {\bibfnamefont
  {R.}~\bibnamefont {Liaw}}, \bibinfo {author} {\bibfnamefont {E.}~\bibnamefont
  {Liang}}, \bibinfo {author} {\bibfnamefont {W.}~\bibnamefont {Paul}},
  \bibinfo {author} {\bibfnamefont {M.~I.}\ \bibnamefont {Jordan}},\ and\
  \bibinfo {author} {\bibfnamefont {I.}~\bibnamefont {Stoica}},\ }\href@noop {}
  {\bibfield  {journal} {\bibinfo  {journal} {CoRR}\ } (\bibinfo {year}
  {2017})},\ \Eprint {https://arxiv.org/abs/1712.05889} {arXiv:1712.05889}
  \BibitemShut {NoStop}%
\bibitem [{\citenamefont {Rodd}\ and\ \citenamefont {Toomey}()}]{NPTFit-Sim}%
  \BibitemOpen
  \bibfield  {author} {\bibinfo {author} {\bibfnamefont {N.}~\bibnamefont
  {Rodd}}\ and\ \bibinfo {author} {\bibfnamefont {M.}~\bibnamefont {Toomey}},\
  }\href {https://github.com/nickrodd/NPTFit-Sim} {\bibinfo {title}
  {{NPTFit-Sim}}}\BibitemShut {NoStop}%
\bibitem [{\citenamefont {Dembinski}\ \emph {et~al.}(2020)\citenamefont
  {Dembinski} \emph {et~al.}}]{iminuit}%
  \BibitemOpen
  \bibfield  {author} {\bibinfo {author} {\bibfnamefont {H.}~\bibnamefont
  {Dembinski}} \emph {et~al.}\ }\href {https://doi.org/10.5281/zenodo.4310361}
  {10.5281/zenodo.4310361} (\bibinfo {year} {2020})\BibitemShut {NoStop}%
\bibitem [{\citenamefont {McKerns}\ \emph {et~al.}(2011)\citenamefont
  {McKerns}, \citenamefont {Strand}, \citenamefont {Sullivan}, \citenamefont
  {Fang},\ and\ \citenamefont {Aivazis}}]{dill}%
  \BibitemOpen
  \bibfield  {author} {\bibinfo {author} {\bibfnamefont {M.~M.}\ \bibnamefont
  {McKerns}}, \bibinfo {author} {\bibfnamefont {L.}~\bibnamefont {Strand}},
  \bibinfo {author} {\bibfnamefont {T.}~\bibnamefont {Sullivan}}, \bibinfo
  {author} {\bibfnamefont {A.}~\bibnamefont {Fang}},\ and\ \bibinfo {author}
  {\bibfnamefont {M.~A.~G.}\ \bibnamefont {Aivazis}},\ }\href
  {http://arxiv.org/abs/1202.1056} {\bibfield  {journal} {\bibinfo  {journal}
  {Proceedings of the 10th Python in Science Conference}\ } (\bibinfo {year}
  {2011})}\BibitemShut {NoStop}%
\bibitem [{\citenamefont {Feroz}\ \emph {et~al.}(2009)\citenamefont {Feroz},
  \citenamefont {Hobson},\ and\ \citenamefont {Bridges}}]{Feroz2009}%
  \BibitemOpen
  \bibfield  {author} {\bibinfo {author} {\bibfnamefont {F.}~\bibnamefont
  {Feroz}}, \bibinfo {author} {\bibfnamefont {M.~P.}\ \bibnamefont {Hobson}},\
  and\ \bibinfo {author} {\bibfnamefont {M.}~\bibnamefont {Bridges}},\ }\href
  {https://doi.org/10.1111/j.1365-2966.2009.14548.x} {\bibfield  {journal}
  {\bibinfo  {journal} {MNRAS}\ }\textbf {\bibinfo {volume} {398}},\ \bibinfo
  {pages} {1601} (\bibinfo {year} {2009})},\ \Eprint
  {https://arxiv.org/abs/0809.3437} {arXiv:0809.3437} \BibitemShut {NoStop}%
\bibitem [{\citenamefont {Buchner}\ \emph {et~al.}(2014)\citenamefont
  {Buchner}, \citenamefont {Georgakakis}, \citenamefont {Nandra}, \citenamefont
  {Hsu}, \citenamefont {Rangel}, \citenamefont {Brightman}, \citenamefont
  {Merloni}, \citenamefont {Salvato} \emph {et~al.}}]{Buchner2014}%
  \BibitemOpen
  \bibfield  {author} {\bibinfo {author} {\bibfnamefont {J.}~\bibnamefont
  {Buchner}}, \bibinfo {author} {\bibfnamefont {A.}~\bibnamefont
  {Georgakakis}}, \bibinfo {author} {\bibfnamefont {K.}~\bibnamefont {Nandra}},
  \bibinfo {author} {\bibfnamefont {L.}~\bibnamefont {Hsu}}, \bibinfo {author}
  {\bibfnamefont {C.}~\bibnamefont {Rangel}}, \bibinfo {author} {\bibfnamefont
  {M.}~\bibnamefont {Brightman}}, \bibinfo {author} {\bibfnamefont
  {A.}~\bibnamefont {Merloni}}, \bibinfo {author} {\bibfnamefont
  {M.}~\bibnamefont {Salvato}}, \emph {et~al.},\ }\href
  {https://doi.org/10.1051/0004-6361/201322971} {\bibfield  {journal} {\bibinfo
   {journal} {A{\&}A}\ }\textbf {\bibinfo {volume} {564}},\ \bibinfo {pages}
  {A125} (\bibinfo {year} {2014})},\ \Eprint {https://arxiv.org/abs/1402.0004}
  {arXiv:1402.0004} \BibitemShut {NoStop}%
\bibitem [{\citenamefont {Zheng}(2011)}]{Zheng2011}%
  \BibitemOpen
  \bibfield  {author} {\bibinfo {author} {\bibfnamefont {S.}~\bibnamefont
  {Zheng}},\ }\href {https://doi.org/10.1007/s13042-011-0031-2} {\bibfield
  {journal} {\bibinfo  {journal} {International Journal of Machine Learning and
  Cybernetics}\ }\textbf {\bibinfo {volume} {2}},\ \bibinfo {pages} {191}
  (\bibinfo {year} {2011})}\BibitemShut {NoStop}%
\bibitem [{\citenamefont {Hatalis}\ \emph {et~al.}(2019)\citenamefont
  {Hatalis}, \citenamefont {Lamadrid}, \citenamefont {Scheinberg},\ and\
  \citenamefont {Kishore}}]{Hatalis2019}%
  \BibitemOpen
  \bibfield  {author} {\bibinfo {author} {\bibfnamefont {K.}~\bibnamefont
  {Hatalis}}, \bibinfo {author} {\bibfnamefont {A.~J.}\ \bibnamefont
  {Lamadrid}}, \bibinfo {author} {\bibfnamefont {K.}~\bibnamefont
  {Scheinberg}},\ and\ \bibinfo {author} {\bibfnamefont {S.}~\bibnamefont
  {Kishore}},\ }\href {http://arxiv.org/abs/1909.12122} {\bibfield  {journal}
  {\bibinfo  {journal} {preprint (arXiv:1909.12122)}\ } (\bibinfo {year}
  {2019})}\BibitemShut {NoStop}%
\end{thebibliography}
%

\vspace{0.5cm}
\appendix
\section*{Appendices}
The following sections contain further details and cross-checks of our results. First, we carry out the exercise of applying both a CNN and \texttt{NPTFit} to a map with a strong unmodeled large-scale asymmetry and to a pixel-shuffled version thereof, as qualitatively discussed in Sec.~\ref{sec:cnns}. Next, we compare the results of our NN-based framework for the \emph{Fermi} map with those of \texttt{NPTFit} when making the same modeling choices (apart from the SCD parameterization). Also, we show the cumulative SCDs for the mismodeling experiment in Sec.~\ref{sec:mismodeling_experiment}, as well as the constraints on $\eta_P$ arising from those estimates. For the recovery of GCE flux artificially injected into the \emph{Fermi} map (see Sec.~\ref{sec:injection}), we present and discuss the results of $g^{\boldsymbol{\varpi}}$ and $h^{\boldsymbol{\nu}}$, which predict the SCDs and constraints on the Poisson flux fraction $\eta_P$, respectively. Then, we provide the analytic likelihood in the case of a homogeneous isotropic PS population considered in Sec.~\ref{sec:constraining_poisson_iso_without_PSF}, and we compare the simple estimator for the Poisson flux fraction presented in Sec.~\ref{sec:constraining_poisson_simple} with the NN estimator $h^{\boldsymbol{\nu}}$ (see Sec.~\ref{sec:constraining_poisson_with_NN}). Finally, we list our priors for the generation of training data and tabulate our NN architectures.

\section{Unmodeled north-south asymmetry: an example}
\label{sec:NS_asymmetry_example}
\begin{figure*}
\centering
  \noindent
   \resizebox{1\textwidth}{!}{
    \includegraphics{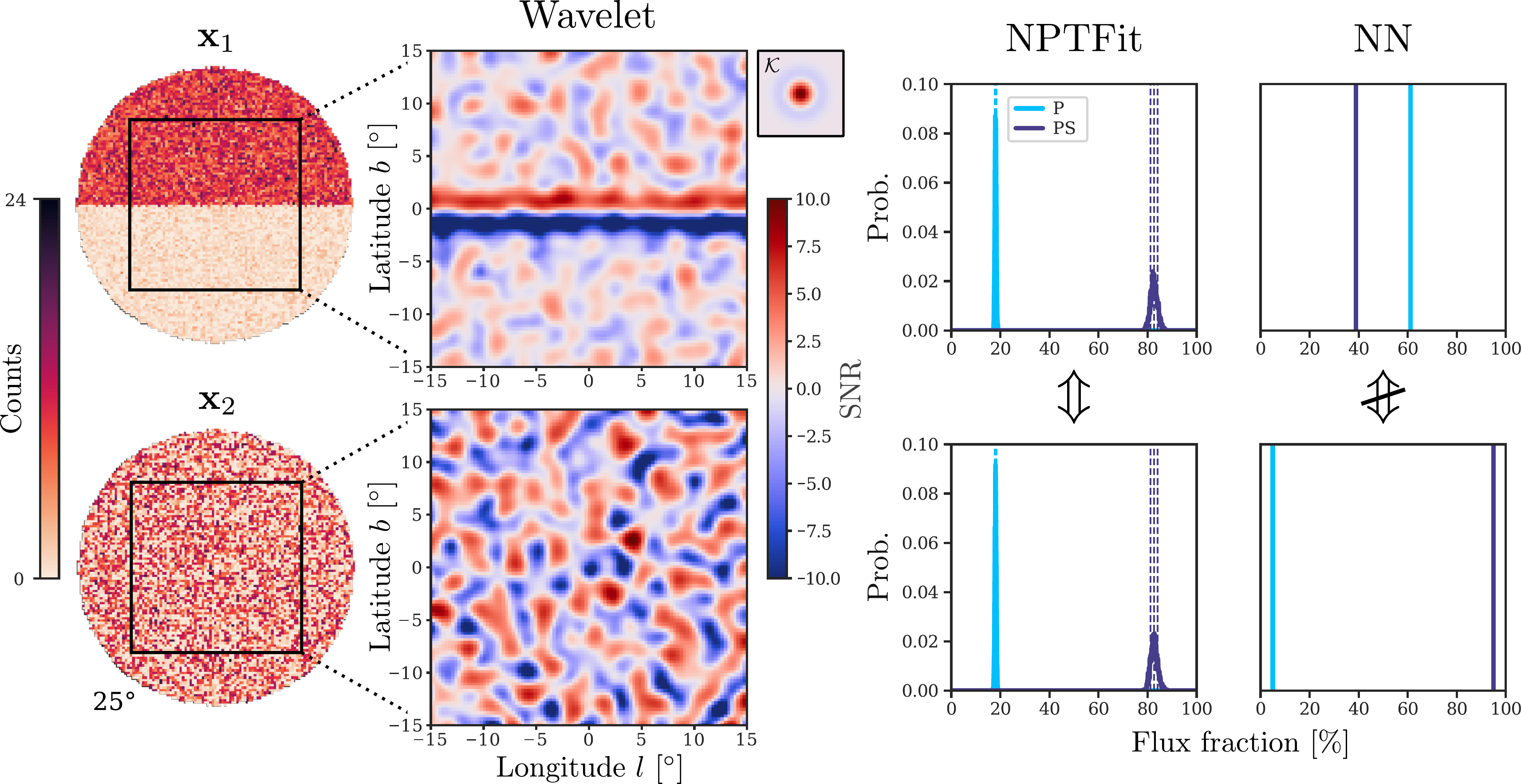}
    }
    \caption{A realization of the large-scale mismodeling scenario with an unmodeled north-south asymmetry discussed in Fig.~\ref{fig:reshuffling_sketch}. The counts in $\mathbf{x}_1$ are drawn from a Poissonian distribution in each pixel with mean $10$ ($1$) in the northern (southern) hemisphere. The map $\mathbf{x}_2$ is a random permutation of the pixels in map $\mathbf{x}_1$. When modeling the map using a spatially constant template, these two maps are indistinguishable for methods that rely on a product likelihood such as NPTF. Therefore, the resulting posteriors for the Poissonian (P) and PS-like flux are identical and attribute the bulk of the flux to PS emission owing to the large pixel-to-pixel variance that arises from the mismodeling. In contrast, the NN finds $\sim60\%$ Poissonian / $40\%$ PS flux in $\mathbf{x}_1$, and close to $100\%$ PS flux in $\mathbf{x}_2$. We also plot the signal-to-noise ratio (SNR) map resulting from convolving the maps with a Mexican hat wavelet kernel $\mathcal{K}$ (see Eq.~(2) in \cite{Bartels2016}), which is dominated by the jump across the equator in map $\mathbf{x}_1$, but otherwise contains higher peaks and deeper troughs for map $\mathbf{x}_2$.}
    \label{fig:asymmetry_comparison}
\end{figure*}
In this appendix, we apply both our NN and \texttt{NPTFit}, the latter of which relies on the product likelihood over the pixels, to a Poissonian map $\mathbf{x}_1$ with an unmodeled north-south asymmetry as discussed in the motivational example in Sec.~\ref{sec:cnns}, and compare the NN prediction for map $\mathbf{x}_1$ to that for a randomly shuffled version $\mathbf{x}_2$. We take the exposure to be constant and do not include a PSF in this example. Figure~\ref{fig:asymmetry_comparison} shows the two maps $\mathbf{x}_1$ and $\mathbf{x}_2$, where $\mathbf{x}_2 = \sigma(\mathbf{x}_1)$ with a random permutation $\sigma$. For illustration purposes, we consider a strong north-south asymmetry in the map $\mathbf{x}_1$, which is taken as a circular region of radius $25^\circ$ with an expected number of counts of 10 and 1 in the northern and southern hemisphere, respectively. We also plot the signal-to-noise ratio (SNR) after projecting the maps to Cartesian images and convolving them with a Mexican hat (or Ricker) wavelet kernel $\mathcal{K}$ with scale $\sigma = 1^\circ$, depicted in the upper right corner. Note that since we consider a small number of counts in this motivational example, the assumption of Gaussianity for the counts is clearly not justified, for which reason the SNR should not be interpreted as the significance for a source at a given location here. Rather, the purpose of the wavelet plot is to provide an intuition for the different outcomes expected for $\mathbf{x}_1$ and $\mathbf{x}_2$ with the wavelet method. We restrict ourselves to the central region so as to avoid boundary effects. 
\par For this illustrative example, we choose a resolution of $n_\text{side} = 128$ and simply train our NN to predict a maximum likelihood estimate for the flux fraction of the Poissonian and PS-like components using an $l^2$ loss function. During the NN training, maps with PS and Poissonian counts corresponding to a uniform spatial template are shown to the NN, implying that the asymmetry is unmodeled when evaluating the trained NN on map $\mathbf{x}_1$. For \texttt{NPTFit}, we assume an isotropic template for the entire map and fit 5 free parameters, namely 1 Poissonian template normalization $A_\text{P}$ and 4 parameters describing the broken power-law SCD of the PS-like component (template normalization $A_\text{NP}$, negative power-law slopes $n_1$ and $n_2$, and the location of the break $S_b$). 
\par The right-hand side of Fig.~\ref{fig:asymmetry_comparison} shows the resulting posterior flux fractions predicted by \texttt{NPTFit} and the NN estimates for maps $\mathbf{x}_1$ and $\mathbf{x}_2$. For \texttt{NPTFit}, the posteriors coincide as anticipated, assigning $> 80\%$ of the flux to PS-like emission. On the other hand, the NN draws different conclusions for the two maps: for the Poissonian map $\mathbf{x}_1$ with a jump across the equator, the NN prefers a mixture between Poissonian ($\sim$~60\%) and PS-like ($\sim$~40\%) emission. In contrast, the highly granular map $\mathbf{x}_2$ causes the NN to assign almost the entire flux to PS-like emission. This simple example illustrates the importance of combining different methods when drawing conclusions about the GCE in the \emph{Fermi} data: each method exhibits a different behavior when mismodeling is at play, as is clearly the case in every GCE analysis to a certain extent, given that models never perfectly describe the reality. We emphasize that we do not attempt to address the intricacies related to the Poisson vs. faint PS degeneracy discussed in the main body or potential biases arising from the default prior parametrization with \texttt{NPTFit} (see the discussion in Ref.~\cite{Collin:2021ufc}) in this experiment, and the results of both methods can be expected to vary depending on the priors and the exact extent of the asymmetry. The key takeaway from this experiment, however, is simply that CNNs respond differently to mismodeling than methods relying on a per-pixel likelihood such as the NPTF.

\section{Comparison with \texttt{NPTFit}}
\label{sec:comparison_nptfit}
\begin{figure}
\centering
  \noindent
   \resizebox{1\columnwidth}{!}{
    \includegraphics{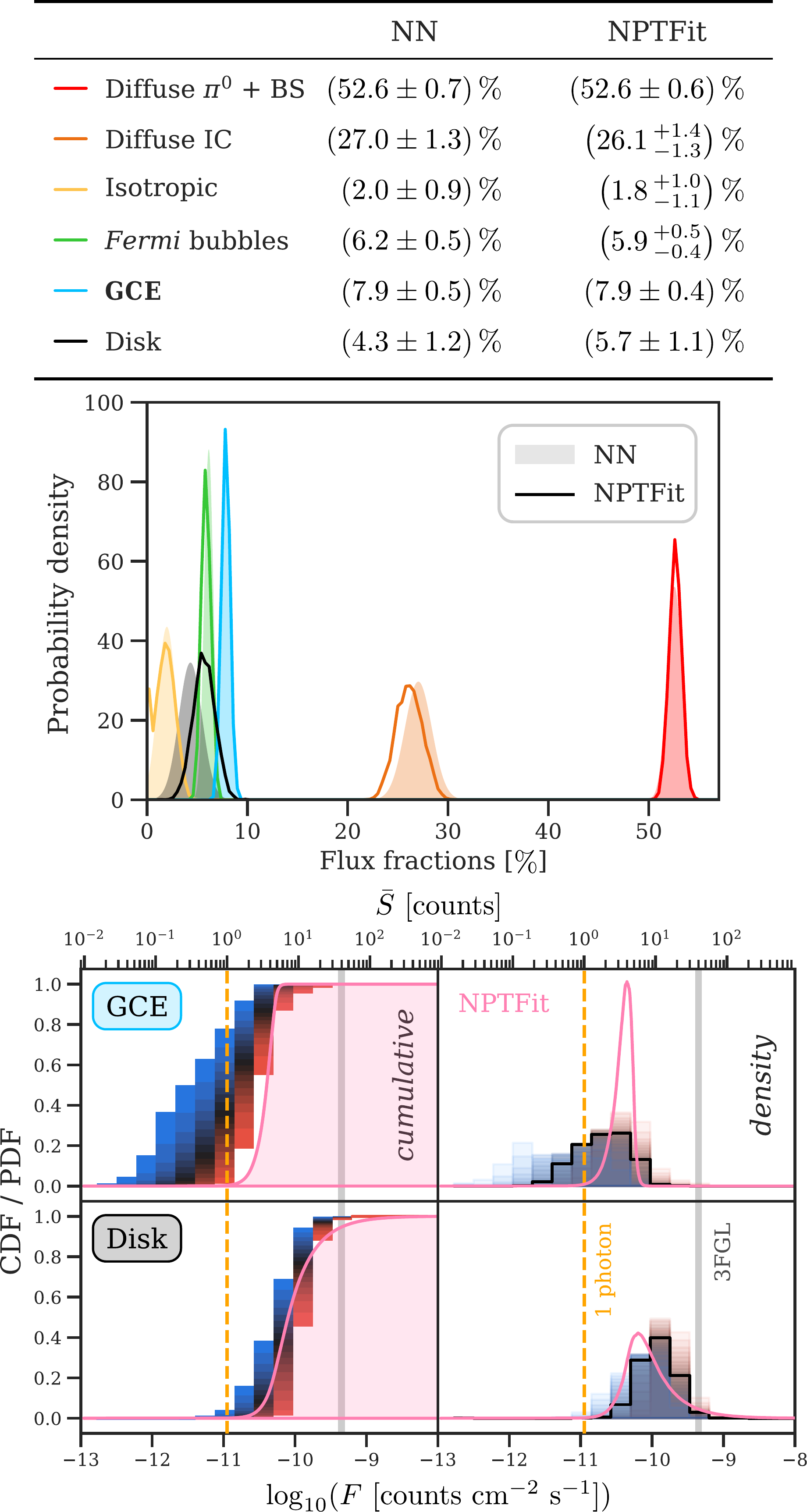}
    }
    \caption{Comparison of our NN results with \texttt{NPTFit}, for the same ROI. Since a Poissonian GCE is described as the limit of ultra-faint PSs in our NN framework (rather than modeling it as a separate component), we do not include a GCE DM template for \texttt{NPTFit} either for the sake of consistency. For \texttt{NPTFit}, we show the median estimates for the SCDs. For most of the templates, the flux fraction posteriors of the NN and \texttt{NPTFit} are in excellent agreement. The NN prefers slightly less disk PS flux and more diffuse IC emission, which is likely a consequence of the spatial degeneracy between these two templates close to the Galactic Plane that can be expected to hamper the distinction between faint disk PSs and diffuse IC flux. This could also explain why the \texttt{NPTFit} estimate for the disk SCD is somewhat fainter than its NN counterpart. However, also for these two templates, the uncertainties for the flux fractions produced by the two methods are consistent. The most striking difference is the GCE SCD: \texttt{NPTFit} places almost no flux below the 1-photon line and favors a narrower distribution than the NN.}
    \label{fig:NPTFit_comparison}
\end{figure}
We present a brief comparison of our NN results for the \emph{Fermi} map to those of \texttt{NPTFit}. To ensure comparability of the results, we use the same ROI for \texttt{NPTFit} (a $25^\circ$ radius circle around the Galactic Center, with $|b| \leq 2^\circ$ and 3FGL sources masked). Also, we use the same templates as for the NN; in particular, we do not include a Poissonian GCE template for the fit such that one would expect a Poissonian GCE in the data to be absorbed by the PS-like GCE template with a very dim SCD function $dN / dF$. We use a resolution parameter of $n_\text{side} = 128$ for \texttt{NPTFit} (instead of $n_\text{side} = 256$ for the NN), to ensure a pixel size larger than the standard deviation of the appropriate \emph{Fermi} instrument PSF (see e.g. Ref.~\cite{Collin:2021ufc}). We parameterize $dN / dF$ as a singly-broken power law for the disk and the GCE templates, giving rise to 4 free parameters for each PS-like template (template normalization $A$, break in terms of counts $S_b$, negative power-law coefficients $n_1$ and $n_2$; see e.g. Ref.~\cite{Mishra-Sharma2017} for details). The only free parameter of the Poissonian models is their template normalization $A$. The prior ranges used for our fit are tabulated in Tab.~\ref{tab:NPTFit_priors}. The templates are normalized to sum up to unity within a ROI radius of $30^\circ$ around the Galactic Center, which anchors the template normalizations $A$.
\par Figure~\ref{fig:NPTFit_comparison} compares the results between our NN and \texttt{NPTFit} for the posterior flux fractions as well as for the relative $F \, dN / d (\log_{10} F)$ SCDs. \texttt{NPTFit} computes the posteriors of the model parameters listed in Tab.~\ref{tab:NPTFit_priors} using the nested sampler \texttt{MultiNest} \cite{Feroz2009, Buchner2014}, which can then be converted to posteriors for the flux fractions and the SCDs. Both the location and the width of the flux fraction posteriors predicted by the NN and \texttt{NPTFit} are very similar. The biggest discrepancy occurs for the disk PS and diffuse IC templates. Both of these templates are bright close to the Galactic Plane, for which reason some cross-talk between faint disk PSs and the diffuse IC template can be expected (see also Sec.~\ref{sec:fermi_example_results_simulated}). Still, the difference in the medians only amounts to 1.4\% and 0.9\% for disk PSs and diffuse IC, respectively, and the estimated uncertainty regions are consistent. The SCD predicted by \texttt{NPTFit} for the disk PSs is somewhat fainter than the NN estimate; however, the differences are modest (particularly when judged by the cumulative distribution, which is the fundamental object on which the NN is trained). For the GCE, we obtain a different picture: both methods roughly agree about the brightest GCE PSs having $5 - 10$ expected counts, but \texttt{NPTFit} favors a much steeper distribution that places almost all the GCE flux above the 1-photon line. In this context, we remark that Ref.~\cite{Chang2019} found in their analysis that median SCDs recovered by \texttt{NPTFit} might be biased toward higher fluxes at the very faint flux end (albeit still within the 95\% region), which is exacerbated when the PS flux is concealed by diffuse emission and by the presence of a PSF (see Fig.~2 in said reference), as is of course the case for the real \emph{Fermi} map. Also, Ref.~\cite{Leane2020a} demonstrated that steep SCDs can arise in \texttt{NPTFit} analyses as artifacts from mismodeling, using a north-south asymmetry of the GCE as an example. However, with regard to the interpretation of our results reported herein, we note that template deficiencies can be expected to bias the recovered SCD to either direction also with our NN approach, as shown in Sec.~\ref{sec:mismodeling_experiment}.
\par Repeating the \texttt{NPTFit} analysis with an additional Poissonian GCE DM template does not appreciably change the GCE SCD: since the \texttt{NPTFit} prefers a PS-like GCE for our choice of priors and ROI, the GCE flux is almost entirely absorbed by the GCE PS template. In fact the Bayes factor in favor of adding the PS component to a purely DM model is $\approx 8 \times 10^3$, although note that this preference can be impacted by the choice of priors \cite{Collin:2021ufc}. With both a PS and a DM template for the GCE, we obtain a flux break of $F_b = 4.8 \times 10^{-11} \ \text{counts} \ \text{cm}^{-2} \ \text{s}^{-1}$ as compared to $F_b = 5.0 \times 10^{-11} \ \text{counts} \ \text{cm}^{-2} \ \text{s}^{-1}$ when omitting the GCE DM template. Also with a GCE DM template, \texttt{NPTFit} identifies a much brighter GCE PS population when replacing Model~O by \texttt{p6v11}, yielding a value of $F_b = 1.4 \times 10^{-10} \ \text{counts} \ \text{cm}^{-2} \ \text{s}^{-1}$, similar to the flux break without a GCE DM template $F_b = 1.5 \times 10^{-10} \ \text{counts} \ \text{cm}^{-2} \ \text{s}^{-1}$ (see Sec.~\ref{sec:results_for_fermi_map}).

\begin{table}
\begin{tabular}{@{}llc@{}}
\toprule
Template & Parameter & Prior range \\ \midrule
Diffuse $\pi^0$ + BS  & $\log_{10} A$ & $[0, 2]$ \\
Diffuse IC            & $\log_{10} A$ & $[0, 2]$ \\
Isotropic             & $\log_{10} A$ & $[-3, 2]$ \\
\emph{Fermi} bubbles  & $\log_{10} A$ & $[-3, 2]$ \\
GCE                   & $\log_{10} A$ & $[-6, 1]$ \\
Disk                  & $\log_{10} A$ & $[-6, 2]$ \\
GCE \& disk           & $n_1$         & $[2.05, 30]$ \\
                      & $n_2$         & $[-5, 1.95]$ \\
                      & $S_b$         & $[0.05, 60]$ \\
\bottomrule                    
\end{tabular}
\caption{Prior ranges used for \texttt{NPTFit} (uniform distribution for all the parameters).}
\label{tab:NPTFit_priors}
\end{table}

\section{Cumulative histograms and constraints on $\eta_P$ for the mismodeling experiment}
\label{sec:mismodeling_cdfs}
\begin{figure}
\centering
  \noindent
   \resizebox{1\columnwidth}{!}{
    \includegraphics{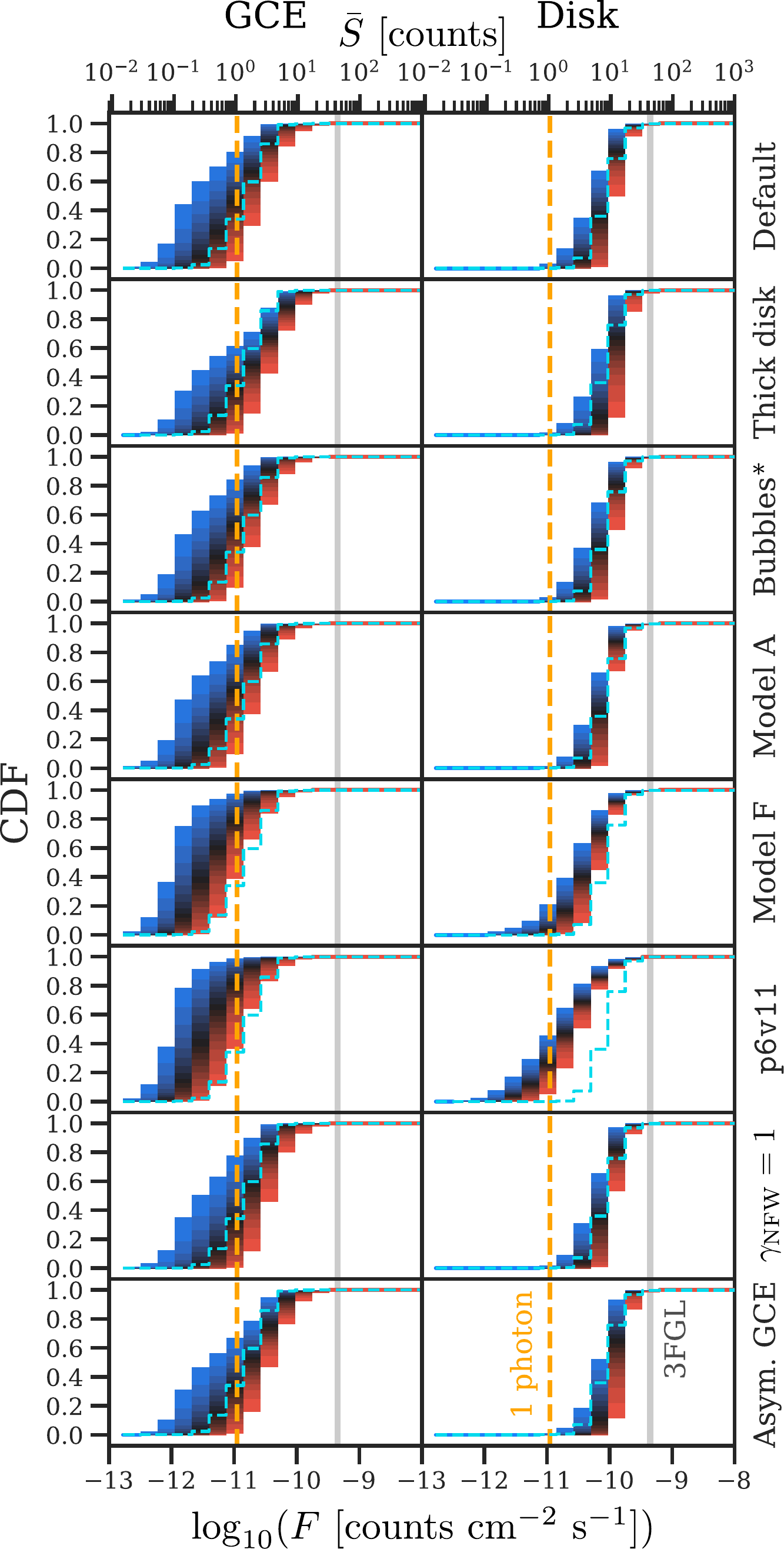}
    }
    \caption{Same as the right two columns in Fig.~\ref{fig:mismodelling}, but for the cumulative histograms.}
    \label{fig:mismodelling_cdf}
\end{figure}
\begin{figure}
\centering
  \noindent
   \resizebox{1\columnwidth}{!}{
    \includegraphics{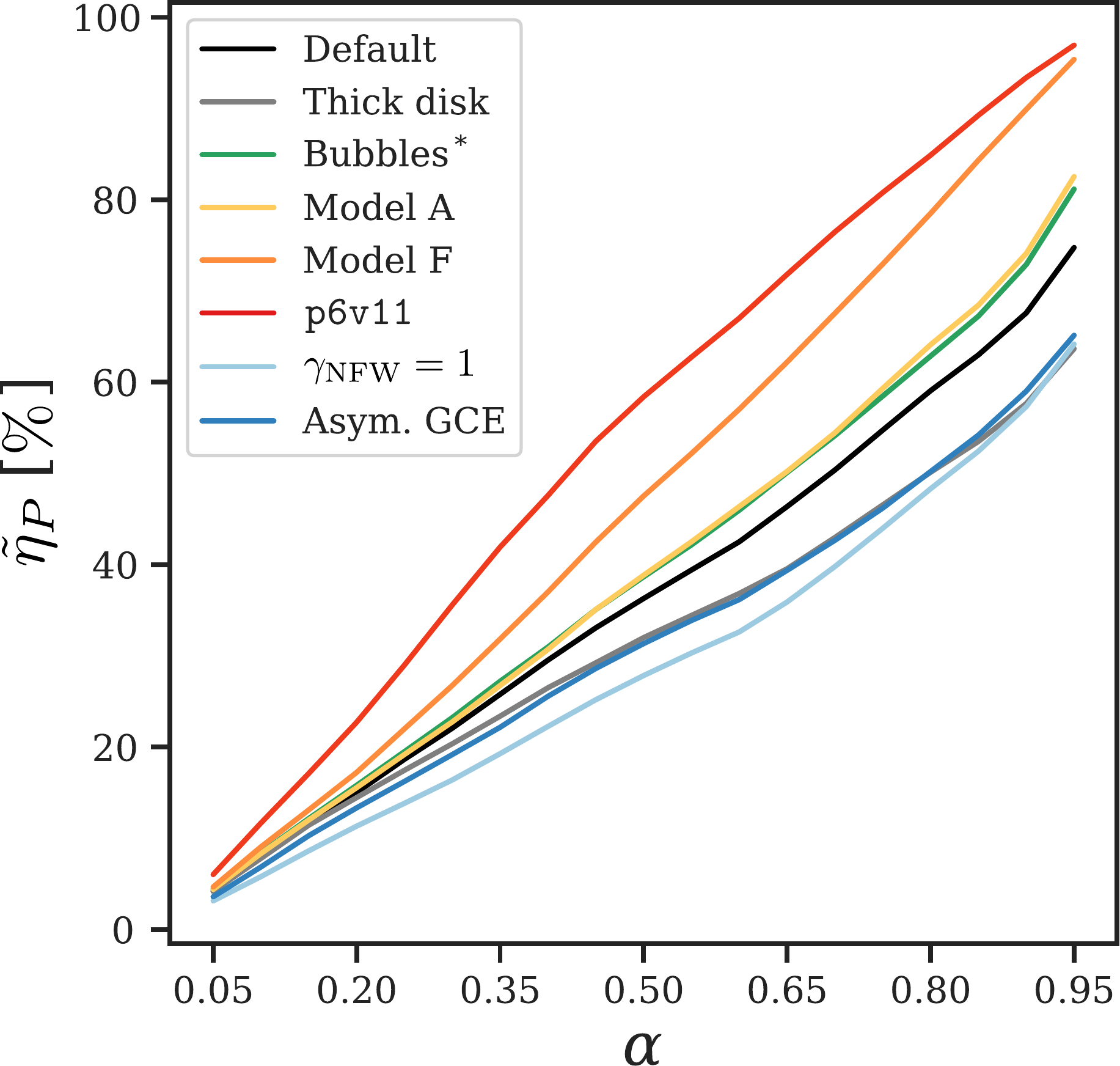}
    }
    \caption{Constraints on the Poisson flux fraction of the GCE $\eta_P$ as a function of the confidence level $\alpha$ in the various mismodeling experiments, obtained from $h^{\boldsymbol{\nu}}$ (median over 256 realizations). For the corresponding flux fractions and SCDs, see Fig.~\ref{fig:mismodelling} in the main body and Fig.~\ref{fig:mismodelling_cdf}.}
    \label{fig:mismodelling_eta}
\end{figure}
In this appendix, we provide the cumulative SCD histograms for the 7 mismodeling scenarios (in addition to the case without mismodeling) shown in Fig.~\ref{fig:mismodelling}, as well as the resulting constraints for $\eta_P$. Figure~\ref{fig:mismodelling_cdf} depicts the median over 256 MC realizations for each scenario, for the GCE and the disk. The colored regions show 5 $-$ 95\% quantiles in steps of 5\%. For all considered discrepancies between the modeled and true morphology of the disk, the \emph{Fermi} bubbles, and the GCE that we consider, the uncertainty regions for the SCD remain consistent with the true SCD, while diffuse mismodeling causes stronger biases (see the main body for a detailed discussion). 
This is also reflected in the constraints for the Poisson flux fraction $\eta_P$, which are shown as a function of the confidence level $\alpha$ in Fig.~\ref{fig:mismodelling_eta}. When the diffuse emission in the maps is generated with Model~F or \texttt{p6v11}, the 95\% confidence constraint obtained from our Model~O-trained analysis pipeline increases to nearly 100\%. In turn, this implies that if the true diffuse emission in the sky departed from Model~O in the direction of either of these two models, the GCE in the \emph{Fermi} data should be expected to be more PS-like than what is shown in Fig.~\ref{fig:fermi_constraints}. As already mentioned in the main body, this is because in this case a fraction of the diffuse flux would be misidentified as dim GCE flux, artificially shifting the SCD to lower fluxes and thus leading to weaker constraints on $\eta_P$. For the other mismodeling cases considered herein, the 95\% confidence constraints move up or down by roughly 10\%, not affecting the conclusion that the existence of a PS-like GCE component is preferred. To further increase the robustness of the constraints, a degree of mismodeling could be incorporated into the NN training, or flexible background models could be constructed (see e.g. \cite{Mishra-Sharma2020}), which we will consider in future work.

\section{SCDs and constraints on $\eta_P$ for the injection experiment}
\label{sec:injection_scds}
\begin{figure*}
\centering
  \noindent
   \resizebox{1\textwidth}{!}{
    \includegraphics{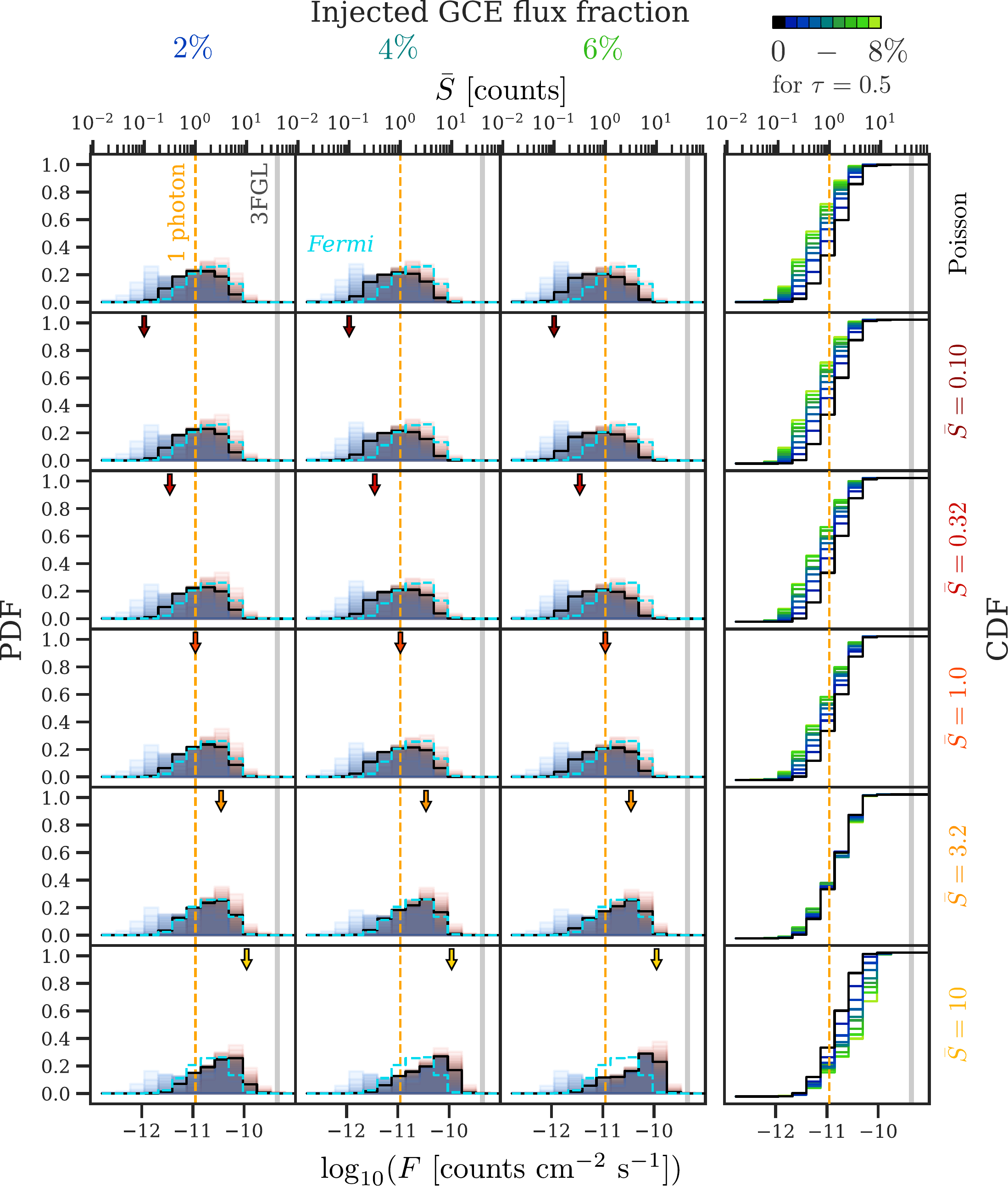}
    }
    \caption{SCD estimates of the GCE for the \emph{Fermi} map with artificially injected GCE flux (median over 64 realizations). The first three columns show the estimated \emph{density} histograms for 2\%, 4\%, and 6\% injected GCE flux (columns), as a function of the brightness of the injected GCE emission (rows). As in Figs.~\ref{fig:fermi_results} and \ref{fig:mismodelling}, the different colors belong to different quantile levels $\tau$ (red to blue, from $5 - 95$\%), and the median is drawn as a black line. The light blue dashed lines show the SCD median estimate for the original \emph{Fermi} map for comparison, and the location of the $dN/dF$ SCD for the injected PS flux is indicated by arrows. The rightmost column shows the \emph{cumulative} median SCD histograms, for injected GCE flux fractions between 0 and 8\% in steps of 1\%.}
    \label{fig:injection_SCD}
\end{figure*}
\begin{figure}
\centering
  \noindent
   \resizebox{1\columnwidth}{!}{
    \includegraphics{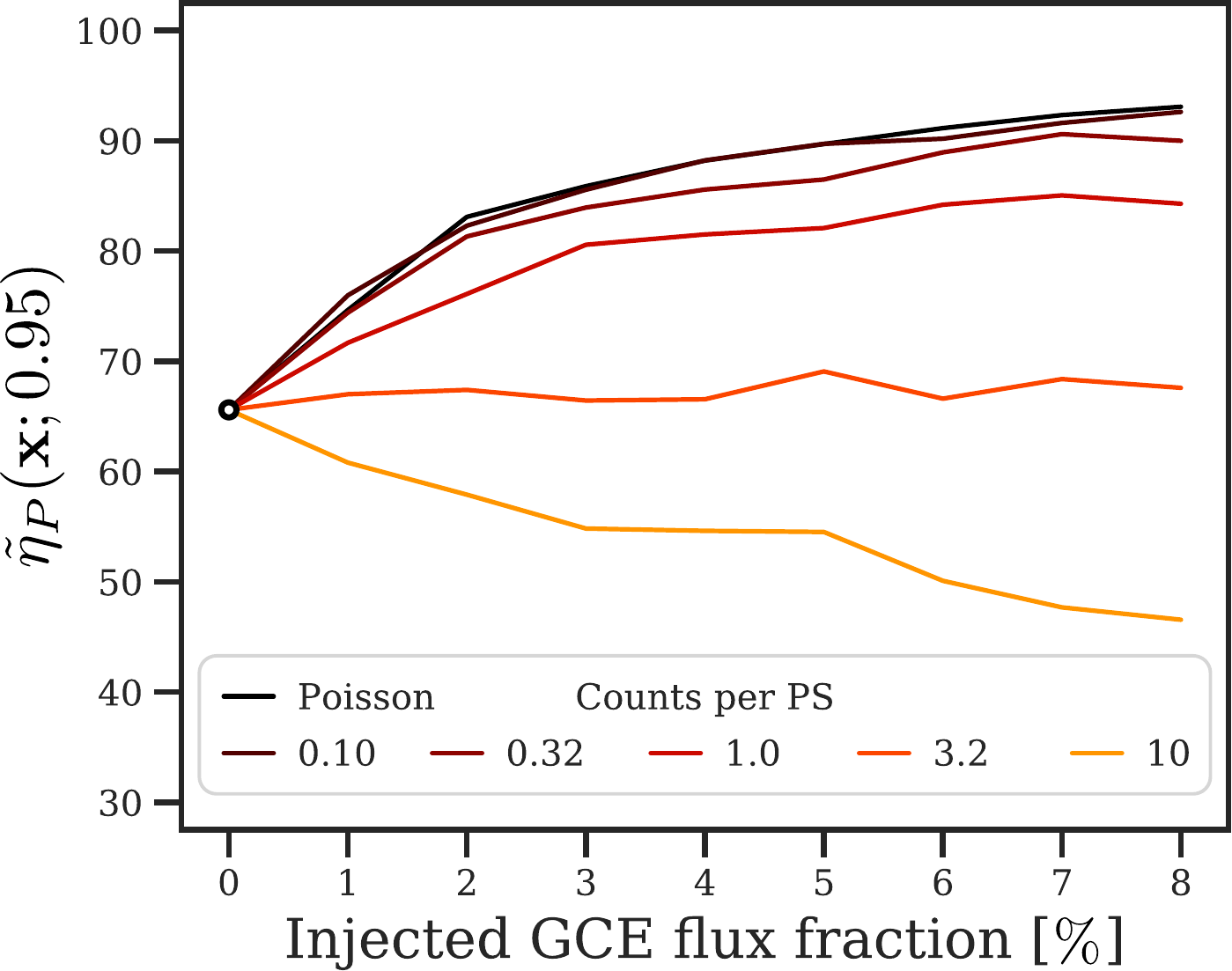}
    }
    \caption{Constraints on the Poisson flux fraction of the GCE $\eta_P$ in the \emph{Fermi} data with artificially injected GCE flux as a function of the injected flux fraction, at 95\% confidence (median over 64 realizations). These constraints are derived by $h^{\boldsymbol{\nu}}$ from the SCD estimates of $g^{\boldsymbol{\varpi}}$ (see Fig.~\ref{fig:mismodelling_cdf}). The line for PS flux with $\bar{S} = 0.1$ expected counts per PS almost coincides with that for Poisson flux, reflecting the Poisson / faint PS degeneracy. For $\bar{S} = 3.2$ expected counts per PS, the constraints remain roughly unchanged, while the constraints gradually become stronger and weaker for brighter and dimmer injected emission, respectively.}
    \label{fig:injection_constraints}
\end{figure}
In Sec.~\ref{sec:injection}, we showed that our NN $f^{\boldsymbol{\omega}}$ accurately recovers artificially injected GCE flux from the \emph{Fermi} data, regardless of whether it is Poissonian or PS-like. In this appendix, we present the SCD estimates of the NN $g^{\boldsymbol{\varpi}}$ and the resulting constraints on the Poisson flux fraction $\eta_P$ for the \emph{Fermi} data with injected GCE flux provided by the NN $h^{\boldsymbol{\nu}}$.
\par Figure~\ref{fig:injection_SCD} shows the predicted SCD for the GCE as a function of the injected GCE flux fraction and the origin of the GCE emission (Poissonian and 5 different homogeneous PS populations). For injected flux of dim PSs each responsible for $\bar{S} \leq 1$ expected count, the SCD estimates move to fainter fluxes as more flux is injected. For $\bar{S} = 3.2$ expected counts per PS, which is approximately the peak of the median SCD for the original \emph{Fermi} map without injection, the predictions are largely unaffected by the injection, and for $\bar{S} = 10$, the SCD moves to higher fluxes, with a peak gradually forming in the corresponding bin. The estimates for the faintest considered PSs are virtually indistinguishable from the Poissonian case, consistent with our unified approach in which Poisson flux is treated as the ultrafaint limit of PS emission. Recall that there is no ``correct'' bin for Poissonian flux (and genuinely Poissonian GCE flux was not included in the training data for $g^{\boldsymbol{\varpi}}$) -- instead, the predictions for maps with a Poissonian GCE characterize the PS flux below which $g^{\boldsymbol{\varpi}}$ is unable to tell which of two PS populations is brighter. Whereas the injection of the brightest considered PSs with $\bar{S} = 10$ expected counts per PS (which is still $\gtrsim$ 3 times fainter than the 3FGL threshold) leads to a localized increase of the SCD in the associated bin, injecting fainter flux does not give rise to narrower SCDs despite the injecting flux following a Dirac delta $dN / dF$. In view of the fact that a fraction of faint PS flux is indistinguishable from Poisson flux to the NN, we suspect that the argument of faint flux affecting several bins partially applies already to faint PS emission. Furthermore, since this experiment uses the real \emph{Fermi} map, rather than simulated MC maps, some interplay with non-GCE templates such as the diffuse foregrounds might also be present. We leave a detailed investigation of this phenomenon for future work.
\par In Fig.~\ref{fig:injection_constraints}, we plot the constraints on the Poisson flux fraction of the GCE $\eta_P$ at confidence level 95\% for each case, as a function of the injected GCE flux. Qualitatively, the constraints are in line with what one expects based on the SCD estimates in Fig.~\ref{fig:injection_SCD}: for $\bar{S} = 3.2$, the median SCD remains largely unchanged, and, accordingly, so does the constraint on $\eta_P$, while the constraints become stronger (weaker) when brighter (fainter) flux is injected. For $\bar{S} = 0.10$, the constraints are nearly the same as in the Poissonian case. As the injected GCE flux increases, the constraints become somewhat weaker than what one would obtain from extrapolating the 95\%-confidence constraint for the original \emph{Fermi} map ($\tilde{\eta}_P = 65.6\%$). For example, adding synthetic Poissonian GCE flux that accounts for 6\% of the total flux in the map (post-injection) to 65.6\% of the GCE flux in the original \emph{Fermi} map yields a GCE that is 81\% Poissonian, while the 95\%-confidence constraint of $h^{\boldsymbol{\nu}}$ is $\tilde{\eta}_P = 91\%$. However, note that the constraining power of $h^{\boldsymbol{\nu}}$ varies depending on the true Poisson flux fraction (e.g., compare the comparatively stronger constraint for $\eta_P = 0$ as compared to $\eta_P = 0.8$ for $\bar{S} = 0.25$ expected counts per PS in Fig.~\ref{fig:constraints_iso_no_PSF}). Also, the size of the uncertainties for the constraint is affected by the injected GCE flux: the IQR between $\alpha = 0.05$ and $0.95$ amounts to a difference in $\eta_P$ of 63\% for the original \emph{Fermi} map, compared with 90\%, 89\%, and 43\% when injecting 8\% Poissonian GCE flux, faint PS flux with $\bar{S} = 0.1$, and bright PS flux with $\bar{S} = 10$, respectively. The growing uncertainties as more faint GCE flux is injected reflect the Poisson vs. faint PS degeneracy. To conclude this experiment, we emphasize again that our NN-based framework is able to accurately identify even small amounts of synthetic GCE flux in the \emph{Fermi} data, and the median SCD estimate gradually moves toward the location of the $dN / dF$ that describes the injected flux.

\section{Constraining the Poisson flux using the analytic likelihood}
\label{sec:frequentist_llh_iso}
Here we provide the analytic likelihood for the example of an isotropically distributed PS population with Dirac delta SCD in the absence of a PSF as considered in Sec.~\ref{sec:constraining_poisson_iso_without_PSF}. In this simple setting we can efficiently evaluate the exact PS likelihood (see \textcite{Collin:2021ufc} for a discussion of the obstructions that arise in more realistic scenarios).
\par In a single pixel and under the above stated assumptions, we can define the likelihood for a model containing both Poissonian and PS flux through the following generating function
\begin{equation}
    \mathcal{P}(t) = \exp \left[ \mu_P(t-1) + N \left( e^{\bar{S} (t-1)} - 1 \right) \right].
    \label{eq:Gen-PPS}
\end{equation}
Here $\mu_P$ is the mean expected Poissonian counts, whereas $N$ and $\bar{S}$ are the expected number of sources per pixel and the expected counts per source, respectively. Given that the generating function of a purely Poissonian model is given by $e^{\mu_P (t-1)}$, we can see that in the limit where each source contributes far less than one count on average, $\bar{S} \ll 1$, the expression in Eq.~\eqref{eq:Gen-PPS} reduces to the Poisson distribution with mean $\mu_P + N\bar{S}$. This formalizes the notion that a population of dim sources becomes exactly degenerate with Poisson emission.
\par Continuing, from Eq.~\eqref{eq:Gen-PPS}, the probability to observe $k$ counts can be determined through successive derivatives of the generating function,
\begin{equation}
    P(k) = \frac{1}{k!} \left. \frac{d^k \mathcal{P}(t)}{dt^k} \right\vert_{t=0}.
\end{equation}
The product of these probabilities across all pixels then specifies the exact likelihood for the common set of model parameters $\boldsymbol{\theta} = \{\mu_P, N, \bar{S}\}$.
\par Our goal is to use this likelihood to establish a limit on the Poissonian flux fraction of the map. In order to do this, we perform a change of coordinates from $\{\mu_P, N, \bar{S}\}$ to $\{\eta_P, N, S_T\}$, where $\eta_P = \mu_P / (\mu_P + N\bar{S})$ is the fraction of counts that is Poissonian, whilst $S_T = \mu_P + N\bar{S}$ is the total number of expected counts per pixel in the map. We can then obtain frequentist limits on $\eta_P$, accounting for $N$ and $S_T$ using the profile likelihood technique, and the results are shown in Fig.~\ref{fig:constraints_iso_no_PSF}.

\section{Constraining the Poisson flux based on the SCD}
\label{sec:systematic_constraints_simple}
\begin{figure}
\centering
  \noindent
   \resizebox{0.85\columnwidth}{!}{
    \includegraphics{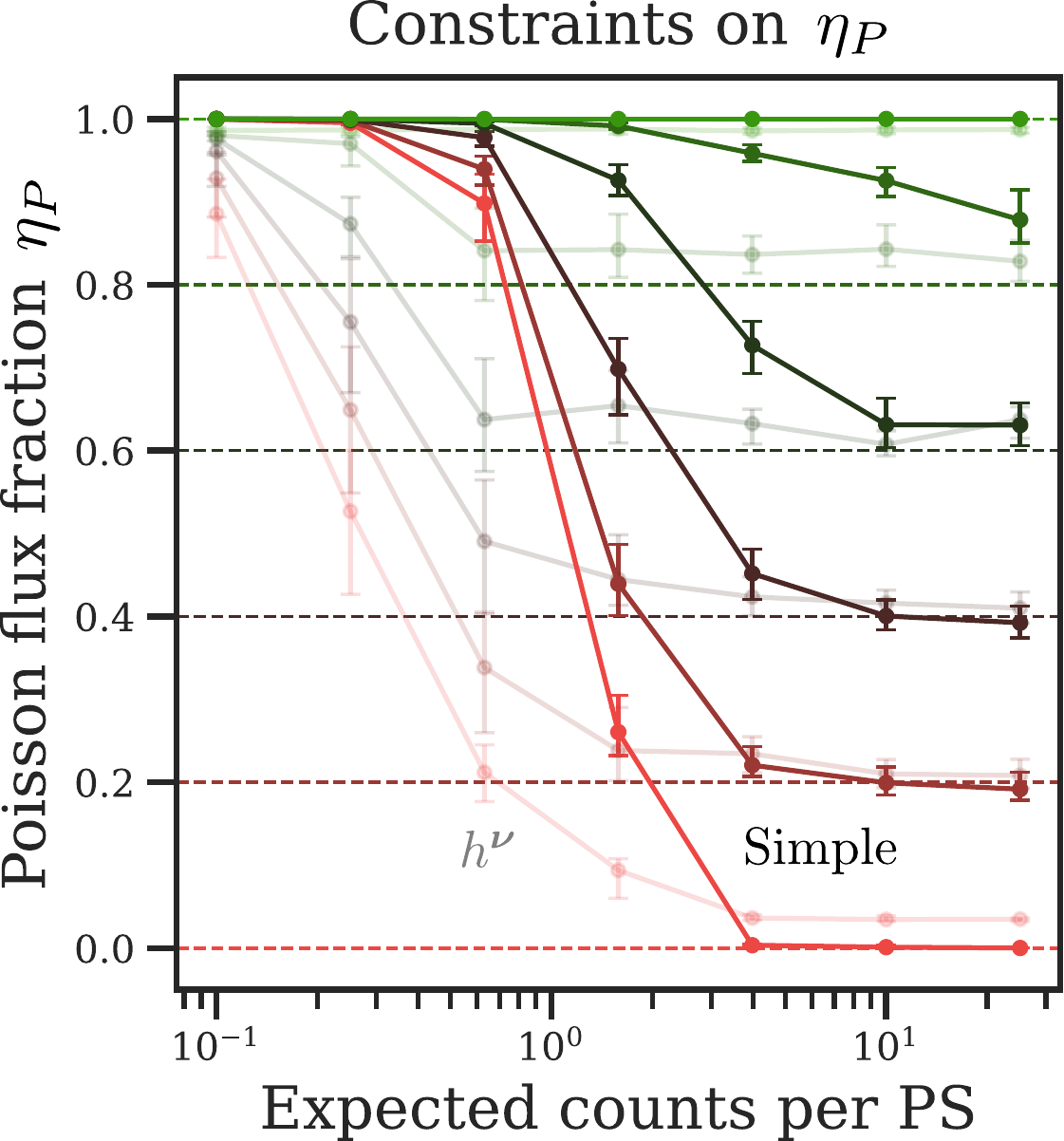}
    }
    \caption{A comparison between the simple estimator that constrains the Poisson flux fraction $\eta_P$ directly from the SCD predicted by the NN $g^{\boldsymbol{\varpi}}$ (see Sec.~\ref{sec:constraining_poisson_simple}) with the constraints produced by the additional NN $h^{\boldsymbol{\nu}}$ (see Sec.~\ref{sec:constraining_poisson_with_NN}), at confidence level $\alpha = 0.95$. The NN $h^{\boldsymbol{\nu}}$ provides much stronger constraints on $\eta_P$ for faint PS populations. For a comparison of the NN $h^{\boldsymbol{\nu}}$ to frequentist constraints derived from the exact likelihood function, we refer to Fig.~\ref{fig:constraints_iso_no_PSF} in the main body.}
    \label{fig:constraints_iso_no_PSF_simple}
\end{figure}
In Sec.~\ref{sec:constraining_poisson_simple}, we introduced a simple estimator for the Poisson flux fraction $\eta_P$ of a flux component, which consists in evaluating the predicted (relative) cumulative median SCD histogram $\tilde{Q}^{\boldsymbol{\varpi}}(\mathbf{x}; 0.5)\vert_{\phi^*(\alpha)}$ at a value $\phi^* = \phi^*(\alpha)$ such that the resulting value \emph{on average} exceeds the true value $\eta_P$ for $(100 \times \alpha)\%$ of the maps in the calibration dataset $\mathcal{X}_\text{cal}$. As mentioned in the main body, the constraining power of this estimator is not sufficient to derive non-trivial constraints on the Poissonian flux component of the GCE in the \emph{Fermi} map. To demonstrate that the NN estimator $h^{\boldsymbol{\nu}}$ provides much tighter constraints, we compare the 95\%-confidence constraints of the simple estimator with those of $h^{\boldsymbol{\nu}}$ for the benchmark example considered in Sec.~\ref{sec:constraining_poisson_iso_without_PSF}. Figure~\ref{fig:constraints_iso_no_PSF_simple} shows the constraints of the simple estimator and the NN $h^{\boldsymbol{\nu}}$ (see Fig.~\ref{fig:constraints_iso_no_PSF} for the same comparison between the frequentist constraints based on the analytic likelihood and $h^{\boldsymbol{\nu}}$). Unlike $h^{\boldsymbol{\nu}}$, the simple estimator is unable to constrain the Poissonian flux component for maps containing counts from PSs below the 1-photon line.

\section{Priors for the training data generation}
\label{sec:NN_priors}
We list the priors that we used for training our NNs $f^{\boldsymbol{\omega}}$ and $g^{\boldsymbol{\varpi}}$ in Tab.~\ref{tab:NN_priors}. For Poissonian templates, the only free parameter is the template normalization $A$, which we draw from a uniform distribution for each map. The prior ranges for the realistic scenario considered in Sec.~\ref{sec:fermi} are chosen such that the values for the \emph{Fermi} map are expected to lie well within the prior cube (see Fig.~\ref{fig:flux_fraction_errors} for the resulting flux fraction ranges). For the PS-like templates (GCE and disk), we draw the total expected flux $F_\text{tot} \ [\text{counts} \ \text{cm}^{-2} \ \text{s}^{-1}]$ in each map from a uniform distribution and take random skew normal distributions for the SCD defined with respect to the flux logarithm $\log_{10}(F)$, whose PDF is given by
\begin{equation}
    f_\text{skew}(x) = \frac{2}{\omega} \ \psi\left(\frac{x - \xi}{\omega}\right) \ \Psi\left(a \left(\frac{x - \xi}{\omega} \right)\right),
\end{equation}
where $\psi(x)$ and $\Psi(x)$ are the standard normal PDF and CDF, respectively. The parameters $\xi$, $\omega$, and $a$ define the location, scale, and skewness of the distribution, respectively. Note that for the mixed Poisson + PS maps used in Sec.~\ref{sec:constraining_Poisson}, $F_\text{tot}$ is not randomly drawn, but defined by the Poissonian flux fraction $\eta_P$ (which, in turn, is drawn from a uniform distribution $U([0, 1])$), together with the desired total expected Poisson + PS flux: for the isotropic example in Sec.~\ref{sec:constraining_poisson_iso_without_PSF}, we take the number of expected Poisson + PS counts in the map to be 50,000, and for the realistic case in Sec.~\ref{sec:constraining_poisson_gce}, the total expected GCE counts correspond to the best-fit prediction of $f^{\boldsymbol{\omega}}$ for the \emph{Fermi} map.

\begin{table}
\emph{Isotropic proof-of-concept example}: \\[0.2cm]
\begin{tabular}{@{}llc@{}}
\toprule
Template & Parameter & Priors \\ \midrule
Isotropic            & $\xi$ & $U([-1, 1.5])$ (Sec.~\ref{sec:toy}) \\
                     & & $U([-2, 1.5])$ (Sec.~\ref{sec:constraining_poisson_iso_without_PSF}) \\
                     & $\omega^2$ & $0.1 \, \chi^2(1)$ \\ 
                     & $a$ & $\mathcal{N}(0, 3)$ \\
                     & $F_\text{tot}$ & $U([1,$ 100,000$])$\footnotemark[1] \\
                     \bottomrule                    
\end{tabular}
\\
\vspace{0.5cm}
\emph{Realistic scenario}: \\[0.2cm]
\begin{tabular}{@{}llc@{}}
\toprule
Template & Parameter & Priors \\ \midrule
Diffuse $\pi^0$ + BS  & $A$ & $U([1.75, 3.5])$ \\
Diffuse IC            & $A$ & $U([1, 2.25])$ \\
Isotropic             & $A$ & $U([0, 0.5])$ \\
\emph{Fermi} bubbles  & $A$ & $U([0, 0.5])$ \\
GCE \& disk           & $\xi$ & $U([-12, -9])$ \\
                      & $\omega^2$ & $0.25 \, \chi^2(1)$ \\ 
                      & $a$ & $\mathcal{N}(0, 3)$ \\
                      & $F_\text{tot}$ & $U([0, 1.4 \times 10^{-7}])$\footnotemark[1] \\
                      \bottomrule                    
\end{tabular}
\footnotetext[1]{The GCE counts in the realistic scenario, as well as the isotropic PS counts in the isotropic proof-of-concept example without PSF in Sec.~\ref{sec:constraining_poisson_iso_without_PSF}, are the sum of \emph{two} template maps. Therefore, the \emph{total} flux of the respective template follows a symmetric triangular distribution between $0$ and $2.8 \times 10^{-7}$ in the realistic scenario, and between $0$ and 200,000 in the isotropic proof-of-concept example for constraining the Poisson flux.}
\caption{Priors used for the training data generation for the NNs $f^{\boldsymbol{\omega}}$ and $g^{\boldsymbol{\varpi}}$, for the isotropic proof-of-concept example (Sec.~\ref{sec:toy}) and the realistic scenario (Sec.~\ref{sec:fermi}). The unit for the flux $F_\text{tot}$ is $\text{counts} \ \text{cm}^{-2} \ \text{s}^{-1}$. The significant difference in the total flux $F_\text{tot}$ between the two examples is due to the difference in the exposure, which we set to $1 \ \text{cm}^{2} \ \text{s}$ in the proof-of-concept example, whereas the \emph{Fermi} mean exposure is $9.1 \times 10^{10} \ \text{cm}^{2} \ \text{s}$ within our ROI.}
\label{tab:NN_priors}
\end{table}

\section{Neural network details and architectures}
\label{sec:NN_architectures}
In Tab.~\ref{tab:NN_arch}, we list the NN architectures of $f^{\boldsymbol{\omega}}$, $g^{\boldsymbol{\varpi}}$ for the realistic scenario with application to the \emph{Fermi} map (Sec.~\ref{sec:fermi}), and for constraining the Poisson flux component of the GCE using NN $h^{\boldsymbol{\nu}}$ (Sec.~\ref{sec:constraining_poisson_gce}). 
\par We improved our NN implementation as compared to \citetalias{List2020b} such that the input maps only consist of the pixels within our ROI, rather than of all the pixels within the coarse $n_\text{side} = 1$ pixel (1 out of 12 that together cover the entire sky) that contains our ROI with zero counts in pixels not belonging to the ROI. This has two consequences: (1) the vertices of the \texttt{DeepSphere} graph utilized for the definition of the convolution via the graph Laplacian operator are given only by the pixels within the ROI at each hierarchy level, (2) for the maximum pooling operation, only pixels within the ROI at the current hierarchy level are taken into account. Consider the $r$-th pixel $p^{n_\text{side} / 2}_{r}$ at resolution $n_\text{side} \, / \, 2$, consisting of the 4 finer pixels $p^{n_\text{side}}_{r, s}$ at resolution $n_\text{side}$, for $s = \{1, 2, 3, 4\}$. If any of the pixels $p^{n_\text{side}}_{r, s}$ lie within the ROI at resolution $n_\text{side}$, the maximum of the outputs of the graph convolutions can be taken over those $s$. Consequently, the output of the convolutional blocks in $p^{n_\text{side} / 2}_{r}$ is defined and is then further processed by the subsequent convolutional blocks, making $p^{n_\text{side} / 2}_{r}$ become part of the ROI at resolution $n_\text{side} / 2$. Thus, the  ``holes'' in the ROI are gradually closed as the resolution decreases when propagating the map through the NN. We take the kernel size of the graph convolutions to be $5$.
\par The output dimension $2 \times 6$ of $f^{\boldsymbol{\omega}}$ corresponds to mean and (log-)variance for each of the 6 templates. The two input channels for $g^{\boldsymbol{\varpi}}$ contain the original map and the Poissonian residual, computed by removing the expected contributions of the purely Poissonian templates (that is, all but disk and GCE) based on the means estimated by $f^{\boldsymbol{\omega}}$. Since we train $f^{\boldsymbol{\omega}}$ and $g^{\boldsymbol{\varpi}}$ consecutively, the residual estimates are accurate already when the training of $g^{\boldsymbol{\varpi}}$ starts. The quantile level $\tau$ of interest, which is drawn uniformly from $[0, 1]$ during the training and can be chosen arbitrarily at evaluation time, is concatenated to the output of the convolutional blocks in Layer XI of $g^{\boldsymbol{\varpi}}$ after mapping it to the interval $[-6, 6]$ via
\begin{equation}
    \tau \mapsto 12 \left(\tau - 0.5\right).
\label{eq:tau_mapping}
\end{equation}
In our implementation, we replace the EMPL loss function in Eq.~\eqref{eq:EM_pinball_loss_discrete} by a smooth variant inspired by Refs.~\cite{Zheng2011, Hatalis2019}, given by
\begin{equation}
    \mathcal{L}^\tau_\beta(\tilde{\mathbf{u}}, \mathbf{u}) = \frac{1}{M} \sum_{j=1}^M \left[\tau \left(\tilde{U}_j - U_j\right) + \beta \operatorname{softplus}\left(\frac{U_j - \tilde{U}_j}{\beta}\right) \right],
    \label{eq:EM_pinball_loss_smooth}
\end{equation}
where we choose the smoothing parameter $\beta = 0.001$. In the limit $\beta \searrow 0$, one finds that $\mathcal{L}^\tau_\beta \to \mathcal{L}^\tau_\text{EMPL}$.

\par The NN architecture of $g^{\boldsymbol{\varpi}}$ for the isotropic proof-of-concept example in Sec.~\ref{sec:toy} is very similar, but the input maps only have a single channel (as there is no residual to compute), and the output is a single SCD histogram with dimension $1 \times 22$, where $N = 22$ is the number of bins. Moreover, we found that replacing batch normalization by instance normalization \cite{Ulyanov2016} led to significantly better generalization from the training to the testing dataset in the isotropic example -- possibly because the noise introduced by the batch-dependent normalization of the means and variances for the weights with batch normalization deters the NN from achieving optimal performance in this simple case. Therefore, the results presented for the isotropic proof-of-concept example use instance normalization. 
\par The NN $h^{\boldsymbol{\nu}}$ that yields constraints on the Poissonian flux fraction $\eta_P$ given an SCD histogram as an input is a simple fully-connected NN with 2 hidden layers. For the confidence level $\alpha$ that plays the role of $\tau$ in the definition of the pinball loss (Eq.~\eqref{eq:pinball_loss}), we use the same mapping as in Eq.~\eqref{eq:tau_mapping} before appending it to the input histogram, and we use a slightly smoothed version of the pinball loss similar to Eq.~\eqref{eq:EM_pinball_loss_smooth}. 
\begin{table*}
$f^{\boldsymbol{\omega}}$ (map $\to$ template flux fractions): \\[0.2cm]
\begin{tabular}{@{}lllcc@{}}
\toprule
Layer & Operations                                 & Output shape & Output $n_\text{side}$ & Trainable parameters \\ \midrule
I     & Input map (normalized)                         & 30,805 $\times$ 1   & 256             & $-$                  \\
II    & ConvBlock & 8,117 $\times$ 32   & 128             & 160 + 32             \\
III   & ConvBlock & 2,199 $\times$ 64   & 64              & 10,240 + 64          \\
IV    & ConvBlock & 598 $\times$ 128    & 32              & 40,960 + 128         \\
V     & ConvBlock & 164 $\times$ 256    & 16              & 163,840 + 256        \\
VI    & ConvBlock & 50 $\times$ 256     & 8               & 327,680 + 256        \\
VII   & ConvBlock & 14 $\times$ 256     & 4               & 327,680 + 256        \\
VIII  & ConvBlock & 4 $\times$ 256      & 2               & 327,680 + 256        \\
IX    & ConvBlock & 1 $\times$ 256      & 1               & 327,680 + 256        \\
X     & Append $\log_{10}(S_\text{tot})$    & 1 $\times$ 257      &                 & $-$                  \\
XI    & ReLU $\circ$ FC                            & 1 $\times$ 2,048    &                 & 526,336 + 2,048      \\
XII   & ReLU $\circ$ FC                            & 1 $\times$ 512      &                 & 1,048,576 + 512      \\
XIII  & Reshape $\circ$ FC                         & 2 $\times$ 6        &                 & 6,144 + 0            \\
XIV   & Softmax (means only)                       & 2 $\times$ 6        &                 & $-$                  \\ \cmidrule(l){5-5}
      &                                            &                     &                 & 3,111,040            \\ \bottomrule
\end{tabular} \\
\vspace{0.5cm}
$g^{\boldsymbol{\varpi}}$ (map $\to$ SCD histograms): \\[0.2cm]
\begin{tabular}{@{}lllcc@{}}
\toprule
Layer & Operations                                 & Output shape & Output $n_\text{side}$ & Trainable parameters \\ \midrule
I     & Input map (normalized)                         & 30,805 $\times$ 2   & 256             & $-$                  \\
II    & ConvBlock & 8,117 $\times$ 32   & 128             & 320 + 32             \\
III   & ConvBlock & 2,199 $\times$ 64   & 64              & 10,240 + 64          \\
IV    & ConvBlock & 598 $\times$ 128    & 32              & 40,960 + 128         \\
V     & ConvBlock & 164 $\times$ 256    & 16              & 163,840 + 256        \\
VI    & ConvBlock & 50 $\times$ 256     & 8               & 327,680 + 256        \\
VII   & ConvBlock & 14 $\times$ 256     & 4               & 327,680 + 256        \\
VIII  & ConvBlock & 4 $\times$ 256      & 2               & 327,680 + 256        \\
IX    & ConvBlock & 1 $\times$ 256      & 1               & 327,680 + 256        \\
X     & Append $\log_{10}(S_\text{tot})$ & 1 $\times$ 257      &                 & $-$                  \\
XI    & Append $\tau$               & 1 $\times$ 258      &                 & $-$                  \\
XII   & ReLU $\circ$ FC                            & 1 $\times$ 2,048    &                 & 528,384 + 2,048      \\
XIII  & ReLU $\circ$ FC                            & 1 $\times$ 512      &                 & 1,048,576 + 512      \\
XIV   & Reshape $\circ$ FC                         & 2 $\times$ 22       &                 & 22,528 + 0           \\
XV    & Normalized softplus                        & 2 $\times$ 22       &                 & $-$                  \\ \cmidrule(l){5-5} 
      &                                            &                     &                 & 3,129,632            \\ \bottomrule
\end{tabular} \\
\vspace{0.5cm}
$h^{\boldsymbol{\nu}}$ (GCE SCD histogram $\to$ Poissonian flux fraction $\eta_P$): \\[0.2cm]
\begin{tabular}{@{}lllc@{}}
\toprule
Layer & Operations                 & Output shape & Trainable parameters \\ \midrule
I     & Input histogram            & 22                  & $-$                  \\
II    & Append $\alpha$            & 23                  & $-$                  \\
III   & ReLU $\circ$ FC & 256                 & 5,888 + 256          \\
IV    & ReLU $\circ$ FC & 256                 & 65,536 + 256         \\
V     & Sigmoid $\circ$ FC & 1                   & 256 + 1              \\ \cmidrule(l){4-4} 
      &                            &                     & 72,193               \\ \bottomrule
\end{tabular}
\caption{NN architectures of $f^{\boldsymbol{\omega}}$, $g^{\boldsymbol{\varpi}}$, and $h^{\boldsymbol{\nu}}$ for the realistic scenario. Each convolutional block consists of ConvBlock $=$ MaxPool $\circ$ ReLU $\circ$ BN $\circ$ GC. The following abbreviations are used: MaxPool (maximum pooling), ReLU (Rectified Linear Unit activation function), BN (batch normalization), GC (graph convolution), and FC (fully-connected layer). We normalize the maps to sum up to unity before feeding them to the NN, and we append the total number of counts in the map $S_\text{tot}$ in Layer X. The combined NN for Step 1 and Step 2 ($f^{\boldsymbol{\omega}}$ and $g^{\boldsymbol{\varpi}}$) has 6,240,672 trainable parameters. The quantile level of interest $\tau$ for the SCD histogram estimation is appended to the output of convolutional blocks in Layer XI of $g^{\boldsymbol{\varpi}}$. The NN for constraining the Poissonian flux fraction $\eta_P$ is a simple fully-connected NN with 2 hidden layers, which takes a histogram and the confidence level $\alpha$ as inputs. The trainable parameters are split up into matrix weights + bias vectors. For example, Layer II of $f^{\boldsymbol{\omega}}$ has $5 \times 1 \times 32 = 160$ matrix weights ($=$ kernel size $\times$ input channels $\times$ output channels) and $32$ weights that compose the bias vector ($=$ output channels). The output dimensions correspond to the means and log-variances for each of the 6 templates for $f^{\boldsymbol{\omega}}$, SCD histograms for the GCE and the disk with $M = 22$ bins for $g^{\boldsymbol{\varpi}}$, and a single estimate $\tilde{\eta}_P$ for $h^{\boldsymbol{\nu}}$.}
\label{tab:NN_arch}
\end{table*}

\end{document}